\documentclass[12pt]{puthesis}

\usepackage{epsfig}
\usepackage{amsfonts}
\usepackage{amssymb}
\usepackage{amsmath}
\usepackage{amsthm}
\usepackage{latexsym}
\usepackage{psfrag}
\usepackage{graphicx}
\usepackage{rotating}
\newcommand{\ol}{\overline}
\newcommand{\ra}{\rangle}
\newcommand{\la}{\langle}
\newcommand{\be}{\begin{equation}}
\newcommand{\ee}{\end{equation}}
\newcommand{\ba}{\begin{eqnarray}}
\newcommand{\ea}{\end{eqnarray}}
\newcommand{\Par}{\parallel}
\newcommand{\Perp}{\perp}
\newcommand{\grad}{\nabla}
\newcommand{\ignore}[1]{}  

\title{Kinetic Effects on Turbulence Driven by the Magnetorotational
Instability in Black Hole Accretion}
\submitted{September 2006}  
\author{Prateek Sharma}

\abstract{
Many astrophysical objects (e.g., spiral galaxies, the solar system,
Saturn's rings, and luminous disks around compact objects) occur in
the form of a disk. One of the important astrophysical problems is
to understand how rotationally supported disks lose angular
momentum, and accrete towards the bottom of the gravitational
potential, converting gravitational energy into thermal (and
radiation) energy.

The magnetorotational instability (MRI), an instability causing
turbulent transport in ionized accretion disks, is studied in the
kinetic regime. Kinetic effects are important because radiatively
inefficient accretion flows (RIAFs), like the one around the
supermassive black hole in the center of our Galaxy, are
collisionless. The ion Larmor radius is tiny compared to the scale
of MHD turbulence so that the drift kinetic equation (DKE), obtained
by averaging the Vlasov equation over the fast gyromotion, is
appropriate for evolving the distribution function. The kinetic MHD
formalism, based on the moments of the DKE, is used for linear and
nonlinear studies. A Landau fluid closure for parallel heat flux,
which models kinetic effects like collisionless damping, is used to
close the moment hierarchy.

We show that the kinetic MHD and drift kinetic formalisms give the
same set of linear modes for a Keplerian disk. The BGK collision
operator is used to study the transition of the MRI from kinetic to
the MHD regime. The ZEUS MHD code is modified to include the key
kinetic MHD terms: anisotropic pressure tensor and anisotropic
thermal conduction. The modified code is used to simulate the
collisionless MRI in a local shearing box. As magnetic field is
amplified by the MRI, pressure anisotropy ($p_\perp>p_\parallel$) is
created because of the adiabatic invariance ($\mu \propto
p_\perp/B$). Larmor radius scale instabilities---mirror,
ion-cyclotron, and firehose---are excited even at small pressure
anisotropies ($\Delta p/p \sim 1/\beta$). Pressure isotropization
due to pitch angle scattering by these instabilities is included as
a subgrid model. A key result of the kinetic MHD simulations is that
the anisotropic stress can be as large as the Maxwell stress.

It is shown, with the help of simple tests, that the centered
differencing of anisotropic thermal conduction can cause the heat to
flow from lower to higher temperatures, giving negative temperatures
in regions with large temperature gradients. A new method, based on
limiting the transverse temperature gradient, allows heat to flow
only from higher to lower temperatures.  Several tests and
convergence studies are presented to compare the different methods.

}

\acknowledgements{
I foremost thank my adviser Greg Hammett, whose guidance made this
thesis possible. His insight, quest for perfection, and passion for
science has always inspired me. He was always patient, and ensured
that I understood every subtle point. Thanks to Eliot Quataert, who
is an inspiring mentor, and initiated me into the fascinating field
of astrophysics. He was so accessible that I never felt that he was
not in Princeton. I am thankful to Jim Stone for his constant
encouragement, and help with numerical methods, especially ZEUS.

Finally, thanks to my loving and supporting family, and wonderful
friends.
Special thanks to my brother Rohit, sister Chubi, and cousins, who
have always brought joy in my life. My wife Asha has been a loving
and caring companion; her suggestion to keep it simple has significantly
improved the thesis.

The thesis work was supported by U.S. DOE contract \#
DE-AC02-76CH03073, and NASA grants NAG5-12043 and NNH06AD01I. Many
thanks to the NASA websites for amazing pictures, some of which I
have used in this thesis.

}

\dedication{To my parents and grandparents}

\begin{document}

\listoftables \listoffigures
\chapter{Introduction}
\label{chap:chap1}
\begin{figure}
\begin{center}
\includegraphics[width=5in,height=3in]{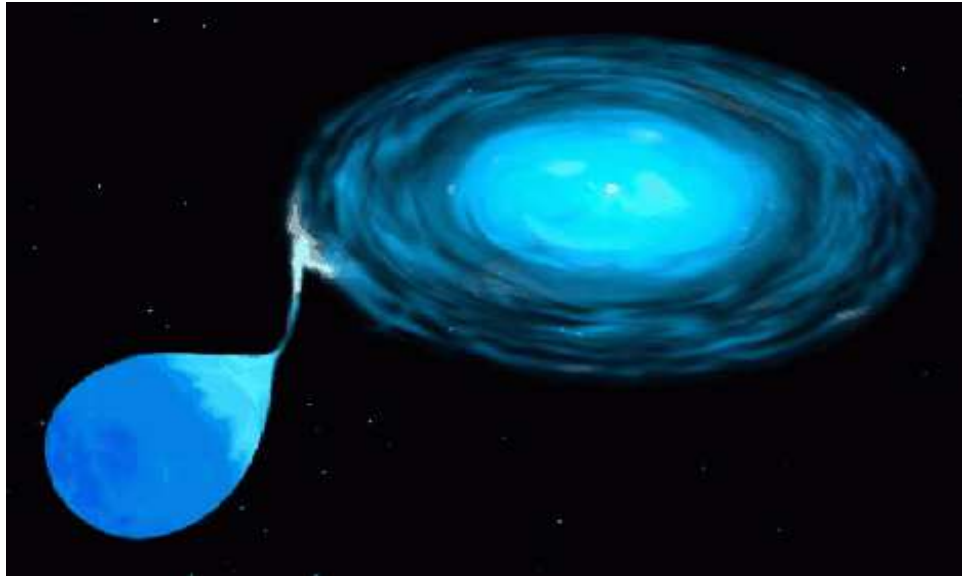}
\caption[An artist's impression of a binary accretion disk]{An
artist's impression of a binary accretion disk. Plasma overflows
from the stellar companion and forms an accretion disk around the
compact object.  Drawing Credit: ST ScI, NASA;
http://antwrp.gsfc.nasa.gov/apod/ap991219.html.
\label{Ch1fig:BinaryDisk}}
\end{center}
\end{figure}

\begin{figure}
\begin{center}
\includegraphics[width=4in,height=3in]{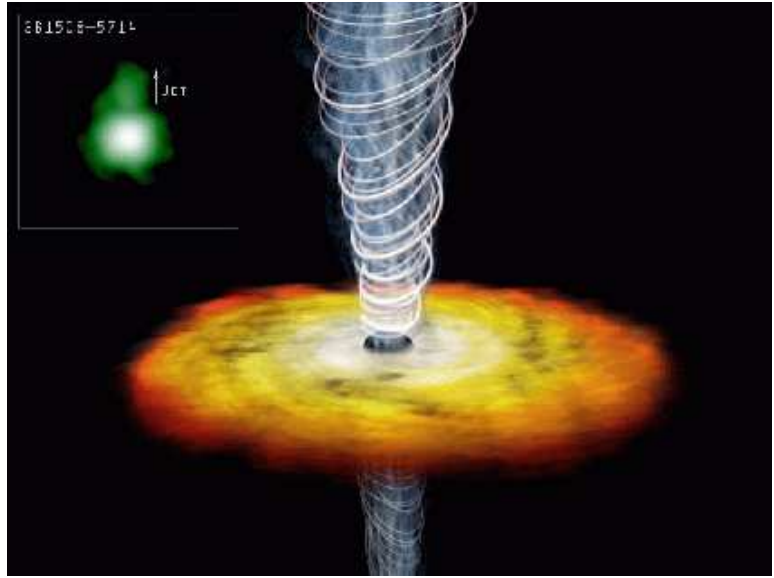}
\caption[Disk and jet associated with a supermassive black
hole]{Inset at upper left shows X-ray emission from energetic
particles in the jet of quasar GB1508+5714. Many accretion disks
have jets associated with them. The illustration shows an accretion
disk surrounding a supermassive black hole, which launches a
collimated jet. Credit: A. Siemiginowska, Illustration by M.Weiss;
http://antwrp.gsfc.nasa.gov/apod/ap031128.html. \label{Ch1fig:AGNDisk}}
\end{center}
\end{figure}

Disks are ubiquitous in astrophysics. Many astrophysical objects,
e.g., Saturn's rings, the solar system, and galaxies, are disk
shaped. A disk is formed when the matter has sufficient angular
momentum for the centrifugal force to balance the attractive
gravitational force; this differs from other systems like stars and
planets where gravitational attraction is balanced by pressure. Key
astrophysical processes, like star and planet formation, and many
sources in high energy astrophysics, are based on an accretion disk.
Accretion refers to the accumulation of matter onto a central
compact object or the center of mass of an extended system. Examples
of accreting systems are: binaries where matter flows from a star to
a compact object like a black hole, a neutron star, or a white dwarf
(see Figure \ref{Ch1fig:BinaryDisk}); Active Galactic Nuclei (AGN)
powered by accretion onto a supermassive black hole in the center of
galaxies (see Figure \ref{Ch1fig:AGNDisk}); and protostellar and
protoplanetary disks, the predecessors of stars and planets.

To accrete, matter has to lose angular momentum. Gravitational
binding energy released because of the infall of matter is a
powerful source of luminosity. Quasars, one of the most luminous
sources in the universe, are powered by accretion
\cite{Lynden-Bell1969}. The central problem in accretion physics is,
how does matter lose rotational support and fall in? In many disks
the mass of the central object is much larger than the disk mass,
resulting in a Keplerian rotation profile ($\Omega \sim R^{-3/2}$).
In principle, the presence of a shear viscosity allows the transport
of angular momentum from the faster inner fluid elements to the
slower outer ones. However, the accretion rate obtained by putting
in a typical number for microscopic (collisional) viscosity is
several orders of magnitude smaller than needed to explain
observations.

Turbulent stress due to interacting large scale ($\approx$ disk
height) eddies is sufficient to provide the needed accretion rates.
For turbulent stress one needs a source to sustain the turbulence;
otherwise the nonlinear motions will be damped due to viscosity.
Hydrodynamic disks with specific angular momentum (angular momentum
per unit mass) increasing outwards (e.g., Keplerian disks) are
linearly stable. A large Reynolds number is not sufficient to
produce nonlinear turbulent motions from small perturbations. A
source to produce and to sustain the turbulence is required. A
linear instability, that can tap the free energy in differential
rotation, can amplify small amplitude fluctuations into large scale
nonlinear motions, and provide such a source. A big advance was made
when Balbus and Hawley realized that the magnetorotational
instability (MRI), an instability of magnetized, differentially
rotating flows, can cause turbulent transport in accretion disks
\cite{Balbus1991,Hawley1995}.

Although, the identification of the MRI as the source of turbulence in
accretion flows was a major step in understanding accretion,
there are several unsolved problems. Although the MRI only requires
a small amount of ionization to work \cite{Blaes1994}, protostellar
disks, from which stars and planets form, are very cold and may have
such a low degree of ionization the MRI does not operate.
Another topic of investigation is, whether the hydrodynamic
Keplerian flow, like the planar shear flow \cite{Lesur2005}, can
become turbulent at large enough Reynolds numbers (discussed more in
subsection \ref{Ch1:Hydro}). Another problem, a motivation for this
thesis, is to understand why some black hole accretion disks are
unusually dim \cite{Narayan1997}. Understanding of the microphysics
and the global structure of accretion flows, in important physical
regimes, is still incomplete. The theoretical disk models have to be
tested against the ever detailed observations.
\section{Accretion as an energy source}
Accretion is a very efficient source of energy. Disk models, based
on accretion of matter from a stellar companion on to a compact
object (first mentioned by \cite{Kuiper1941}), were used to explain
novae outbursts \cite{Crawford1956}, and later compact X-ray sources
\cite{Prendergast1968}. The release of thermonuclear energy from
stars is insufficient to account for the high luminosity, and
significant X-ray (non-blackbody) luminosity. This section,
including the subsections on the Eddington limit and the emitted
spectrum, are based on Chapter 1 of \cite{Frank2002}.

To illustrate the enormous power of accretion consider the following
example from \cite{Frank2002}. For a body of mass $M_*$ and radius
$R_*$, the gravitational energy released by accretion of mass $m$ on
to its surface is $\Delta E_{acc} = GM_*m/R_*$, where $G$ is the
gravitational constant. This energy is expected to
be released mainly in the form of electromagnetic radiation.
Luminosity, the energy radiated per unit time, is proportional to
the ratio $M_*/R_*$ and $\dot{M}$, the mass accretion rate.

Writing in terms of the rest mass energy, $\Delta E_{acc} = 0.15
(M_*/M_\odot)(10$ km$/R_*) mc^2$, where $M_\odot$ is the solar mass.
If the accreting body is a neutron star with $R_* \sim 10$ km and
$M_* \sim M_\odot$, then the efficiency of accretion is $0.15$. For
comparison, the nuclear energy released on burning hydrogen to helium is
$\Delta E_{nuc} = 0.007 mc^2$, about one twentieth of the accretion
yield. Thus, accretion is an even more efficient energy source than
fusion (in fact by a factor of few tens)!

Since black holes have no surface, $R_*$ refers to the radius beyond
which matter does not radiate. This radius depends on black hole
spin, which is difficult to measure. Our ignorance of $R_*$ can be
parameterized by an efficiency $\eta$, with $\Delta E_{acc}=\eta
mc^2$. Relativistic calculations give an efficiency of $6\%$ for a
non-rotating Schwarzchild black hole, and $42.3\%$ for a maximally
rotating Kerr black hole~\cite{Misner1973} (see Appendix
\ref{app:BHefficiency} for a discussion of the efficiency of black
hole accretion).

For a white dwarf with $M_* \sim M_\odot$, $R_* \sim 10^9 $ cm,
nuclear burning is more efficient than accretion by factors $40-50$.
Although the efficiency for nuclear burning for white dwarfs is much
higher, in many cases the reaction tends to `run away' to produce an
event of great brightness but short duration, a nova outburst, in
which available nuclear fuel is rapidly exhausted. For almost all of
its lifetime no nuclear burning occurs, and the white dwarf may
derive its entire luminosity from accretion. Whether accretion or
nuclear fusion dominates depends on $\dot{M}$, the accretion rate.
\subsection{The Eddington limit}
At high luminosity, the accretion flow is affected by the outward
momentum transferred from radiation to the accreting matter by
scattering and absorption. We derive an upper limit on luminosity of
an accretion disk by considering spherical, steady state accretion.
Assume the accreting matter to be fully ionized hydrogen plasma. If
$S$ is radiant energy flux (erg cm$^{-2}$ sec$^{-1}$), and
$\sigma_T=6.7 \times 10^{-25}$cm$^2$ is the electron Thomson
scattering cross section, the outward radial force is $\sigma_T
S/c$. The effective cross section can exceed $\sigma_T$ if photons
are absorbed by spectral lines. Because of the charge neutrality of
plasma, radiative force on electrons couples to protons. If $L$ is
the luminosity of the accreting source, $S=L/4\pi r^2$, net inward
force on proton is $(GM_*m_p - L\sigma_T/4\pi c)/r^2$. The limiting
luminosity for which the radial force vanishes, the Eddington limit,
is $L_{Edd}=4\pi G M_* m_p c/\sigma_T \cong 1.38 \times 10^{38}
(M_*/M_\odot)$ erg s$^{-1}$. At greater luminosities, the radiation
pressure will halt accretion. The Eddington limit is a crude
estimate of the upper limit on the steady state disk luminosity.

If all the kinetic energy of accretion is given up at the stellar
surface, $R_*$, then the luminosity is $L_{acc}=GM_*\dot{M}/R_*$.
For accretion powered objects, the Eddington limit implies an upper
limit on the accretion rate, $\dot{M} \lesssim \dot{M}_{Edd}=4\pi
R_*m_pc/\sigma_T=9.5 \times 10^{11} R_*$ g s$^{-1}$. The Eddington
limit applies only for uniform, steady accretion; e.g., photon
bubble instability \cite{Arons1992,Gammie1998,Blaes2001}, a
compressive instability of radiative disks that opens up optically
thin ``holes" through which radiation can escape, can allow for
super-Eddington luminosity \cite{Turner2005,Begelman2006}.
\subsection{The emitted spectrum}
Order of magnitude estimates of spectral range of the emission from
compact accreting objects can be made. The continuum spectrum can be
characterized by a temperature $T_{rad}=h\overline{\nu}/k$ of
emitted radiation, where $\overline{\nu}$ is the frequency of a
typical photon. For an accretion disk with luminosity $L_{acc}$, one
can define a blackbody temperature as $T_{b}=(L_{acc}/4\pi
R_*^2\sigma)^{1/4}$, where $\sigma$ is the Stefan-Boltzmann
constant. Thermal temperature, $T_{th}$, is defined as the
temperature material would reach if its gravitational potential
energy is converted entirely into the thermal energy. For each
proton-electron pair accreted, the potential energy released is
$GM_*(m_p+m_e)/R_* \cong GM_*m_p/R_*$, and the thermal energy is $2
\times (3/2) kT$; therefore $T_{th} = GM_*m_p/3kR_*$. The virial
temperature, $T_{vir}=T_{th}/2$, is also used frequently. If the
accretion flow is optically thick, photons reach thermal equilibrium
with the accreted material before leaking out to the observer and
$T_{rad}=T_{b}$. Whereas, if accretion energy is converted directly
into radiation which escapes without further interaction (i.e., the
intervening material is optically thin), $T_{rad}=T_{th}$. In
general, the observed radiation temperature is expected to lie
between the two limits, $T_{b} \lesssim T_{rad} \lesssim T_{th}$.

Applying these limits to a solar mass neutron star radiating at the
Eddington limit gives, 1 keV $\lesssim h \overline{\nu} \lesssim$ 50
MeV; similar results would hold for stellar mass black holes. Thus
we can expect the most luminous accreting neutron star and black
hole binary disks to appear as medium to hard X-ray emitters, and
possibly as $\gamma$-ray sources. Similarly for white dwarf
accretion disks with $M_*=M_\odot$, $R_*= 10^9$ cm, we obtain 6 eV
$\lesssim h \overline{\nu} \lesssim$ 100 keV. Consequently,
accreting white dwarfs should be optical, UV, and possibly X-ray
sources. Observations are mostly consistent with these estimates.

Nonthermal emission mechanisms also operate in disks. Examples are:
synchrotron emission by relativistic electrons spiraling around
magnetic field lines and inverse Compton up-scattering of photons by
relativistic electrons. Line emission because of electronic
transition between energy levels provides a useful diagnostic of
density, temperature, and velocities in the emitting region.

Accretion disks, being efficient sources of energy, can be very
luminous. Their spectra are also very rich, extending all the way
from radio to X-ray and $\gamma$-ray frequencies. In order to
interpret the radiative signatures, one needs to understand
transport and radiation processes in accretion disks.
\section{Accretion disk phenomenology}
Much of the phenomenology of accretion disks was developed in
mid-1970's when two influential papers, by Shakura and Sunyaev
\cite{Shakura1973}, and Lynden-Bell and Pringle
\cite{Lynden-Bell1974}, appeared. It was shown that in the presence
of a shear viscosity, an infinitesimal mass can carry away all the
angular momentum of the inner fluid elements, facilitating mass
accretion \cite{Lynden-Bell1974,Pringle1981}. The structure (thick
or thin) and radiation spectrum (luminous or radiatively
inefficient) of a disk depends mainly on the rate of matter inflow,
$\dot{M}$ \cite{Shakura1973}. Of course the overall luminosity and
accretion time scale depends on $M_*$,the mass of the central
object.

A binary system consisting of a star and a compact object (black
hole, neutron star, or white dwarf) is likely to be very common in
the Galaxy. The outflow of matter from the star's surface---the
stellar wind---is significant ($\sim 10^{-5} M_\odot$ /yr) for
massive O-stars and Wolf-Rayet stars ($M \gtrsim 20 M_\odot$). In
binary systems, an additional strong matter outflow connected with
the Roche limiting surface is possible. The Roche surface is the
surface around the star beyond which the gravitational influence of
the compact object dominates. When a star leaves the main sequence
at later stages of its evolution, it can increase in size and fill
its Roche volume, giving rise to an intensive outflow of matter
mostly through the inner Lagrangian point (an unstable equilibrium
point between the star and compact object) \cite{Frank2002}. Figure
\ref{Ch1fig:BinaryDisk} shows an artist's impression of a binary
accretion system, with a star filling its Roche lobe and accreting
on to a compact object via a disk. Accretion in AGN is likely to be
fed by the winds from nearby massive stars, or the infall of
intergalactic gas. Once an accretion disk is formed around the
compact object, the subsequent evolution does not depend on the
source of matter.
\begin{figure}
\begin{center}
\includegraphics[width=4in,height=3in]{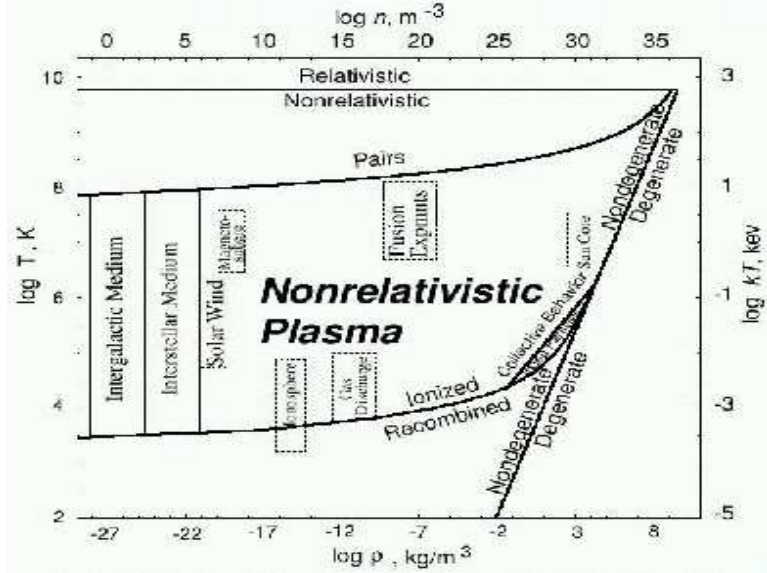}
\caption[Density-temperature diagram for hydrogen]{ The
density-temperature diagram for hydrogen in the regime where it
behaves as a non-relativistic plasma. Many astrophysical systems
including accretion disks are in the plasma state. The figure is
taken from lecture notes by Niel Brandt,
http://www.astro.psu.edu/users/niel/astro485/lectures/lecture08-overhead07.jpg
\label{Ch1fig:DensityTemperature}}
\end{center}
\end{figure}

Figure \ref{Ch1fig:DensityTemperature} shows the density-temperature
phase diagram for hydrogen. This shows that the accretion disks with
temperatures exceeding a few eV (and reasonable densities) are fully
ionized. Most accretion disks, except possibly protostellar and
protoplanetary disks, are sufficiently ionized for the plasma
description to be valid \cite{Blaes1994,Balbus1998}. Even relatively
cold gas disks may have enough ionization by cosmic rays
\cite{Gammie1996}, X-rays \cite{Igea1999}, and radioactivity
\cite{Sano2000} to be sufficiently conducting for the MHD-like
phenomena to occur.

Magnetohydrodynamics (MHD) is a good approximation for a magnetized
plasma when the mean free path is much smaller than the scales of
interest, e.g., in efficiently radiating, dense, thin disks. This is
not always the case; the radiatively inefficient accretion flows
(RIAFs), a motivation for this thesis, are believed to be
collisionless with the mean free path comparable to (or even larger
than) the disk size (see Table \ref{Ch1tab:SgrA} for plasma parameters in the
Galactic center disk). Ideal MHD, where resistive effects are
negligible and the field is frozen into the plasma, is a good
approximation for large scale dynamics of almost all astrophysical
plasmas; as the dynamical scales are orders of magnitude larger than
the resistive scale or the gyroradius scale. Even with a large
separation between the dynamical and resistive/viscous scales,
dissipation cannot be ignored---energy cascades from large scales to
smaller scales, terminating at the dissipative scales, where it is
dissipated in shocks and reconnection. In the inertial range of
isotropic, homogeneous turbulence, energy dissipation rate balances
the rate at which energy is injected, independent of resistivity and
viscosity \cite{Frisch1995,Biskamp2003}.

In rest of the section we closely follow the review article by
Balbus and Hawley to use the conservation of mass, energy, and
angular momentum to derive the transport properties of disks. The
widely used $\alpha$ model for turbulent stress is introduced
\cite{Shakura1973}.
\subsection{Governing equations}
Following \cite{Balbus1998}, the conservation of total energy in
magnetohydrodynamics (MHD), gives \be \label{Ch1eq:TotalEnergy}
\frac{\partial}{\partial t} \left ( \frac{1}{2}\rho V^2 +
\frac{3}{2} p + \rho \Phi + \frac{B^2}{8\pi} \right ) + {\bf \nabla
\cdot} \left [~ \right] = -{\bf \nabla \cdot F_{rad}}, \ee where
$\rho$, ${\bf V}$, $p$, ${\bf B}$, and $\Phi$ are density, fluid
velocity, pressure, magnetic field, and gravitational potential,
respectively. The term on the right represents radiative losses. The
conservative flux term ${\bf \nabla \cdot} [~]$ consists of a
dynamic contribution \be \label{Ch1eq:EnergyFlux} {\bf v} \left (
\frac{1}{2}\rho V^2 + \frac{5}{2} p + \rho \Phi \right ) +
\frac{{\bf B}}{4\pi} \times ({\bf V} \times {\bf B}), \ee and a
viscous contribution, \be \label{Ch1eq:ViscousFlux} -\eta_V \left (
{\bf \nabla} \frac{V^2}{2} + \frac{{\bf V}}{3} {\bf \nabla \cdot V}
\right ) + \frac{\eta_B}{4\pi} ( {\bf \nabla} \times {\bf B} )
\times {\bf B}, \ee where, $\eta_V$ is the microscopic kinematic
shear viscosity, and $\eta_B$ the microscopic resistivity. Here we
use a uniform, isotropic viscosity for simplicity. The Braginskii
viscosity is highly anisotropic as will be discussed later in the
thesis. The dynamic flux in Eq.~(\ref{Ch1eq:EnergyFlux}) consists of
an advective flux of kinetic and thermal energy (the first term),
and the Poynting flux of electromagnetic energy (the second term).
The equation for angular momentum conservation in cylindrical,
$(R,\phi,z)$, coordinate system is given by \ba
\label{Ch1eq:AngularMomentum} \nonumber \frac{\partial}{\partial t}
(\rho R V_\phi) &+& {\bf \nabla \cdot} R \left [ \rho V_\phi {\bf V}
- \frac{B_\phi}{4 \pi} {\bf B_p} + \left ( p +  \frac{B^2}{8\pi}
\right )
{\bf \hat{\phi}} \right ] \\
&-& {\bf \nabla \cdot } \left [ \frac{R \eta_V}{3} ({\bf \nabla
\cdot V}) {\bf \hat{\phi}} + \eta_V R^2 {\bf \nabla}
\frac{V_\phi}{R} \right ] = 0, \ea where ${\bf \hat{\phi}}$ is the
unit vector in the azimuthal direction, the subscript ${\bf p}$
refers to the poloidal magnetic field components (the $R$ and
$z$ components). In an accretion disk, there is a net flux of energy
and angular momentum in the radial direction, so the divergence
terms in Eqs.~(\ref{Ch1eq:TotalEnergy})
and~(\ref{Ch1eq:AngularMomentum}) are dominated by the radial
derivatives of radial fluxes.
\subsection{Fluctuations}
A fiducial disk system consists of a point mass potential situated
at the center of the disk, with the gas going around in a Keplerian
rotation, $\Omega^2=GM_*/R^3$. The fluctuation velocity is given by
$V_R$, $\delta V_\phi=V_\phi-R\Omega$, and $V_z$. When the azimuthal
velocity $R\Omega$ much exceeds the isothermal sound speed
$c_s=\sqrt{p/\rho}$, the disk is thin; the vertical structure is
determined by hydrostatic balance, with the disk height scale
$H=c_s/\Omega \ll R$. In this section we consider only thin disks
because they are simpler, for the vertical dynamics and pressure
forces do not play a significant role. For thick disks, where
thermal forces are equally important and vertical motion is coupled
to the motion in plane, there is no universally accepted standard
model \cite{Narayan1994,Quataert2000,Blandford1999}.

The radial flux of angular momentum from
Eq.~(\ref{Ch1eq:AngularMomentum}) is $R \left [ \rho V_R (R\Omega +
\delta V_\phi) - B_R B_\phi/4\pi \right ]$. Taking an azimuthal
average, integrating over height, and averaging over a narrow range
$\Delta R$ in $R$, one obtains, $\Sigma R [ R \Omega \la  V_R
\ra_\rho + \la V_R \delta V_\phi - V_{AR} V_{A\phi} \ra_\rho]$,
where the surface density $\Sigma=\int \rho dz$, and for any $X$,
$\la X \ra_\rho = 1/(2\pi\Sigma \Delta R) \int X \rho d\phi dR dz$.
The notation $V_{AR}$, etc. denotes the Alfv\'en velocity, ${\bf
V_A} = {\bf B}/\sqrt{4\pi \rho}$. The first term in the radial
angular momentum flux is the direct inflow of angular momentum due
to radially inward accretion of matter; the second term represents
an outward component of flux due to turbulent transport because of
statistical correlations in the velocity and magnetic stress
tensors \cite{Tennekes1972}. The $R\phi$ component of the stress, responsible for
angular momentum transport (see Eq. \ref{Ch1eq:AngularMomentum}), is
$W_{R \phi} \equiv \la V_R \delta V_\phi - V_{AR} V_{A\phi}
\ra_\rho$.

In steady state, the angular momentum flux must be divergence free,
and thus vary as $1/R$, i.e., $\Sigma R^2( R\Omega \la V_R \ra_\rho
+ W_{R\phi})$ is independent of $R$. The condition of vanishing
stress at the inner edge ($R_*$) gives, $ \Sigma (\Omega R \la V_r
\ra_\rho + W_{R\phi}) = \Sigma_* \Omega_* R_* \la V_{r*} \ra_\rho (
R_*/R )^2 $. Expressing in terms of the constant accretion rate,
$\dot{M} = -2\pi R \Sigma \la V_R \ra_\rho$, leads to $-\dot{M}
R\Omega/2\pi + \Sigma R W_{R\phi} = - \dot{M} R_*^2 \Omega_* /2\pi
R$. This gives an expression for the variation of stress with radius
as, \be \label{Ch1eq:Stress} W_{R\phi} = \frac{\dot{M}\Omega}{2\pi
\Sigma} \left [ 1- \left ( \frac{R_*}{R} \right)^{1/2} \right ]. \ee

Keeping only the second order terms in the energy flux in
Eq.~(\ref{Ch1eq:EnergyFlux}), one gets $\rho V_R(\Phi +
R^2\Omega^2/2 + R\Omega \delta V_\phi) - (R\Omega/4\pi)B_R B_\phi$.
Upon averaging, height integrating, and using the Keplerian
potential $\Phi=-R^2\Omega^2$, energy flux becomes $ F_E =
\dot{M}R\Omega^2/4\pi + \Sigma R \Omega W_{R\phi} $. Substituting
for $\Omega$ and using Eq.~(\ref{Ch1eq:Stress}) for the stress
tensor, this reduces to \be \label{Ch1eq:FluxEnergy} F_E = \frac{3 G
M_* \dot{M}}{4\pi R^2} \left [ 1 - \frac{2}{3} \left (
\frac{R_*}{R}\right )^{1/2} \right ]. \ee The energy deposited by
this flux is the source of disk's luminosity. Minus the divergence
of the flux gives the disk surface emissivity (energy per unit area
per unit time), $Q$. Dividing by a factor of two for each side of
the disk gives \be \label{Ch1eq:Emissivity} Q = \frac{3 G M_*
\dot{M}}{8\pi R^3} \left [ 1- \left ( \frac{R_*}{R} \right )^{1/2}
\right ]. \ee The $Q-\dot{M}$ relation depends on local energy
conservation and is, as expected, independent of the form of the
stress tensor (see \cite{Shakura1973,Pringle1981}).
Eliminating $\dot{M}$ between Eqs. (\ref{Ch1eq:Stress}) and
(\ref{Ch1eq:Emissivity}) yields \be
\label{Ch1eq:FluctuationDissipation} Q = \frac{3}{4}\Sigma \Omega
W_{R\phi} = \frac{3}{4} \Sigma \Omega \la V_R \delta V_\phi -
V_{AR}V_{A\phi} \ra_\rho, \ee a kind of fluctuation-dissipation
relation for accretion disks~\cite{Balbus1994}. From Eqs.
(\ref{Ch1eq:Stress}) and (\ref{Ch1eq:FluctuationDissipation}), it is
clear that the correlation of velocity (and magnetic field)
fluctuation components is responsible for much of the disk
transport and luminosity.

Above discussion is valid only for a cold, thin disk where pressure
can be ignored. For a radiatively inefficient, hot, thick disk the
pressure term $(5/2) p V_R$ should be included in the radial energy
flux; ${\bf \nabla \cdot} F_E \approx 0$ in absence of radiation,
and the gravitational energy released from accretion is converted
into thermal and kinetic energies.

Total luminosity emitted from $R_*$ to $R$ is, $L(R_*<R) = 2\pi
[R_*F_E(R_*)-RF_E(R)] = (GM_*\dot{M}/2R_*) [1-3R/R_* + (R_*/R)^{3/2}
]$. In the limit $R \rightarrow \infty$, $L = GM_*\dot{M}/2R_*$,
which shows that half the binding energy of the innermost orbit is
converted to radiation. The other half is retained as kinetic
energy. The fate of the residual energy depends on the nature of
central accretor. If a stellar surface is present, remaining energy
will be radiated in a boundary layer; if the central object is a
black hole, the energy may be swallowed and lost.
\subsection{$\alpha$ disk models}
Although the relationship between disk's surface emissivity $Q$ and
the mass accretion rate $\dot{M}$ is independent of stress tensor,
most other relations involve a dependence on $W_{R \phi}$.
Recognizing the central importance of $W_{R \phi}$ and its
computational inaccessibility, Shakura and Sunyaev
\cite{Shakura1973} suggested a natural scaling for the stress
tensor, $W_{R\phi} = \alpha c_s^2$, where $\alpha \lesssim 1$ is a
parameter. The idea behind the $\alpha$ prescription is that the
turbulent velocities, whose correlation determines $W_{R\phi}$, are
limited by the sound speed $c_s$, as supersonic velocities will be
quickly dissipated in shocks. The $\alpha$ formalism bypasses the
thorny issue of disk turbulence, and can be thought as a closure for
the stress tensor. The $\alpha$ formalism can be thought of as
equivalent to a ``turbulent viscosity" \be \label{Ch1eq:AlphaModel}
\nu_t = \alpha c_s H \ee that is similar to microscopic viscosity in
Navier Stokes equation. The role of random particle velocity is
played by $c_s$, and the scale height $H$ is the effective mean free
path (eddy size). This is a closure based on plausibility arguments
and is not rigorous like the Chapman-Enskog procedure
\cite{Chapman1970}. The
$\alpha$
formalism is the basis of much of observationally driven disk
phenomenology.
The radial dependence of various physical quantities, like
temperature, density, height, etc., can be obtained in terms of
parameters $\dot{M}$ and $\alpha$ \cite{Frank2002}.
\section{MRI: the source of disk turbulence}
\label{Ch1sec:MRI}
A breakthrough occurred when Balbus and Hawley proposed the
magnetorotational instability (MRI), an instability of
differentially rotating flows, as the source for turbulence and
transport in accretion disks~\cite{Balbus1991}. Before this, a
robust mechanism to sustain turbulent angular momentum transport in
accretion disks was unknown. Although the instability was described
in its global form for magnetized Couette flow by
Velikhov~\cite{Velikhov1959} and
Chandrasekhar~\cite{Chandrasekhar1960}, its importance for accretion
disks was not recognized. In his classic book
\cite{Chandrasekhar1961}, Chandrasekhar points out the essential
feature of the MRI, ``in the limit of zero magnetic field, a
sufficient condition for stability is that the angular speed,
$|\Omega|$, is a monotonic increasing function of $r$. At the same
time, any adverse gradient of angular velocity can be stabilized by
a magnetic field of sufficient strength."

Both local \cite{Hawley1995} and global
\cite{Armitage1998,Hawley2000,Stone2001} numerical simulations have
confirmed that the MRI can amplify small perturbations to nonlinear
turbulent motions. Correlations between the radial and azimuthal
fields results in a sustained turbulent stress corresponding to
$\alpha \equiv W_{R\phi}/p \sim 0.001-0.5$, enough to account for
typical disk luminosities. Next, we discuss the inadequacy of the
hydrodynamic models, followed by the linear and nonlinear
characteristics of the MRI.
\subsection{Insufficiency of hydrodynamics}
\label{Ch1:Hydro} In the Boussinesq approximation (${\bf \nabla
\cdot V} = 0$ in the equation of motion), if we ignore pressure,
then a fluid element disturbed slightly from its Keplerian orbit
will execute retrograde epicycles at a frequency $\kappa$ ($
\kappa^2 \equiv \frac{1}{R^3} \frac{d(R^4 \Omega^2)}{dR}$), as seen
by an observer in an unperturbed Keplerian orbit. The criterion for
local linear stability is simply $\kappa > 0$, i.e., specific
angular momentum increases outwards, the Rayleigh
criterion.\footnote{The Rayleigh criterion applies only for
axisymmetric disturbances.} Therefore, a Keplerian disk with
specific angular momentum $R^2\Omega \sim R^{1/2}$, increasing
outwards is linearly stable, unable to produce (and sustain)
nonlinear turbulent stress.

A rotating shear flow is different from a planar shear flow because
of the coriolis force. Coriolis force is responsible for stable
epicyclic oscillations in Keplerian flows, whereas planar shear
flows are marginally stable in the linear regime. Nonlinear local
shearing box simulations show that while planar shear flows can be
nonlinearly unstable and become turbulent even at relatively small
Reynolds numbers \cite{Bayley1988}, $10^3-10^4$ (orders of magnitude
smaller than the true Reynolds number for disks), Keplerian disks are nonlinearly
stable and give no turbulence over the same range of Reynolds
numbers \cite{Balbus1996,Hawley1999}. This is because stable
epicycles prevent nonlinear instabilities to develop in Keplerian
flows. Whether turbulence and transport can occur in hydrodynamic
Keplerian flows, is still not universally agreed. There are
experimental claims that the Keplerian disks are nonlinearly
unstable~\cite{Richard1999}, but recent experiments, with more
carefully controlled boundary conditions (especially Eckman flows),
which directly measure the Reynolds stress,
show otherwise~\cite{Burin2006}. Also, there is some recent work on
transient amplification in the linear regime, that can give rise to
nonlinear amplitudes (and maybe turbulence) in hydrodynamic
differentially rotating
flows~\cite{Chagelishvili2003,Afshordi2005,Umurhan2004}.

Convective turbulence was also proposed as a source of enhanced
shear viscosity \cite{Lin1980}. Convection is believed to arise from
heating due to energy dissipated in the disk midplane. The hope was
that somehow convective blobs can cause nonlinear correlations to
produce non-vanishing stress. However, the linear analysis of
convective instability in Keplerian flows gives a wrong sign of
stress \cite{Ryu1992}, with inward transport of angular momentum.
Three dimensional simulations of convectively unstable disk show
very small angular momentum transport ($\alpha \sim -10^{-4}$), and
in the opposite direction \cite{Stone1996a}. None of the local
hydrodynamic mechanisms to date are able to give sufficient angular
momentum transport.\footnote{Global modes like Papaloizou-Pringle
instability \cite{Papaloizou1984,Goldreich1986}, and spiral shocks
can in principle cause turbulence and transport, however, their role
as a universal transport mechanism for Keplerian disks is not
clear \cite{Balbus1998}.} This launches us
to the study of the dramatic effect of magnetic fields on accretion
disk stability.
\subsection{MHD accretion disks: Linear analysis}
The ideal MHD equations are \ba \label{Ch1eq:MHD1} && \frac{\partial
\rho}{\partial t} + \nabla \cdot \left(\rho {\bf V}\right)=0,
\\
\label{Ch1eq:MHD2} && \rho \frac{\partial {\bf V}}{\partial t} +
\rho\left({\bf V} \cdot \nabla\right) {\bf V}= \frac{\left(\nabla
\times {\bf B}\right) \times {\bf B}}{4\pi} - {\bf \nabla} p + {\bf F_g},\\
\label{Ch1eq:MHD3} && \frac{\partial {\bf B}}{\partial t}= \nabla
\times \left({\bf V} \times {\bf
B}\right), \\
\label{Ch1eq:MHD4} && \frac{\partial e}{\partial t} + {\bf \nabla}
\cdot ( e {\bf V}) = -p {\bf \nabla \cdot V} , \ea
 where, ${\bf F_g}$ is the gravitational force, and $e=p/(\gamma-1)$
 relates internal energy density and pressure ($\gamma=5/3$ in a 3-D
 non-relativistic plasma). Making ${\bf B}=0$ in the MHD equations
 gives the hydrodynamic equations.
\subsubsection{WKB analysis in a Keplerian disk}
The linear response of a Keplerian hydrodynamic flow is stable
epicyclic motion, however, addition of weak magnetic fields renders
it unstable. Before considering Keplerian flows, it is useful to
study waves in a homogeneous, non-rotating equilibrium. Linear waves
in MHD and hydrodynamics are quite different. MHD is richer in waves
with fast, Alfv\'en, and slow modes, compared to hydrodynamics with
only an isotropically propagating sound wave \cite{Kulsrud2005}. As
the name suggests, the fast mode has the fastest phase speed and
propagates isotropically. The Alfv\'en mode is intermediate with
propagation along the field lines, and the slow mode with the
smallest phase speed also propagates along the field lines. In
addition to these, there is a non-propagating entropy mode with
anticorrelated density and temperature fluctuations.

Consider a differentially rotating disk threaded by a magnetic field
with a vertical component $B_z$ and an azimuthal component $B_\phi$.
Consider WKB perturbations of the form $\exp i({\bf k \cdot
r}-\omega t)$, $kR \gg 1$. Notation is the standard one: ${\bf k}$
is the wave vector, ${\bf r}$ the position vector, $\omega$ the
angular frequency, and $t$ the time. Linear perturbations are
denoted by $\delta$. Only a vertical wave number is considered,
${\bf k}=k {\bf \hat{z}}$. The local linear equations are \ba
\label{Ch1eq:LMHD1}
&& -\omega \frac{\delta \rho}{\rho} + k \delta V_z = 0, \\
\label{Ch1eq:LMHD2} && -i\omega \delta V_R - 2\Omega \delta V_\phi - i
\frac{k B_z}{4\pi
\rho} \delta B_R = 0, \\
\label{Ch1eq:LMHD3} && -i \omega \delta V_\phi +
\frac{\kappa^2}{2\Omega} \delta V_R -i
\frac{kB_z}{4\pi \rho} \delta B_\phi = 0, \\
\label{Ch1eq:LMHD4}
 && - \omega \delta V_z + k \left ( \frac{\delta
p}{\rho} + \frac{B_\phi
\delta B_\phi}{4\pi \rho} \right ) = 0, \\
\label{Ch1eq:LMHD5}
&&-\omega \delta B_R = k B_z \delta V_R, \\
\label{Ch1eq:LMHD6}
 && -i \omega \delta B_\phi = \delta B_R
\frac{d\Omega}{d \ln R} +i k
B_z \delta V_\phi - B_\phi i k \delta V_z, \\
\label{Ch1eq:LMHD7}
&& \delta B_z = 0,\\
\label{Ch1eq:LMHD8} && \frac{\delta p}{p} = \frac{5}{3} \frac{\delta
\rho}{\rho}. \ea The resulting dispersion relation is (Eq. (99) in
\cite{Balbus1998}) \ba \label{Ch1eq:Dispersion} \nonumber [ \omega^2
&-& ({\bf k\cdot V_A})^2 ] [ \omega^4 - k^2 \omega^2 (a^2+V_A^2) +
({\bf k \cdot V_A})^2 k^2 a^2]
\\  \nonumber &-& \left[ \kappa^2 \omega^4 -  \omega^2 \left ( k^2 \kappa^2
(a^2+V_{A\phi}^2) + ({\bf
k\cdot V_A})^2 \frac{d\Omega^2}{d \ln R} \right ) \right ] \\
&-& k^2 a^2 ({\bf k \cdot V_A})^2 \frac{d\Omega^2}{d \ln R} = 0,\ea
where $a^2= (5/3) P/\rho$, ${\bf V_A} = {\bf B}/\sqrt{4\pi \rho}$.
Only the first term in the dispersion relation, Eq.
(\ref{Ch1eq:Dispersion}), is non-zero in the non-rotating limit; the
roots of the dispersion relation correspond to the fast, Alfv\'en,
and slow modes.

The effect of Keplerian rotation on the three MHD modes is shown in
Fig. (15) of Balbus and Hawley's review article \cite{Balbus1998},
with $\omega^2$ plotted as a function of $\Omega^2$ (also see Figure
\ref{Ch3fig:compare}). It shows that $\omega^2$ becomes negative for
the slow mode when $d\Omega^2/d \ln R >({\bf k \cdot V_A})^2$, i.e.,
slow modes becomes unstable. This destabilized MHD slow mode in
differentially rotating flows is the MRI. For a fixed wave number
there is an upper limit on the field strength for the MRI to exist,
i.e., it is a weak field instability.
\subsubsection{Spring model of the MRI}
A simple physical description of the MRI, based on the spring model
of Balbus and Hawley, is presented; the discussion closely follows
\cite{Balbus1998}. It is useful to study the instability in the
Boussinesq limit, where fast waves are eliminated. The simplest
model to think is of axisymmetric perturbations on uniform vertical
magnetic field in a Keplerian disk. If a fluid element is displaced
from its circular orbit by ${\bf \xi}$, with a spatial dependence
$\mbox{e}^{ikz}$, induction equation leads to ${\bf \delta B}=ikB
{\bf \xi}$. Magnetic tension force is then $ikB {\bf \delta B}/{4\pi
\rho} = -( {\bf k\cdot V_A})^2 {\bf \xi}$. In an incompressible,
pressure free limit, the equations of motion become \ba
\label{Ch1eq:Spring1} \ddot{\xi}_R - 2\Omega \dot{\xi}_\phi &=& -
\left ( \frac{d\Omega^2}{d\ln R} + ({\bf k \cdot V_A})^2 \right )
\xi_R,
\\ \label{Ch1eq:Spring2} \ddot{\xi}_\phi + 2 \Omega \dot{\xi}_R
&=& -({\bf k \cdot V_A})^2 \xi_\phi . \ea As before, $2\Omega$ and
$d\Omega^2/d\ln R$ terms represent coriolis and tidal forces,
respectively. These equations also describe two orbiting point
masses connected with a spring of spring constant $({\bf k \cdot
V_A})^2$.
\begin{figure}
\begin{center}
\psfrag{A}{$M_*$} \psfrag{B}{$m_i$} \psfrag{C}{$m_o$}
\psfrag{T}{$T$} \psfrag{P}{$\phi$} \psfrag{R}{$R$} \psfrag{z}{$z$}
\psfrag{H}{$H$}
\includegraphics[width=5in,height=3in]{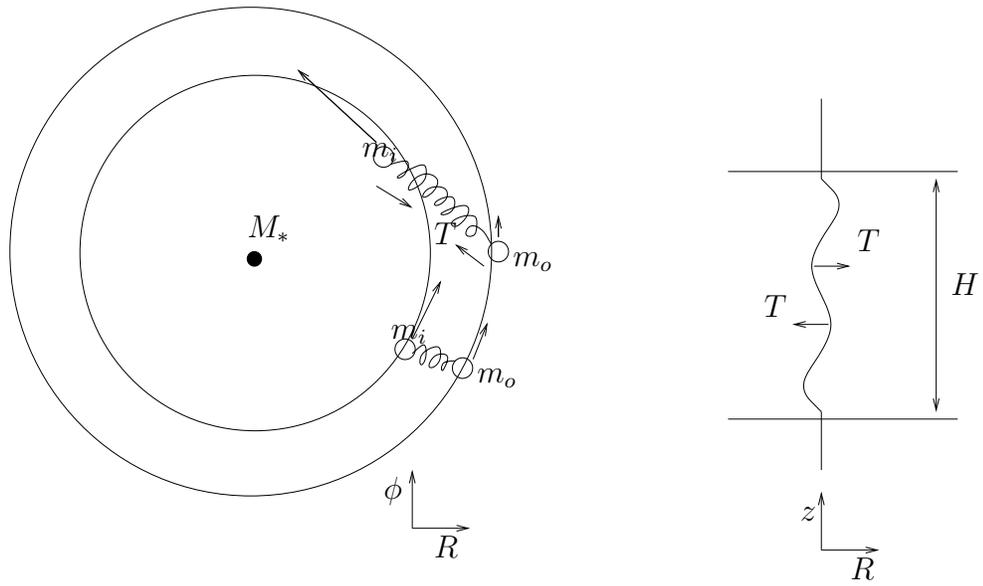}
\caption[Spring model of the MRI]{Spring model of the MRI. Left part
shows a top view of inner and outer point masses $m_i$ and $m_o$
connected by a spring. Mass $m_i$ is moving faster than $m_o$
because velocity decreases outwards in a Keplerian flow. Spring
force slows down $m_i$, and makes $m_o$ go faster. Inner mass falls
in as it loses angular momentum to the outer one, which moves out.
Right part shows a side view of a perturbed field line that results
in a restoring spring force. The field strength should be weak
enough for an unstable mode to fit within a disk height scale, $H$.
\label{Ch1fig:SpringMRI}}
\end{center}
\end{figure}

Consider two point masses, initially at the same orbital location,
displaced slightly in the radial direction, as shown in Figure
\ref{Ch1fig:SpringMRI}. The inner mass $m_i$ at radius $R_i$ is
connected via a spring to outer mass $m_o$ at $R_o$. In a Keplerian
disk, the inner mass rotates faster than the outer one. In the
absence of a spring both execute stable epicycles. However, the
spring stretches and builds up a tension $T$. $T$ pulls backward on
$m_i$ and forward on $m_o$. Thus, $m_i$ slows down and loses angular
momentum to $m_o$ which gains speed. This means that the slower
$m_i$ (compared to the local Keplerian velocity) cannot remain in
orbit at $R_i$ and must drop to a yet lower orbit. Similarly, $m_o$
acquire too much angular momentum to stay at $R_o$ and must move
outwards. The separation widens, the spring stretches yet more, $T$
goes up, and the process runs away. This is the essence of weak
field instability in differentially rotating flows. The presence of
other field components does not affect this picture, as by selecting
${\bf k} = k_z {\bf \hat{z}}$ we have ensured that only vertical
field couples dynamically. It is very crucial that the spring be
weak; if spring is very stiff, there are many stable vibrations in
an orbital time and no net transport of angular momentum. The right
side of Eq. (\ref{Ch1eq:Spring1}) reproduces the stability criterion
for the slow mode, $({\bf k \cdot V_A})^2 > - d\Omega^2/d\ln R$. One
can always choose a small enough $k$ to make a Keplerian disk
unstable. Thus, the necessary and sufficient condition for the
stability of a magnetized differentially rotating disk is
$d\Omega^2/d\ln R > 0$.

Just how large a wavelength is permitted? In order for the WKB
approximation to be valid, at least a half wavelength needs to fit
in the box height $H$. The stability criterion for a Keplerian disk
becomes $V_A^2 > \frac{H^2}{\pi^2} \frac{d\Omega^2}{d\ln R} \sim
(6/\pi^2) c_s^2$, i.e., the Alfv\'en speed must significantly exceed
the sound speed, if all the modes in a disk thickness are to be
stable. The MRI is called a weak field instability because it
requires pressure to exceed magnetic energy ($\beta = 8\pi p/B^2
\gtrsim 1$). It is interesting to note that there is no lower limit
on the strength of the magnetic field for the instability to exist
if dissipation scales are arbitrarily small \cite{Krolik2006b}.

The dispersion relation from Eqs. (\ref{Ch1eq:Spring1}) and
(\ref{Ch1eq:Spring2}), on assuming ${\bf \xi} \sim \exp(-i\omega
t)$, is \be  \label{Ch1eq:IncompressibleDispersion} \omega^4 -
\omega^2 [\kappa^2 + 2 ({\bf k \cdot V_A})^2 ] + ({\bf k\cdot
V_A})^2 \left ( (k\cdot V_A)^2 + \frac{d\Omega^2}{d \ln R} \right )
= 0, \ee which is precisely the $a\rightarrow \infty$ Boussinesq
limit of Eq. (\ref{Ch1eq:Dispersion}). Eq.
(\ref{Ch1eq:IncompressibleDispersion}) is a quadratic in $\omega^2$
which can be solved easily. The fastest growth rate for a Keplerian
disk is $\gamma_{max}=(3/4)\Omega$, and occurs at $({\bf k\cdot
V_A})_{max}=(\sqrt{15}/4) \Omega$. This is a very fast instability
that would cause amplification by $\sim 10^4$ in energy, per orbit.
The instability is very robust, independent of the magnetic field
orientation. In presence of a toroidal field the MRI is
non-axisymmetric for the perturbations to couple to the field
\cite{Balbus1992}. Nonlinear correlations resulting from this
instability can provide a sizeable stress to explain fast angular
momentum transport in disks, as we see in the next subsection.

\subsection{MHD accretion disks: Nonlinear simulations}
Tremendous progress has been made in the understanding of growth and
saturation of the MRI. Numerical studies started with unstratified
local shearing box simulations \cite{Hawley1991,Hawley1995} using
the ZEUS MHD code \cite{Stone1992a,Stone1992b}. In the shearing box
limit, equations are written in a frame rotating with the mean flow.
There is shear in a Keplerian box with $d\Omega/d\ln R = -3/2$.
Boundary conditions are periodic in $y$- (azimuthal) and $z-$
(vertical) directions, and shearing periodic in $x$- (radial
direction).\footnote{Shearing periodic means that periodic boundary
conditions are applied after a time dependent remap in $y$-
direction at the $x$- boundaries \cite{Hawley1995,Balbus1998}. There
is a jump of $-(3/2) \Omega L_x$ in $V_y$ between the inner and
outer radial faces to take differential rotation into account. There
are similarities between this and the boundary conditions used in
fusion energy research to handle sheared magnetic fields in
flux-tube simulations of turbulence
\cite{Cowley1991,Hammett1993,Beer1995}.}

\begin{figure}
\begin{center}
\includegraphics[width=4in,height=3in]{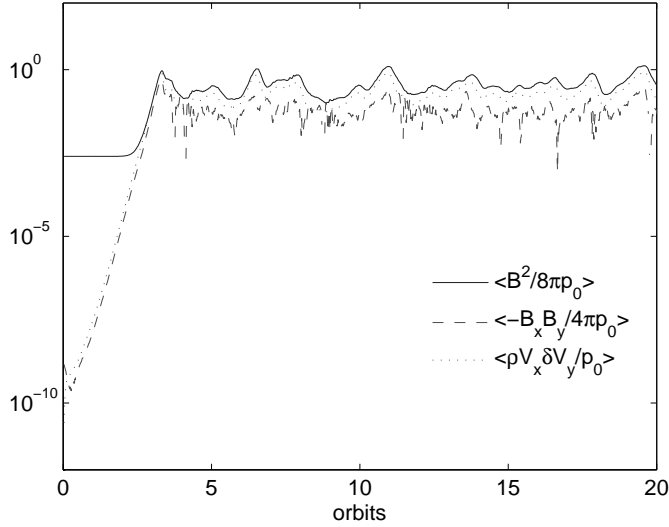}
\caption[Magnetic energy and stresses for an MHD simulation with
net vertical flux]{Volume averaged magnetic energy, Maxwell stress,
and Reynolds stress normalized to the initial pressure, for an
initial $\beta=400$ vertical field. The Maxwell stress is $\sim 3$
larger than the Reynolds stress. Time and volume averaged values in
the turbulent state are: $\alpha=0.286$ and $\beta=2.56$ (case $ZMl$
in Table \ref{Ch4tab:tab1} from Chapter \ref{chap:chap4}).
\label{Ch1fig:NonlinearMRI}}
\end{center}
\end{figure}
Shearing box simulations start with a random white noise imposed on
an initial equilibrium. In simulations with a net vertical flux,
magnetic energy increases exponentially until the channel solution
(the nonlinear form of the fastest growing mode with $k_z
V_{Az}/\Omega \sim 1$) becomes unstable to secondary
Kelvin-Helmholtz type instabilities \cite{Goodman1994}. Magnetic
energy increases by several orders of magnitude before secondary
instabilities break the channel solutions into turbulence. Magnetic
energy saturates at sub-equipartition ($\beta = 8\pi p_0/B^2 \sim
1-100$), with $\alpha=\la\la (\rho V_x \delta V_y - B_x B_y/4\pi)
\ra\ra/p_0 \sim 0.001-0.5$, where ``$\la \la \ra \ra$" represents a
box and time average in the turbulent state. Figure
\ref{Ch1fig:NonlinearMRI} shows the time evolution of magnetic
energy, and Maxwell and Reynolds stress for a simulation with an
initial vertical field with $\beta=400$. Magnetic energy is
dominated by the toroidal component. All variables show large
fluctuations from the mean in the turbulent state. Magnetic and
kinetic energy power spectra are peaked at low wave numbers,
indicating significant energy at scales comparable to the box size.

Nonlinear simulations of an initially toroidal field observe that
the growth rates are smaller than the vertical field runs
\cite{Hawley1995,Matsumoto1995}. The growth rate of the
non-axisymmetric mode is fastest for largest vertical wave number
$k_z$ \cite{Balbus1992}, but in  simulations, these wave numbers are
damped because of a finite resolution. In the saturated state
dominated by large wave numbers $\alpha \sim 0.01$, smaller than the
net vertical flux cases. Simulations with no net flux, $\la {\bf B}
\ra =0$, also result in sustained MHD turbulence and transport at
large scales. However, both magnetic energy and turbulent stress are
smaller by a factor of $10-100$ compared with the net vertical field
case \cite{Hawley1996}. Toroidal fields and the Maxwell stress
dominate the other components, irrespective of the initial field
configuration. Total stress is roughly proportional to the magnetic
energy for all cases.

Stratified shearing boxes with vertical gravity were simulated to
closely model a real accretion disk in the local limit
\cite{Brandenburg1995,Stone1996b}. Vertical stratification allow the
possibility of vertical motions driven by magnetic buoyancy.
Stratified simulations are not very different from the unstratified
ones because the Mach number (the ratio of fluid velocity and sound
speed, $V/c_s$) is much less than unity for MRI turbulence. This
ensures that the MRI timescale ($1/\Omega=H/c_s$) is much faster
than the time scale for buoyant motions at large scales ($H/V$).
Results are similar for the adiabatic and isothermal equations of
state \cite{Stone1996b}. While the $R-\phi$ dynamics is dominated by
the MRI, vertical stratification can result in significant mixing in
the $z-$ direction. Stratified simulations show the eventual
emergence of a magnetically dominated corona stable to the MRI
because of buoyantly rising magnetic fields \cite{Miller2000}. It is
reassuring that irrespective of initial fields geometry, equation of
state, boundary conditions, vertical stratification, numerical
methods, etc., MHD turbulence and efficient transport of angular
momentum always ensues. But the question of the exact saturation
level and its dependence on physical and numerical parameters, such
as net vertical flux, box size, or dissipation mechanisms, remain a
topic of continued research
\cite{Hawley1996,Sano2004,DeVilliers2006,Blackman2006,Pessah2006}.

Local shearing boxes have been used extensively to understand MRI
turbulence in presence of other physical effects, e.g., resistivity
\cite{Fleming2000}, ambipolar diffusion \cite{Hawley1998}, Hall
effect \cite{Sano2002}, radiation and the photon bubble instability
\cite{Turner2003,Turner2005}, and the thermal instability
\cite{Piontek2005}.
\begin{figure}
\begin{center}
\includegraphics[width=5in,height=4in]{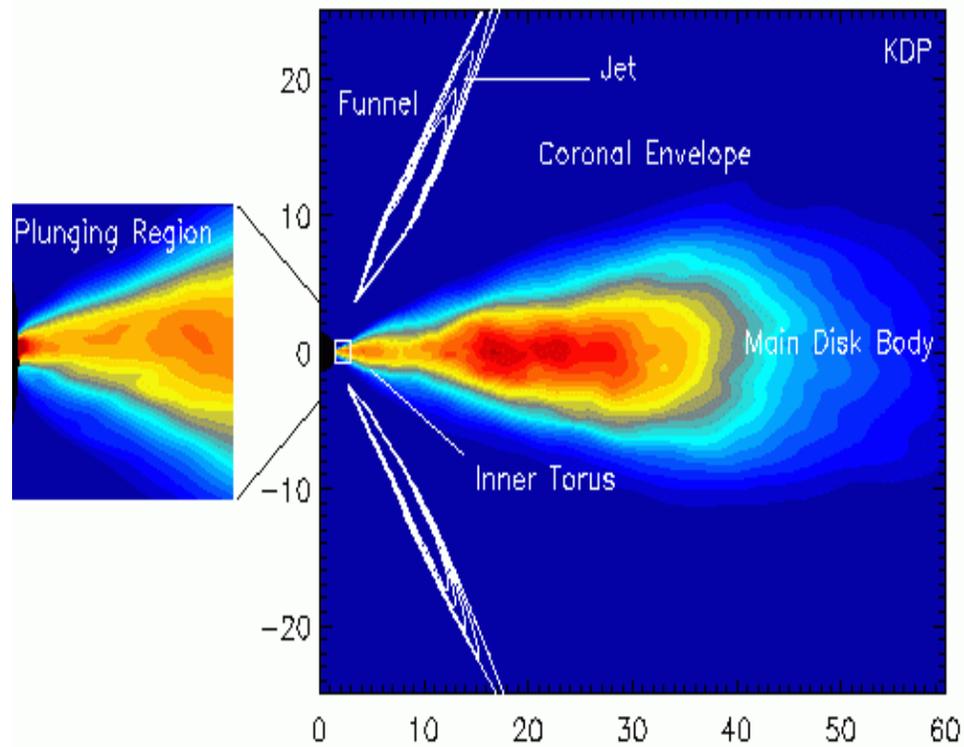}
\caption[Inner regions of a black hole accretion disk from a
GRMHD simulation]{The inner regions of an accretion disk around a
black hole calculated in a GRMHD simulation (Figure 3 of
\cite{DeVilliers2003}). The black hole is at the origin with an
event horizon of radius unity. The accretion disk rotates around the
vertical direction. Color contours show the density distribution,
with red representing highest density and dark blue the lowest.
There is a hot magnetized corona above the disk, and between the
corona and the rotation axis there is an ejection of mildly
relativistic plasma. This example shows a non-radiating, thick
disk.\label{Ch1fig:GlobalDisk}}
\end{center}
\end{figure}

The effect of MRI turbulence was first observed in global 2-D MHD
simulations of Shibata and Uchida with a net vertical flux
\cite{Shibata1986}, but the reason for the disruption of the flow
was not understood. Starting from the 2-D simulations
\cite{Shibata1986,Stone1994}, tremendous progress in computer
hardware and algorithms has made it possible to simulate realistic
disks around rotating Kerr black holes with general relativistic MHD
(GRMHD) in 3-D \cite{DeVilliers2003,Krolik2006a}. Figure
\ref{Ch1fig:GlobalDisk} shows the structure of a disk from a GRMHD
simulation \cite{DeVilliers2003}. In addition to the efficient
angular momentum transport in disks due to the MRI, global
simulations allows one to study angular momentum extraction by
global mechanisms such as magnetic braking and winds
\cite{Blandford1982}, and extraction of black hole spin energy in
form of jets
\cite{Blandford1977,McKinney2004,Komissarov2005,Krolik2006a}. Global
simulations have also been used to understand the structure of thick
disks in radiatively inefficient accretion flows (RIAFs, see Fig.
\ref{Ch1fig:GlobalDisk}), the subject of the next section
\cite{Stone1999,Stone2001,Hawley2002,Proga2003b}.
\section{Radiatively inefficient accretion flows}
\label{Ch1sec:RIAFs}
\begin{figure}
\begin{center}
\includegraphics[width=4in,height=4in]{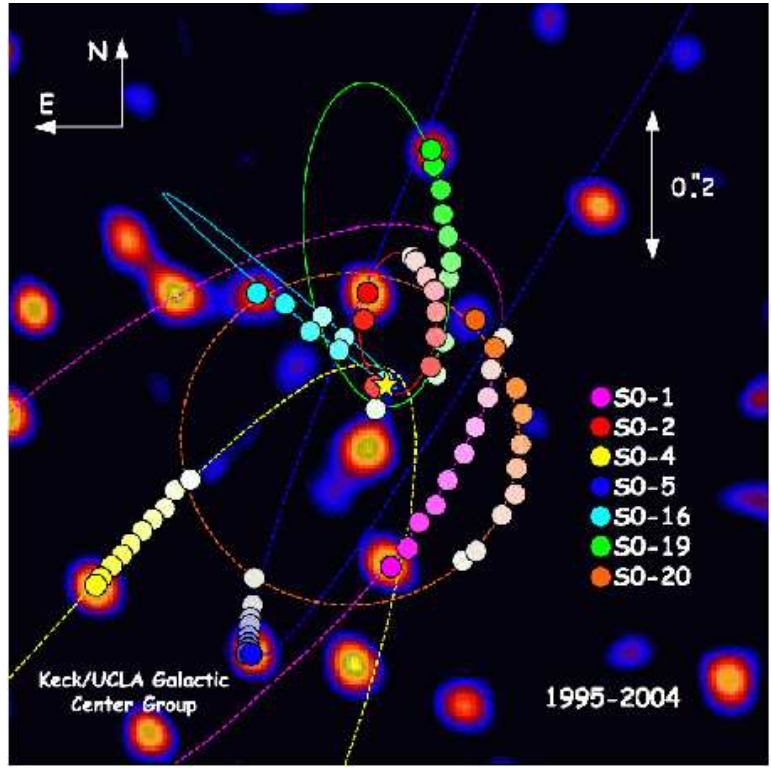}
\caption[Infrared observations of stellar orbits around the Galactic
center black hole]{Keck observations of stellar orbits in the
central $1\times 1$ arcsecond ($0.13$ light years) of our Galaxy are
shown. Stars show significant motion over a period of 9 years.
Changing stellar locations with time, and best fitting Keplerian
orbits are indicated. The orbital parameters confirm the presence of
a $ 4.1 \pm 0.6 \times 10^6 M_\odot$ black hole in the center of our
Galaxy. Source:
http://www.astro.ucla.edu/\~{ }ghezgroup/gc/pictures/orbitsOverImage04.shtml
\label{Ch1fig:CentralStars}}
\end{center}
\end{figure}
This section borrows heavily from an unpublished document on the
motivation for studying radiatively inefficient accretion flow
(RIAF) regimes, by E. Quataert. There is growing observational
evidence for the presence of supermassive black holes (SMBHs) in
galactic nuclei. High resolution imaging of the stellar orbits
around a dark object in the Galactic center, using adaptive optics,
provides a compelling evidence for a $ 4.1 \pm 0.6 \times 10^6 M_\odot
M_\odot$ SMBH~\cite{Schodel2002,Ghez2003} (see Figure
\ref{Ch1fig:CentralStars}).
Very large baseline interferometry
(VLBI) observations of water masers in NGC 4258 show gas in a
Keplerian orbit about a SMBH~\cite{Miyoshi1995}. More generally,
stellar motions and radiation from hot gas in the central regions of
nearby galaxies have shown that SMBHs are present in nearly every
galaxy with a bulge component \cite{Magorrian1998,Gebhardt2000,Ferrarese2000}.
\footnote{The bulge component
of a galaxy is the central roughly spherical region with old stars.}
\begin{figure}
\begin{center}
\includegraphics[width=3in,height=3in]{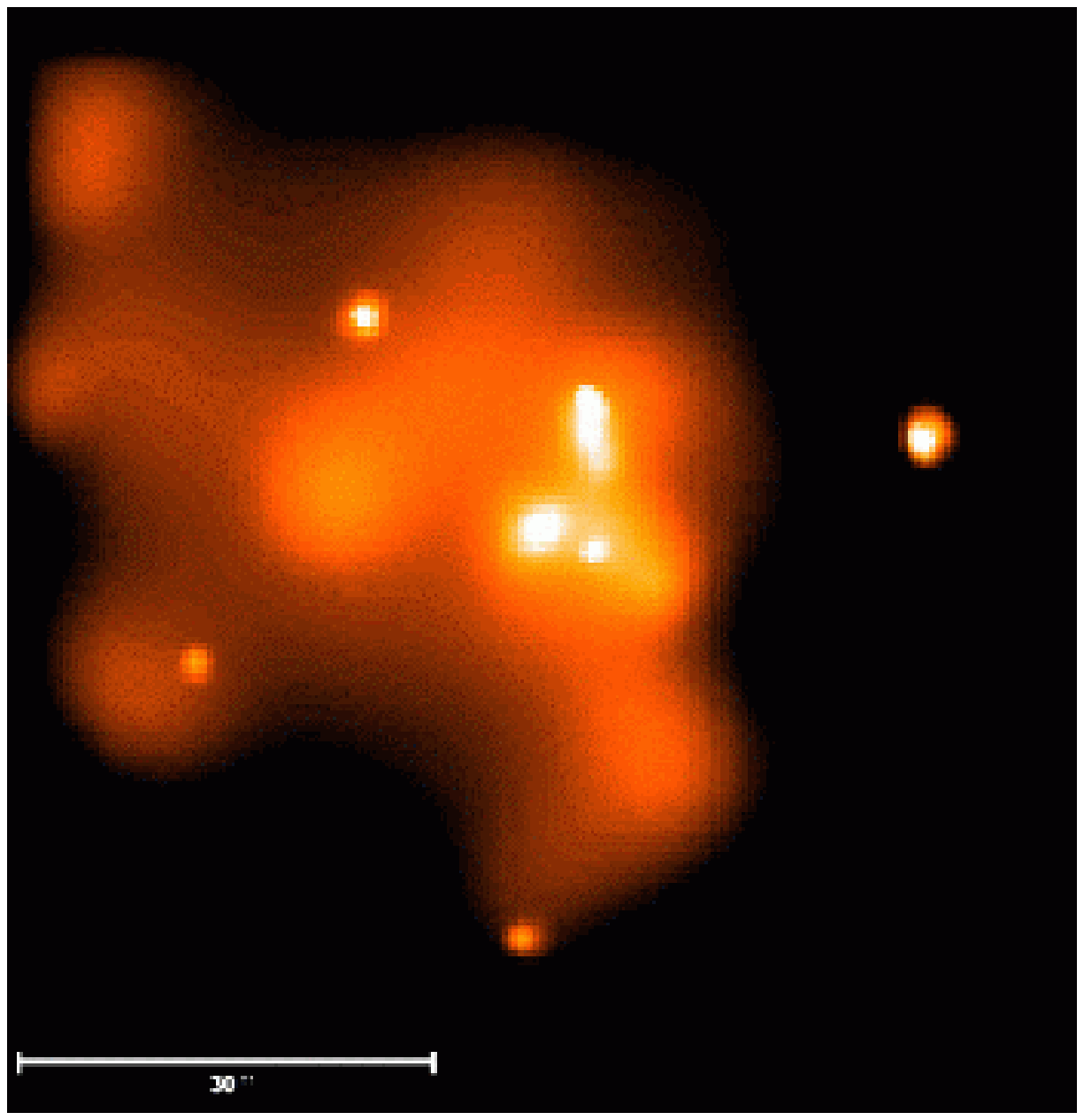}
\caption[X-ray image of the Galactic center]{Chandra X-ray image of
the innermost 10 light years ($\approx 100$ times larger than Figure
\ref{Ch1fig:CentralStars}) at the center of our Galaxy. The image
shows an extended cloud of hot gas surrounding the supermassive
black-hole Sagittarius A* (larger white dot at the very center of
the image---a little to the left and above the smallest white dot).
This gas glows in X-rays as it has been heated to a temperature of
millions of degrees by shock waves produced by winds from young
massive stars (and perhaps by supernova explosions). Source:
http://chandra.harvard.edu/photo/2000/0204/index.html
\label{Ch1fig:GCChandra}}
\end{center}
\end{figure}

One of the puzzles about many SMBHs is their extreme low luminosity,
despite their gas rich environments. In contrast, the Active
Galactic nuclei (e.g., quasars), which are also powered by accretion
onto SMBHs, are luminous enough to outshine the rest of the galaxy.
Our Galactic center (GC) is the canonical example of low luminosity
accretion (see Figure \ref{Ch1fig:GCChandra}) \cite{Narayan2003}.
The winds from massive stars in the central $\sim 0.1$ pc of the
Galactic center feed the black hole at an estimated Bondi accretion
rate of $\dot{M}_{\rm Bondi} \approx 10^{-5} M_\odot$
/yr (see Appendix \ref{app:Bondi} for Bondi accretion model)
\cite{Baganoff2003,Melia1992,Quataert2004}.\footnote{The hot wind from 
the X-ray source IRS 13E1 alone supplies $\approx 10^{-3} M_\odot$ 
yr$^{-1}$ \cite{Najarro1997}; however, much of the hot gas in the GC 
is gravitationally unbound,
leaving only a small fraction to be accreted by the central black
hole.} If this gas were to
accrete onto the black hole with $\approx 10 \%$ efficiency (typical
of the Active Galactic Nuclei), the luminosity would be $\approx
10^{41}$ erg s$^{-1}$, five orders of magnitude larger than the
observed luminosity (see Table \ref{Ch1tab:RIAFs})
\cite{Quataert2003}.
\begin{table}[hbt]
\begin{center}
\caption{Dim SMBHs in the Galactic center and nearby galaxies
\label{Ch1tab:RIAFs}} \vskip0.05cm
\begin{tabular}{|c|c|c|c|c|c|c|c|c|} \hline
Galaxy & $M_{\rm SMBH}$ & $\dot M_{\rm Bondi}$ & ${L_{\rm
Bondi}}^{a}$ &
${L_{\rm X}}^{b}$ & $L_X/L_{\rm Bondi}$ \\
& $10^8$ M$_{\odot}$ & M$_{\odot}$ yr$^{-1}$ & erg s$^{-1}$ & erg
s$^{-1}$ & \\ \hline Milky Way$^{c}$ & $0.03$ & $10^{-5}$ & $6
\times 10^{40}$ & $2 \times 10^{33}-10^{35}$ & $3 \times
10^{-8}-10^{-6}$ \\ \hline NGC 1399$^{d}$ & 10.6 & $4 \times
10^{-2}$ & $2 \times 10^{44}$ & $\lesssim 10^{39}$ & $\lesssim 5
\times 10^{-6}$\\  \hline NGC 4472$^{d}$ & 5.6 & $8 \times 10^{-3}$
& $5
\times 10^{43}$ & $\lesssim 10^{39}$ & $\lesssim 2 \times 10^{-5}$ \\
\hline NGC 6166$^{e}$ & $10$ & $3 \times 10^{-2}$ & $2 \times
10^{44}$ & $10^{40}$ & $5 \times 10^{-5}$ \\ \hline NGC 4636$^{d}$ &
0.8 & $8 \times 10^{-5}$ & $5 \times 10^{41}$ & $\lesssim 3  \times
10^{38}$ & $\lesssim 6 \times 10^{-4}$\\ \hline
\end{tabular}
\end{center}
$^a$ $0.1 \dot M_{\rm Bondi} c^2$ \ \ $^b$ $2-10$ keV luminosity or
an upper limit \ \ $^c$ \cite{Baganoff2001,Baganoff2003} \ \ $^d$
\cite{Lowenstein2001} \ \ $^e$ \cite{DiMatteo2001} \\
Sgr A$^*$ shows $\approx 100$ times larger X-ray luminosity in the
flaring state as compared to the quiescent state. The total RIAF
luminosity ($L_{\rm Tot}$) is dominated by the radio emission, which
is 2-3 orders of magnitude larger than the quiescent X-ray output in
case of Sgr A$^*$, so that $L_{\rm Tot}/L_{\rm Bondi} \sim 10^{-5}$
is still surprisingly small \cite{Quataert2003}.
\end{table}

The Chandra X-ray Observatory, with its excellent spatial resolution
(0.5 arcseconds), has put stringent constraints on the nuclear
emission in a large number of nearby galaxies
\cite{Ho2001,Lowenstein2001,DiMatteo2001}. Table \ref{Ch1tab:RIAFs}
gives some examples.  In addition to the observed X-ray luminosity
$L_X$ and the mass of the SMBH, the table lists the observationally
inferred Bondi accretion rate and ``Bondi luminosity.''  Bondi rate
is the accretion rate calculated from the density and temperature in
the vicinity of the black hole (measured on $\sim 1''$ scales, which
is $\sim 10^5-10^6$ Schwarzschild radii for the systems in Table
\ref{Ch1tab:RIAFs}), and assuming spherical hydrodynamic accretion
(see Appendix \ref{app:Bondi}). Bondi luminosity is the luminosity
if the ambient gas accretes onto the SMBH at the Bondi rate and
emits with $\approx 10 \%$ efficiency. For all cases in Table
\ref{Ch1tab:RIAFs}, $L_X$ is much less than the Bondi luminosity
(which is orders of magnitude smaller than the Eddington limit for
these systems). Thus, the observed luminosities are orders of
magnitude smaller than simple theoretical predictions. Moreover,
these discrepancies are not unique to X-ray observations, but are
present in high resolution observations from the radio to the
gamma-rays \cite{Ho1999}.
\subsection{RIAF models}
With compelling evidence  for low luminosity SMBHs in the Galactic
center and nearby galaxies, one needs to account for their extreme
dimness. The explanation for their low luminosity must lie in how
the surrounding gas accretes onto the central black hole. The
standard accretion disk model is that of a geometrically thin,
optically thick disk \cite{Shakura1973}, applied extensively to
luminous accreting sources in X-ray binaries and AGN
\cite{Koratkar1999,Fabian2000}. Low luminosity disks are
fundamentally different; radiatively inefficient disks retain most
of the accretion energy as thermal motion and puff up to become
thick. Also, RIAFs show no significant black body component in their
spectra in infrared-UV \cite{Lasota1996,Ho1999,Quataert1999}; this
emission is seen in luminous sources such as Seyferts and quasars
\cite{Koratkar1999}. Most low luminosity disk models have appealed
to modes other than thin disks. Accretion disks where very little of
the gravitational potential energy of the accreting gas is radiated
away is referred to as radiatively inefficient accretion flows
(RIAFs).

The plasma in RIAFs is hot and dilute because the gravitational energy
released from accretion is stored as thermal energy. Because of the
low densities and high temperatures, Coulomb collisions are
inefficient at exchanging energy between the electrons and protons
(see Table \ref{Ch1tab:SgrA}). If protons and electrons are heated
to their respective virial temperatures without exchanging energy,
then protons will be hotter than electrons by their mass ratio
$m_p/m_e \approx 2000$. But the temperatures depend on how energy
released from accretion is dissipated into electrons and ions, which
remains poorly understood. Most RIAF models assume that protons
($\sim 10^{12}$ K) are much hotter than the electrons ($\sim
10^{10}-10^{12}$ K) \cite{Quataert2003}. The electron temperature is
not well constrained but crucial as it determines the radiation that
we see. The hot RIAFs are thus very different from the thin
accretion disks, which are much cooler ($\sim 10^5-10^6$ K) and
denser. In addition, because of the different physical conditions in
the accretion flow, thin disk and RIAF models predict very different
multiwavelength spectra (e.g., RIAFs are optically thin and do not
produce blackbody emission).

Two ways to make a disk radiatively inefficient are:  1) energy
released from accretion at Bondi rate is channeled preferentially
into poorly radiating ions, which are eventually swallowed (with
their energy) by the hole; and 2) instead of accreting all the
available gas supply, processes like winds and outflows, and
convection can constrict the net accretion ($\dot{M}\ll M_{\rm
Bondi}$) onto the black hole.

The original RIAF models  by Ichimaru (1977) and Rees et. al. (1982;
the ``ion torus'' model) \cite{Ichimaru1977,Rees1982} were based on
the first approach. These models were revived in the 1990s, and
extensively applied to observed systems, under the name
advection-dominated accretion flows (ADAFs), by Narayan, Abramowicz,
and others \cite{Narayan1994,Narayan1995,Abramowicz1995}. In ADAF
models, the gas accretes at about the Bondi rate, but the radiative
efficiency is $\ll 10 \%$, providing a possible explanation for the
very low luminosity of most galactic nuclei
\cite{Rees1982,Fabian1995}. The radiative efficiency is very low
because it is assumed that the electrons, which produce the
radiation we see, are much colder than the ions which are advected
(with their thermal energy) on to the hole. Thus, instead of energy
release in the form of radiation like in the cool, thin disks,
energy is lost forever to the black hole in ADAF models.

The past few years have seen new steps in the theoretical
understanding of RIAFs.  In particular, hydrodynamic and MHD
numerical simulations of RIAFs have been performed
\cite{Stone1999,Igumenshchev1999,
Stone2001,Hawley2001,Hawley2002,Igumenshchev2003,Pen2003,
Proga2003a,Proga2003b}. The hydrodynamic simulations based on
$\alpha$ model for stress (e.g.,
\cite{Stone1999,Igumenshchev1999,Proga2003a}) found that convection
can stall accretion, with density varying like $\rho \sim r^{-1/2}$
with radius, as compared to a steeper $\sim r^{-3/2}$ dependence
in ADAFs and Bondi accretion (see \cite{Narayan1994} and Appendix
\ref{app:Bondi}). These simulations motivated analytical
self-similar models known as convectively dominated accretion flows
(CDAFs). The reason for a less steep dependence of density on radius
is that the mass accretion rate in CDAFs decreases as we move in towards
the black hole, $\dot{M}/\dot{M}_{\rm ADAF} \sim (r/r_{acc})$. The low
luminosity
in CDAFs is not because of low efficiency of accretion ($\eta \sim
0.1$), but because of the reduction of mass accretion due to
convection. In global MHD simulations strong magnetic fields ($\beta
\lesssim 10$) are generated by MHD turbulence driven by the MRI, and
convection is unimportant
\cite{Stone2001,Hawley2001,Hawley2002,Igumenshchev2003,Pen2003,Proga2003b}.
Numerical simulations by different groups (using different codes and
boundary conditions) lead to the same conclusion---magnetically
driven outflows prevent most of the mass supplied at outer regions
to accrete. Outflows are natural outcome of hot RIAFs and have been
incorporated in theoretical models to account for low accretion
rates \cite{Narayan1994,Blandford1999,Blandford2004}; this adiabatic
inflow-outflow solution (ADIOS) model also predicts a smaller
accretion rate in the inner regions, ($\dot{M}/\dot{M}_{\rm ADAF} \sim (r/r_{acc})^p$,
with $0 \leq p \leq 1$), and a gentle dependence of density on
radius ($\rho \sim r^{-3/2+p})$ compared to an ADAF.

The ADIOS/CDAF models look very different from ADAF models; very
little of the mass supplied at large radii actually accretes into
the black holes. The accretion rate can be smaller than the Bondi
estimate (e.g., Table \ref{Ch1tab:RIAFs}) by a factor of $\sim
R_{acc}/R_{S} \sim 10^5$, where $R_{S}$ and $R_{acc}$ are the inner
($\sim 2GM_*/c^2$, the Schwarzschild radius) and the outer ($\sim
r_{acc}=2GM_*/a^2$, the Bondi accretion radius) radii of the
accretion flow. This very low accretion rate may explain the low
luminosity of most galactic nuclei.
\subsection{The Galactic center}
\label{Ch1subsec:SgrA} Following Baganoff et al.
\cite{Baganoff2003}, we apply the models discussed in the previous
subsection to Sgr A$^*$, the RIAF in the Galactic center (GC). The
high resolution Chandra X-ray observations have enabled the
detection of X-rays in the vicinity of Sgr A$^*$, unpolluted by the
emission from other X-ray sources in the region \cite{Baganoff2003}.
The X-rays arise because of thermal bremsstrahlung at larger radii,
and synchrotron and Compton processes near the SMBH (these processes
need very hot electrons). By assuming a thermal bremsstrahlung model
for X-ray observations at 10$^{\prime\prime}$, the ambient
temperature is estimated to be $ T(\infty) \approx 1.3$ keV and the
plasma number density to be $n (\infty) \approx 26$ cm$^{-3}$.
Quataert \cite{Quataert2004} has argued that the $10^{\prime\prime}$
observation probes the gas being driven out of the central star
cluster, while the $1.^{\prime \prime}5$ observation probes the gas
which is gravitationally captured by the black hole; we use
$1.^{\prime\prime}5$ observations ($n\approx130$ cm$^{-3}$ and
$T\approx 2$ keV) to estimate the accretion rate and to make Table
\ref{Ch1tab:SgrA}.

We will use the ambient conditions and different RIAF models to
estimate physical conditions in accretion flow of Sgr A$^*$. The
Bondi capture radius is given by $r_{acc}=2GM/a^2 \approx 0.072$ pc
($1.^{\prime\prime}8$), where $a$ is the sound speed (see Appendix
\ref{app:Bondi}). The Bondi accretion rate is given by (see Eq.
\ref{Ch1eq:MdotNum}) \be \label{Ch1eq:BondiSgr} \dot{M}_{\rm Bondi}
\approx 3 \times 10^{-6} \left ( \frac{n}{130~{\rm cm}^{-3}}\right )
\left ( \frac{kT}{2~{\rm keV}} \right)^{-3/2} M_\odot~{\rm
yr}^{-1}.\ee This is an order of magnitude smaller than what is
estimated from the amount of gas available from stellar winds (see
Table \ref{Ch1tab:RIAFs}). The ADAF model gives $\dot{M}_{\rm ADAF}
\sim \alpha \dot{M}_{\rm Bondi}$, where $\alpha$ is the
Shakura-Sunyaev viscosity parameter \cite{Narayan1995}. The mass
accretion rate as a function of radius for ADIOS/CDAF models is
$\dot{M}_{\rm ADIOS/CDAF} \sim \alpha \dot{M}_{\rm
Bondi}(R_S/r_{acc})^p$. Using $p=1$ corresponding to a CDAF (or a
CDAF-like ADIOS), the accretion rate is \be \label{Ch1eq:CDAF}
\dot{M}_{\rm CDAF/ADAF} \approx 1.2 \times 10^{-12} \left (
\frac{\alpha}{0.1} \right ) \left ( \frac{n}{130~{\rm
cm}^{-3}}\right ) \left ( \frac{kT}{2~{\rm keV}} \right)^{-3/2}
M_\odot~{\rm yr}^{-1}, \ee much smaller than the Bondi estimate.
Consistent with the CDAF/ADIOS models, the detection of linear
polarization of radio emission from Sgr A$^*$ (see
\cite{Aitken2000,Bower2003}) implies a small Faraday rotation
(indicating a small density and magnetic field) and places a
stringent upper limit on $\dot{M} \lesssim 10^{-8} M_\odot$
yr$^{-1}$ \cite{Agol2000,Quataert2000b}.

\begin{table}[hbt]
\begin{center}
\caption{Plasma parameters for Sgr A$^*$ \label{Ch1tab:SgrA}}
\vskip0.05cm
\begin{tabular}{|c|c|c|c|} \hline
Parameter & $r=r_{acc}$ & $r=\sqrt{r_{acc}R_S}$ & $r=R_S$ \\
& $2.2 \times 10^{17}$ cm & $4.2 \times 10^{14}$ cm & $7.8 \times
10^{11}$ cm \\
\hline $\Omega_K = \sqrt{GM_*/r^3}$ (s$^{-1}$) & $1.84 \times
10^{-10}$ & $2.2 \times 10^{-6}$ & $0.028$ \\
\hline $T \sim r^{-1}$ keV & 2 & 1048 & $5.7 \times 10^5$ \\
\hline $n_{\rm ADAF} \sim r^{-3/2}$ (cm$^{-3}$) & 130 & $1.56 \times 10^6$ & $1.95 \times 10^{10}$ \\
\hline $n_{\rm CDAF} \sim r^{-3/2+p}$ (cm$^{-3}$) & 130 & 3000 & $7 \times 10^4$ \\
\hline $B^a_{\rm ADAF} \sim r^{-5/4}$ (G) & 0.0012 & 2.93 & $7.6\times 10^3$ \\
\hline $B^a_{\rm CDAF} \sim r^{-5/4+p/2}$ (G) & 0.0012 & 0.13 & $14.4$ \\
\hline $\nu_{i,\rm ADAF}/\Omega_K \sim r^{3/2}$ & 11.4 & $9.4 \times 10^{-4}$ & $7.6 \times 10^{-8}$ \\
\hline $\nu_{i,\rm CDAF}/\Omega_K \sim r^{3/2+p}$ & 11.4 & $1.81 \times 10^{-6}$ & $2.62 \times 10^{-13}$ \\
\hline $\rho_{i,\rm ADAF}/H \sim r^{-1/4}$ & $2\times 10^{-11}$ & $9.94 \times 10^{-11}$ & $4.59 \times 10^{-10}$ \\
\hline $\rho_{i,\rm CDAF}/H \sim r^{-1/4-p/2}$ & $2\times 10^{-11}$ & $2.23 \times 10^{-9}$ & $2.48 \times 10^{-7}$ \\
\hline
\end{tabular}
\end{center}
$^a$ equipartition field, \ \ $H \approx 0.87 r$, \ \ $p=0$ for
ADAF, $p=1$ for ADIOS/CDAF, \\
$\nu_i$ is the ion collision frequency, \ \ $\rho_i$ the ion
gyroradius \\
$\nu_e=\nu_i(m_i/m_e)^{1/2}(T_e/T_i)^{-3/2}$, Coulomb logarithm
($\ln \Lambda$) chosen to be 30, \\
$\rho_e=\rho_i(T_e/T_i)^{1/2}(m_i/m_e)^{1/2}$
\end{table}

Table \ref{Ch1tab:SgrA} shows different physical variables; the
number density $n$, temperature $T$, equipartition magnetic field
$B$, etc. at three radial locations ($r_{acc}$, $\sqrt{r_{acc}
R_S}$, and $R_S$) using an ADAF (equivalent to the Bondi model for
$\alpha \sim 1$) and CDAF/ADIOS model with $p=1$. At radii smaller than
$r_{acc}$, the
mean free path is much larger than the disk height scale $H =
c_s/\Omega \sim r$; the Larmor radius is many orders of magnitudes smaller
than the disk height. This motivates us to investigate the role of
plasma kinetic effects in the physics of RIAFs, as we discuss in the
next section.

\section{Motivation}
As discussed in the previous section, there is ample evidence that
RIAFs are collisionless, with the Coulomb collision time much longer
than accretion time (see Table \ref{Ch1tab:SgrA}). However, most
studies of the MRI have used
ideal MHD equations, which are formally valid only for collisional,
short mean free path plasmas. A collisionless analysis should use
the Vlasov equation \cite{Krall1986} which describes the time
evolution of the distribution function of a collisionless plasma in
a 6-D phase space. In cases when the scales of interest are much
larger than the ion Larmor radius (e.g., in RIAFs ion Larmor radius
is $\sim 10^8$ times smaller than the disk height scale, the
scale of largest eddies in MRI turbulence), one can average over the
fast gyromotion to obtain the drift kinetic equation (DKE)
describing the distribution function in a 5-D phase space
\cite{Kulsrud1983,Snyder1997}. Collisionless plasmas are different
from the collisional MHD plasmas, in that the pressure is
anisotropic with respect to the magnetic field, and rapid thermal
conduction can occur along the field lines.

Quataert and coworkers \cite{Quataert2002}, used the DKE to study
the collisionless MRI in the linear regime. They found that with an
equal vertical and azimuthal fields, the fastest growing mode is
twice as fast as in MHD and occurs at a much larger length scale.
The aim of the thesis is to follow up their work with numerical
simulations of the MRI in the kinetic regime. A method based on the
DKE that evolves the distribution function in a 5-D phase space is
more expensive than the 3-D MHD simulations. A less expensive
approach (and equivalent to the DKE in the linear regime) is to use
the kinetic MHD (KMHD) equations with Landau fluid closure for
parallel heat flux \cite{Snyder1997}. We started by showing the
equivalence of linear modes in the drift kinetic and KMHD formalisms
(the stable fast, Alfv\'en, slow, and entropy modes, and the
unstable MRI) in a Keplerian disk~\cite{Sharma2003}. This was
followed up by nonlinear KMHD simulations in a local shearing box
\cite{Sharma2006}.

Transition of the MRI from collisionless to collisional regime was
studied linearly, using a BGK collision operator \cite{Sharma2003}.
Transition from kinetic to the Braginskii regime occurs as the mean
free path becomes short compared to the parallel wavelength,
$\lambda_{\rm mfp} \ll \lambda_\parallel \equiv 2\pi/k_\parallel$;
for the fastest growing mode this corresponds to $\nu \lesssim
\Omega \sqrt{\beta}$, where $\nu$ is the collision frequency. As
collision frequency is increased further ($\nu \gtrsim \Omega
\beta$), anisotropic stress becomes negligible compared to the
Maxwell stress, and transition to MHD occurs. Differences between the
kinetic and MHD regimes is striking at large $\beta$'s. A crucial
difference from MHD is the presence of damped modes, indicating a
possibility of wave-particle interactions in form of Landau and
Barnes damping \cite{Landau1946,Barnes1966,Stix1992}. Balbus and
Islam \cite{Balbus2004,Islam2005} have studied MRI in the weakly
collisional Braginskii regime and found agreement with our results.

The ZEUS MHD code \cite{Stone1992a,Stone1992b} is modified to
include anisotropic pressure, and parallel thermal conduction based
on Landau fluid closure \cite{Snyder1997}. Nonlinear KMHD
simulations are done in a local shearing box limit
\cite{Hawley1995}. The adiabatic invariant, $\mu=p_\perp/B$, is
conserved in collisionless plasmas, as a result, pressure anisotropy
($p_\perp>p_\parallel$) is created as magnetic field is amplified by
the MRI. Pressure anisotropy cannot become large
($p_\perp/p_\parallel - 1 \lesssim \mbox{few}/\beta_\perp$), as
mirror and ion-cyclotron instabilities will isotropize the pressure
by pitch angle scattering. Subgrid models of pitch angle scattering
by these instabilities at the Larmor radius scale have been
included. Pressure anisotropy gives rise to a stress in addition to
the usual Maxwell and Reynolds stress in MHD \cite{Sharma2006}.
Pressure anisotropy driven instabilities are expected to arise in
any collisionless plasma, when the field strength changes in a
$\beta \gtrsim 1$ plasma, e.g., the solar wind
\cite{Marsch1982,Kasper2002}, magnetosphere
\cite{Tsurutani1982,Gary1993}, and galaxy clusters
\cite{Schekochihin2005}.

The next step is to include collisionless effects in global MHD
simulations \cite{Stone2001}. Anisotropic thermal conduction is
expected to change the convective stability criterion from an
outward increasing entropy to an outward increasing temperature
\cite{Balbus2000,Balbus2001,Parrish2005}. Self-similar solutions
using saturated (isotropic) conduction have shown significant
differences from standard non-conducting ADAF models
\cite{Menou2005,Tanaka2006}. While implementing anisotropic thermal
conduction in global MHD simulations we discovered that the centered
differencing of anisotropic thermal conduction can give rise to heat
flowing from lower to higher temperatures, causing the temperature
to become negative in regions with large temperature gradients,
e.g., disk corona interface. We have developed a new numerical
method that uses slope limiters to ensure that the temperature
extrema are not amplified by anisotropic conduction
\cite{Sharma2007}. Global numerical simulations with anisotropic
thermal conduction can tell us about the global structure of RIAFs.
These, combined with the insights on local energy dissipation in
disks from the local KMHD simulations, can shed light on their low
luminosity.
\section{Overview}
The main body of the thesis (chapters \ref{chap:chap3},
\ref{chap:chap4}, and \ref{chap:chap5}) is based on three papers,
the first on the transition of the MRI from collisionless to the
collisional regime \cite{Sharma2003}, the second on the shearing box
simulations of the collisionless MRI \cite{Sharma2006}, and the
third on numerical implementation of anisotropic conduction in
presence of large temperature gradients \cite{Sharma2007}.

Chapter \ref{chap:chap2} introduces the kinetic MHD formalism. We
begin with the Vlasov description of a collisionless plasma, and
derive the drift kinetic equation (DKE) in the limit of length
scales much larger than the Larmor radius, and frequencies much
smaller than the gyrofrequency. Moments of the DKE with an
anisotropic pressure tensor are called the kinetic MHD (KMHD)
equations. The KMHD equations are closed by the `3+1' Landau fluid
closure for heat flux along the field lines. Landau closure is
equivalent to a Pad\'e approximation to the drift kinetic linear
response function. We discuss different ways to implement the
nonlocal closure in a numerical simulation. We show that the moment
equations with a BGK collision operator recover the Braginskii
equations in the high collisionality regime.

Chapter \ref{chap:chap3}, which is based on \cite{Sharma2003},
describes the transition of the MRI from collisionless to the
collisional regime. Linear modes of a magnetized, collisionless
plasma in a Keplerian rotation are derived using both the DKE and
the KMHD equations with Landau closure for the heat flux. The two
methods agree very well for the real and imaginary parts for the
frequency response; this motivates kinetic MHD simulations of the
collisionless MRI. The presence of damped modes in the collisionless
regime can cause waves to be damped by Landau/Barnes damping at
large scales, instead of being damped only at small scales as in MHD
with small resistivity and viscosity. A BGK collision operator is
used to study the transition of the MRI from collisionless to the
collisional (MHD) regime; the transition from kinetic to the
Braginskii regime occurs when the mean free path becomes small
compared to the wavelength, for the fastest growing mode this
corresponds to $\nu \gtrsim \Omega \sqrt{\beta}$.

Chapter \ref{chap:chap4} presents results from the nonlinear
shearing box simulations of the collisionless MRI. Pressure
anisotropy ($p_\perp>p_\parallel$) is created because of adiabatic
invariance ($\mu=p_\perp/B$), as magnetic field is amplified by the
MRI. The effect of $p_\perp>p_\parallel$ is to make the field lines
stiffer. If the pressure anisotropy is allowed to become arbitrarily
large, the stiff field lines (because of $p_\perp>p_\parallel$) can
result in the stabilization of all the MRI modes into small
amplitude anisotropic Alfv\'en waves. However, at large pressure
anisotropies ($p_\perp/p_\parallel -1 > ({\rm a~few})/\beta$),
mirror and ion-cyclotron instabilities are expected to arise.
Although the mirror instability is  present in the kinetic MHD
approximation, the resolution (and hence the growth rate) is not
enough to keep the pressure anisotropy within the marginal
anisotropy. The ion-cyclotron instability is ordered out of the
drift kinetic ordering. Therefore, subgrid models for pitch angle
scattering due to these instabilities are included. Pitch angle
scattering due to microinstabilities imposes an MHD-like dynamics on
collisionless plasmas, this is the reason MHD provides a good
approximation for many collisionless plasmas in astrophysics, e.g.,
the solar wind, the magnetosphere, and the interstellar medium.
Modifications to the ZEUS MHD code, to include the kinetic MHD
terms, are described. The key result of the collisionless MRI
simulations is that anisotropic stress, a qualitatively new
mechanism to transport angular momentum, is as important as the
dominant Maxwell stress in MHD. This can also affect the energetics;
in particular the rate at which anisotropic pressure (collisionless
damping is included in it) heats ions and electrons can be
comparable. If electron heating is comparable to ion heating, it will be
difficult to maintain $T_e \ll T_i$ as required by some RIAF models.

Chapter \ref{chap:chap5} is the result of our attempts to carry out
global non-radiative disk simulations. The aim was to include the
effect of anisotropic conduction on global MHD disk simulations
\cite{Stone2001}. The initial condition consists of a constant
angular momentum torus surrounded by a hot, low density corona. We
were running into numerical difficulties with this initial set up;
the temperature was becoming negative at some grid points near the
disk corona interface. This motivated us to investigate the effect
of anisotropic conduction in regions of high temperature gradient.
Chapter \ref{chap:chap5} describes simple tests where centered
differencing of anisotropic thermal conduction results in heat
flowing from lower to higher temperatures, resulting in negative
temperature at large temperature gradients. We introduce a new
numerical method based on limiting the transverse temperature
gradient; this ensures that heat flows from higher to lower
temperatures and the temperature extrema are not amplified. Many
tests and convergence studies are described.

Chapter \ref{chap:chap6} concludes the thesis with an outline of
possible future work. Future work include global disk simulations
with anisotropic thermal conduction, local simulations with more
sophisticated models for non-local anisotropic thermal conduction,
and more accurate drift kinetic simulations evolving the distribution
function in a 5-D phase space.

Appendix \ref{app:app1} describes the efficiency of black hole accretion
based on a simple pseudo-Newtonian potential which captures key general
relativistic effects \cite{Paczynski1980}. Also presented is the
derivation of spherically symmetric, steady accretion.

Appendix \ref{app:app2} shows the derivation of closures for
$p_\parallel$ and $p_\perp$, using the DKE with a BGK collision
operator, in both high and low collisionality limit. These have been
used in Chapter \ref{chap:chap2} to show the equivalence of the
drift kinetic formalism and the kinetic MHD approximation with
Landau closure for heat flux.

Appendix \ref{app:app3} describes the modifications to the ZEUS MHD
code to include the kinetic MHD terms, anisotropic pressure and
anisotropic thermal conduction based on Landau fluid closure. This
also includes some tests of the collisionless aspects of the code,
e.g., damping of a linear fast mode, mirror instability in an
initially anisotropic plasma ($p_\perp>p_\perp$), shear generated
pressure anisotropy and firehose instability driven by the
anisotropy ($p_\parallel > p_\perp$).

Appendix \ref{app:app4} describes the error analysis of a time
series where the sampling time is smaller than the correlation time. For
such a data, all entries are not independent and the standard deviation
is no a correct measure of uncertainty. This method based on \cite{Nevins2005}
is used to put error bars on the time and volume
averaged quantities derived from the shearing box simulations (in
Chapter \ref{chap:chap4}).

\chapter{Description of collisionless plasmas}
\label{chap:chap2}
Macroscopically collisionless plasmas, with collision mean free path
comparable to the system size, are ubiquitous in astrophysics, e.g.,
the solar wind, earth's magnetotail, radiatively inefficient
accretion flows (RIAFs), and X-ray clusters. Fluid theories, such as
hydrodynamics and MHD, are applicable only when the mean free path
is much smaller than the system size, but are routinely used even
when the plasma is collisionless. While fluid theories are sometimes
useful even outside of their rigorous regime of validity,
collisionless plasmas can be quite different from MHD plasmas. For
example, whereas, viscous and resistive dissipation at small scales
are the only ways to dissipate kinetic and magnetic energies into
thermal motion in MHD, collisionless damping at large scales is an
important source of heating in collisionless plasmas. Although
kinetic instabilities may enforce an MHD-like behavior,
collisionless effects can be crucial, especially to understand
energetics and particle acceleration.

In this chapter we discuss several descriptions of collisionless
plasmas valid in different approximations. We start with the Vlasov
equation, the most detailed description of a collisionless plasma,
which describes the time evolution of the distribution function in a
6-D phase space. The drift kinetic equation (DKE) is obtained from
the Vlasov equation, in the limit when length scales are much larger
than the Larmor radius and time scales much longer than the
gyroperiod. Moments of the DKE result in kinetic MHD (KMHD)
equations, where pressure is anisotropic with respect to the
magnetic field direction. Landau fluid closure for heat flux along
the field lines, which recovers the correct kinetic response in the
linear regime, is described.

\section{The Vlasov equation}
A complete statistical description of the species `$s$' in a
collisionless plasma involves a distribution function $F_s$ in a
$6N_s$ dimensional phase space, where $N_s$ is the total number of
particles of species `$s$'. The distribution function, $F_s$,
satisfies the Liouville equation for an N-body system
\cite{Goldstein2002}, \be \label{Ch2eq:Liouville} \frac{DF_s}{Dt}
\equiv \frac{\partial F_s}{\partial t} + \sum_{i=1}^{N_s} {\bf v}_i
\cdot \frac{\partial F_s}{\partial {\bf x}_i} + \sum_{i=1}^{N_s}
{\bf a}_i \cdot \frac{\partial F_s}{\partial {\bf v}_i} = 0, \ee
corresponding to the conservation of probability, where ${\bf x}$,
${\bf v}$, and ${\bf a}$ are position, velocity, and acceleration
respectively, and $D/Dt$ is the Lagrangian derivative in the $6N_s$
dimensional phase space \cite{Krall1986}. Reduced distributions are
obtained by integrating $F_s$ over all but one, two, three, etc.,
particles.

Evolution for the single particle distribution function $f_s({\bf
x},{\bf v},t)$ is obtained by integrating Eq. \ref{Ch2eq:Liouville}
over all but one particle's phase space, \be \label{Ch2eq:Vlasov}
\frac{\partial f_s}{\partial t} + {\bf v \cdot \nabla} f_s + \left [
\frac{q_s}{m_s} ({\bf E} + {\bf v} \times {\bf B}) + \frac{{\bf
F_g}}{m_s} \right ] \cdot \nabla_{\bf v} f_s = 0, \ee where all
terms of order the plasma parameter, $g\equiv 1/n_s\lambda_D^3 \ll
1$ \cite{Montgomery1964}, are neglected. The plasma parameter is the
inverse of the number of particles in a Debye sphere. The Debye
length is the length scale over which plasma establishes
quasineutrality, $\lambda_D=\sqrt{k T_s/4\pi n_s q_s^2}$, where
$n_s$, $T_s$ are number density and temperature \cite{Krall1986}.
For an ideal plasma, with effective shielding, $n_s \lambda_D^3 \gg
1$ or $g \ll 1$. The force of gravity is denoted by ${\bf F_g}$, and
the electromagnetic fields are governed by the Maxwell equations \ba
\label{Ch2eq:DivE} {\bf \nabla \cdot E} &=& 4\pi \sum_s q_s \int f_s
d{\bf v}, \\
\label{Ch2eq:DivB}
{\bf \nabla \cdot B} &=& 0, \\
\label{Ch2eq:CurlE} \frac{\partial {\bf B}}{\partial t} &=& -c {\bf
\nabla} \times {\bf E}, \\
\label{Ch2eq:CurlB} {\bf \nabla} \times {\bf B} &=& 4\pi \sum_s
\frac{q_s}{c} \int f_s {\bf v} d {\bf v} + \frac{1}{c}
\frac{\partial {\bf E}}{\partial t}.
 \ea

The higher order terms (in $g$) that we neglect in the derivation of
Eq. \ref{Ch2eq:Vlasov}--negligible compared to the collective force
due to plasma--arise because of scattering due to microscopic fields
of nearby particles. Eq. \ref{Ch2eq:Vlasov} is the Vlasov equation
(also known as the collisionless Boltzmann equation) that describes
the distribution function $f_s({\bf x},{\bf v},t)$, the probability
of finding a particle of species `$s$' in an interval $d{\bf x}d{\bf
v}$ at $({\bf x},{\bf v})$ in phase space at time $t$. The
Vlasov-Maxwell equations are more complicated than the fluid
equations as they involve seven independent variables $t,{\bf
x},{\bf v}$ rather than four in MHD, $t,{\bf x}$. A collision
operator, that takes into account the microscopic fields due to
individual charges, can be added on the right side of Eq.
\ref{Ch2eq:Vlasov} to obtain the Boltzmann equation. Going from
$6N_s$ to 6 variables in phase space simplifies the description
considerably for an ideal plasma with $g \ll 1$. Further
simplifications can be made as we show in the following sections.

\section{The drift kinetic equation}
\label{Ch2sec:DKE}
The Vlasov equation can be simplified further if the Larmor radius
($\rho_s$) is much smaller than the spatial scales ($\rho_s/L \ll 1$),
and the gyroperiod ($2\pi/\Omega_s$) much smaller than the time
scales ($\Omega_s \gg \omega$). An asymptotic expansion in $\rho_s/L
= (m_s/q_s) (cv/BL) \ll 1$ can reduce the number of variables by
two; the gyration phase is irrelevant, and the perpendicular
velocity is governed by the adiabatic invariant, $\mu = v_\perp^2/2
B$ \cite{Chew1956}.

To the lowest order, all particles drift with an ${\bf E} \times
{\bf B}$ velocity perpendicular to the field lines, and a parallel
motion along the field lines. The parallel electric field is small,
$E_\parallel \sim O(1/q_s) \sim O(\epsilon)$, as charges streaming
along the magnetic field lines will short it out. Lowest order term
in the expansion of $f_s$, $f_{0s}$, is independent of the
gyrophase. We avoid the messy details of the derivation
\cite{Kulsrud1962,Kulsrud1983,Rosenbluth1959,Kruskal1958}, and
simply state the kinetic equation for the zeroth-order distribution
function given by (we follow Kulsrud's derivation
\cite{Kulsrud1962,Kulsrud1983}) \be \label{Ch2eq:DKE} \frac{\partial
f_{0s}}{\partial t} + ({\bf V}_E + v_\parallel {\bf \hat{b}}) \cdot
\nabla f_{0s} + \left ( -{\bf \hat{b}} \cdot \frac{D{\bf V}_E}{Dt} -
\mu {\bf \hat{b}} \cdot \nabla B + \frac{1}{m_s} (q_s E_\parallel +
F_{g\parallel}) \right ) \frac{\partial f_{0s}}{\partial
v_\parallel} = 0, \ee where ${\bf \hat{b}}={\bf B}/B$ is the unit
vector in magnetic field direction, ${\bf V}_E=c({\bf E} \times {\bf
B})/B^2$ is the drift velocity independent of species, and $D/Dt
\equiv
\partial/\partial t + ({\bf V}_E + v_\parallel {\bf \hat{b}}) \cdot
\nabla$ is the comoving derivative in phase space. The Maxwell
equations to the lowest order gives the charge neutrality condition,
\ba
\label{Ch2eq:ChargeNeutrality1} \sum_s \int q_s f_{0s} d {\bf v} &=& 0, \\
\label{Ch2eq:ChargeNeutrality2} \sum_s \int q_s f_{0s} {\bf v} d {\bf v}
&=& 0. \ea

Some remarks on the drift kinetic equation (DKE) are in order. Eq.
\ref{Ch2eq:DKE} can be interpreted as the conservation of
probability in a 5-D phase space (${\bf x}, \mu, v_\parallel$) with
characteristics, $d{\bf x}/dt = {\bf V}_E + v_\parallel {\bf
\hat{b}}$, and $dv_\parallel/dt = -{\bf \hat{b}} \cdot \frac{D{\bf
V}_E}{Dt} - \mu {\bf \hat{b}} \cdot \nabla B + \frac{1}{m_s} (q_s
E_\parallel + F_{g\parallel})$. Only the ${\bf E} \times {\bf B}$
drift shows up in the perpendicular drift, other drifts--curvature,
$\nabla B$, etc., $\propto 1/q_s$--are higher order in $\epsilon$ in
the drift-kinetic ordering. The force along the field lines consists
of the fluid inertial force ($-{\bf \hat{b}} \cdot \frac{D{\bf
V}_E}{Dt}$), the magnetic mirror force ($-\mu {\bf \hat{b}} \cdot
\nabla B$), the parallel electric force ($q_s E_\parallel$), and the
parallel gravitational force ($F_{g\parallel}$). Although
$E_\parallel \ll E_\perp$ and can be dropped in the Ohm's law, it
needs to be kept in the parallel particle dynamics where it ensures
quasineutrality. The condition that determines $E_\parallel$ will be
given in Eq. \ref{Ch2eq:Epar}. The DKE evolves the distribution
function in a 5-D phase space (${\bf x},\mu,v_\parallel$), with
$\mu$ a parameter. A term with $\partial/\partial \mu$ does not
appear in Eq. \ref{Ch2eq:DKE} as $d\mu/dt=0$ along the
characteristics.

\section{Kinetic MHD equations}
\label{Ch2sec:KMHD}
In the drift-kinetic approximation, particles free stream along the
field lines, but move with the field lines in the perpendicular
direction. This fluid-like behavior in the perpendicular plane
restores the possibility of a fluid description of a $\rho_s/L \ll
1$ plasma.

Moments of the Vlasov equation combined with the Maxwell equations
in the non-relativistic limit (ignoring the displacement current,
$\partial E/\partial t $, in Eq. \ref{Ch2eq:CurlB}) yields the
kinetic MHD equations \cite{Kulsrud1962,Kulsrud1983},  \ba
\label{Ch2eq:KMHD1} \frac{\partial \rho}{\partial t} & + &
\nabla \cdot \left(\rho {\bf V}\right)=0, \\
\label{Ch2eq:KMHD2} \rho \frac{\partial {\bf V}}{\partial t} & + &
\rho\left({\bf V} \cdot \nabla\right) {\bf V}= \frac{\left(\nabla
\times {\bf B}\right) \times {\bf B}}{4\pi}
- \nabla \cdot {\bf P},\\
\label{Ch2eq:KMHD3} \frac{\partial {\bf B}}{\partial t} &=& \nabla
\times \left({\bf V} \times {\bf B}\right), \\
\label{Ch2eq:KMHD4}
 {\bf P}&=& p_{\Perp} {\bf I} + \left(p_{\Par}- p_{\Perp}\right){\bf
\hat{b}\hat{b}}, \\
\label{Ch2eq:KMHD5} p_\perp &=& \sum_s m_s \int f_{0s}
\frac{v_\perp^2}{2} d{\bf v}, \\
\label{Ch2eq:KMHD6} p_\parallel &=& \sum_s m_s \int f_{0s}
(v_\parallel - {\bf V \cdot \hat{b}})^2 d{\bf v}, \ea where ${\bf V}
={\bf V}_E + V_\parallel {\bf \hat{b}}$ is the fluid velocity.
Kinetic MHD, like MHD, is a single fluid description of plasma
obtained by combining the moments of all species. Kinetic MHD
appears similar to MHD, except the pressure is an anisotropic tensor
unlike an isotropic pressure in MHD; the pressure tensor is
determined by moments of the solution of the DKE, unlike the
equation of state in MHD. The asymptotic ordering in $1/q_s$ leads to
the ideal Ohm's law \cite{Krall1986}. In principle, equations for
$p_\parallel$ and $p_\perp$ can be derived from the moments of the
DKE. However, because of the inherent complexity of a phase space
description, fluid approximations for closure are usually employed.

The simplest and the oldest approximation is the double adiabatic
(CGL) approximation, where heat flux is assumed to vanish
\cite{Chew1956}, \ba \label{Ch2eq:CGL1} \frac{d}{dt} \left (
\frac{p_\perp}{\rho B} \right ) &=& 0, \\ \frac{d}{dt} \left (
\frac{p_\parallel B^2}{\rho^3}\right ) &=& 0.  \ea The assumption
that the heat flux vanishes is valid only if the phase speed,
$\omega/k_\parallel$, is much larger than electron and ion thermal
speeds, a cold plasma criterion almost never satisfied for slow and
fast magnetoacoustic waves at high $\beta$.\footnote{This cold
plasma criterion is also not satisfied for Alfv\'en waves, though
the heat flux is zero for linear Alfv\'en waves in a uniform
plasma.} Furthermore, the CGL equations are non-dissipative,
incapable of modeling collisionless damping. The CGL closure is also
known to give an incorrect marginal stability criterion for the
mirror instability, an instability that regulates pressure
anisotropy in collisionless plasmas \cite{Kulsrud1983,Snyder1997}.

\section{Landau fluid closure}
\label{Ch2sec:LFC} In this section we describe a fluid closure that
maintains the simplicity of the CGL model, while including kinetic
effects like Landau damping \cite{Snyder1997}. We also include a
simple BGK collision operator, which conserves number, momentum, and
energy. Fluid closures that incorporated kinetic effects like
collisionless damping were first derived in the electrostatic limit
for nonlinear studies of drift-wave instabilities
\cite{Hammett1990,Hammett1992,Dorland1993}. Snyder et al.
\cite{Snyder1997} extended the closure to electromagnetic
bi-Maxwellian (anisotropic) plasmas; similar closures were obtained
by \cite{Chang1992}. Fluid closures that capture kinetic effects are
somewhat analogous to the flux limited diffusion methods used in
radiation transport \cite{Levermore1981,Mihalas1985}.

\subsection{The moment hierarchy}
\label{Ch2subsec:Moments} Multiplying Eq. \ref{Ch2eq:DKE} by $B$,
and using Eq. \ref{Ch2eq:KMHD3} leads to the kinetic equation in a
conservative form, \be \label{Ch2eq:DKECons}
\frac{\partial}{\partial t} f_s B + {\bf \nabla \cdot} [ f_s
B(v_\parallel {\bf \hat{b}} + {\bf V}_E) ] +
\frac{\partial}{\partial v_\parallel} \left [ f_s B \left ( -{\bf
\hat{b}} \cdot \frac{D{\bf V}_E}{Dt} - \mu {\bf \hat{b} \cdot
\nabla} B + \frac{q_s}{m_s} E_\parallel  \right ) \right ] = B
C(f_s), \ee where subscript `$0$' has been suppressed. The term on
the right is a BGK collision operator \cite{Bhatnagar1954} \be
C(f_j) = - \sum_k \nu_{jk} (f_j - F_{Mjk}), \ee where $\nu_{j,k}$ is
the collision rate of species $j$ with $k$. The collisions cause
$f_j$ to relax to a shifted Maxwellian with effective temperature of
the species $j$ and fluid velocity of the species $k$, \be F_{Mjk} =
\frac{n_j}{(2 \pi T_j/m_j)^{3/2}} \exp \left [ -
\frac{-m_j(v_\parallel - V_{\parallel k})^2}{2T_j} - \frac{m_j \mu
B}{T_j} \right ], \ee where $T_j=(T_{\parallel j}+2T_{\perp
j})/3$.\footnote{A variant of this model is required to handle the
large differences between energy and momentum relaxation rates that
can occur in some cases, but this simpler model is sufficient for
the case at hand, where $V_{\parallel,e}=V_{\parallel,i}$ to lowest
order.}

We define the velocity moments as: \ba \nonumber n_s &=& \int f_s
d{\bf v}, \hspace{0.5in} n_s V_{\parallel s} = \int f_s v_\parallel
d{\bf v}, \\  \nonumber p_{\parallel s} &=& m \int f_s (v_\parallel
- V_\parallel)^2 d{\bf v}, \hspace{0.5in} p_{\perp s} = m \int f_s
\mu B d{\bf v}, \\ \nonumber q_{\parallel s} &=& m \int f_s
(v_\parallel - V_\parallel)^3 d{\bf v}, \hspace{0.5in} q_{\perp s} =
m \int f_s \mu B (v_\parallel-V_\parallel)d{\bf v}, \\ \nonumber
r_{\parallel,\parallel s} &=& m \int f_s (v_\parallel-V_\parallel)^4
d{\bf v}, \hspace{0.5in} r_{\parallel,\perp s} = m\int f_s \mu B
(v_\parallel-V_\parallel)^2 d{\bf v}, \\ \nonumber r_{\perp,\perp s
} &=& m\int f_s \mu^2 B^2 d{\bf v}. \ea Specializing to the case of
an electron-proton plasma, and using the charge neutrality condition
(Eqs. \ref{Ch2eq:ChargeNeutrality1} and
\ref{Ch2eq:ChargeNeutrality2}), $n=n_e=n_i$ and
$V_\parallel=V_{\parallel e}=V_{\parallel i}$. In this limit when
electrons and protons drift at equal velocity, the only role of
collisions is to isotropize the distribution function. Taking
appropriate moments of Eq. \ref{Ch2eq:DKECons}, \ba
\label{Ch2eq:SingleFluid1} \frac{\partial n}{\partial t} &+& {\bf
\nabla \cdot} (n {\bf V}) = 0, \\
\nonumber \label{Ch2eq:SingleFluid2} \frac{\partial
V_\parallel}{\partial t} &+& {\bf V \cdot \nabla} V_\parallel + {\bf
\hat{b} \cdot} \left ( \frac{\partial {\bf V}_E}{\partial t} + {\bf
V \cdot \nabla V}_E \right ) + \frac{{\bf \nabla} \cdot ( {\bf
\hat{b}} p_{\parallel s} )}{n m_s} -
\frac{p_{\perp s} {\bf \nabla \cdot \hat{b}}}{n m_s} \\
&-& \frac{q_s E_\parallel}{m_s} = 0, \\
\label{Ch2eq:SingleFluid3} \nonumber \frac{\partial p_{\Par
s}}{\partial t} &+& \nabla \cdot (p_{\Par s} {\bf V}) + \nabla \cdot
{\bf q_{\Par s}} + 2 p_{\Par s} {\bf \hat{b}} \cdot \nabla {\bf V}
\cdot {\bf
\hat{b}} - 2q_{\Perp s} \nabla \cdot {\bf \hat{b}} \\
&=& -\frac{2}{3} \nu_s (p_{\Par s} - p_{\Perp s}), \\
\nonumber \label{Ch2eq:SingleFluid4} \frac{\partial p_{\Perp
s}}{\partial t}  &+& \nabla \cdot (p_{\Perp s} {\bf V}) + \nabla
\cdot {\bf  q_{\Perp s}} + p_{\Perp s} \nabla \cdot {\bf V} -
p_{\Perp s} {\bf \hat{b}} \cdot \nabla {\bf V} \cdot {\bf \hat{b}}
\\ &+& q_{\Perp s} \nabla \cdot {\bf \hat{b}}
= -\frac{1}{3} \nu_s (p_{\Perp s} - p_{\Par s}), \\ \nonumber
\label{Ch2eq:SingleFluid5} \frac{\partial q_{\parallel s}}{\partial
t} &+& {\bf \nabla \cdot} ({\bf V} q_{\parallel s}) + {\bf \nabla
\cdot ( \hat{b}} r_{\parallel,\parallel s}) + 3 q_{\parallel s} {\bf
\hat{b} \cdot \nabla V \cdot \hat{b}} \\ &-& \frac{3 p_{\parallel
s}}{n m_s} {\bf \hat{b} \cdot \nabla} p_{\parallel s} + 3 \left (
\frac{p_{\perp s} p_{\parallel s}}{n m_s} - \frac{ p_{\parallel
s}^2}{n m_s} - r_{\parallel, \perp s} \right ) {\bf \nabla \cdot
\hat{b}} = -\nu_s q_{\parallel s}, \\ \nonumber
\label{Ch2eq:SingleFluid6} \frac{\partial q_{\perp s}}{\partial t}
&+& {\bf \nabla \cdot} ({\bf V} q_{\perp s}) + {\bf \nabla \cdot (
\hat{b}} r_{\parallel,\perp s}) + q_{\perp s} {\bf \nabla \cdot}
(V_\parallel {\bf \hat{b}}) - \frac{p_{\perp s}}{n m_s} {\bf \hat{b}
\cdot \nabla} p_{\parallel s}
\\ &+& \left ( \frac{p_{\perp s}^2}{n m_s} - \frac{p_{\perp s}
p_{\parallel s}}{n m_s} - r_{\perp, \perp s} + r_{\parallel,\perp s}
\right ) {\bf \nabla \cdot \hat{b}} = -\nu_s q_{\perp s}, \ea where
$\rho = n(m_i+m_e)$, ${\bf V} = {\bf V}_E + V_\parallel {\bf
\hat{b}}$, $\nu_i=\nu_{ii}+\nu_{ie}$ and $\nu_e=\nu_{ee}+\nu_{ei}$,
and ${\bf q_{\parallel s}}={\bf \hat{b}} q_{\parallel s}$ and ${\bf
q_{\perp s}}={\bf \hat{b}} q_{\perp s}$ are thermal fluxes of
$p_{\parallel s}$ and $p_{\perp s}$ along the field lines;
perpendicular heat flux vanishes as $\rho_s/L \ll 1$. The
perpendicular equation of motion is given by the perpendicular
component of Eq. \ref{Ch2eq:KMHD2}, whose parallel component is
equivalent to Eq. \ref{Ch2eq:SingleFluid2}. The condition
$V_{\parallel i}=V_{\parallel e}$, and Eq. \ref{Ch2eq:SingleFluid2}
gives \cite{Kulsrud1983}, \be \label{Ch2eq:Epar} E_\parallel =
\frac{ \sum_s (q_s/m_s) {\bf \hat{b}} \cdot \nabla \cdot {\bf
P_s}}{\sum_s (n_s q_s^2/m_s)}. \ee Eqs.
\ref{Ch2eq:SingleFluid1}-\ref{Ch2eq:SingleFluid4}, like Eqs.
\ref{Ch2eq:KMHD1}-\ref{Ch2eq:KMHD4}, are not complete, and need a
closure equation for $q_{\parallel s}$ and $q_{\perp s}$. In the
next subsection we introduce Landau fluid closure for heat fluxes.

\subsubsection{Conservation properties}
The moment equations (Eqs.  \ref{Ch2eq:KMHD1}, \ref{Ch2eq:KMHD2},
\ref{Ch2eq:SingleFluid3} and \ref{Ch2eq:SingleFluid4}) conserve
momentum and total energy irrespective of the closure for higher
moments. Combining Eqs. \ref{Ch2eq:KMHD1} and \ref{Ch2eq:KMHD2}
gives the momentum conservation equation, \be
\label{Ch2eq:MomentumConservation} \frac{\partial}{\partial t} (\rho
{\bf V}) = - {\bf \nabla \cdot} \left [ \rho {\bf V V} + \left (
\frac{B^2}{8\pi} {\bf I} - \frac{{\bf B B}}{4\pi} \right )+ {\bf P}
\right ], \ee where $p_\parallel = p_{\parallel i} + p_{\parallel
e}$ and $p_\perp = p_{\perp i} + p_{\perp e}$.

Total energy (the sum of kinetic, magnetic, and thermal energies),
defined as $\Gamma = \rho V^2/2 + B^2/8\pi + p_\perp +
p_\parallel/2$, is also conserved as \be
\label{Ch2eq:EnergyConservation} \frac{\partial \Gamma}{\partial t}
= -{\bf \nabla \cdot} \left [ \left ( \frac{1}{2} \rho V^2 + p_\perp
+ \frac{1}{2} p_\parallel \right ) {\bf V} \right ] - {\bf \nabla
\cdot} \left [ \frac{{\bf B} \times ({\bf V} \times {\bf B}) }{4\pi}
\right ] - {\bf \nabla \cdot ( V \cdot P}) - {\bf \nabla \cdot q},
\ee where ${\bf q} = (q_\perp+q_\parallel/2) {\bf \hat{b}}$, and
$q_\parallel = q_{\parallel i} + q_{\parallel e}$ and $q_\parallel =
q_{\parallel i} + q_{\parallel e}$.

\subsection{The 3+1 Landau closure}

A simple model which evolves $p_\parallel$ and $p_\perp$, and
truncates the moment hierarchy with Eqs. \ref{Ch2eq:SingleFluid3}
and \ref{Ch2eq:SingleFluid4}, using closure approximations for
$q_\parallel$ and $q_\perp$ is called a ``3+1 model," as it evolves
3 parallel moments $(n,u_\parallel,p_\parallel)$ and 1 perpendicular
moment $(p_\perp)$ \cite{Snyder1997}.

The $3+1$ closure is derived by writing $q_\parallel$ and $q_\perp$
in terms of the lower moments and $\delta B$, and solving for
coefficients by matching with the linear kinetic response. This
gives \cite{Snyder1997} \ba \label{Ch2eq:q_parallel} q_{\parallel s}
&=& - n \sqrt{\frac{8}{\pi}} v_{t \parallel s}
\frac{i k_\parallel T_{\parallel s}}{|k_\parallel|}, \\
\label{Ch2eq:q_perp} q_{\perp s} &=& - n \sqrt{\frac{2}{\pi}} v_{t
\parallel s} \frac{i k_\parallel T_{\perp s}}{|k_\parallel|} + n
\sqrt{\frac{2}{\pi}} v_{t \parallel s} T_{\perp s} \left ( 1-
\frac{T_{\perp s}}{T_{\parallel s}} \right ) \frac{i k_\parallel
\delta B}{|k_\parallel|B}, \ea where $v_{t \parallel
s}=\sqrt{T_{\parallel s}/m_s}$, and $k_\parallel$ is the parallel
wavenumber of small perturbation. The second term in the closure for
$q_{\perp s}$ vanishes in the electrostatic limit or if pressure is
isotropic, and is needed to conserve $\mu$ linearly
\cite{Snyder1997}.

Substituting the closures, Eqs. \ref{Ch2eq:q_parallel} and
\ref{Ch2eq:q_perp}, into Eqs.
\ref{Ch2eq:SingleFluid1}-\ref{Ch2eq:SingleFluid4} yields the density
response \be \label{Ch2eq:DensityResponse} \delta n_s = \frac{i
n}{k_\parallel T_{\parallel s}} q_s E_\parallel R_3(\zeta_s) + n
\frac{\delta B}{B} \left [ 1 - \frac{T_{\perp s}}{T_{\parallel s}}
R_3(\zeta_s) \right ], \ee and the perpendicular pressure response
\be \label{Ch2eq:PperpResponse} p_{\perp s} = - \frac{i p_{\perp
s}}{k_\parallel T_{\parallel s}} q_s E_\parallel R_3(\zeta_s) + 2
p_{\perp s} \frac{\delta B}{B} \left [ 1 - \frac{T_{\perp
s}}{T_{\parallel s}} \left ( \frac{R_3(\zeta_s}{2} +
\frac{R_1(\zeta_s}{2} \right ) \right ], \ee where $\zeta_s =
\omega/\sqrt{2} |k_\parallel| v_{t \parallel s}$, and $R_3(\zeta_s)$
is the three-pole Pad\'e approximation of the electrostatic response
function \be R_3(\zeta_s) =
\frac{2-i\sqrt{\pi}\zeta_s}{2-3i\sqrt{\pi} \zeta_s - 4 \zeta_s^2 + 2
i \sqrt{\pi} \zeta_s^3},\ee and $R_1(\zeta_s)$ is a one-pole model
of $R(\zeta_s)$, $R_1(\zeta_s) = 1/(1-i\sqrt{\pi} \zeta_s)$. The
electrostatic response function, $R(\zeta_s)=1+\zeta_s Z(\zeta_s)$,
where $Z(\zeta)=(1/\sqrt{\pi}) \int dt \exp(-t^2)/(t-\zeta)$, arises
frequently in linearized moments of the Vlasov equation (or the
DKE). The $3+1$ model recovers the fully kinetic response function
in both asymptotic limits, $\zeta_s \ll 1$ and $\zeta_s \gg 1$, and
provides a good approximation in the intermediate regime. Figs.
(1)-(4) in \cite{Snyder1997} show that Landau closure is a good
approximation for the linear response function from the DKE. While
the linear response function in CGL (and MHD) approximation shows no
imaginary part in frequency, Landau closure gives collisionless
damping rates consistent with the DKE.

The complete $3+1$ system of equations is given by Eqs.
\ref{Ch2eq:KMHD1}-\ref{Ch2eq:KMHD4}, and Eqs.
\ref{Ch2eq:SingleFluid3} and \ref{Ch2eq:SingleFluid4}, closed by the
inverse Fourier transform of Eqs. \ref{Ch2eq:q_parallel} and
\ref{Ch2eq:q_perp}. In Section \ref{Ch2sec:LFCImplementation} we discuss
ways of computing the heat fluxes in coordinate space from the Fourier space
expressions.

\section{Collisional effects}
Collisions serve two roles, first they isotropize the pressure
tensor, and second they reduce the heat fluxes. With the BGK
collision operator, a pitch angle scattering term that isotropizes
pressure appears in equations for $p_\parallel$ and $p_\perp$ (terms
on the right side of Eqs. \ref{Ch2eq:SingleFluid3} and
\ref{Ch2eq:SingleFluid4}). Certain collisional effects, such as
perpendicular diffusion, resistive effects, etc., are not included
because of the drift kinetic ordering ($\Omega_s \gg \omega$,
$\nu$); also not included is the collisional heat transfer from one
species to another.

To extend Landau closure to the collisional regime, it is useful to
write Eqs. \ref{Ch2eq:SingleFluid3} and \ref{Ch2eq:SingleFluid4} in
a form similar to Braginskii's equations \cite{Braginskii1965}. This
is done by defining an average pressure, $p_s = (p_{\parallel
s}+2p_{\perp s})/3$, a differential pressure, $\delta p_s=
p_{\parallel s} - p_{\perp s}$, and a heat flux, $q_s = q_{\parallel
s}/2 + q_{\perp s}$. The pressure tensor, ${\bf P}_s$, can be
divided into an isotropic part, $p_s {\bf I}$, and an anisotropic
stress, ${\bf \Pi_s} = -\delta p_s {\bf I}/3 + \delta p_s {\bf
\hat{b}\hat{b}}$, with $\delta p_s = (p_{\parallel s} - p_{\perp
s})$. Combining Eqs. \ref{Ch2eq:SingleFluid3} and
\ref{Ch2eq:SingleFluid4}, then gives \cite{Snyder1997} \ba
\label{Ch2eq:Braginskii1} \frac{d p_s}{dt} &+& \frac{5}{3} p_s {\bf
\nabla \cdot V} = - \frac{2}{3} {\bf \nabla \cdot ( \hat{b}} q_s) -
\frac{2}{3} {\bf \Pi_s : \nabla V}, \\ \nonumber
\label{Ch2eq:Braginskii2} \frac{d \delta p_s}{dt} &+& \frac{5}{3}
\delta p_s {\bf \nabla \cdot V} + {\bf \Pi_s : \nabla V} + 3 p_s
{\bf \hat{b} \cdot \nabla V \cdot \hat{b}} - p_s {\bf \nabla \cdot
V} - 3 q_\perp {\bf \nabla \cdot V}
\\ &+& {\bf \nabla \cdot [ \hat{b}} (q_{\parallel s} - q_{\perp s})
] = - \nu_s \delta p_s. \ea

\subsection{The high collisionality limit}
\label{Ch2subsec:Braginskii} In the high collisionality limit, $\nu
\gg \omega$, the above equations yield approximation to the
Braginskii transport equations in the $\nu \ll \Omega_s$ regime, as
required by the initial ordering. An expansion in $1/\nu_s$ of Eqs.
\ref{Ch2eq:SingleFluid5}, \ref{Ch2eq:SingleFluid6}, and
\ref{Ch2eq:Braginskii2} implies that $q_{\parallel 0 s} = q_{\perp 0
s} = \delta p_{0 s}=0$. Combining this with Eq.
\ref{Ch2eq:Braginskii2} gives, to next order, \be \delta p_{1 s} = -
\frac{p_{0s}}{\nu_s} (3 {\bf \hat{b} \cdot \nabla V \cdot \hat{b}} -
{\bf \nabla \cdot V} ). \ee The expression for ${\bf \Pi_s}$ is the
same as Braginskii's result, if $\nu_s$ is the inverse of
Braginskii's collision time ($\nu_i^{-1}=0.96 \tau_{i, Brag}$ and
$\nu_e^{-1}=0.73 \tau_{e, Brag}$; see \cite{Huba2000}).

Similarly, a heat flux matching Braginskii's result can be obtained
by taking the high collisionality limit of the equations evolving
$q_\parallel$ and $q_\perp$ (Eqs. \ref{Ch2eq:SingleFluid5} and
\ref{Ch2eq:SingleFluid6}), which to the lowest order in $1/\nu_s$
gives \be {\bf \nabla \cdot} \left [ {\bf \hat{b}} \left (
\frac{r_{\parallel,\parallel 0 s}}{2} + r_{\parallel, \perp 0 s}
\right ) \right ] - \frac{5}{2} \frac{p_{0 s}}{n_0 m_s} {\bf \hat{b}
\cdot \nabla} p_{0 s} - \left ( \frac{r_{\parallel,\perp 0 s}}{2} +
r_{\perp, \perp 0 s} \right ) {\bf \nabla \cdot \hat{b}} = - \nu_s
q_s. \ee In the collisional limit $r_0$'s will take their
collisional values, $r_{\parallel, \parallel 0 s} = 3 p_{\parallel
0}^2/m_s n_0$, $r_{\parallel, \perp 0 s} = p_{\parallel 0}^2/m_s
n_0$, and  $r_{\parallel, \parallel 0 s} = 2 p_{\parallel 0}^2/m_s
n_0$. Substituting in the above equation gives \be q_{s} = -
\frac{5}{2} \frac{p_0}{\nu_s m_s} \nabla_\parallel T_{0 s}, \ee
which matches Braginskii's parallel heat flux (within factors of
order unity).

\subsection{3+1 closure with collisions}
In principle, it is possible to use a kinetic response with the
collision terms and to choose Landau closures that match the
collisional linear response. Collisional heat fluxes can also be
derived by using a higher moment model (e.g., a 4+1 model) and
reducing the number of moments by taking a low frequency limit of
the highest moment equations, with the collision terms included (see
\cite{Snyder1997}). Without giving the details of derivation, we
state the results for 3+1 closures that include the effects of
collisions, \ba \label{Ch2eq:q_parallelCollisions} q_{\parallel s}
&=& - 8 n v_{t
\parallel s}^2 \frac{ i k_\parallel T_{\parallel s}}{(\sqrt{8\pi}|k_\parallel|
v_{t \parallel s} + (3\pi-8)\nu_s )}, \\ \nonumber
\label{Ch2eq:q_perpCollisions} q_{\perp s} &=& - \frac{n v_{t
\parallel s}^2 i k_\parallel T_{\perp s} }{\left(  \sqrt{\frac{\pi}{2}}
|k_\parallel| v_{t \parallel s} + \nu_s \right )} \\
&+& \left ( 1 - \frac{T_{\perp s}}{T_{\parallel s}}\right ) \frac{n
v_{t \parallel s}^2 T_\perp i k_\parallel \delta B}{B \left (
\sqrt{\frac{\pi}{2}} |k_\parallel| v_{t \parallel s} + \nu_s
\right)}. \ea These closures allow a smooth transition from
collisionless regime where collisionless damping is important, to
the collisional regime with only viscous (collisional) damping.
These closures give results similar to those derived from the DKE
with a collision operator (in the linear regime), as shown for the
case of MRI in Chapter \ref{chap:chap3}. Thus, Landau models can be
used to study collisionless and marginally collisional ($\omega \sim
\nu$) regimes. However, accurate modeling of all the collisional
effects, particularly those involving momentum exchange between
species, requires a Braginskii formalism in highly collisional
regime ($\nu \gg \omega$) or extension of the BGK model to use a
velocity dependent collision frequency.

\section{Nonlinear implementation of closure}
\label{Ch2sec:LFCImplementation}
Landau fluid closure
for heat fluxes (Eqs. \ref{Ch2eq:q_parallel} and \ref{Ch2eq:q_perp})
involve terms containing $i k_\parallel/|k_\parallel|$. Numerical
implementation of these in $k$-space is straightforward for
electrostatic problems, as magnetic perturbations vanish and a
simple Fourier transform along the magnetic field direction is
needed. However, in more general problems, heat fluxes need to be
calculated along the total (equilibrium+perturbation) magnetic
field, and so $k_\parallel$ involves Fourier transforms along the
perturbed field lines. Linear approximation of parallel heat flux,
$q_\parallel \propto {\bf \hat{b_0} \cdot \nabla} \delta T_\parallel
+ {\bf \delta \hat{b} \cdot \nabla} T_\parallel$, has a contribution
due to perturbed field lines. In an incompressible, ideally
conducting plasma, temperature is constant along a field line
($q_\parallel=0$), but temperature gradient along the unperturbed
field lines gives a nonzero result. Thus, the calculation of heat
fluxes along the perturbed field lines is non-trivial.

For fully nonlinear, electromagnetic calculations one can use a
Lagrangian coordinate system moving with the field lines, and
coordinates aligned with the magnetic field. Then the standard
fast Fourier transform (FFT) algorithm along the coordinate can be
used to evaluate closures. While Lagrangian methods are useful for
fusion plasma simulations where magnetic field fluctuations are
small, in most astrophysical cases fields are turbulent with $B
\lesssim \delta B$, making a grid aligned with field lines extremely
difficult to implement. Alternatively, in an Eulerian grid, one
needs to map $T_\parallel$ from the simulation grid to a field line
following coordinate system, carry out the FFT, and then remap the
result back to the simulation grid. FFTs can be avoided by working
with the real space form of closures. This involves convolutions in
one direction [$O(N^4)$ operations for $N$ grid points in each
direction], rather than the FFT algorithm [$O(N^3 \ln N)$
operations]. For example, the real-space form of the collisionless
3+1 closure for $q_{\parallel s}(z)$, Eq.
\ref{Ch2eq:q_parallelCollisions}, is the convolution  \be
\label{Ch2eq:NonLocal} q_{\parallel s} = -n \left( \frac{2}{\pi}
\right)^{3/2} v_{t
\parallel s} \int_0^\infty d z' \frac{T_{\parallel s}(z+z') -
T_{\parallel s}(z-z')}{z'} g(\frac{z'}{\lambda_{mfp}}) , \ee where
$g(z'/\lambda_{mfp})=1$ for $z'$ small compared to the mean free
path, but $g$ falls off rapidly for $z'$ large compared to the mean
free path (see Eq. 51 of \cite{Snyder1997}).  In a very low
collisionality plasma, exact evaluation of the heat flux requires
integrating a very long distance along a magnetic field line, but in
practice the integral can be cut off at a few correlation lengths.
Truncating the integral at $z' = L$ essentially means that Landau
damping is applied to modes with $k_\| > 1/L$, while Landau damping
is ignored for large scale $k_\| < 1/L$ modes.  Choice of an
appropriate $L$ could be made by convergence studies.

\subsubsection{A crude closure for heat fluxes}
\begin{figure}
\begin{center}
\includegraphics[width=2.95in,height=2.5in]{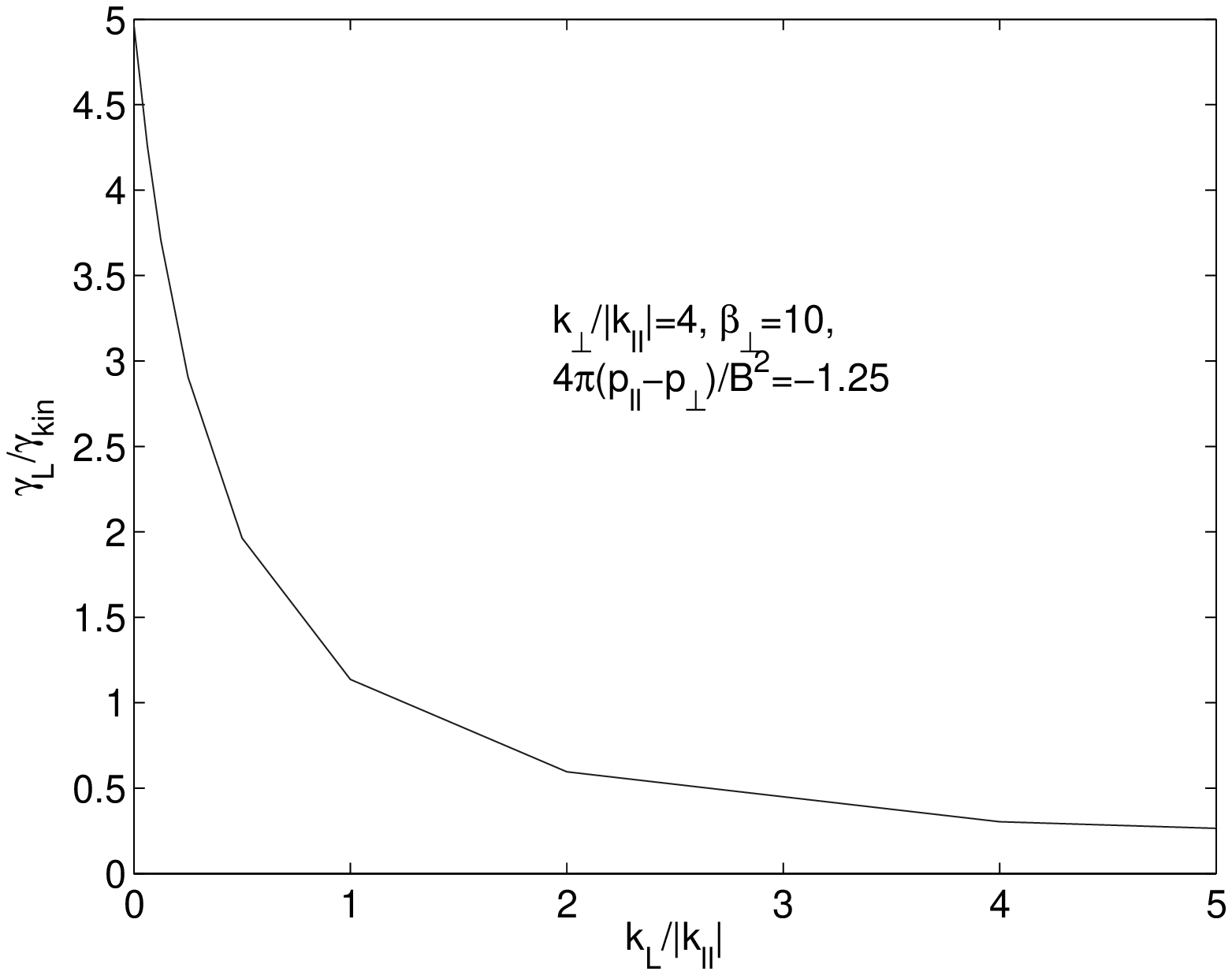}
\includegraphics[width=2.95in,height=2.5in]{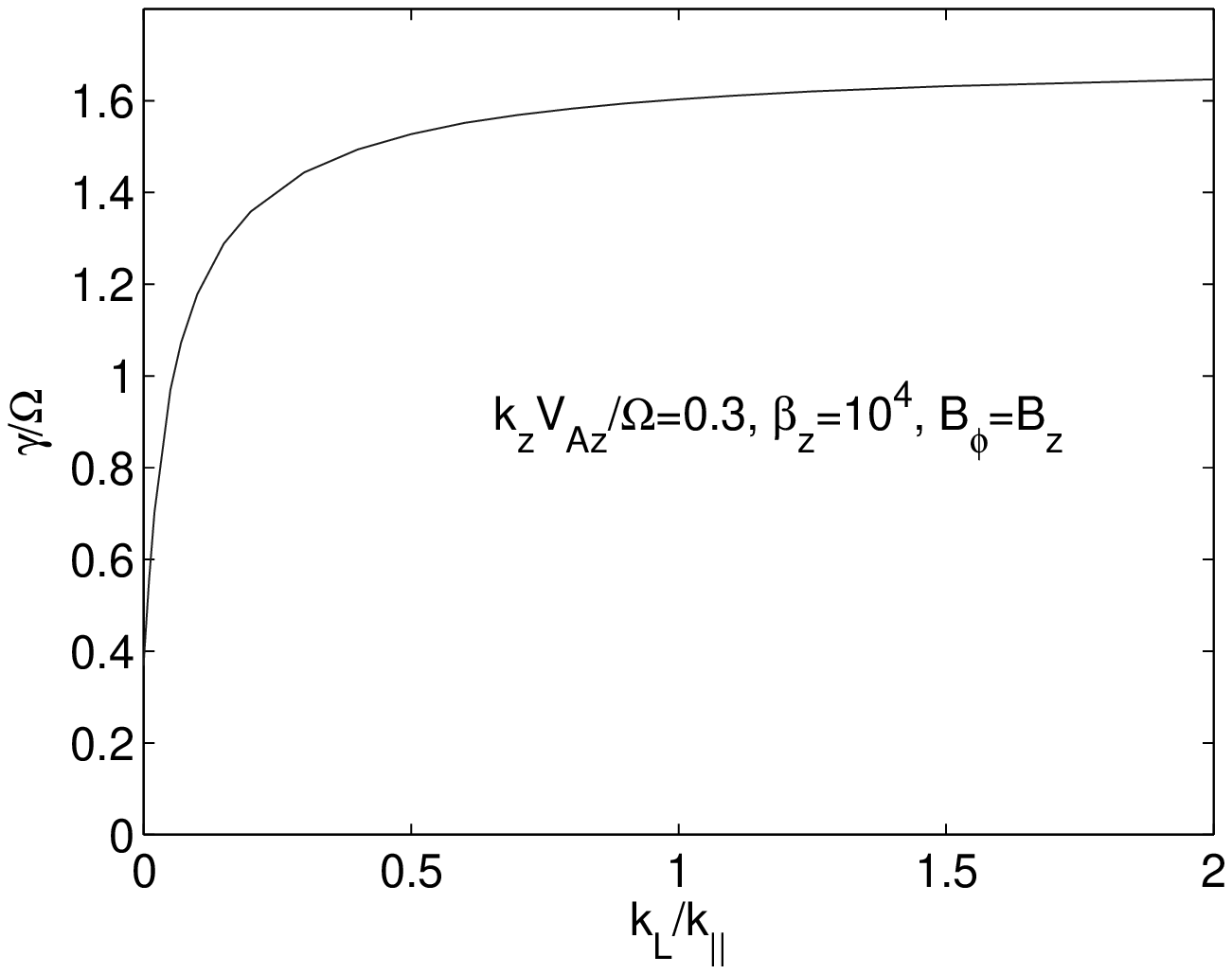}
\caption[Dependence of the growth rate of a linear mode on
$k_L$]{The dependence of the kinetic MHD growth rate, for a mirror
mode and an MRI mode, on the assumed $k_L$ Landau damping parameter
in the heat fluxes. The optimal 3-pole approximation to the plasma
$Z$ function is recovered if $k_L/k_\Par = 1$.  The plot on the left
shows that the growth rate of a mirror instability is fairly
sensitive to the assumed $k_L$. On the other hand, the plot on the
right shows that the growth rate of an MRI mode is not very
sensitive to the assumed value of $k_L$ (for these parameters). Note
that $k_L = 0$ corresponds to an isothermal limit and $k_L
\rightarrow \infty$ corresponds to a CGL limit where parallel heat
fluxes are ignored. The existence of anomalous pitch-angle
scattering by velocity-space microinstabilities may further reduce
the sensitivity of the nonlinear MRI results to the assumed $k_L$
parameter. \label{Ch2fig:Crude}}
\end{center}
\end{figure}
A crude closure for parallel heat fluxes is obtained by using a
local approximation where $|k_\parallel|$ in the denominator of Eqs.
\ref{Ch2eq:q_parallelCollisions} and \ref{Ch2eq:q_perpCollisions} is
replaced by a parameter $k_L$, i.e., \ba
\label{Ch2eq:q_parallelLocal}  q_{\parallel s} &=& - 8 n v_{t
\parallel s}^2 \frac{ i k_\parallel T_{\parallel
s}}{(\sqrt{8\pi}k_L v_{t \parallel s} + (3\pi-8)\nu_s )}, \\
\label{Ch2eq:q_perpLocal} q_{\perp s} &=& - \frac{n v_{t
\parallel s}^2 i k_\parallel T_{\perp s} }{\left(  \sqrt{\frac{\pi}{2}}
k_L v_{t \parallel s} + \nu_s \right )} + \left
( 1 - \frac{T_{\perp s}}{T_{\parallel s}}\right ) \frac{n
v_{t \parallel s}^2 T_\perp i k_\parallel \delta B}{B \left (
\sqrt{\frac{\pi}{2}} k_L v_{t \parallel s} + \nu_s \right)}. \ea
These are readily calculable local expressions for parallel heat
fluxes that recovers the correct growth/damping rate for a mode with
wavenumber $k_L$. Modes with $|k_\parallel|>k_L$
($|k_\parallel|<k_L$) have a faster (slower) heat conduction rate
than collisionless Landau damping, but the final impact on the
growth or damping rate of a mode depends on the type of mode (see
Figure \ref{Ch2fig:Crude} for examples of mirror and MRI modes).
Note that heat fluxes in Eqs. \ref{Ch2eq:q_parallelLocal} and
\ref{Ch2eq:q_perpLocal} with a constant $k_L$ are diffusive, like
Braginskii's heat fluxes. The
shearing box simulations of the MRI in the collisionless regime use
these local expressions \cite{Sharma2006}; the nonlinear results are
not very sensitive to the choice of $k_L$, but do show some
dependence (as shown in Chapter \ref{chap:chap4}). However, there
may be velocity-space microinstabilities that enhance the effective
pitch-angle scattering rate, which may make the nonlinear results
less sensitive to assumptions about $k_L$ than one might at first
think.

To improve on this in future work, there are several possible
approaches that could be taken, such as a direct evaluation of the
non-local heat flux expressions like Eq. \ref{Ch2eq:NonLocal}, along
field lines to some maximum length $L$. Another would be to use
better Pad\'e approximations to the $k$-space operator corresponding
to the Landau-damping operator.  For example, in Eqs.
\ref{Ch2eq:q_parallel} and \ref{Ch2eq:q_perp} the heat flux is
proportional to $i k_\Par / |k_\Par|$, which at present we
approximate as $i k_\Par / k_L$ and then  Fourier transform to real
space to get the local operator $(1/k_L) \grad_\Par$.  A next order
Pad\'e approximation to $i k_\Par / |k_\Par$ would be of the form
$\alpha_0 i k_\Par / (1 + \beta_2 k_\Par^2)$. Fourier transforming
this gives the operator $(1 - \beta_2 \grad_\Par^2)^{-1} \alpha_0
\grad_\Par$ \cite{Dimits1994}. If a fast iterative Krylov or
Multigrid solver could be developed to invert the $(1 - \beta_2
\grad_\Par^2)$ operator (which would be non-trivial because it is an
anisotropic operator corresponding to diffusion only along field
lines), then this could be a faster way to approximate the non-local
heat flux operator that would be relatively good over a range of
$k_\|$ instead of only near $k_\|=k_L$.  This procedure could be
made more accurate by using higher order Pad\'e approximations.

Another way to improve the calculation of the heat flux while
retaining a local approximation could be by keeping more fluid
moments before introducing a closure approximation, and modifying
the closure approximations to correspond to hyper-collisions that
selectively damp fine scales in velocity space. This would reduce
the number of fluid moments needed.  Keeping higher order fluid
moments before closing is related to a kinetic calculation that uses
higher order Hermite polynomial basis functions in velocity space
\cite{Hammett1993,Smith1997}.  While this is possible in principle,
the rate of convergence as more terms are added and its
computational cost relative to other options have not been
evaluated.

Finally, another way to improve the calculation would be to do a
direct 5-D calculation of the drift-kinetic equation.  This would be
computationally challenging, but would be feasible for a range of
problems.  It would be similar to 5-D gyrokinetic simulations
recently developed in fusion energy research
 that have made significant
contributions to understanding drift-wave turbulence in fusion
devices \cite{Dorland2000,Jenko2000,Candy2003}.

\subsection{The effects of small-scale anisotropy-driven
instabilities} \label{Ch2subsec:nu_eff} The MRI acts as a dynamo
that amplifies the magnetic field. Conservation of the magnetic
moment $\mu = v_\Perp^2 / (2 B)$ means that as the magnetic field
fluctuates, the perpendicular pressure $p_\Perp$ will change,
creating pressure anisotropies.  As we will discuss in more detail
in Chapter \ref{chap:chap4}, if these pressure anisotropies exceed a
certain threshold, they can drive velocity space instabilities (the
mirror, cyclotron, and firehose instabilities) that have very fast
growth rates at small scales. These instabilities can drive
gyro-radius scale fluctuations that break adiabatic invariance and
cause scattering to reduce the pressure anisotropy back to
threshold.

To estimate the magnitude of this enhanced scattering rate, consider
Eq. \ref{Ch2eq:Braginskii2} in an incompressible limit, $\grad \cdot
{\bf V} = 0$ (as might be expected at high $\beta$ for low Mach
number MRI-driven flows):
$$
d \delta p_s / dt + (3 p_s + \delta p_s) {\bf \hat{b}} \cdot \grad
{\bf V} \cdot {\bf \hat{b}} + \grad \cdot [ {\bf \hat{b}}
(q_{\parallel,s} - q_{\Perp,s}) ] = - \nu \delta p_s.
$$
Expanding the magnetic field evolution equation in the
incompressible limit, $\partial {\bf B} / \partial t = \grad \times
({\bf V} \times {\bf B}) = {\bf B \cdot \grad V} - {\bf V \cdot B}$,
and dotting it with ${\bf B}$ gives
$$
\frac{\partial}{\partial t} \left ( \frac{1}{2} B^2 \right) + \grad
\cdot \left( {\bf V} \frac{1}{2} B^2 \right) = B^2 {\bf \hat{b}
\cdot \grad V \cdot \hat{b}}.
$$
Substituting this in the pressure anisotropy equation, gives:
$$
d \delta p_s / dt = - (3 p_s + \delta p_s) \left [ \frac{\partial
\log{B}}{\partial t} + \grad \cdot \left( {\bf V} \log{B} \right)
\right ] - \grad \cdot [ {\bf \hat{b}} (q_{\parallel,s} - q_{\Perp,s}) ] -
\nu \delta p_s.
$$
The first term on the RHS represents the rate at which pressure
anisotropies are driven due to adiabatic invariance in a changing
magnetic field, which we will estimate as of order $3 p_s \partial
\log{B} / \partial t \sim 3 p_s \gamma_{MRI}$, where $\gamma_{MRI}$
is the growth rate for the dominant MRI modes in the simulation
(this might be modified in the nonlinear state).  In steady state,
this will be balanced by the last term in this equation, which
represents isotropization due to scattering at rate $\nu$, due
either to binary collisions (which are negligible for the regimes we
focus on) or due to gyro-scale velocity-space instabilities.  The
growth rate of velocity-space instabilities is very rapid if the
threshold for instability is exceeded, so a simple model for the
effect of these instabilities is that they will cause just enough
scattering $\nu_{eff}$ to keep the pressure anisotropy $\delta
p_s/p_s = (p_{\Par s} - p_{\Perp s})/p_s$ close to the threshold
value, which for the mirror instability is of order $7/ \beta$
(further details of this will be discussed in Chapter 4). Thus,
balancing the first and last terms on the RHS, we estimate the
effective scattering rate by velocity-space instabilities to be of
order \be \label{Ch2eq:nu_eff} \nu_{eff} \sim 3 p_s \gamma_{MRI} /
\delta p_s \sim \gamma_{MRI} \beta . \ee The mean free path
associated with this is \be \label{Ch2eq:mfp_eff} \lambda_{mfp,eff}
\sim v_t / \nu_{eff} \sim L_{MRI}/\sqrt{\beta}, \ee assuming
$\gamma_{MRI} \sim k_{MRI} V_{Alfv\acute{e}n} \sim V_{Alfv\acute{e}n} / L_{MRI}$,
where $L_{MRI}$ is of order the wavelength of a typical MRI mode in
the system. There are factors of $2$, $\pi$, etc. uncertainties in
these estimates, but they suggest that, at very high $\beta$, the
effective mean free path due to scattering by velocity-space
instabilities might be short compared to the dominant MRI wavelength.
This would reduce the sensitivity of the results to the assumed value
of the $k_L$ Landau damping parameter. However, there are intermittency
issues that may complicate the picture, as we discuss next.

\subsubsection{Intermittency of pitch angle scattering}
\begin{figure}
\begin{center}
\includegraphics[width=2.95in,height=2.5in]{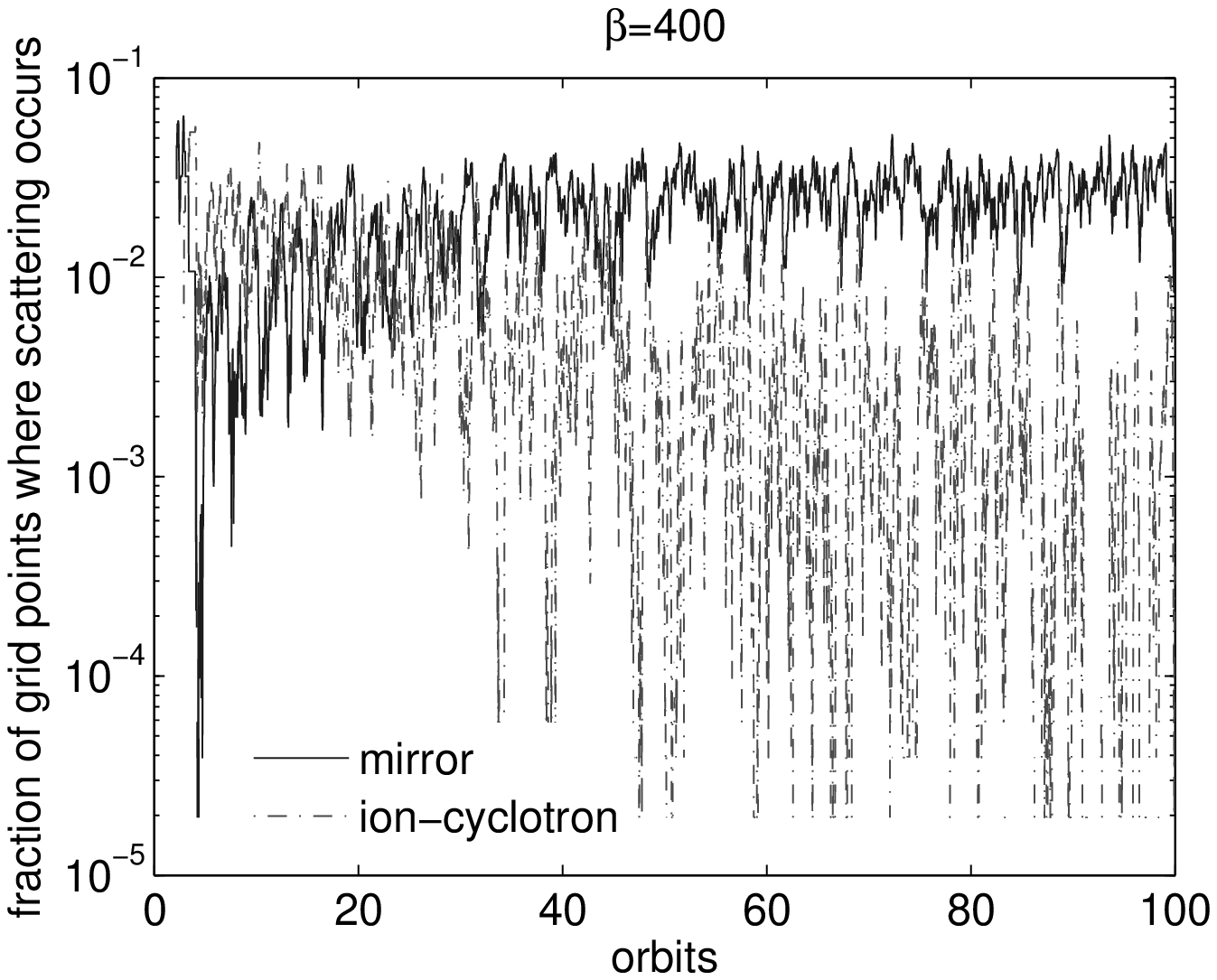}
\includegraphics[width=2.95in,height=2.5in]{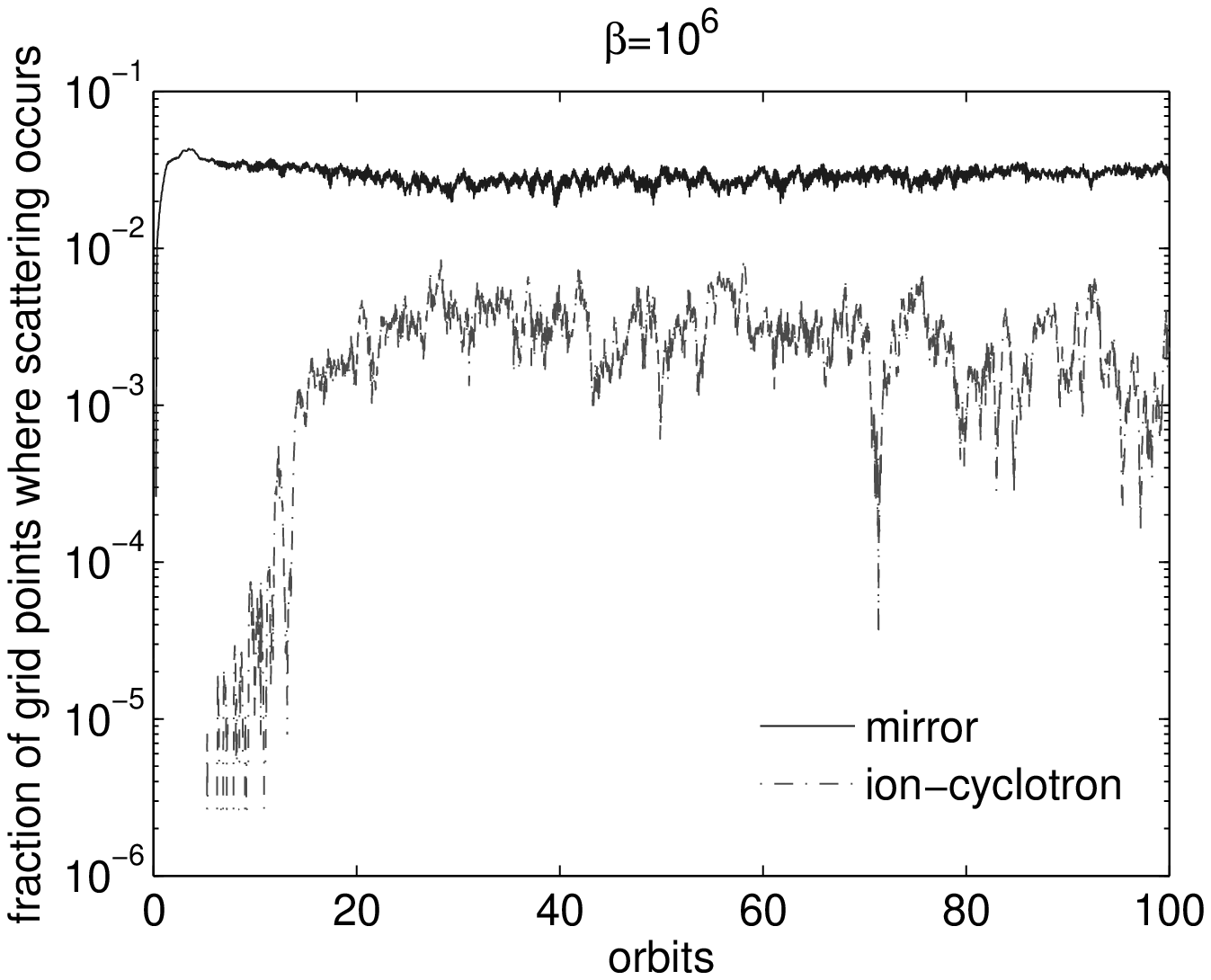}
\includegraphics[width=2.95in,height=2.5in]{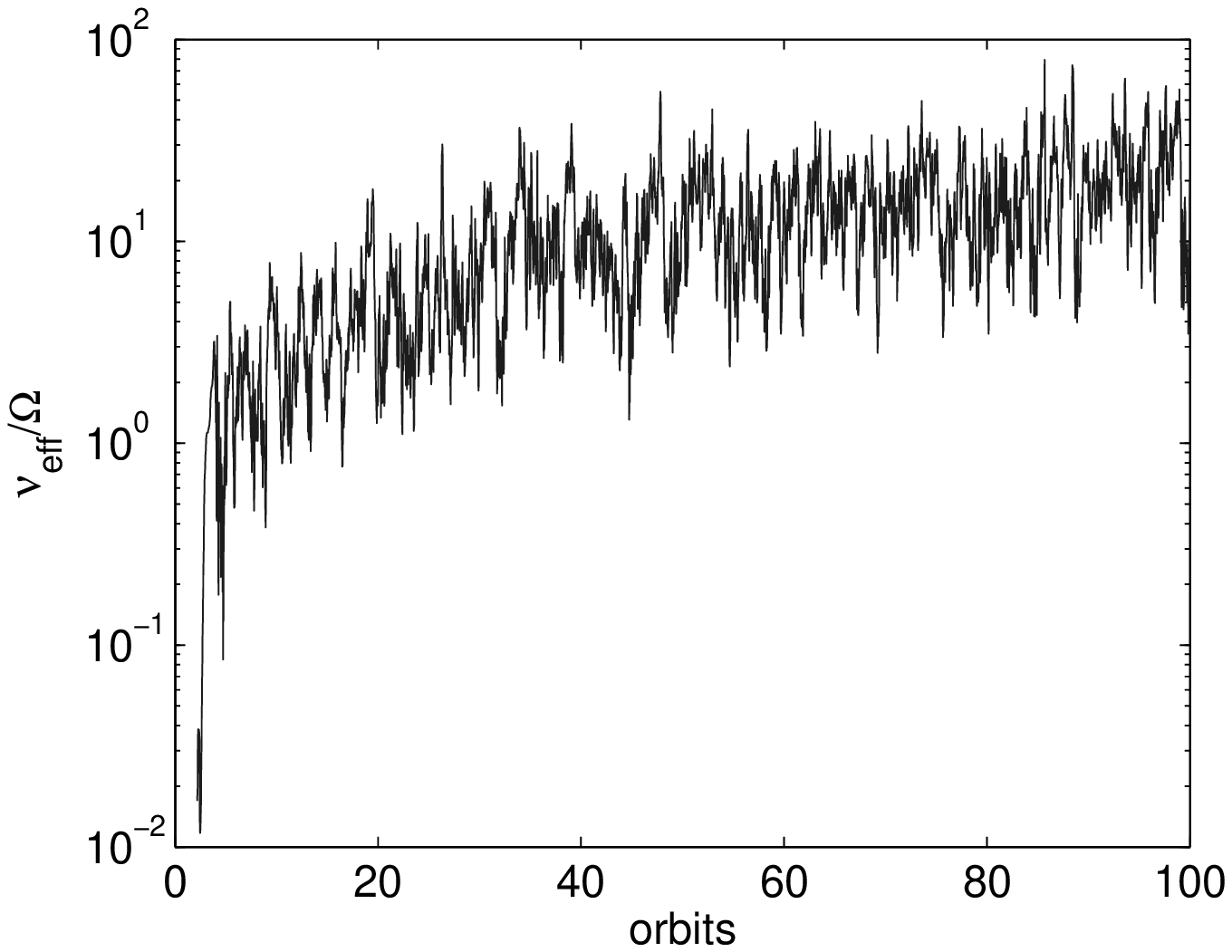}
\includegraphics[width=2.95in,height=2.5in]{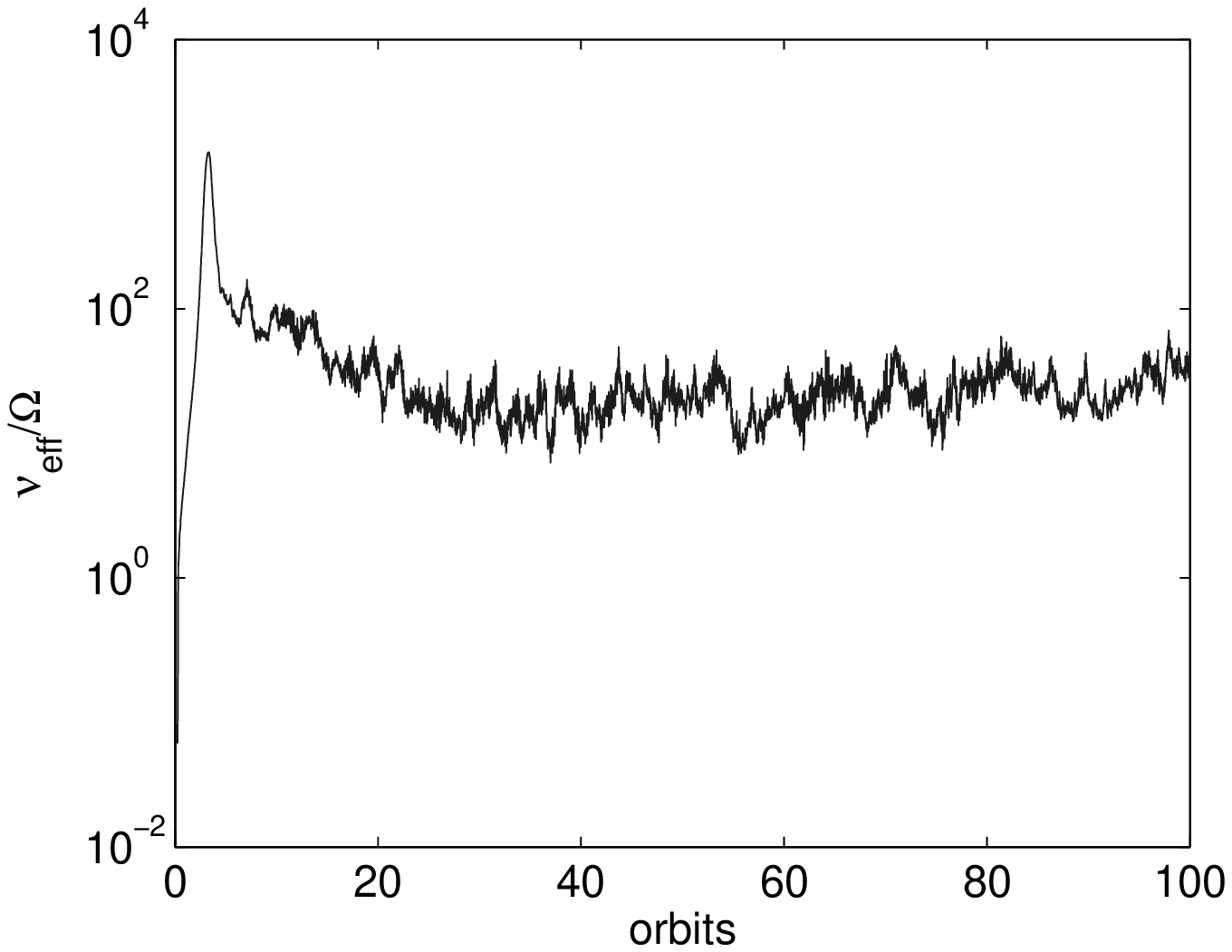}
\caption[Scattering fraction and effective collisionality due to
pitch angle scattering]{Top two plots show the fraction of grid
points undergoing pitch angle scattering due to mirror (solid line)
and ion-cyclotron (dot-dashed line) instabilities for $\beta=400$
(left; low resolution run $Zl4$) and $10^6$ (right; high resolution run 
$KZ4h$). Effective collision frequency due to pitch angle scattering
($\nu_{eff}/\Omega$) for $\beta=400$ (left) and $\beta=10^6$. Pitch
angle scattering is not uniform in space, and occurs only in small
volume-fraction of the box ($\sim 0.01-0.1$).
\label{Ch2fig:figure2}}
\end{center}
\end{figure}

Figure \ref{Ch2fig:figure2} shows that the fraction of the box where pitch
angle scattering occurs is small ($\sim 0.01-0.1$) for both
$\beta=400$ and $\beta=10^6$ simulations (runs $Zl4$ and $KZ4l$ in
Chapter \ref{chap:chap4}). 
The density of scattering regions (and hence the effective mean free
path) is very similar for $\beta=400$ and $\beta=10^6$ simulations
(see Figure \ref{Ch2fig:figure2}).
The volume averaged effective
collision frequency $\nu_{eff}$ is also shown in Figure \ref{Ch2fig:figure2};
at late times both initial
$\beta=400$ and initial $\beta=10^6$ simulations give similar values for
$\nu_{eff}$ because $\beta$ at late times for the two simulations 
are comparable ($\beta \sim 500-1000$), roughly consistent with the $\beta$
scaling of Eq. \ref{Ch2eq:nu_eff}, but an order of magnitude smaller than
the estimate of Eq. \ref{Ch2eq:nu_eff}. 
Eqs. \ref{Ch2eq:nu_eff} and \ref{Ch2eq:mfp_eff} assume that
pitch angle scattering occurs roughly uniformly everywhere in the 
box. However,
nonlinear simulations show that pitch angle scattering is not
uniform but is concentrated in small volumes in the box. 
Because of the sparsity of scattering regions the
true mean free path of most particles will be much longer than
$H/\sqrt{\beta}$. The true mean free path will be some average
measure of how far particles have to move along field lines before
they find one of the isolated regions where rapid scattering is
occurring; this may be comparable to the box size (or larger).

Further studies are
required to understand the role of the distribution of intermittent
scattering regions on thermal conduction and momentum transport, and
to what extent does pitch angle scattering lead to MHD-like
dynamics.

As we will find in Chapter \ref{chap:chap4}, the limits on
anisotropy provided by these velocity-space instabilities cause the
nonlinear kinetic MHD simulations of the MRI to be qualitatively
closer to regular MHD simulations of the MRI. However, even with
these limits on anisotropy, the anisotropic pressure component of
the angular momentum transport is found to be competitive with the
usual Maxwell and Reynolds stress transport mechanisms.  The
enhanced scattering by these velocity-space instabilities may alter
the relative electron/ion heating in MRI turbulence, a topic we
leave for future research.

The enhanced scattering by velocity-space instabilities can also
cause an increase in the effective Reynolds number (and thus a
reduction of the effective magnetic Prandtl number, the ratio of 
viscosity and resistivity) of high $\beta$ MHD
turbulence in general. The possible implications of this are beyond
the focus of this thesis, but they have been discussed in a recent
paper \cite{Schekochihin2005}, on which I was a co-author.

To summarize, in this chapter we began with the most detailed Vlasov
description of collisionless plasmas, and motivated the drift
kinetic equation (DKE) in the $k \rho_s \gg 1$, $\omega \ll
\Omega_s$ regime. Further simplification was introduced in the form
of fluid closures for parallel heat fluxes that reproduce correct
kinetic behavior, and capture collisionless damping. Fourier space
representation of the heat fluxes, and the nonlocal integral
expression in coordinate space (and the ways to numerically compute
it) were indicated. A generalization to include the collisional
effects, which reduces to Braginskii's result in $\nu \gg
\omega$ regime, was given. A crude, local approximation for the heat
fluxes was presented, in which the parameter $k_L$ represents a
typical wavenumber of the mode. Only the local approximation has
been used in the local shearing box simulations of the collisionless
MRI, leaving sophisticated treatments for the future. Pitch-angle
scattering by velocity space instabilities might provide a reduction 
of the effective mean-free-path, which may contribute to
reducing the sensitivity of the results to the assumed $k_L$, but the
scattering is found to be very intermittent spatially, so the reduction
in the mean free path might be less than one might expect at first.

\chapter{Transition from collisionless to collisional MRI}
\label{chap:chap3}
This chapter is based on our paper on the transition of the MRI from
the collisionless to the collisional regime \cite{Sharma2003}.
Calculations by Quataert and coworkers \cite{Quataert2002} found
that the growth rates of the magnetorotational instability (MRI) in
a collisionless plasma can differ significantly from those
calculated using MHD, particularly at long wavelengths. This can be
important in hot accretion flows around compact objects (see Section
\ref{Ch1sec:RIAFs} for a review). In this chapter we study the
transition from the collisionless kinetic regime to the collisional
MHD regime, mapping out the dependence of the MRI growth rate on
collisionality. The Landau fluid closure for parallel heat flux,
which recovers kinetic effects like Landau/Barnes damping, is used
and the effect of collisions is included via a BGK operator.
The kinetic MHD equation of motion has three forces: the isotropic
pressure force, the magnetic
force, and the anisotropic pressure force. For $\beta \gtrsim 1$ the
transition from
collisionless to Braginskii regime occurs as the anisotropic pressure
becomes small compared to the isotropic pressure ($\nu \gtrsim \Omega
\sqrt{\beta}$); and the transition
from Braginskii to MHD occurs when anisotropic pressure force becomes
negligible compared to the magnetic force ($\nu \gtrsim \Omega \beta$).
In the weak magnetic field regime where the Alfv\'en and MRI
frequencies $\omega$ are small compared to the sound wave frequency
$k_{\Par} c_0$, the dynamics are still effectively collisionless
even if $\omega \ll \nu$, so long as the collision frequency $\nu
\lesssim k_{\Par} c_{0}$ (i.e., so long as the mean free path is
long compared to a wavelength); for an accretion flow this requires
$\nu \lesssim \Omega \sqrt{\beta}$. The low collisionality regime
not only modifies the MRI growth rate, but also introduces
collisionless Landau or Barnes damping of long wavelength modes,
which may be important for heating of electrons and protons. The
fastest growth rate in the collisionless regime is $\approx$ twice
faster than in MHD; moreover, the fastest growing mode occurs at
large length scales compared to the fastest growing MHD mode.
\section{Introduction}
Balbus and Hawley \cite{Balbus1991} showed that the
magnetorotational instability~(MRI), a local instability of
differentially rotating magnetized plasmas, is the most efficient
source of angular momentum transport in many astrophysical accretion
flows~(see Section \ref{Ch1sec:MRI} for a review). The MRI may also be
important for dynamo generation of galactic and stellar magnetic
fields. Most studies of the MRI have employed standard MHD equations
which are appropriate for collisional, short mean free path plasmas,
but it is not obvious that this instability is relevant for collisionless,
low luminosity accretion flows (see Section \ref{Ch1sec:RIAFs}; Table
\ref{Ch1tab:SgrA} shows the collsion frequency is small compared to
the rotation frequency).
Quataert and coworkers (\cite{Quataert2002}; hereafter
QDH) studied the MRI in the collisionless regime using the kinetic
results of Snyder, Hammett $\&$ Dorland \cite{Snyder1997}. They showed
that the
MRI persists as a robust instability in a collisionless plasma, but
that at high $\beta \gg 1$~(ratio of plasma pressure to magnetic
pressure), the physics of the instability is quite different and the
kinetic growth rates can differ significantly from the MHD growth
rates.

One motivation for studying the MRI in the collisionless regime is
to understand radiatively inefficient accretion flows onto compact
objects. An example of non-radiative accretion is the radio and
x-ray source Sagittarius A$^*$, which is powered by
gas accreting onto a supermassive black hole at the center of our
galaxy (see Subsection \ref{Ch1subsec:SgrA} for a review).  In radiatively
inefficient accretion flow models, the accreting gas is a hot, low
density, plasma, with the proton temperature large compared to the
electron temperature~($T_p \approx 10^{12}$ K $\gg T_e \approx
10^{10}-10^{12}$ K).  In order to maintain such a two-temperature
configuration, the accretion flow must be collisionless in the sense
that the timescale for electrons and protons to exchange energy by
Coulomb collisions is longer than the inflow time of the gas (for
models of Sagittarius A*, the collision time close to the black hole
is $\approx 7$ orders of magnitude longer than the inflow time, see
Table \ref{Ch1tab:SgrA}).

In this chapter we extend the kinetic results of QDH to include
collisions; we study the behavior of the MRI in the transition from
the collisionless regime to the collisional MHD regime.  Instead of
using a more accurate (but very complicated) Landau or Balescu-Lenard
collision operator, we use the simpler Bhatnagar-Gross-Krook (BGK)
collision operator \cite{Bhatnagar1954} that conserves number,
momentum and energy.

There are several reasons for studying the MRI with a varying
collision frequency: (1) to gain additional understanding of the
qualitatively different physics in the MHD and kinetic regimes, (2)
the key difference between the kinetic and MHD regimes is that the
pressure is anisotropic (with respect to the local magnetic field)
in a collisionless plasma (see Section \ref{Ch2sec:KMHD}). Even if particle
collisions are negligible, high frequency waves with frequencies
$\sim$ the proton cyclotron frequency can isotropize the proton
distribution function (see Subsection \ref{Ch4subsec:Isotropization}). Our
treatment of ``collisions'' can qualitatively describe this process
as well; and (3) the transition from the collisional to the kinetic
MRI could be dynamically interesting if accretion disks undergo
transitions from thin disks to hot radiatively inefficient flows (as
has been proposed to explain, e.g., state changes in X-ray binaries;
\cite{Esin1997}). There can be associated changes in the rate of
angular momentum transport ($\alpha$), as disk transitions from
collisional to collisionless state, and vice versa.

We begin with the linearized drift kinetic equation with a BGK
collision operator and derive the exact closures for $\delta
p_\parallel$ and $\delta p_\perp$; these are used to close the
linearized kinetic MHD equations. We use Landau fluid closure for
parallel heat fluxes, and show that they are equivalent to the
kinetic closures in both low and high collisionality regimes; Landau
fluids are considered because they are easier to implement
computationally and we use them for local, nonlinear MHD disk
simulations described in Chapter \ref{chap:chap4}. The kinetic MHD
linear analysis shows the presence of damped modes at all scales
(see Figure \ref{Ch3fig:compare}), a feature absent in MHD.
\section{Linearized kinetic MHD equations}
The analysis is restricted to fluctuations that have
wavelengths much larger than proton Larmor radius and frequencies
well below the proton cyclotron frequency.  In this limit, a plasma
can be described by the following kinetic MHD
equations (see Section \ref{Ch2sec:KMHD}): \ba
\label{Ch3eq:MHD1} && \frac{\partial \rho}{\partial t} + \nabla \cdot
\left(\rho {\bf V}\right)=0,
\\
\label{Ch3eq:MHD2} && \rho \frac{\partial {\bf V}}{\partial t} +
\rho\left({\bf V} \cdot \nabla\right) {\bf V}= \frac{\left(\nabla
\times {\bf B}\right) \times {\bf B}}{4\pi} - \nabla \cdot {\bf
P} + {\bf F_g},\\
\label{Ch3eq:MHD3} && \frac{\partial {\bf B}}{\partial t}= \nabla
\times \left({\bf V} \times {\bf
B}\right), \\
\label{Ch3eq:MHD4} && {\bf P}= p_{\Perp} {\bf I} + \left(p_{\Par}-
p_{\Perp}\right){\bf \hat{b}\hat{b}}, \ea where $\rho$ is the mass
density, ${\bf V}$ is the fluid velocity, ${\bf B}$ is the magnetic
field, ${\bf F_g}$ is the gravitational force, ${\bf \hat{b}}={\bf
B}/|{\bf B}|$ is a unit vector in the direction of the magnetic
field, and ${\bf I}$ is the unit tensor. In equation~(\ref{Ch3eq:MHD3})
an ideal Ohm's law is used, neglecting effects such as resistivity.
${\bf P}$ is the pressure tensor that has different
perpendicular~($p_{\Perp}$) and parallel~($p_{\Par}$) components
with respect to the background magnetic field (unlike in MHD, where
there is only a scalar pressure). The pressures are determined by
solving the drift kinetic equation given below.  ${\bf P}$ should in
general be a sum over all species but in the limit where ion
dynamics dominate and electrons just provide a neutralizing
background, the pressure can be interpreted as the ion pressure.
This is the case for hot accretion flows where $T_p \gg T_e$.

We assume that the background~(unperturbed) plasma is described by a
non-relativistic Maxwellian distribution function with equal
parallel and perpendicular pressures~(temperatures). Although the
equilibrium pressure is assumed to be isotropic, the perturbed
pressure is not. We take the plasma to be differentially rotating,
but otherwise uniform (we neglect temperature and density
gradients). Equilibrium state for equation~(\ref{Ch3eq:MHD2}) in
presence of a subthermal magnetic field with vertical~($B_z=B_0 \sin
\theta$) and azimuthal~($B_{\phi}=B_0 \cos \theta$) components gives
a Keplerian rotation~($\Omega \propto R^{-3/2}$) profile.

In a differentially rotating plasma, a finite $B_R$ is sheared to
produce a time-dependent $B_{\phi}$, which complicates the kinetic
analysis~(unlike in MHD, where a time-dependent $B_{\phi}$ does not
couple to axisymmetric disturbances; \cite{Balbus1991}); we
therefore set $B_R=0$. For linearization we consider fluctuations of
the form $\exp[-i\omega t+i {\bf k}\cdot{\bf x}]$, with ${\bf k}=k_R
\hat{R}+k_z \hat{z}$, i.e., axisymmetric modes; we also restrict our
analysis to local perturbations for which $|{\bf k}|R \gg 1$.
Writing $\rho=\rho_0+ \delta \rho$, ${\bf B}={\bf B_0} + \delta {\bf
B}$, $p_{\Perp}=p_0+\delta p_{\Perp}$, and $p_{\Par}= p_0+\delta
p_{\Par}$, ${\bf V} = \hat{\phi} \Omega R + \delta {\bf V}$ (with
Keplerian rotation $\Omega(R)$), and working in cylindrical
coordinates, the linearized versions of
equations~(\ref{Ch3eq:MHD1})-(\ref{Ch3eq:MHD3}) become (see QDH) \ba
\label{Ch3eq:lin1}
&& \omega \delta \rho= \rho_0 {\bf k} \cdot \delta{\bf V}, \\
\label{Ch3eq:lin2} && -i \omega \rho_0 \delta V_R - \rho_0 2\Omega
\delta V_{\phi}= -\frac{i k_R}{4 \pi}\left(B_z \delta B_z+B_{\phi}
\delta B_{\phi}\right)+\frac{i k_z B_z
\delta B_R}{4 \pi} - i k_R \delta p_{\Perp}, \\
\label{Ch3eq:lin3} && -i \omega \rho_0 \delta V_{\phi} + \rho_0 \delta
V_R \frac{\kappa^2} {2 \Omega} = \frac{i k_z B_z \delta B_{\phi}}{4
\pi} - i k_z \sin \theta \cos
\theta[ \delta p_{\Par} - \delta p_{\Perp}], \\
\label{Ch3eq:lin4} && -i \omega \rho_0 \delta V_z= - \frac{i k_z
B_{\phi} \delta B_{\phi}}{4 \pi} - i k_z[ \sin^2 \theta \delta
p_{\Par} + \cos^2 \theta \delta
p_{\Perp}], \\
\label{Ch3eq:lin5}
&& \omega \delta B_R= - k_z B_z \delta V_R, \\
\label{Ch3eq:lin6} && \omega \delta B_{\phi}= - k_z B_z \delta V_{\phi}
- \frac{i k_z B_z}{\omega} \frac{d \Omega}{d \ln R} \delta V_R +
B_{\phi} {\bf k} \cdot
\delta {\bf V}, \\
\label{Ch3eq:lin7} && \omega \delta B_z=k_R B_z \delta V_R, \ea where
$\kappa^2=4\Omega^2 + d\Omega^2/d \ln R$ is the epicyclic frequency.
To complete this system of equations and derive the dispersion
relation for linear perturbations, we need expressions for $\delta
p_{\Perp}$ and $\delta p_{\Par}$ in terms of lower moments. These can
be obtained by taking
moments of the linearized and Fourier transformed drift-kinetic
equation that includes a linearized BGK collision operator
(see Section \ref{Ch2sec:DKE}).

The drift-kinetic equation for the distribution function $f$,
including the effects of gravity is (see Section \ref{Ch2sec:DKE} for
details) \be \frac{\partial f}{\partial t} + \left(v_{\Par}{\bf
\hat{b}}+{\bf V}_E\right)\cdot \nabla f + \left(-{\bf\hat{b}} \cdot
\frac{D {\bf V}_E}{Dt} -\mu {\bf \hat{b}} \cdot \nabla B +
\frac{e}{m} (E_{\Par}+ F_{g \Par}/e)\right) \frac{\partial
f}{\partial v_{\Par}}=C\left(f\right), \label{Ch3eq:DKE} \ee where
${\bf V}_E=c\left({\bf E} \times {\bf B}\right)/B^2$, $\mu=({\bf
v}_{\Perp}-{\bf V}_E)^2/2B$ is the magnetic moment (conserved in our
approximations in the absence of collisions), $F_{g \Par}=GM_*m_p
\hat{R} \cdot \hat{\bf b}/R^2$, and $D/Dt=\partial /\partial t +
\left(v_{\Par} {\bf \hat{b}} + {\bf V}_E\right) \cdot \nabla$.
The fluid velocity ${\bf V} = {\bf V}_E + {\bf\hat{b}} V_\Par$, where
the $E \times B$ drift ${\bf V}_E$ is determined by the
perpendicular component of equation (\ref{Ch3eq:MHD2}).
The parallel component of the
gravitational force, $F_{g \Par}$, is included as it is of the same
order as the parallel electric force.
Notice the addition of a collision operator on the right hand side
to allow for generalization to collisional regimes. In the next
section we derive the linearly-exact kinetic expressions for $\delta
p_{\Par}$ and $\delta p_{\Perp}$ using the BGK collision operator in
equation (\ref{Ch3eq:DKE}).  We then compare these with Landau
closure approximations from Snyder et al. \cite{Snyder1997}.
\section{Kinetic closure including collisions}
In this section we use a simple BGK collision
operator~\cite{Bhatnagar1954} to calculate $\delta p_{\Par}$ and
$\delta p_{\Perp}$ from equation (\ref{Ch3eq:DKE}). Since we
consider only ion-ion collisions (see Subsection \ref{Ch2subsec:Moments} for
multiple species), the BGK operator is
$C_K\left(f\right)=-\nu\left(f-F_M\right)$ where $\nu$ is the
ion-ion collision frequency and $F_M$ is a shifted Maxwellian with
the same density, momentum, and energy as $f$ (so that collisions
conserve number, momentum, and energy). The integro-algebraic BGK
operator greatly simplifies the calculations while adequately
modeling many of the key properties of the full integro-differential
collision operator.  In some situations, the effects of weak
collisions can be enhanced in a more complete collision operator due
to sharp velocity gradients in the distribution function; we ignore
such effects in the present analysis.

In this section, we calculate the linearization of the drift-kinetic
equation around an accretion disk equilibrium, including equilibrium
flows and gravity.  A number of complicated intermediate terms end
up canceling, and the final forms of the closures used (from
equations (\ref{Ch3eq:closure_perp}-\ref{Ch3eq:closure_par})
onwards) are identical to what one would get from perturbing around
a slab equilibrium with no flows. We carried out the more detailed
calculation to verify that there were no missing terms in the final
closures.

The equilibrium distribution function $f_0$ is given by \be
f_0=\frac{n_0}{(2 \pi T_0/m)^{3/2}} \exp\left(-\frac{m}{2T_0}|{\bf
v}- {\bf V}_{0}|^2 \right), \ee where ${\bf V}_{0}={\bf
V}_{E0}+V_{\Par 0} {\bf \hat{b}}_0$ is the Keplerian rotation
velocity in the $\hat{\phi}$ direction. Since $|{\bf v}-{\bf
V}_{0}|^2=(v_{\Par}-V_{\Par 0})^2 + 2 \mu B_0$, $f_0$ can be
expressed in terms of $\left(\mu,v_\Par\right)$ as \be
f_0=\frac{n_0}{\left(2\pi T_0/m\right)^{3/2}}
\exp\left(-\frac{m}{2T_0}\left((v_\Par-V_{\Par 0})^2+ 2\mu
B_0\right)\right) . \label{Ch3eq:equilibrium} \ee We shall linearize
the drift-kinetic equation and the BGK collision operator. The
distribution function is given as $f=f_0+\delta f$ where $\delta f$
is the perturbation in the distribution function. The shifted
Maxwellian that appears in the BGK collision operator is given by
\be F_{M}=\frac{N_M}{\left(2\pi T_M/m\right)^{3/2}}
\exp{\left\{-\frac{m}{2 T_M}\left(\left(v_\Par - V_{\Par M}\right)^2
+ 2 \mu B\right)\right\} }. \label{Ch3eq:BGK} \ee $F_{M}$ has three
free parameters ($N_M$, $V_{\Par M}$, $T_M$) which are to be chosen
so as to conserve number, parallel momentum, and energy.  When
taking moments of the BGK operator, it is important to note that
$\int d^3v=\int 2 \pi \left(B_0+\delta B\right) d\mu d v_\Par$. From
equation~(\ref{Ch3eq:BGK}) and conservation of number, momentum, and
energy it follows that \ba N_M & = & n_0 + \delta n \approx n_0
\left(1+ \frac{\delta B}{B_0}\right)+ 2 \pi B_0 \int d \mu d v_\Par
\delta f, \label{Ch3eq:number_conservation}
\\
N_M V_{\Par M} & = & N_M(V_{\Par 0} + \delta V) \approx n_0V_{\Par
0} \left(1+ \frac{\delta B}{B_0}\right) + 2 \pi B_0 \int d \mu d
v_\Par \delta f v_\Par, \label{Ch3eq:norm_momentum}
\\
N_M T_M & = & p_0 + \delta p = p_0 + (\delta p_\Par + 2 \delta
p_\Perp)/3, \label{Ch3eq:norm_energy}
\\
\delta p_\Par & \approx & p_0 \delta B / B_0 + 2 \pi B_0 \int d \mu
d v_\Par \delta f m (v_\Par-V_{\Par 0})^2, \label{Ch3eq:norm_ppar}
\\
\delta p_\Perp & \approx & 2 p_0 \delta B / B_0 + 2 \pi B_0 \int d
\mu d v_\Par \delta f \mu m B_0, \label{Ch3eq:norm_pperp} \ea where
the approximate expressions retain only linear terms in perturbed
quantities. Linearizing the expression for the relaxed Maxwellian in
equation~(\ref{Ch3eq:BGK}) about $f_0$, the drift-kinetic BGK
collision operator is given by \ba \nonumber C_K\left(\delta
f\right) &=& -\nu \delta f + \nu f_0 \times \left \{
\left(\frac{\delta n}{n_0}-\frac{3\delta T}{2T_0}\right) \right . \\
&+& \left . \frac{m}{T_0}\left(\left(v_{\Par}-V_{\Par
0}\right)\delta u +\left(v_{\Par}-V_{\Par 0}\right)^2\frac{\delta
T}{2T_0}\right) - \frac{m \mu B_0}{T_0}\left(\frac{\delta B}{B_0} -
\frac{\delta T}{T_0}\right)\right\}. \label{Ch3eq:linearized_BGK}
\ea The drift-kinetic equation including the BGK operator can be
linearized to obtain the following equation for $\delta f$ \ba
\nonumber \delta f &=& V_{\phi 0}(v_\Par - V_{\Par 0})f_0
\sin{\theta} \frac{\left( \delta B_{\phi} \sin{\theta} - \delta B_z
\cos{\theta}\right) m}{T_0 B_0}+ \frac{m\left(v_\Par-V_{ \Par
0}\right) f_0}{T_0 \left(-i\omega+i
k_\Par\left(v_\Par- V_{\Par 0} \right)+\nu\right)} \times \\
\nonumber && \left(-i k_\Par \mu \delta B +\frac{\left(eE_\Par+F_{g
\Par}\right)}{m}\right) + \frac{\nu f_0}{\left(-i\omega+i
k_\Par\left(v_\Par-V_{\Par 0} \right)+\nu\right)}
\times \\
&& \left(\frac{\delta n}{n_0}- \frac{3}{2}\frac{\delta T}{T_0} +
\frac{m(v_\Par-V_{\Par 0}) \delta u}{T_0}+ \frac{m (v_{\Par}-V_{\Par
0})^2}{2 T_0} \frac{\delta T}{T_0}+\frac{m \mu B_0}{T_0}\frac{\delta
T}{T_0} -\frac{m \mu \delta B}{T_0}\right),
\label{Ch3eq:linearized_DKE} \ea where $F_{g \Par}=GM_*m_p\delta
B_R/B_0R^2$ is the component of gravitational force in the direction
of magnetic field. Choosing a compact notation where $- i \omega
\sin{\theta} (\delta B_{\phi} \sin{\theta}-\delta B_z \cos{\theta})
m V_{\phi 0}/e B_0$ $ +F_{g \Par}/e + E_\Par \rightarrow E_\Par$,
the moments of the perturbed distribution function $\delta f$ in
drift coordinates $(v_\Par, \mu)$, $\int\left(1,2 \mu
B_0,(v_{\Par}-V_{\Par 0})^2\right) \delta f 2 \pi B_0 d\mu dv_\Par$
give \ba \nonumber \frac{\delta n}{n_0} &=& \frac{\delta
B}{B_0}\left(1-R\right)+\frac{e E_{\Par}} {i k_\Par T_0} R -
\zeta_2\left\{\left(\frac{\delta n}{n_0}-\frac{3}{2}\frac{\delta
T}{T_0}\right)Z+\left(\frac{\delta
T}{T_0}-\frac{\delta B}{B_0}\right)Z \right. \\
&+& \left. \sqrt{2}\frac{\delta V}{c_0} R + \left(\frac{\delta
T}{T_0}+2 i \sin{\theta} \frac{k_\Par V_{\phi 0}}{\nu} \frac{\left(
\delta B_{\phi}\sin{\theta}-\delta B_z
\cos{\theta}\right)}{B_0}\right)\zeta
R\right\},\\
\nonumber \frac{\delta p_{\Perp}}{p_0} &=& 2\frac{\delta
B}{B_0}\left(1-R\right)+\frac{e E_{\Par}} {i k_\Par T_0} R -
\zeta_2\left\{\left(\frac{\delta n}{n_0}-\frac{3}{2}\frac{\delta
T}{T_0}\right)Z+2\left(\frac{\delta T}{T_0}- \frac{\delta
B}{B_0}\right)Z
\right.  \\
&+& \left. \sqrt{2}\frac{\delta V}{c_0}R+ \left(\frac{\delta
T}{T_0}+2 i \sin{\theta} \frac{k_\Par V_{\phi 0}}{\nu} \frac{\left(
\delta B_{\phi}\sin{\theta} -\delta B_z
\cos{\theta}\right)}{B_0}\right) \zeta R\right\},
\\
\nonumber \frac{\delta p_{\Par}}{p_0} &=& -2\frac{\delta
B}{B_0}\zeta^2 R + \frac{ e E_{\Par}}{i k_\Par T_0}\left(1+2 \zeta^2
R\right) - \zeta_2\left\{2\left(\frac{\delta
n}{n_0}-\frac{3}{2}\frac{\delta T}{T_0}\right)\zeta R \right. \\
\nonumber &+& 2\left(\frac{\delta T} {T_0}-\frac{\delta
B}{B_0}\right) \zeta R + \sqrt{2} \frac{\delta V}{c_0}\left(1+ 2
\zeta^2 R\right) \\
&+& \left. \left(\frac{\delta T}{T_0}+2 i \sin{\theta} \frac{k_\Par
V_{\phi 0}}{\nu} \frac{\left( \delta B_{\phi}\sin{\theta}-\delta B_z
\cos{\theta}\right)}{B_0}\right)\zeta\left(1+2 \zeta^2 R\right)
\right\}. \ea The parallel electric field, $E_\Par$, can be
eliminated by taking appropriate combinations of these three
equations, giving \be \frac{\delta \rho}{\rho_0}-\frac{\delta
p_{\Perp}}{p_0} =-\frac{\delta B}{B_0}\left(1-R\right)+\zeta_2
Z\left(\frac{\delta T}{T_0}- \frac{\delta B}{B_0}\right),
\label{Ch3eq:closure_perp} \ee and \be
\left(1+2\zeta^2R\right)\frac{\delta \rho}{\rho_0}-R\frac{\delta
p_{\Par} }{p_0}=\frac{\delta B}{B_0}\left(1+2\zeta^2 R-R\right)
-\zeta_2\left(Z-2 \zeta R\right)\left(\frac{\delta
\rho}{\rho_0}-\frac{\delta T}{2 T_0}- \frac{\delta B}{B_0}\right),
\label{Ch3eq:closure_par} \ee where $\delta T=\left(2\delta
T_{\Perp}+\delta T_{\Par}\right)/3$, $\delta B={\bf \hat{b}_0} \cdot
\delta{\bf B}$, $\zeta= \left(\omega+ i \nu\right)/\sqrt{2}
|k_{\Par}| c_0$, $\zeta_2=i \nu/\sqrt{2} |k_{\Par}| c_0$,
$k_{\Par}={\bf \hat{b}_0 \cdot k}$, $T_{\Par,\Perp}=m
p_{\Par,\Perp}/\rho$, and $c_0=\sqrt{T_0/m}$ is the isothermal sound
speed of the ions. In equations~(\ref{Ch3eq:closure_perp})
and~(\ref{Ch3eq:closure_par}), $R=1+\zeta Z$ is the plasma response
function, where \be Z\left(\zeta\right)=\frac{1}{\sqrt{\pi}} \int dx
\frac{\exp[-x^2]}{x-\zeta} \label{Ch3eq:response_function} \ee is
the plasma dispersion function~\cite{Huba2000}. Equations
(\ref{Ch3eq:closure_perp}) and (\ref{Ch3eq:closure_par}) can be
substituted into the linearized fluid equations
\ref{Ch3eq:lin1}-\ref{Ch3eq:lin7} to derive the dispersion relation
for the plasma.  The full closures are, however, very complicated,
so it is useful to consider several simplifying limits that isolate
much of the relevant physics.  In addition, the solution of
linearized kinetic MHD equations fully kinetic closures will give an
implicit equation for the growth rate (involving the $Z$ function)
that has to be solved numerically.

The closure equations can be simplified in two limits, $|\zeta| \ll
1$, the collisionless limit, and $|\zeta| \gg 1$, the high
collisionality limit.  The derivation of the asymptotic solution for
the closure equations in these two limits is given in Appendix
\ref{app:app2}. In the high collisionality limit, \be \frac{\delta
p_{\Perp}}{p_0}=\frac{5}{3}\frac{\delta \rho}{\rho_0}+
\frac{\zeta_1}{\zeta_2}\left(\frac{4}{3}+\frac{5}{9
\zeta_1^2}\right) \frac{\delta \rho}{\rho_0}-
2\frac{\zeta_1}{\zeta_2} \frac{\delta B}{B_0},
\label{Ch3eq:high_closure_perp} \ee and \be \frac{\delta
p_{\Par}}{p_0}=\frac{5}{3}\frac{\delta \rho}{\rho_0} +
\frac{\zeta_1}{\zeta_2}\left(-\frac{2}{3}+\frac{5}{9
\zeta_1^2}\right)\frac{\delta \rho}{\rho_0}
+\frac{\zeta_1}{\zeta_2}\frac{\delta B}{B_0},
\label{Ch3eq:high_closure_par} \ee where $\zeta_1=\omega/\sqrt{2}
|k_{\Par}| c_0$. Notice that in the limit that the collision
frequency is very high, $\zeta_2 \rightarrow \infty$, one recovers
the MHD result that the perturbations are adiabatic and isotropic:
$\delta p_{\Par}/p_0= \delta p_\Perp / p_0 = 5\delta \rho/3 \rho_0$.

For low collisionality, $|\zeta| \ll 1$, to second order in $\zeta$,
\be \frac{\delta p_{\Perp}}{p_0}=\frac{\delta \rho}{\rho_0}-i
\sqrt{\pi} \zeta_1 \frac{\delta B}{B_0}-\frac{\pi \zeta_1
\zeta_2}{3}\frac{\delta
\rho}{\rho_0}+\zeta_1\zeta_2\left(2-\frac{\pi}{3}\right)\frac{\delta
B}{B_0}, \label{Ch3eq:low_closure_perp} \ee and \ba \nonumber
\frac{\delta p_{\Par}}{p_0}&=&\frac{\delta \rho}{\rho_0}-i
\sqrt{\pi} \zeta_1\left(\frac{\delta \rho}{\rho_0} - \frac{\delta
B}{B_0}\right)+ \frac{\delta \rho}{\rho_0}\left(4\zeta_1 \zeta_2-\pi
\zeta_1^2-\frac{7 \pi
\zeta_1\zeta_2}{6}\right)+ \\
&&\frac{\delta B}{B_0}\left(\sqrt{\pi}\zeta_1\zeta_2 -\frac{\pi
\zeta_1\zeta_2}{6}-2\zeta^2 - 4 \zeta_2 \zeta\right).
\label{Ch3eq:low_closure_par} \ea To first order, there is no effect
of collisions on the growth rate of the MRI; the results above are
then exactly same as equations (20) and (21) in QDH (who neglected
collisions entirely). Collisional effects modify the closure only at
order $\zeta^2$, though one has to go to this order to find the
first order dependence of $\omega$ on $\nu$ in the dispersion
relation.
\section{Comparison with Landau fluid closure}
The results from last section provide expressions for $\delta
p_\Perp$ and $\delta p_\Par$ in both low and high collisionality
regimes, $|\zeta| \ll 1$ and $|\zeta| \gg 1$, but it would be
convenient to have a single set of equations that can provide a
robust transition between these two regimes.  The Landau fluid
closure \cite{Snyder1997}, which we discuss in Section \ref{Ch2sec:LFC}, can
do this.

The second order moments of the drift kinetic equation (Eq.
\ref{Ch3eq:DKE}) yield evolution equations for $\delta p_\Perp$ and
$\delta p_\Par$ (see, e.g., Eqs. \ref{Ch2eq:SingleFluid3} and
\ref{Ch2eq:SingleFluid4}). The linearized versions of these
equations, including a BGK collision operator, are given by
\footnote{A comparison of our equations~(\ref{Ch3eq:p_par})
and~(\ref{Ch3eq:p_perp}) with equations~(30) and~(31) in Snyder et
al.\ (linearized version of Eqs. \ref{Ch2eq:SingleFluid3} and
\ref{Ch2eq:SingleFluid4}) shows that our equations have an extra
term proportional to the Keplerian rotation frequency; this is
because \cite{Snyder1997} did not include gravitational effects and
Keplerian rotation in their linearized equations.} \be -i\omega
\delta p_{\Par}+p_0 i{\bf k \cdot \delta v} + i k_{\Par} q_{\Par} +
2 p_0 i k_{\Par} \delta v_{\Par} - 3 p_0 \Omega \cos{\theta} \frac{
\delta B_R}{B_0}=-\frac{2}{3} \nu\left(\delta p_{\Par}-\delta
p_{\Perp}\right), \label{Ch3eq:p_par} \ee and \be -i\omega \delta
p_{\Perp}+2p_0 i{\bf k \cdot \delta v} + i k_{\Par} q_{\Perp} - p_0
i k_{\Par} \delta v_{\Par} + \frac{3}{2} p_0 \Omega \cos{\theta}
\frac{\delta B_R}{B_0}=-\frac{1}{3} \nu\left(\delta p_{\Perp}
-\delta p_{\Par}\right). \label{Ch3eq:p_perp} \ee As is usual with
moment hierarchies, the above equations for $\delta p_{\Par,\Perp}$
depend on third moments of the distribution function, $q_\Par$ and
$q_\perp$, the parallel and perpendicular heat fluxes.\footnote{It is
important to note that $q_\parallel$ and $q_\perp$ are the fluxes of
$p_\parallel$ and $p_\perp$ along the field lines; thermal
conduction perpendicular to field lines vanishes as the Larmor
radius is tiny.} Snyder et al. \cite{Snyder1997} introduced closure
approximations for $q_\Par$ and $q_\perp$ that determine $\delta
p_\Perp$ and $\delta p_\Par$ without solving the full kinetic
equation of the previous section (see Section \ref{Ch2sec:LFC} for a
review). These Landau-fluid approximations ``close''
equations~(\ref{Ch3eq:MHD1})-(\ref{Ch3eq:MHD4}) and allow one to
solve for the linear response of the plasma.

The linearized heat fluxes of parallel and perpendicular pressures
are given by \be q_{\Perp}=-p_0 c_0^2\frac{i k_{\Par} \left(\delta
p_{\Perp}/p_0-\delta \rho /\rho_0\right)}{\left(\sqrt{\pi/2}
|k_{\Par}| c_0+\nu\right)} \label{Ch3eq:heat_perp} \ee and \be
q_{\Par}=-8 p_0 c_0^2 \frac{i k_{\Par}\left(\delta
p_{\Par}/p_0-\delta \rho/\rho_0\right)}{\left(\sqrt{8 \pi}|k_{\Par}|
c_0+\left(3 \pi -8\right) \nu\right)}. \label{Ch3eq:heat_par} \ee As
discussed in earlier work \cite{Snyder1997, Hammett1992,
Hammett1993, Smith1997}, Landau-fluid closure approximations provide
n-pole Pad\'e approximations to the exact plasma dispersion function
$Z(\zeta)$ that appears in the kinetic plasma response (see
Section \ref{Ch2sec:LFC}). These Pad\'e approximations are thus able to
provide robust results that capture kinetic effects such as Landau
damping, and that can also smoothly transition between the high and
low $\zeta$ regimes.\footnote{The approximations are fairly good
near or above the real $\zeta$ axis, though they will have only a
finite number of damped roots, corresponding to the finite number of
poles in the lower half of the complex plane, while the full
transcendental $Z(\zeta)$ function has an infinite number of damped
roots.}  We have found, not surprisingly, that the fluid
approximations remain robust when collisions are included.  That is,
in all of the numerical tests we have carried out, we have found
good agreement between the results from
equations~(\ref{Ch3eq:p_par})-(\ref{Ch3eq:heat_par}) and the
asymptotic kinetic results from the previous section for the low and
high collisionality regimes. All plots in this chapter are
calculated with the Landau-fluid closure
equations~(\ref{Ch3eq:p_par})-(\ref{Ch3eq:heat_par}).

The Landau-Fluid closure approximations provide a useful way to
extend existing non-linear MHD codes to study key kinetic effects
(see Chapter \ref{chap:chap4}). The closure approximations are
independent of the frequency (or the $Z$ function), so are
straightforward to implement in a nonlinear initial value code
(though, as discussed in Chapter \ref{chap:chap2} they do require
FFT's or non-local heat flux integrals to evaluate some
terms\cite{Snyder1997, Hammett1992}; however, in nonlinear
simulations discussed in Chapter \ref{chap:chap4} we use a simple
local form for heat flux). But one should remember that they are
approximations and so do not accurately model all kinetic effects in
all regimes, particularly near marginal stability
(\cite{Mattor1992,Smith1997,Dimits2000}), though it is generally
found that they work fairly well in strong turbulence regimes
(\cite{Hammett1993,Parker1994,Smith1997,Dimits2000}).

As an aside, we note that the double adiabatic (CGL) closure
\cite{Chew1956}, which is a simpler closure approximation that sets
$q_\Par = q_\Perp = 0$ in equations (\ref{Ch3eq:p_par}) and
(\ref{Ch3eq:p_perp}), generally does a poor job of reproducing the
full kinetic calculations.  This is because the perturbations of
interest have $\omega \ll |k_\Par| c_0$ and are thus far from
adiabatic (see also QDH); moreover, the CGL approximation excludes
kinetic effects like Landau damping.
\section{Collisionality dependence of the MRI growth rate}
\begin{figure}
\begin{center}
\includegraphics[width=2.95in,height=2.5in]{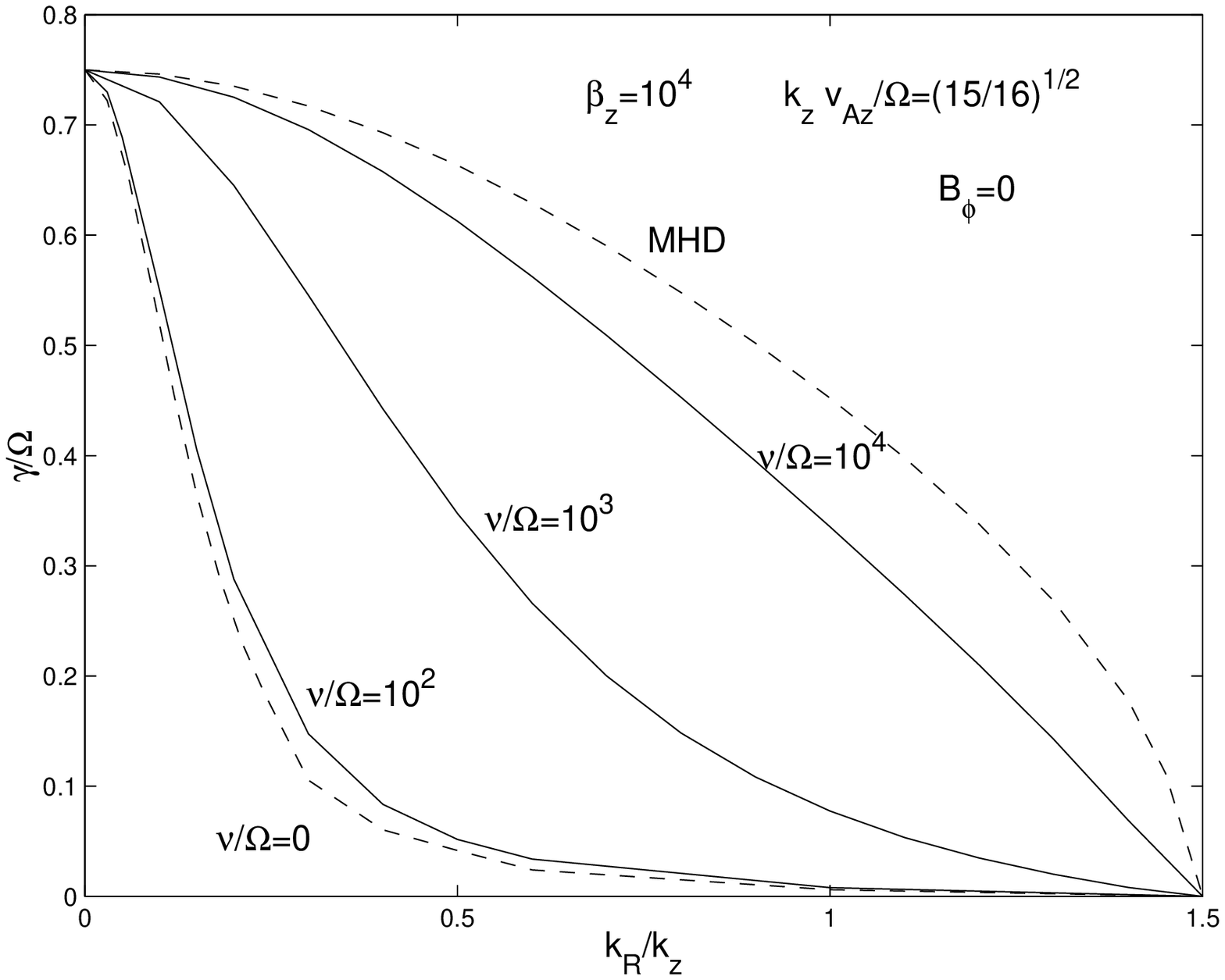}
\includegraphics[width=2.95in,height=2.5in]{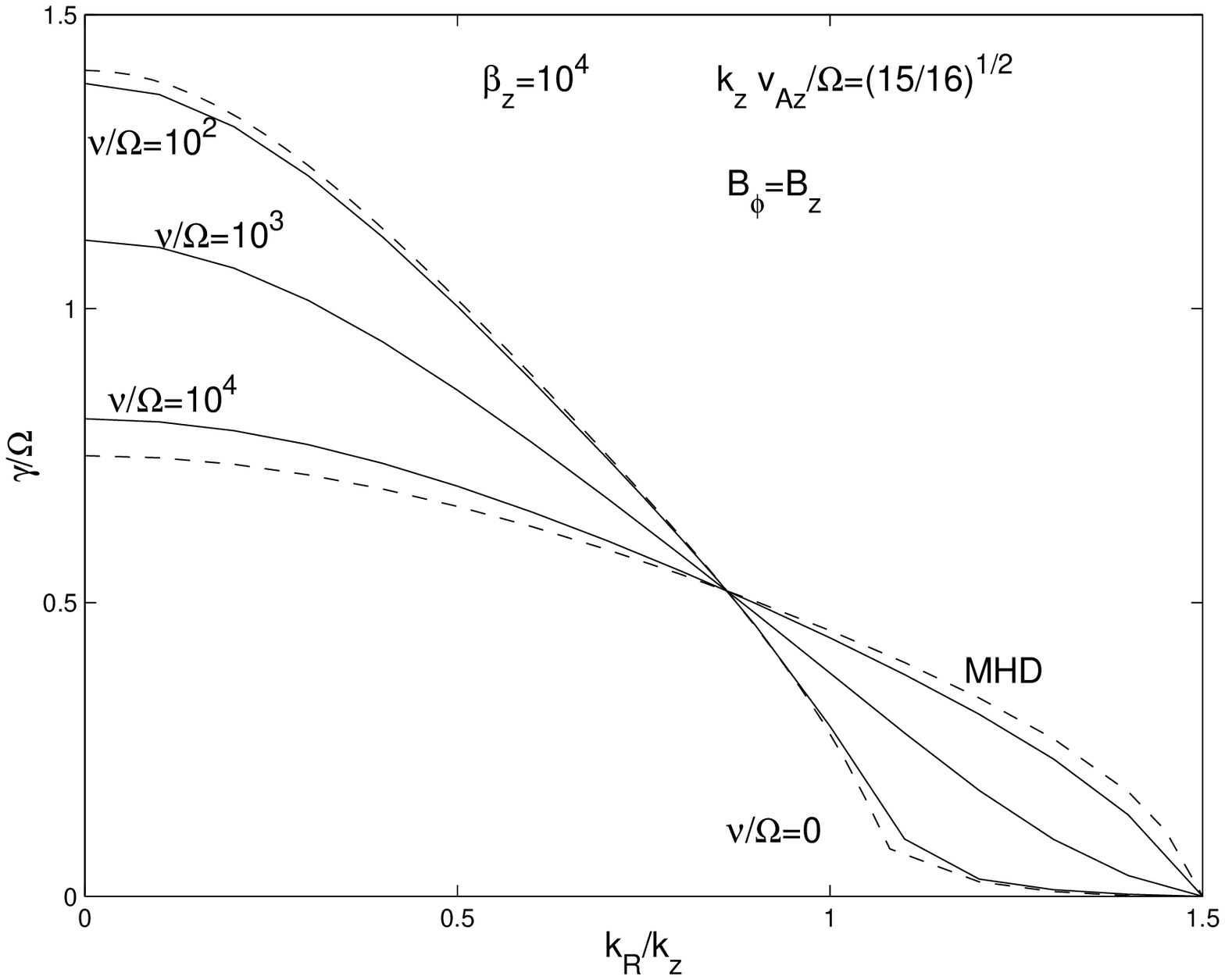}
\caption[Kinetic MRI growth rates with $k_R/k_z$ for $B_\phi=0$ and
$B_\phi=B_z$]{Growth rates of the MRI as a
function of $k_R/k_z$ for different collision frequencies;
$\beta_z=10^4$, $B_{\phi}=0$ for the plot on the left, and
$B_{\phi}=B_z$ for the plot on right. For $\nu/\Omega \geq 10^4$
(=$\beta$; this is the transition to MHD) the growth rates are very
close to the MHD values, while for $\nu/\Omega \leq 10^2$
($=\sqrt{\beta}$; this is the transition to Braginskii regime) they
are quite similar to the collisionless limit. The enhancement of the
growth rate in the collisionless regime for small $k_R$ is the
result of pressure anisotropy.\label{Ch3fig:Fig2}}
\end{center}
\end{figure}
\begin{figure}
\begin{center}
\includegraphics[width=2.95in,height=2.5in]{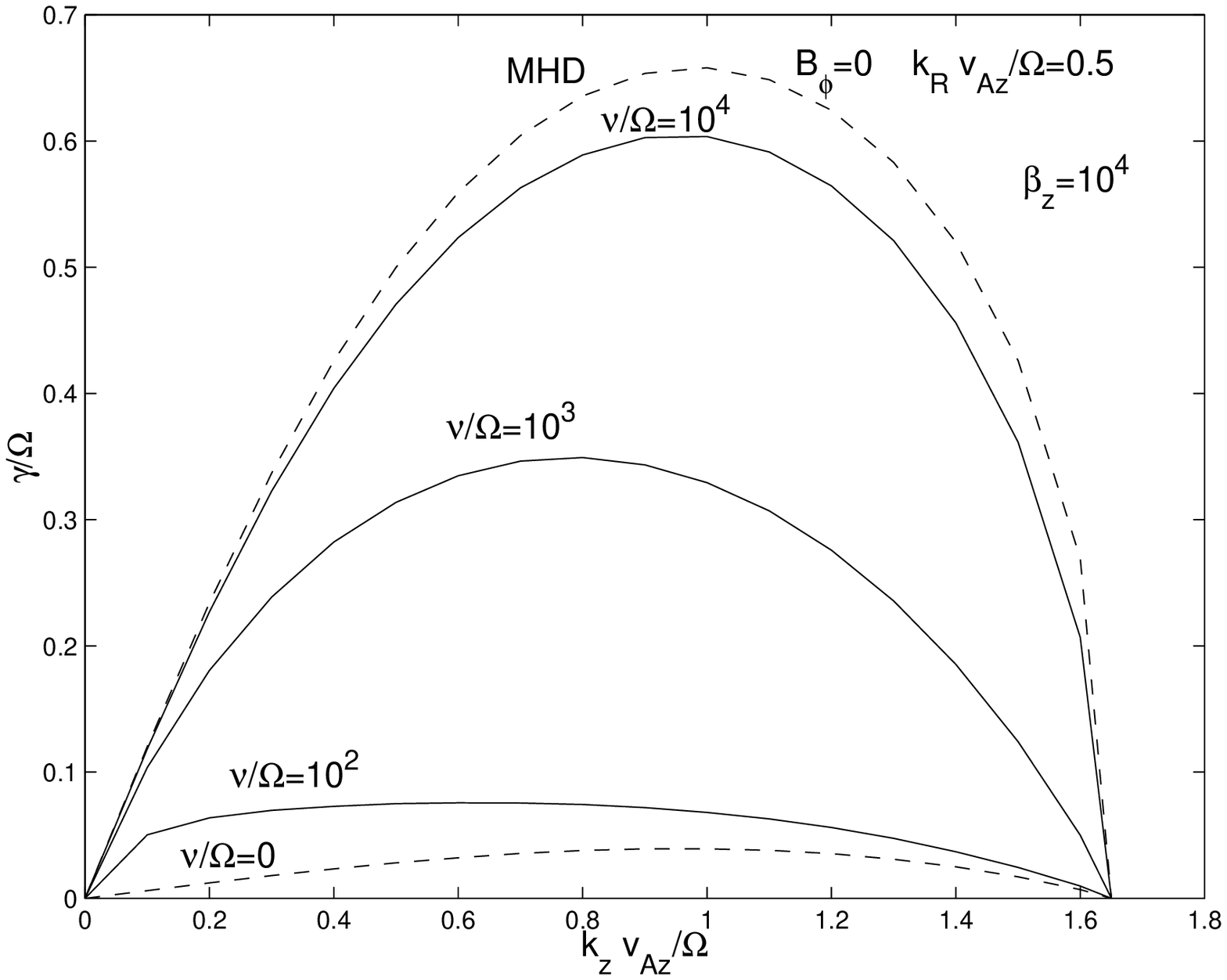}
\includegraphics[width=2.95in,height=2.5in]{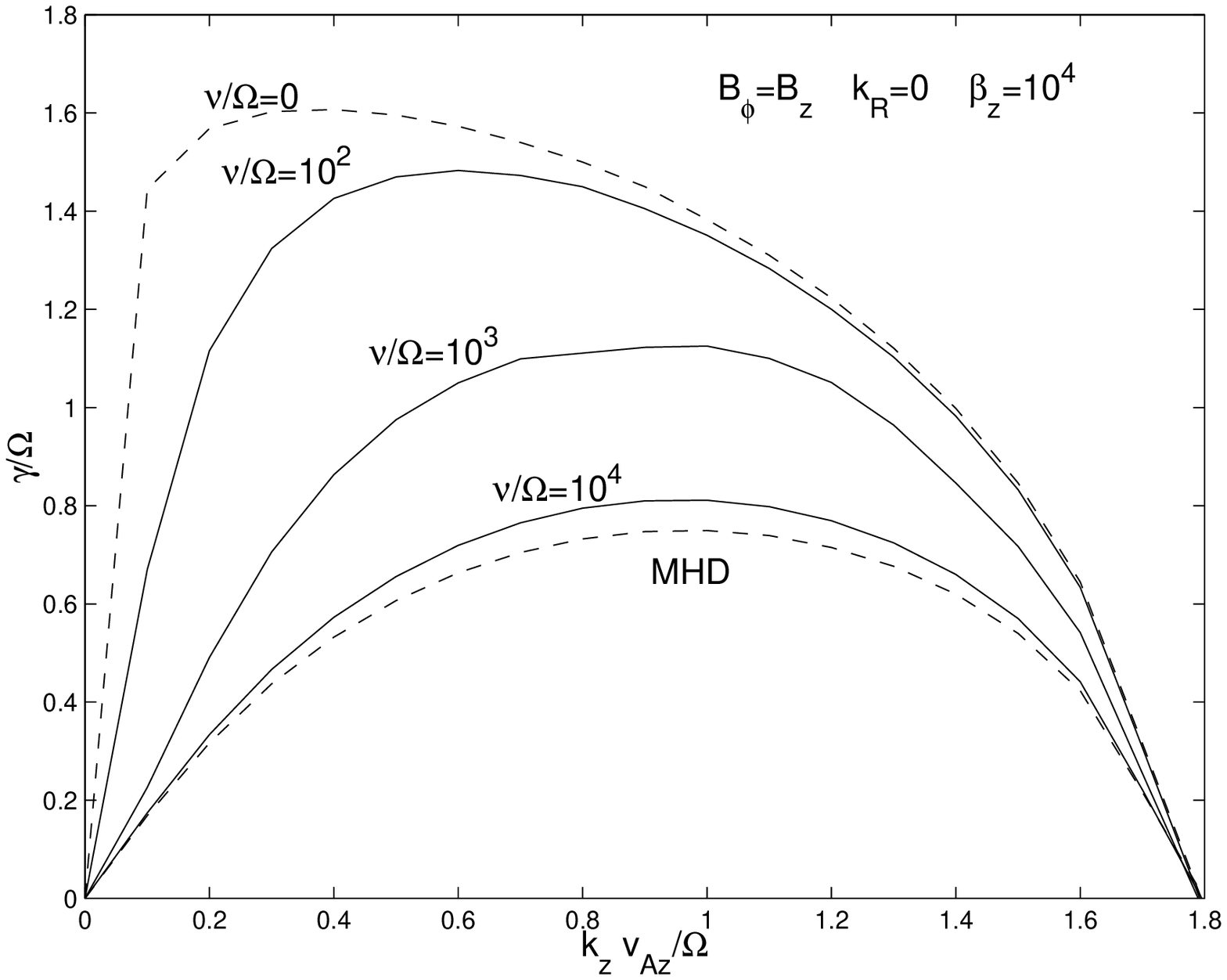}
\caption[Kinetic MRI growth rates with $k_zV_{Az}/\Omega$ for
$B_\phi=0$ and $B_\phi=B_z$]{Growth rates of the
MRI as a function of $k_zV_{Az}/\Omega$ for different collision
frequencies; $\beta_z=10^4$, $B_{\phi}=0$ for the plot on left, and
$B_{\phi}=B_z$ for the plot on right. For $\nu/\Omega \geq \beta$
the growth rates are very close to the MHD values, while for
$\nu/\Omega \leq \sqrt{\beta}$ they are quite similar to the
collisionless limit. Notice that the fastest growing mode in the
collisionless regime is $\approx$ twice faster than the fastest
growing mode in MHD, and also occurs at a much larger length scale.
\label{Ch3fig:Fig4}}
\end{center}
\end{figure}

Figures \ref{Ch3fig:Fig2} and \ref{Ch3fig:Fig4} show the growth rate
of the MRI for intermediate values of collisionality, in addition to
the limits of zero and infinite collision frequency (the MHD limit;
the latter two cases were shown in QDH).  To produce these plots, we
have used equations~(\ref{Ch3eq:lin1})-(\ref{Ch3eq:lin7}) and
(\ref{Ch3eq:p_par})-(\ref{Ch3eq:heat_par}).  These equations were
solved both with a linear initial value code to find the fastest
growing eigenmode, and with MATHEMATICA to find the complete set of
eigenvalues $\omega$.

\begin{figure}
\begin{center}
\includegraphics[width=4in,height=3in]{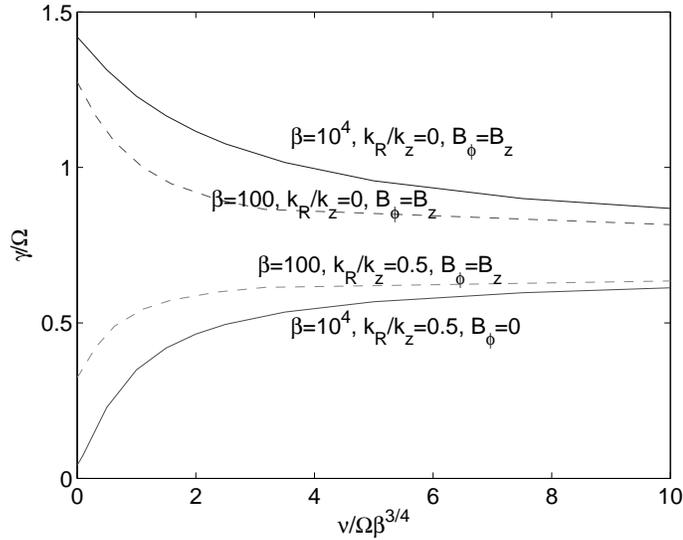}
\caption[MRI growth rate as a function of the collision
frequency]{Variation of the MRI growth rate with collisionality for
$k_R=0$, $B_{\phi}=B_z$ (top curves) and $k_R/k_z=0.5$, $B_{\phi}=0$
(bottom curves). Collisions isotropize the distribution function and
can increase the growth rate in some regimes and decrease it in
others. Solid lines correspond to $\beta=10^4$ and dotted lines to
$\beta=10^2$. \label{Ch3fig:newfig}}
\end{center}
\end{figure}

Figures \ref{Ch3fig:Fig2} and \ref{Ch3fig:Fig4} show that the
transition from the MHD to the collisionless regime is fairly smooth
and occurs, for these particular parameters, in the vicinity of
$\nu/\Omega \sim 10^3$, which corresponds to $\nu \sim \beta^{1/4} k
c_0$, or $k \lambda_{mfp} \sim \beta^{-1/4}$, where $\lambda_{mfp} =
c_0/\nu$ is the mean free path. Figure \ref{Ch3fig:newfig} shows the
growth rate versus collisionality for $\beta_z = 100$ and $\beta_z =
10^4$, and for $B_{\phi}=B_z$, $k_R=0$ and $B_{\phi}=0$,
$k_R/k_z=0.5$.

It is clear from these figures that the transition from the
collisionless to the collisional MRI takes place at far higher
collision rates than $\nu \sim \Omega \sim \omega$.  That is, $\nu >
\omega$ is not a sufficient criterion to be in the collisional
regime. The transition from collisionless to collisional regime can
be understood in terms of the forces in equation of motion: the
isotropic pressure force ($\sim \rho c_0^2$), the anisotropic pressure 
force ($\sim \frac{{\bf \hat{b} \hat{b}:}\grad V}{\nu} \rho c_0^2$), and the
magnetic force ($\sim \rho V_A^2$). For $\nu=0$ and $\beta \gg 1$, the anisotropic
pressure force is comparable to the isotropic pressure and is much
larger than the magnetic force. As $\nu$ is increased, the
anisotropic pressure is reduced in comparison to the isotropic
pressure and the transition to the Braginskii regime occurs when
$\nu \gtrsim k_\parallel c_0$. Transition to MHD occurs on further
increasing the collisionality, as anisotropic pressure becomes
negligible compared to the magnetic force $\nu \gtrsim \beta k_\parallel
V_A = k_\parallel \sqrt{\beta} c_0$. Using $k_\parallel V_A \sim
\Omega$ for the MRI, these transitions are given in terms of the
rotation frequency and $\beta$; collisionless to Braginskii when
$\nu \gtrsim \Omega \sqrt{\beta}$, and Braginskii to MHD when $\nu
\gtrsim \Omega \beta$. Figure \ref{Ch3fig:newfig} clearly shows that
the transition from collisionless to MHD regime occurs roughly when
$\nu \gtrsim \Omega \beta^{3/4}$, the geometric mean of the two
transition collision frequencies.

At high $\beta (\gg 1)$, the Alfv\'en and MRI frequencies are small
compared to the sound wave frequency, and there exists a regime
$\omega \ll \nu \lesssim k_\Par c_0$ where the collisionless results
still hold, despite the collision time being shorter than the growth
rate of the mode.  Physically, this is because in order to wipe out
the pressure anisotropy, that is crucial to the MRI in a
collisionless plasma (see QDH), the collision frequency must be
greater than the sound wave frequency, rather than the (much slower)
growth rate of the mode.  This can also be seen by comparing Figures
\ref{Ch3fig:Fig2} and \ref{Ch3fig:Fig4} with the corresponding
figures in QDH: the effect of increasing collisions (decreasing
pressure anisotropy) is similar to that of decreasing $\beta_z$
(decreasing pressure force relative to magnetic forces). From the
point of view of Snyder et al.'s fluid approach, the weak dependence
of growth rate on collisionality, even if $\nu$ is as large as
$\omega$, is because the terms proportional to $\omega$ and $\nu$ in
Eqs.~(\ref{Ch3eq:p_par}) and~(\ref{Ch3eq:p_perp}) are both much
smaller than the dominant terms involving convection, heat
conduction, and magnetic forces. So the relative magnitudes of
$\omega$ and $\nu$ are not that important, and it is not until $\nu$
is large enough to be relevant in Eqs.
\ref{Ch3eq:p_par}-\ref{Ch3eq:heat_par}, that collisional effects
become noticeable.

\begin{figure}
\begin{center}
\includegraphics[width=5in,height=4in]{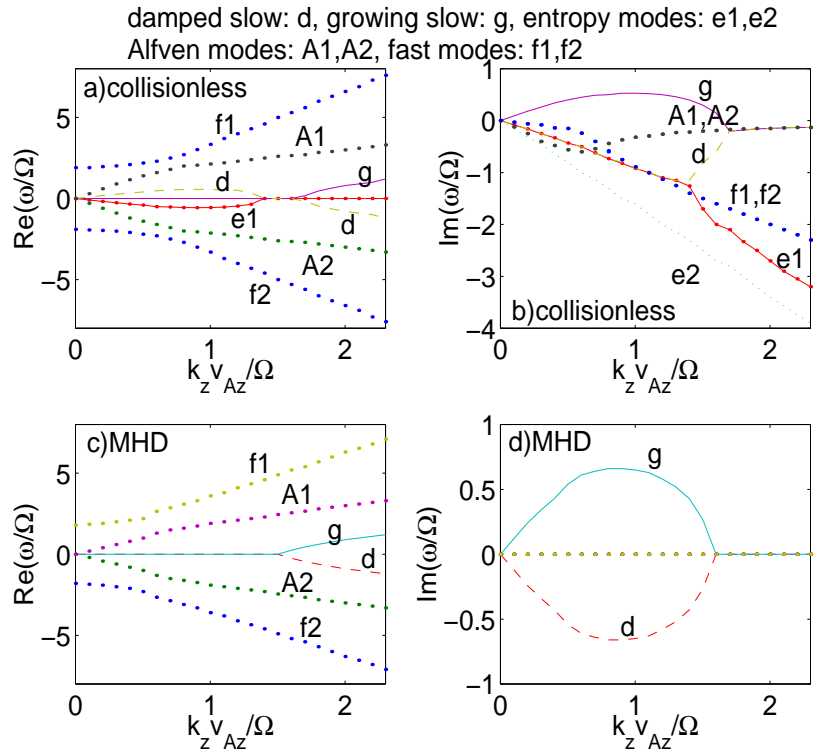}
\caption[Linear modes of a Keplerian disk in collisionless and
collisional regimes]{The real and imaginary parts of the mode
frequency as a function of $k_z$, using collisionless Landau fluid
closures~(a,b) and MHD~(c,d), are shown~($\nu=0$, $k_R
v_{Az}/\Omega=0.5$, $\beta_z=10$, $B_\phi=0$).
\label{Ch3fig:compare}}
\end{center}
\end{figure}

Figure \ref{Ch3fig:compare} shows the complete spectrum of eigenmode
frequencies as $k_z$ is varied, including the propagating and damped
modes, in addition to the unstable MRI branch. We show all the waves
present in collisionless Landau fluid and MHD calculations for a
general choice of wavenumbers and a moderate $\beta_z (=10)$. The
MRI is operational at lower $k_z$, while at high $k_z$ the
eigenfrequencies eventually approach the uniform plasma limit.

Focusing first on the MHD solutions at high $k_z$, we see the
standard set of 3 MHD waves: in order of descending frequency these
are the fast magnetosonic wave, the shear Alfv\'en wave, and the
slow wave. Eqs. \ref{Ch3eq:lin1}-\ref{Ch3eq:lin7} with an MHD
adiabatic pressure equation $\omega \delta p = p_0 {\bf k} \cdot
\delta{\bf v}$ is a set of 8 equations with 8 eigenvalues for
$\omega$.  The standard 3 MHD waves provide 6 of the eigenvalues
($\pm \omega$ for oppositely propagating waves). The remaining roots
are zero frequency modes (not shown in the plot). One is an entropy
mode, corresponding to $\delta \rho/\rho_0 = -\delta T /T_0$ so that
$\delta p=0$.  The other solution corresponds to an unphysical
fluctuation that violates $\grad \cdot {\bf B} =0$, which is
eliminated by imposing the proper initial condition $\grad \cdot
{\bf B} =0$.  At lower $k_z$ in the MHD plots in Figure
\ref{Ch3fig:compare}, the slow mode is destabilized to become the
MRI, as discussed in \cite{Balbus1998}.

Turning next to the collisionless limit in Figure
\ref{Ch3fig:compare}, there are two roots plotted in addition to the
three ``MHD-like'' modes; this is because the single pressure
equation of MHD is replaced by separate equations for the parallel
and perpendicular pressure, so that there are now two entropy-like
modes, both of which have non-zero frequencies but which are also
strongly damped by collisionless heat conduction (which is neglected
in MHD).\footnote{We should point out that while our equations using
the 3+1 Landau-fluid closure approximations have 8 eigenfrequencies,
the equations using the more accurate 4+2 Landau-fluid closure
approximations have 10 eigenfrequencies, with 2 additional strongly
damped roots.  If the exact kinetic response were used, one would
find an infinite number of strongly damped eigenmodes because the
$Z(\zeta)$ function is transcendental.  These strongly damped modes
are related to ``ballistic modes'' and transients in the standard
analysis of Landau damping.}

The fast, Alfv\'en, and slow waves in the collisionless calculation
can again be identified in order of decreasing (real) frequency at
high $k_z$. At lower $k_z$, one of the slow modes becomes
destabilized to become the MRI, as in MHD. Unlike in MHD, however,
the fast magnetosonic waves are strongly Landau damped since the
resonance condition $\omega \sim k_\Par c_0$ is easily satisfied. In
addition, it is interesting to note that both the shear Alfv\'en and
slow waves have some collisionless damping at the highest $k_z$ used
in this plot, though the damping will approach zero for very high
$k_z$.  In a uniform plasma the shear Alfv\'en wave is undamped
unless its wavelength is comparable to the proton Larmor radius or
its frequency is comparable to the proton cyclotron frequency
(neither of which is true for the modes considered here).  By
contrast, the slow mode is strongly damped unless $k_\Perp \ll
k_\Par$ (because $\delta v_\parallel \propto (k_\perp/k_\parallel)
\ll c_0$ in this regime; e.g., \cite{Barnes1966,Foote1979}). The
damping of small $k_z$ shear Alfv\'en waves in Figure
\ref{Ch3fig:compare} is because our background plasma is rotating so
the uniform-plasma modes are mixed together. Thus the well-known
dissipation of the slow mode by transit-time damping also leads to
damping of what we identify as the shear Alfv\'en wave (based on its
high $k_z$ properties).
\section{Summary and Discussion}
In this chapter we have extended the linear, axisymmetric, kinetic
magnetorotational instability (MRI) calculation of QDH to include
the effect of collisions. As the collision frequency is increased,
the MRI transitions from collisionless to Braginskii regime ($\nu
\gtrsim \Omega \sqrt{\beta}$), and eventually to the MHD regime
($\nu \gtrsim \Omega \beta$). Interestingly, the collisionless MRI
results hold not only if $\nu \ll \omega$, but even when $\omega \ll
\nu \ll k_{\Par} c_0$. This intermediate regime can exist in $\beta
\gtrsim 1$ plasmas because the MRI growth rate is slow compared to
the sound wave frequency, $\omega \sim k_\Par V_A = k_\Par c_0
\sqrt{2/\beta} \ll k_\Par c_0$. The fastest growing collisionless
MRI mode is $\approx$ twice faster than the fastest growing MHD
mode, and occurs at a much larger length scale; thus, MRI in the
collisionless regime can result in fast MHD dynamo at large scales
(not much smaller than the disk height scale).

If we consider the application of our results to accretion flows,
the collisionless limit will be applicable so long as $\nu/\Omega
\lesssim \sqrt{\beta}$.  This condition is amply satisfied for
proton-proton and proton-electron collisions in all hot radiatively
inefficient accretion flow models (see Table \ref{Ch1tab:SgrA}),
suggesting that the collisionless limit is always appropriate.
However, high frequency waves such as ion-cyclotron waves can
isotropize the proton distribution function and thus provide an
effective ``collision'' term crudely analogous to the one considered
here (see Subsection \ref{Ch4subsec:Isotropization}). In the drift kinetic
limit, when the Larmor radius is small compared to the dynamical
length scales, the adiabatic invariant $\mu=p_\perp/B$ is conserved.
Nonlinear simulations described in Chapter \ref{chap:chap4} show
that the MRI results in fast growth of magnetic fields resulting in
an anisotropic plasma ($p_\perp>p_\perp$). Fairly large pressure
anisotropies ($\Delta p/p \sim ({\rm a~few})/\beta$) are created at
the dynamical timescales and small scale instabilities---mirror and
ion-cyclotron---are excited. Pressure isotropization due to these
instabilities imposes an MHD like dynamics on a formally
collisionless plasma. However, selective heating of
resonant electrons and ions may result in different electron and ion 
temperatures, and spectral signatures different from MHD.

One might anticipate that the linear differences between the
collisionless and collisional MRI highlighted here and in QDH will
imply differences in the nonlinear turbulent state in hot accretion
flows (see, e.g., \cite{Hawley2002,Igumenshchev2003} for global MHD
simulations of such flows). Not only are there differences in the
linear growth rates of the instability that drives turbulence, but
the spectrum of damped modes is also very different. In particular,
in the kinetic regime there exist modes at all scales in $|{\bf k}|$
that are subject to Landau/Barnes collisionless damping, while in
the MHD regime the only sink for turbulent energy is due to
viscosity/resistivity at very small scales (very high $|{\bf k}|$).
Indeed, as we have shown, even long wavelength Alfv\'en waves can be
damped by collisionless effects because of the mixture of
uniform-plasma modes in the differentially rotating accretion flow
(Figure \ref{Ch3fig:compare}).  Whether these differences are
important or not may depend on how efficiently nonlinearities couple
energy into the damped modes.  These could modify the nonlinear
saturated turbulent spectrum (e.g., the efficiency of angular
momentum transport) or the fraction of electron vs.\ ion heating
(the heating may also be anisotropic), which in turn determines the
basic observational signatures of hot accretion flows (the accretion
rate and the radiative efficiency).  One approach for investigating
nonlinear collisionless effects would be to extend existing MHD
codes to include anisotropic pressure, the fluid closure
approximations for kinetic effects \cite{Snyder1997}, and the BGK
collision operator considered here. By varying the collision
frequency, one can then scan from the collisionless kinetic to the
collisional MHD regime, and assess any differences in the nonlinear
turbulent state. The nonlinear simulations of the collisionless MRI,
based on Landau fluid closure for heat fluxes, are described in the
next chapter.

\chapter{Nonlinear Simulations of kinetic MRI}
\label{chap:chap4}
In this chapter we describe local shearing box simulations of
turbulence driven by the magnetorotational instability (MRI) in a
collisionless plasma. Collisionless effects may be important in
radiatively inefficient accretion flows, such as near the black hole
in the Galactic center (see Section \ref{Ch1sec:RIAFs}). The ZEUS MHD code
is modified to evolve an anisotropic pressure tensor. A Landau-fluid
closure approximation is used to calculate heat conduction along
magnetic field lines. The anisotropic pressure tensor provides a
qualitatively new mechanism for transporting angular momentum in
accretion flows (in addition to the Maxwell and Reynolds stresses).
We estimate limits on the pressure anisotropy due to pitch angle
scattering by kinetic instabilities. Such instabilities provide an
effective ``collision'' rate in a collisionless plasma and lead to
more MHD-like dynamics. We find that the MRI leads to efficient
growth of the magnetic field in a collisionless plasma, with
saturation amplitudes comparable to those in MHD.  In the saturated
state, the anisotropic stress is comparable to the Maxwell stress,
implying that the rate of angular momentum transport may be
moderately enhanced in a collisionless plasma. More importantly,
heating due to anisotropic stress is comparable to the numerical
energy loss in updating magnetic fields; this can have important
consequences for electron and ion heating.

\section{Introduction}
Following the seminal work of Balbus and Hawley~\cite{Balbus1991},
numerical simulations have demonstrated that magnetohydrodynamic
(MHD) turbulence initiated by the magnetorotational instability
(MRI) is an efficient mechanism for transporting angular momentum in
accretion disks (see Section \ref{Ch1sec:MRI} for a review). For a broad
class of astrophysical accretion flows, however, the MHD assumption
is not directly applicable.  In particular, in radiatively
inefficient accretion flow (RIAF) models for accretion onto compact
objects, the accretion proceeds via a hot, low density,
collisionless plasma with the proton temperature larger than the
electron temperature \cite{Narayan1998,Quataert2003} (see
Section \ref{Ch1sec:RIAFs} for a review). In order to maintain such a
two-temperature flow the plasma must be collisionless, with the
Coulomb mean-free path many orders of magnitude larger than the
system size (see Table \ref{Ch1tab:SgrA} for plasma parameters in
Sgr A$^*$). Motivated by the application to RIAFs, this chapter
studies the nonlinear evolution of the collisionless MRI in the
local shearing box limit.

Quataert, Dorland, \& Hammett (2001; hereafter QDH) and Sharma,
Hammett, \& Quataert (2003; hereafter SHQ) showed that the linear
dynamics of the MRI in a collisionless plasma can be quite different
from that in MHD (see Chapter \ref{chap:chap3}).  The maximum growth
rate is a factor of $\approx 2$ larger and, perhaps more
importantly, the fastest growing modes can shift to much longer
wavelengths, giving direct amplification of long wavelength modes.
Dynamical instability exists even when the magnetic tension forces
are negligible because of the anisotropic pressure response in a
collisionless plasma.  In related work using Braginskii's
anisotropic viscosity, the collisionless MRI is studied as the
``magnetoviscous'' instability \cite{Balbus2004,Islam2005}.

We are interested in simulating the dynamics of a collisionless
plasma on length-scales ($\sim$ disk height) and time-scales ($\sim$
orbital period) that are very large compared to the microscopic
plasma scales (such as the Larmor radius and the cyclotron period).
Since the ratio of the size of the accretion flow to the proton
Larmor radius is $\sim 10^{8}$ for typical RIAF models (see Table
\ref{Ch1tab:SgrA}), direct particle methods such as PIC (particle in
a cell), which need to resolve both of these scales, are
computationally challenging and require simulating a reduced range
of scales. Instead, we use a fluid-based method to describe the
large-scale dynamics of a collisionless plasma (``kinetic MHD,''
described in Section \ref{Ch2sec:KMHD}). The key differences with respect to
MHD are that the pressure is a tensor rather than a scalar,
anisotropic with respect to the direction of the local magnetic
field, and that there are heat fluxes along magnetic field lines
(related to Landau damping and wave-particle interactions). The
drawback of our fluid-based method is, of course, that there is no
exact expression for the heat fluxes if only a few fluid moments are
retained in a weakly collisional plasma (the ``closure problem'').
We use results from Snyder, Hammett, \& Dorland (1997; hereafter
SHD) who have derived approximations for the heat fluxes in terms of
nonlocal parallel temperature and magnetic field gradients. These
heat flux expressions can be shown to be equivalent to multi-pole
Pad\'e approximations to the $Z$-function involved in Landau damping
(see Section \ref{Ch2sec:LFC}).  This approach can be shown to converge as
more fluid moments of the distribution function are kept
{\cite{Hammett1993}}, just as an Eulerian kinetic algorithm
converges as more grid points in velocity space are kept. These
fluid-based methods have been applied with reasonable success to
modeling collisionless turbulence in fusion plasmas, generally
coming within a factor of 2 of more complicated kinetic calculations
in strong turbulence regimes
\cite{Dimits2000,Parker1994,Hammett1993,Scott2005}, though there can
be larger differences in weak turbulence regimes
\cite{Hammett1993,Dimits2000}. The simulations we report here use an
even simpler local approximation to the heat flux closures than
those derived in \cite{Snyder1997} (see ``the crude closure" in
Section \ref{Ch2sec:LFC}). While not exact, these closure approximations
allow one to begin to investigate kinetic effects with relatively
fast modifications of fluid codes; whereas, solving the full drift
kinetic equation (see Section \ref{Ch2sec:DKE}) is considerably slower and
requires code development and testing from scratch.

In a collisionless plasma the magnetic moment, $\mu=v_\perp^2/2B$,
is an adiabatic invariant. Averaged over velocity space, this leads
to conservation of $\langle \mu \rangle =p_\Perp/(\rho B)$. As a
result, pressure anisotropy with $p_\Perp>p_\Par$ is created as the
MRI amplifies the magnetic field in the accretion flow.  This
pressure anisotropy creates an anisotropic stress (like a
viscosity!) which can be as important for angular momentum transport
as the magnetic stress. It is interesting to note that for cold
disks, the mean free path is negligible compared to the disk height
resulting in a viscosity insufficient to account for efficient
transport; but hot, thick accretion flows are collisionless with
large viscosity, and viscous stress is quite efficient in
transporting angular momentum. However, it is important to emphasize
that an anisotropic viscosity in a collisionless, magnetized plasma
is very different from an isotropic viscosity (since viscosity
perpendicular to the field lines is vanishingly small). Although,
the Reynolds number ($Re \equiv VL/\eta_V$) based on parallel
viscosity is small, $O(1)$, the plasma is turbulent; this would not
be true if the Reynolds number based on an isotropic viscosity is so
small.

The pressure anisotropy cannot, however, grow without bound because
high frequency waves and kinetic microinstabilities feed on the free
energy in the pressure anisotropy, effectively providing an enhanced
rate of collisions that limit the pressure tensor anisotropy
(leading to more MHD-like dynamics in a collisionless plasma).  We
capture this physics by using a subgrid model to restrict the
allowed amplitude of the pressure anisotropy.  This subgrid model
(described in \S 2.3) is based on existing linear and nonlinear
studies of instabilities driven by pressure anisotropy
\cite{Hasegawa1969,Gary1997}.

The remainder of this paper is organized as follows. We begin with
Kulsrud's formulation of kinetic MHD (KMHD) and our closure model
for the heat fluxes in a collisionless plasma. We also include a
linear analysis of the MRI in the presence of a background pressure
anisotropy and describe limits on the pressure anisotropy set by
kinetic instabilities.  Next, we describe our modifications to the
ZEUS code to model kinetic effects. We present our primary results
on the nonlinear evolution of the MRI in a collisionless plasma. At
the end we discuss these results, their astrophysical implications,
and future work required to understand the global dynamics of
collisionless accretion disks.

\section{Governing equations}
In the limit that all fluctuations of interest are at scales larger
than the proton Larmor radius and have frequencies much smaller than
the proton cyclotron frequency, a collisionless plasma can be
described by the following magnetofluid equations
\cite{Kulsrud1983,Snyder1997} (see Section \ref{Ch2sec:KMHD} for details):
\ba \label{Ch4eq:MHD1} \frac{\partial \rho}{\partial t} & + & \nabla
\cdot \left(\rho {\bf V}\right)=0,
\\
\label{Ch4eq:MHD2} \rho \frac{\partial {\bf V}}{\partial t} & + &
\rho\left({\bf V} \cdot \nabla\right) {\bf V}= \frac{\left(\nabla
\times {\bf B}\right) \times {\bf B}}{4\pi}
- \nabla \cdot {\bf P} + {\bf F_g},\\
\label{Ch4eq:MHD3} \frac{\partial {\bf B}}{\partial t} &=& \nabla
\times \left({\bf V} \times {\bf
B}\right), \\
\label{Ch4eq:MHD4}
 {\bf P}&=& p_{\Perp} {\bf I} + \left(p_{\Par}- p_{\Perp}\right){\bf
\hat{b}\hat{b}} = p_\Perp {\bf I} + {\bf \Pi}, \ea where $\rho$ is
the mass density, ${\bf V}$ is the fluid velocity, ${\bf B}$ is the
magnetic field, ${\bf F_g}$ is the gravitational force, ${\bf
\hat{b}}={\bf B}/|{\bf B}|$ is a unit vector in the direction of the
magnetic field, and ${\bf I}$ is the unit tensor. In
equation~(\ref{Ch4eq:MHD3}) an ideal Ohm's law is used, neglecting
resistivity.  In equation (\ref{Ch4eq:MHD4}), ${\bf P}$ is the pressure
tensor with different perpendicular~($p_{\Perp}$) and
parallel~($p_{\Par}$) components with respect to the background
magnetic field, and ${\bf \Pi} = {\bf \hat{b} \hat{b}} (p_\Par -
p_\Perp)$ is the anisotropic stress tensor. (Note that ${\bf \Pi}$
is not traceless in the convention used here.) ${\bf P}$ should in
general be a sum over all species but in the limit where ion
dynamics dominate and electrons just provide a neutralizing
background, the pressure can be interpreted as the ion pressure.
This is the case for hot accretion flows in which $T_p \gg T_e$.

The exact pressures $p_\Par$ and $p_\Perp$ can be rigorously
determined by taking moments of the drift kinetic equation (see
Section \ref{Ch2sec:DKE}), \be \label{Ch4eq:DKE} \frac{\partial
f_s}{\partial t} + (v_\Par {\bf \hat{b} + V_E}) \cdot \nabla f_s +
\left [ -{\bf \hat{b}} \cdot \frac{D {\bf V_E}}{Dt} - \mu {\bf
\hat{b}} \cdot \nabla B + \frac{e_s}{m_s} \left ( E_\Par +
\frac{F_{g\Par}}{e_s} \right ) \right ] \frac{\partial f_s}{\partial
v_\Par} = C(f_s), \ee which is the asymptotic expansion of the
Vlasov equation for the distribution function $f_s({\bf x}, \mu,
v_\Par, t)$ for species `$s$' with mass $m_s$ and charge $e_s$ in
the limit $\rho_s/L \ll 1$, $\omega/\Omega_s \ll 1$, where $\rho_s$
and $\Omega_s$ are the gyroradius and gyrofrequency, respectively.
In equation (\ref{Ch4eq:DKE}), ${\bf V_E}=c({\bf E} \times {\bf
B})/B^2$ is the perpendicular drift velocity, $\mu=({\bf v_\Perp}
-{\bf V_E})^2/2B$ is the magnetic moment (a conserved quantity in
the absence of collisions), $F_{g \Par}$ is the component of the
gravitational force parallel to the direction of the magnetic field,
and $D/Dt=\partial/\partial t + (v_\Par {\bf \hat{b}+V_E}) \cdot
\nabla$ is the particle Lagrangian derivative in the phase space.
The fluid velocity ${\bf V}= {\bf V_E + \hat{b}} V_\Par$, so the
${\bf E} \times {\bf B}$ drift is determined by the perpendicular
component of equation~(\ref{Ch4eq:MHD2}).  Other drifts such as grad
B, curvature, and gravity $\times {\bf B}$ drifts are higher order
in the drift kinetic ordering and do not appear in this equation. In
equation (\ref{Ch4eq:DKE}), $C(f_s)$ is the collision operator to
allow for generalization to collisional regimes. Collisions can also
be used to mimic rapid pitch angle scattering due to high frequency
waves that break $\mu$ invariance. The parallel electric field is
determined by $E_\| = \sum_s (e_s/m_s) {\bf \hat{b}} \cdot \grad
\cdot {\bf P}_s/\sum_s (n_s e_s^2/m_s) $, which insures
quasineutrality (see Subsection \ref{Ch2subsec:Moments}).

Separate equations of state for the parallel and perpendicular
pressures can be obtained from the moments of the drift kinetic
equation \cite{Chew1956}.  Neglecting the collision term these are:
\ba \label{Ch4eq:CGL1} \rho B \frac{D}{Dt} \left(
\frac{p_\Perp}{\rho B} \right) &=& - \nabla \cdot
{\bf  q_\Perp} - q_\Perp \nabla \cdot {\bf \hat{b}}, \\
\label{Ch4eq:CGL2} \frac{\rho^3}{B^2} \frac{D}{Dt} \left( \frac{p_\Par
B^2}{\rho^3} \right) &=& -\nabla \cdot {\bf q_\Par} + 2 q_\Perp
\nabla \cdot {\bf\hat{b}}, \ea where $D/Dt = \partial/\partial t +
{\bf V} \cdot \nabla$ is the fluid Lagrangian derivative and ${\bf
q_{\Par,\perp}} = q_{\Par, \Perp} {\bf \hat{b}}$ are the heat fluxes
(flux of $p_\Par$ and $p_\Perp$) parallel to the magnetic field.
The equation for the magnetic moment density $\rho \langle \mu
\rangle =p_\Perp/B$ can be written in a conservative form: \be
\label{Ch4eq:mucon} \frac{\partial}{\partial t} \left(
\frac{p_\Perp}{B} \right) + \nabla \cdot \left(\frac{p_\Perp}{B}
{\bf V} \right) = - \nabla \cdot \left( \frac{q_\Perp}{B}{\bf
\hat{b}} \right), \ee  If the heat fluxes are neglected (called the
CGL or double adiabatic limit), as the magnetic field strength ($B$)
increases, $p_\Perp$ increases ($p_\Perp \propto \rho B$), and
$p_\Par$ decreases ($p_\Par \propto \rho^3/B^2$). Integrating
equation~(\ref{Ch4eq:mucon}) over a finite periodic (even a shearing
periodic) box shows that $\langle p_\Perp/B \rangle$ is conserved,
where $\langle \rangle$ denotes a volume average. This implies that
even when $q_{\Par,\Perp} \ne 0$, $p_\Perp$ increases in a volume
averaged sense as the magnetic energy in the box increases. This
means that that for a collisionless plasma, pressure anisotropy
$p_\Perp >(<)\,p_\Par$ is created as a natural consequence of
processes that amplify (reduce) $B$. This pressure anisotropy is
crucial for understanding magnetic field amplification in
collisionless dynamos.

To solve the set of equations (\ref{Ch4eq:MHD1}-\ref{Ch4eq:MHD4}),
(\ref{Ch4eq:CGL1}-\ref{Ch4eq:CGL2}) in a simple fluid based
formalism, we require expressions for $q_\Par$ and $q_\Perp$ in
terms of lower order moments. No simple, exact expressions for
$q_\Par$ and $q_\Perp$ exist for nonlinear collisionless plasmas.
Although simple, the double adiabatic or CGL approximation (where
$q_\Par= q_\Perp=0$) does not capture key kinetic effects such as
Landau damping.  In the moderately collisional limit ($\rho_i <$
mean free path $<$ system size), where the distribution function is
not very different from a local Maxwellian, one can use the
Braginskii equations to describe anisotropic transport
(\cite{Braginskii1965}; see \cite{Balbus2000,Balbus2004} for
astrophysical applications). However, in the hot RIAF regime, the
mean free path is often much larger than the system size and the
Braginskii equations are not formally applicable, though they are
still useful as a qualitative indication of the importance of
kinetic effects. The collisional limit of the kinetic MHD equations
can be shown to recover the dominant anisotropic heat flux and
viscosity tensor of Braginskii (see Subsection \ref{Ch2subsec:Braginskii}).
The local approximation to kinetic MHD that we use here leads to
equations that are similar in form to Braginskii MHD, but with
separate dynamical equations for parallel and perpendicular
pressures.  We also add models for enhanced pitch angle scattering
by microinstabilities, which occur at very small scales and high
frequencies beyond the range of validity of standard kinetic MHD.
\footnote{This would also be needed when using Braginskii equations,
because they are not necessarily well posed in situations where the
anisotropic stress tensor can drive arbitrarily small scale
instabilities\cite{Schekochihin2005}.}

Hammett and collaborators have developed approximate fluid closures
(called Landau fluid closure) for collisionless plasmas
\cite{Hammett1990,Hammett1992,Snyder1997} that capture kinetic
effects such Landau damping. SHD \cite{Snyder1997} give the
resulting expressions for parallel heat fluxes ($q_\Par$, $q_\Perp$)
to be used in equations (\ref{Ch4eq:CGL1}) and (\ref{Ch4eq:CGL2}).
Landau closures are based on Pad\'e approximations to the full
kinetic plasma dispersion function that reproduce the correct
asymptotic behavior in both the adiabatic ($\omega/k_\Par c_\Par \gg
1$) and isothermal ($\omega/k_\Par c_\Par \ll 1$) regimes (and
provide a good approximation in between), where $\omega$ is the
angular frequency, $k_\Par$ is the wavenumber parallel to the
magnetic field, and $c_\Par=\sqrt{p_\Par/\rho}$ is the parallel
thermal velocity of the particles. In Fourier space, the linearized
heat fluxes can be written as equations (39) \& (40) in SHD, \ba
q_\Par &=& -\sqrt{\frac{8}{\pi}} \rho_0 c_{\Par 0}
\frac{ik_\Par \left( p_\Par/\rho \right)}{|k_\Par|}, \\
q_\Perp &=& -\sqrt{\frac{2}{\pi}} \rho_0  c_{\Par 0}
   \frac{ik_\Par \left (p_\Perp/\rho \right)}{|k_\Par|}
   + \sqrt{\frac{2}{\pi}}  c_{\Par 0} \frac{p_{\Perp 0}}{B_0}
   \left ( 1-  \frac{p_{\Perp 0}}{p_{\Par 0} }
   \right) \frac{ik_\Par B}{|k_\Par|},
\ea where $`0'$ subscripts indicate equilibrium quantities.  Real
space expressions are somewhat more cumbersome and are given by
convolution integrals (see Section \ref{Ch2sec:LFC}) \ba
  \label{Ch4eq:nonlocal1}
q_\Par &=& - \left( \frac{2}{\pi}\right)^{3/2} n_0  c_{\Par 0 }
    \int_0^\infty \delta z^\prime
    \frac{T_\Par(z+z^\prime)-T_\Par(z-z^\prime)}{z^\prime}, \\
\label{Ch4eq:nonlocal2}  \nonumber q_\Perp &=& - \left(
\frac{2}{\pi^3} \right)^{1/2} n_0 c_{\Par 0} \int_0^\infty \delta
z^\prime \frac{T_\Perp(z+z^\prime)-T_\Perp(z-z^\prime)}{z^\prime} \\
&+& \left( \frac{2}{\pi^3} \right)^{1/2} c_{\Par 0} \left(1-
\frac{p_{\Perp 0}}{p_{\Par 0}}  \right) \frac{  p_{\Perp 0} }{B_0}
\times \int_0^\infty \delta z^\prime
\frac{B(z+z^\prime)-B(z-z^\prime)}{z^\prime}, \ea where $n_0$ is the
number density, $T_\Par=p_\Par/n$, and $T_\Perp=p_\Perp/n$ are the
parallel and perpendicular temperatures, and $z^\prime$ is the
spatial variable along the magnetic field line. In the previous
chapter (based on \cite{Sharma2003}) we have shown that these fluid
closures for the heat fluxes accurately reproduce the kinetic linear
Landau damping rate for all MHD modes (slow, Alf\'ven, fast and
entropy modes). The growth rate of the MRI using the Landau closure
model is also very similar to that obtained from full kinetic
theory. As noted in the introduction, in addition to reproducing
linear modes/instabilities, Landau fluid closures have also been
used to model turbulence in fusion plasmas with reasonable success.

These closure approximations were originally developed for
turbulence problems in fusion energy devices with a strong guide
magnetic field, where the parallel dynamics is essentially linear
and FFTs could be easily used to quickly evaluate the Fourier
expressions above.  In astrophysical problems with larger amplitude
fluctuations and tangled magnetic fields, evaluation of the heat
fluxes become somewhat more complicated.  One could evaluate the
convolution expressions, equations (\ref{Ch4eq:nonlocal1}) and
(\ref{Ch4eq:nonlocal2}) (with some modest complexity involved in
writing a subroutine to integrate along magnetic field lines),
leading to a code with a computational time $T_{cpu} \propto N_x^3
N_\Par$, where $N_x^3$ is the number of spatial grid points and
$N_\Par$ is the number of points kept in the integrals along field
lines.  (In some cases, it may be feasible to map the fluid
quantities to and from a field-line following coordinate system so
that FFTs can reduce this to $T_{cpu} \propto N_x^3 \log N_\Par$.)
While this is more expensive than simple MHD where $T_{cpu} \propto
N_x^3$, it could still represent a savings over a direct solution of
the drift kinetic equation, which would require $T_{cpu} \propto
N_x^3 N_{v_\Par} N_{v_\Perp}$, where $N_{v_\Par} N_{v_\Perp}$ is the
number of grid points for velocity space.\footnote{On the other
hand, an effective hyperdiffusion operator in velocity space may
reduce the velocity resolution requirements, and recent direct
kinetic simulations of turbulence in fusion devices have found that
often one does not need very high velocity resolution.  This may
make a direct solution of the drift kinetic equation tractable for
some astrophysical kinetic MHD problems. Furthermore, a direct
solution of the drift kinetic equation involves only local
operations, and thus is somewhat easier to parallelize than the
convolution integrals.}

As a first step for studying kinetic effects, in this paper we pick
out a characteristic wavenumber $k_L$ that represents the scale of
collisionless damping and use a local approximation for the heat
fluxes in Fourier space (see the ``crude closure for thermal
conduction" in Section \ref{Ch2sec:LFC}), with a straightforward assumption
about the nonlinear generalization: \ba \label{Ch4eq:qpar1} q_\Par
&=& -\sqrt{\frac{8}{\pi}} \rho  c_{\Par}
\frac{\nabla_\Par \left( p_\Par/\rho \right)}{k_L}, \\
\label{Ch4eq:qperp1} q_\Perp &=& -\sqrt{\frac{2}{\pi}} \rho c_{\Par}
\frac{\nabla_\Par \left( p_\Perp/\rho \right)}{k_L} +
\sqrt{\frac{2}{\pi}} c_\Par  p_\Perp \left( 1-
\frac{p_\Perp}{p_\Par} \right) \frac{\nabla_\Par B}{k_LB}.  \ea Note
that this formulation of the heat flux is analogous to a Braginskii
heat conduction along magnetic field lines.  For linear modes with
$|k_\Par| \sim k_L$, these approximations will of course agree with
kinetic theory as well as the Pad\'e approximations shown in SHD.
One can think of $k_L$ as approximately controlling the heat
conduction rate, though this does not necessarily affect the
resulting Landau damping rate of a mode in a monotonic way, since
this sometimes exhibits impedance matching behavior,i.e., some modes
are weakly damped in both the small and large (isothermal) heat
conduction limits.  We vary $k_L$ to investigate the sensitivity of
our results to this parameter.

\subsection{Linear modes}
\label{Ch4sec:linmodes}

Since pressure anisotropy arises as a consequence of magnetic field
amplification in a collisionless plasma, it is of interest to repeat
the linear analysis of the collisionless MRI done in Chapter
\ref{chap:chap3}, but with a background pressure anisotropy
($p_{\Par 0} \ne p_{\Perp 0}$). We consider the simple case of a
vertical magnetic field. This analysis provides a useful guide to
understanding some of our numerical results.

We linearize equations (\ref{Ch4eq:MHD1})-(\ref{Ch4eq:MHD4}) for a
differentially rotating disk (${\bf V_0}= R \Omega(R) {\bf
\hat{\phi}}$) with an anisotropic pressure about a uniform
subthermal vertical magnetic field (${\bf B_0} = B_z{\bf \hat{z}}$).
We assume that the background (unperturbed) plasma is described by a
bi-Maxwellian distribution ($p_{\Par 0} \neq p_{\Perp 0}$). We also
assume that the perturbations are axisymmetric, of the form exp[$-i
\omega t +i {\bf k \cdot x}$] with ${\bf k} = k_R {\bf \hat{R}} +
k_z {\bf \hat{z}}$. Writing $\rho=\rho_0+\delta \rho$, ${\bf B} =
{\bf B_0} + {\bf \delta B}$, $p_\Perp = p_{\Perp 0} + \delta
p_\Perp$, $p_\Par = p_{\Par 0} + \delta p_\Par$, working in
cylindrical coordinates and making a $|k|R \gg 1$ assumption, the
linearized versions of equations (\ref{Ch4eq:MHD1})-(\ref{Ch4eq:MHD3})
become: \ba \label{Ch4eq:lin1}
\omega \delta \rho &=& \rho_0 {\bf k} \cdot \delta{\bf v}, \\
\label{Ch4eq:lin2} \nonumber -i \omega \rho_0 \delta v_R &-& \rho_0
2\Omega \delta v_{\phi} = -\frac{i k_R}{4 \pi}B_z \delta B_z \\
&+& i k_z \left( \frac{B_z}{4 \pi} - \frac{(p_{\Par 0}
-p_{\Perp 0})}{B_z} \right) \delta B_R - i k_R \delta p_{\Perp}, \\
\label{Ch4eq:lin3} -i \omega \rho_0 \delta v_{\phi} &+& \rho_0 \delta
v_R \frac{\kappa^2} {2 \Omega} = i k_z \left( \frac{B_z}{4
\pi}-\frac{(p_{\Par 0}-p_{\Perp 0})}{B_z}
\right) \delta B_\phi, \\
\label{Ch4eq:lin4} -i \omega \rho_0 \delta v_z &=& -i k_R \left(
p_{\Par 0}-p_{\Perp 0} \right)
\frac{\delta B_R}{B_z} - i k_z \delta p_{\Par}, \\
\label{Ch4eq:lin5}
\omega \delta B_R &=& - k_z B_z \delta v_R, \\
\label{Ch4eq:lin6} \omega \delta B_{\phi} &=& - k_z B_z \delta v_{\phi}
- \frac{i k_z
B_z}{\omega} \frac{d \Omega}{d \ln R} \delta v_R, \\
\label{Ch4eq:lin7} \omega \delta B_z &=& k_R B_z \delta v_R, \ea
where $\kappa^2=4\Omega^2 + d\Omega^2/d \ln R$ is the epicyclic
frequency. Equations~(\ref{Ch4eq:lin1})-(\ref{Ch4eq:lin7}) describe
the linear modes of a collisionless disk with an initial pressure
anisotropy about a vertical magnetic field.  This corresponds to the
$\theta=\pi/2$ case of Chapter \ref{chap:chap3}, but with an
anisotropic initial pressure. Equations (\ref{Ch4eq:lin2}) \&
(\ref{Ch4eq:lin3}) show that an initial anisotropic pressure
modifies the Alfv\'en wave characteristics, so we expect a
background pressure anisotropy to have an important effect on the
MRI.  One way of interpreting equations (\ref{Ch4eq:lin2}) \&
(\ref{Ch4eq:lin3}) is that $p_\Perp>p_\Par$ ($p_\Par>p_\Perp$) makes
the magnetic fields more (less) stiff; as a result, this will shift
the fastest growing MRI mode to larger (smaller) scales.

The linearized equations for the parallel and perpendicular pressure
response are given by Eqs. \ref{Ch3eq:p_par} and \ref{Ch3eq:p_perp}
from Chapter \ref{chap:chap3}. We present them here for the sake of
completeness. \ba \label{Ch4eq:ppar} -i \omega \delta p_\Par &+&
p_{\Par 0} i {\bf k} \cdot {\bf \delta v} + i k_z
q_\Par + 2 p_{\Par 0} i k_z \delta v_z  = 0, \\
\label{Ch4eq:pperp} -i \omega \delta p_\Perp &+& 2 p_{\Perp 0} i
{\bf k} \cdot {\bf \delta v} + i k_z q_\Perp - p_{\Perp 0} i k_z
\delta v_z = 0, \ea where the heat fluxes can be expressed in terms
of lower moments using \ba \label{Ch4eq:qperp} q_\Perp &=&
-\sqrt{\frac{2}{\pi}} c_{\Par 0}  \frac{i k_z}{|k_z|} (\delta
p_\Perp -  c_{\Par 0}^2 \delta \rho) + \sqrt{\frac{2}{\pi}}  c_{\Par
0} p_{\Perp 0} \left( 1 - \frac{p_{\Perp 0}}{p_{\Par 0}} \right)
\frac{i k_z}{|k_z|} \frac{\delta B}{B_z}  , \\
\label{Ch4eq:qpar} q_\Par &=& -\sqrt{\frac{8}{\pi}} c_{\Par 0}
    \frac{i k_z}{|k_z|} (\delta p_\Par
-  c_{\Par 0} ^2 \delta \rho), \ea where $c_{\Par 0}=\sqrt{p_{\Par
0}/\rho_0}$ and $\delta B=|{\bf \delta B}|$.

\begin{figure}
\begin{center}
\includegraphics[width=4in,height=3in]{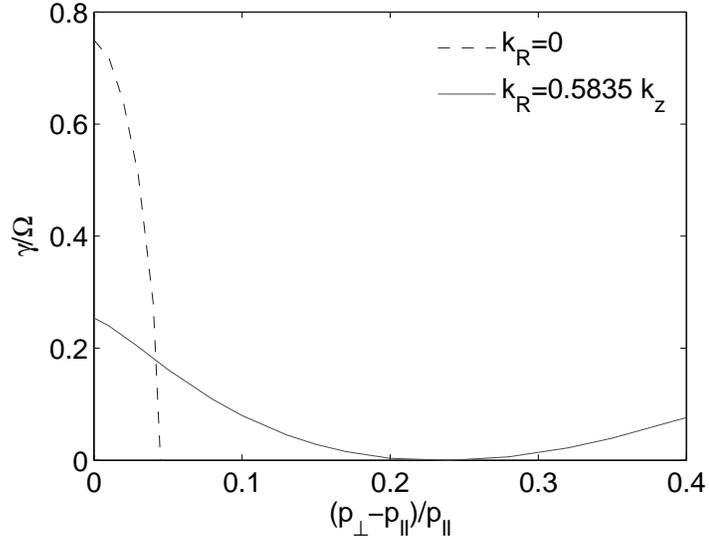}
\caption[Kinetic MRI growth rate as a function of pressure
anisotropy]{Normalized growth rate ($\gamma/\Omega$) of the MRI
versus normalized pressure anisotropy, $(p_\Perp-p_\Perp)/p_\Par$
for $\beta=100$, $k_z V_{Az}/\Omega=\sqrt{15/16}$, and two different
$k_R$'s.  Note that even a small anisotropy can stabilize the
fastest growing MRI mode. The growth at large pressure anisotropy
for $k_R \ne 0$ is due to the mirror mode. \label{Ch4fig:figure1}}
\end{center}
\end{figure}

Figure~\ref{Ch4fig:figure1} shows the MRI growth rate as a function
of pressure anisotropy for two values of $k_R$ for $\beta=100$. This
figure shows that the fastest growing MHD mode ($k_R=0$) is
stabilized for $(p_{\Perp 0}-p_{\Par 0})/p_{\Par 0} \sim 4/\beta$;
modes with $k_R \ne 0$ modes require larger anisotropy for
stabilization.  For $\beta \gg 1$, these results highlight that only
a very small pressure anisotropy is required to stabilize the
fastest growing MRI modes. Growth at large pressure anisotropies in
Figure \ref{Ch4fig:figure1} for $k_R \neq 0$ mode is because of the
mirror instability that is discussed below.  The physical
interpretation of the stabilization of the MRI in Figure
\ref{Ch4fig:figure1} is that as the pressure anisotropy increases
($p_{\Perp 0}>p_{\Par 0}$), the field lines effectively become
stiffer and modes of a given $k$ can be stabilized (though longer
wavelength modes will still be unstable). In a numerical simulation
in which the pressure anisotropy is allowed (unphysically, as we see
in Subsection \ref{Ch4subsec:Isotropization}) to grow without bound as the
magnetic field grows, this effect is capable of stabilizing all of
the MRI modes in the computational domain at very small amplitudes
(see Figure \ref{Ch4fig:figure6}).

\subsection{Isotropization of the pressure tensor in collisionless
plasmas} \label{Ch4subsec:Isotropization} Pressure anisotropy
($p_\perp \ne p_\Par$) is a source of free energy that can drive
instabilities which act to isotropize the pressure, effectively
providing an enhanced ``collision'' rate in a collisionless plasma
\cite{Gary1997}.  In order to do so, the instabilities must break
magnetic moment conservation, and thus must have frequencies
comparable to the cyclotron frequency and/or parallel wavelengths
comparable to the Larmor radius.  Because of the large disparity in
timescales between $\mu$-breaking microinstabilities and the MRI
($\omega_{micro}/\Omega \sim 10^8$), one can envision the
microinstabilities as providing a ``hard wall'' limit on the
pressure anisotropy; once the pressure anisotropy exceeds the
threshold value where microinstabilities are driven and cause rapid
pitch angle scattering, the pressure anisotropy nearly
instantaneously reduces the anisotropy to its threshold value (from
the point of view of the global disk dynamics). In this section we
review the relevant instabilities that limit the pressure anisotropy
in high $\beta$ collisionless plasmas---these are the firehose,
mirror, and ion cyclotron instabilities.  We then discuss how we
have implemented these estimated upper bounds on the pressure
anisotropy in our numerical simulations.

\subsubsection{Maximum anisotropy for $p_\Par > p_\Perp$}

Plasmas with $p_\Par > p_\Perp$ can be unstable to the firehose
instability, whose dispersion relation for parallel propagation is
given by equation (2.12) of \cite{Kennel1967}: \be \omega^2 - \omega
\Omega_i k_\Par^2 \rho_i^2 + \Omega_i^2 k_\Par^2 \rho_i^2 \left( 1 -
\frac{p_\Perp}{p_\Par} - \frac{2}{\beta_\Par} \right)=0, \ee where
$\beta_\Par=8\pi p_\Par/B^2$, $\rho_i$ is the ion Larmor radius,
$\Omega_i$ is the ion cyclotron frequency, and $k_\Par$ is the
wavenumber parallel to the local magnetic field direction. Solving
for $\omega$ gives \be \omega = k_\Par^2\rho_i^2 \frac{\Omega_i}{2}
\pm i k_\Par c_{\Par 0} \left(1 - \frac{p_\Perp}{p_\Par} -
\frac{2}{\beta_\Par} - \frac{k_\Par^2 \rho_i^2}{4} \right)^{1/2} \ee
For long wavelengths, the firehose instability requires
$p_\Par>p_\Perp+B^2/4\pi$, and is essentially an Alfv\'en wave
destabilized by the pressure anisotropy. The maximum growth rate
occurs when $k_\Par^2 \rho_i^2 = 2 (1-p_\Perp/p_\Par-2/\beta_\Par)$
and is given by $\Omega_i(1-p_\Perp/p_\Par-2/\beta_\Par)$. We use an
upper limit on $p_\Par>p_\Perp$ corresponding to
$1-p_\Perp/p_\Par-2/\beta_\Par<1/2$, which is an approximate
condition for the growth of modes that will violate $\mu$
conservation and produce rapid pitch angle scattering (when $\omega
\sim \Omega_i$ and $k_\parallel \rho_i \sim 1$).

\subsubsection{Maximum anisotropy for $p_\Perp > p_\Par$}
\label{Ch4sec:mirror_scattering}

For $p_\Perp > p_\Par$ there are two instabilities that act to
isotropize the pressure, the mirror instability and the ion
cyclotron instability \cite{Gary1997}.  A plasma is unstable to the
mirror instability when $p_\Perp/p_\Par - 1 > 1/\beta_\Perp$,
although as discussed below only for somewhat larger anisotropies is
magnetic moment conservation violated.  Formally, a plasma with any
nonzero pressure anisotropy can be unstable to the ion cyclotron
instability \cite{Stix1992}.  However, there is an effective
threshold given by the requirement that the unstable modes grow on a
timescale comparable to the disk rotation period.

The growth rate of the mirror instability is given by (Eq. (36) of
\cite{Hasegawa1969}) \be \gamma = \left ( \frac{2}{\pi} \right
)^{1/2} \left ( \frac{T_\perp}{T_\parallel} \right)^{3/2}
k_\parallel c_{\perp 0} \left [ \frac{T_\perp}{T_\parallel} -1 -
\frac{1}{\beta_\perp} \left (1+ \frac{k_\parallel^2}{k_\perp^2}
\right) \frac{\exp(\lambda)}{I_0(\lambda)-I_1(\lambda)} \right ],\ee
where $c_{\perp 0}=\sqrt{p_\perp/\rho}$, $\lambda=(k_\parallel
c_{\perp 0}/\Omega_i)^2$, and $I_0$ and $I_1$ are modified Bessel
functions of order 0 and 1. Minimizing the growth rate with respect
to $k_\parallel$ and $k_\perp$ gives equations (43$^\prime$) \&
(44$^\prime$) of \cite{Hasegawa1969}, which give the wavenumber for
the fastest growing mirror mode, \ba
&&\frac{k_\Par}{k_\Perp}=\sqrt{\frac{(D-1)}{4}}, \\ &&k_\Perp \rho_i
= \sqrt{\frac{(D-1)}{6}}, \ea where $D=\beta_\Perp
(p_\Perp/p_\Par-1)$, $\beta_\Perp=8\pi p_\Perp/ B^2$. To estimate
the pressure anisotropy at which $\mu$ conservation is broken and
thus pitch angle scattering is efficient, we calculate $D$ for which
$k_\Par \rho_i \sim k_\Perp \rho_i \sim 1$. This implies $D \approx
7$, or that $\mu$ conservation fails (and pitch angle scattering
occurs) if the pressure anisotropy satisfies \be
\label{Ch4eq:mirror_thresh} \frac{p_\Perp}{p_\Par} - 1 >
\frac{7}{\beta_\Perp}. \ee

The ion cyclotron instability can be also be excited when
$p_\Perp>p_\Par$. Gary and collaborators have analyzed the ion
cyclotron instability in detail through linear analysis and
numerical simulations \cite{Gary1997,Gary1994}. They calculate the
pressure anisotropy required for a given growth rate $\gamma$
relative to the ion cyclotron frequency $\Omega_i$ \be
\label{Ch4eq:gary_thresh} \frac{p_\Perp}{p_\Par} - 1 >
\frac{S^\prime}{\beta_\Par^p} \ee where $S^\prime=0.35$ and $p=0.42$
are fitting parameters quoted in equation (2) of \cite{Gary1994} for
$\gamma/\Omega_i=10^{-4}$. Moreover, for $\gamma \ll \Omega_i$ the
threshold anisotropy depends only very weakly on the growth rate
$\gamma$. As a result, equation (\ref{Ch4eq:gary_thresh}) provides a
reasonable estimate of the pressure anisotropy required for pitch
angle scattering by the ion cyclotron instability to be important on
a timescale comparable to the disk rotation period.

\subsection{Pressure anisotropy limits}
\label{Ch4sec:anisotropy}

Motivated by the above considerations, we require that the pressure
anisotropy satisfy the following inequalities in our simulations (at
each grid point and at all times): \ba \label{Ch4eq:pitch1} &&
\frac{p_\Perp}{p_\Par}-1 +\frac{2}{\beta_\Par} >\frac{1}{2},\\
\label{Ch4eq:pitch2}
&&\frac{p_\Perp}{p_\Par}-1<\frac{2\xi}{\beta_\Perp},\\
\label{Ch4eq:pitch3} &&\frac{p_\Perp}{p_\Par}-1<S
\left(\frac{2}{\beta_\Par}\right)^{1/2}, \ea where $S$ and $\xi$ are
constants described below.  It is important to note that the
fluid-based kinetic theory utilized in this paper can correctly
reproduce the existence and growth rates of the firehose and mirror
instabilities (though not the ion cyclotron
instability).\footnote{The double adiabatic limit ($q_\Perp = q_\Par
= 0$) predicts an incorrect threshold and incorrect growth rates for
the mirror instability \cite{Snyder1997}. Thus it is important to
use the heat flux models described in \S 2 to capture the physics of
the mirror instability.}  However, it can only do so for long
wavelength perturbations that conserve $\mu$. The relevant modes for
pitch angle scattering occur at the Larmor radius scale, which is
very small in typical accretion flows and is unresolved in our
simulations.  For this reason we must impose limits on the pressure
anisotropy and cannot simultaneously simulate the MRI and the
relevant instabilities that limit the pressure anisotropy. The
algorithm to impose the pressure anisotropy limits is explained in
Appendix~\ref{app:pthresh}.

In Eq. \ref{Ch4eq:pitch2}, the parameter $\xi$ determines the
threshold anisotropy above which the mirror instability leads to
pitch angle scattering. A value of $\xi = 3.5$ was estimated in
Section \ref{Ch4sec:mirror_scattering}.  We take this as our fiducial value,
but for comparison also describe calculations with $\xi = 0.5$,
which corresponds to the marginal state for the mirror instability.
We compare both models because the saturation of the mirror
instability is not well understood, particularly under the
conditions appropriate to a turbulent accretion disk.  Eq.
\ref{Ch4eq:pitch3} is based on the pitch angle scattering model used
by~\cite{Birn2001} for simulations of magnetic reconnection in
collisionless plasmas; following them we choose $S=0.3$. Eq.
\ref{Ch4eq:pitch3} with $S=0.3$ gives results which are nearly
identical (for the typical range of $\beta$ studied here) to the
pressure anisotropy threshold for the ion cyclotron instability
discussed in Section \ref{Ch4sec:mirror_scattering} (Eq.
\ref{Ch4eq:gary_thresh}).

In our simulations we find that for typical calculations, if $\xi =
0.5$ then Eq. \ref{Ch4eq:pitch2} (the ``mirror instability'')
dominates the isotropization of the pressure tensor, while if $\xi =
3.5$ then Eq. \ref{Ch4eq:pitch3} (the ``ion cyclotron instability'')
dominates.  We also find that our results are insensitive to the
form of the $p_\Par > p_\Perp$ threshold (Eq. \ref{Ch4eq:pitch1});
e.g., simulations with $1 - p_\Perp/p_\Par < 2/\beta_\Par$ (the
marginal state of the firehose mode) instead of equation
(\ref{Ch4eq:pitch1}) give nearly identical results.  Fully kinetic
simulations of the mirror, firehose, and ion cyclotron instabilities
will be useful for calibrating the pitch angle scattering models
used here.

\section{Kinetic MHD simulations in shearing box}

In this section we discuss the shearing box equations that we solve
numerically, and the modifications made to ZEUS to include kinetic
effects.

\subsection{Shearing box}

The shearing box is based on a local expansion of the tidal forces
in a reference frame rotating with the disk (see HGB for details). A
fiducial radius $R_0$ in the disk is picked out and the analysis is
restricted to a local Cartesian patch such that $L_x,L_y,L_z \ll
R_0$ (where $x=r-R_0$, $y=\phi$ and $z=z$). In this paper only the
radial component of gravity is considered, and vertical gravity and
buoyancy effects are ignored. We also assume a Keplerian rotation
profile. With these approximations, the equations of Landau MHD
(kinetic MHD combined with Landau closure for parallel heat fluxes)
in the shearing box are: \ba \label{Ch4eq:SB1}
\frac{\partial \rho}{\partial t} &+& \nabla \cdot (\rho {\bf V}) = 0, \\
\label{Ch4eq:SB2} \nonumber \frac{\partial {\bf V}}{\partial t}  &+& {\bf V}
\cdot \nabla {\bf V} = -\frac{1}{\rho} \nabla \left( p_\Perp +
\frac{B^2}{8\pi} \right) + \frac{{\bf B} \cdot \nabla {\bf B}}{4\pi
\rho} - \frac{1}{\rho} \nabla \cdot {\bf \Pi} \\
&-& 2 {\bf \Omega} \times {\bf V} + 3 \Omega^2 x {\bf \hat{x}}, \\
\label{Ch4eq:SB3} \frac{\partial {\bf B}}{\partial t} &=& \nabla \times
({\bf V} \times
{\bf B}), \\
\label{Ch4eq:SB4} \frac{\partial p_\Par}{\partial t} &+& \nabla
\cdot (p_\Par {\bf V}) + \nabla \cdot {\bf q_\Par} + 2 p_\Par {\bf
\hat{b}} \cdot \nabla {\bf V} \cdot {\bf \hat{b}} - 2q_\Perp \nabla
\cdot {\bf \hat{b}} = -\frac{2}{3} \nu_{eff} (p_\Par - p_\Perp), \\
\label{Ch4eq:SB5} \nonumber \frac{\partial p_\Perp}{\partial t}  &+&
\nabla \cdot (p_\Perp {\bf V}) + \nabla \cdot {\bf  q_\Perp}  +
p_\Perp \nabla \cdot {\bf V} - p_\Perp {\bf \hat{b}} \cdot \nabla
{\bf V}
\cdot {\bf \hat{b}} + q_\Perp \nabla \cdot {\bf \hat{b}} \\
&=& -\frac{1}{3} \nu_{eff} (p_\Perp - p_\Par), \\
\label{Ch4eq:SB6}
q_\Par &=& - \rho \kappa_\Par \nabla_\Par \left( \frac{p_\Par}{\rho}\right), \\
\label{Ch4eq:SB7} q_\Perp &=& - \rho \kappa_\Perp \nabla_\Par \left(
\frac{p_\perp}{\rho}\right) + \kappa_m {\bf B} \cdot \nabla B, \ea
where ${\bf q_\Par} = q_\Par {\bf \hat{b}}$ and ${\bf q_\Perp} =
q_\Perp {\bf \hat{b}}$ are the heat fluxes parallel to the magnetic
field, $\nu_{eff}$ is the effective pitch-angle scattering rate
(includes microinstabilities, see Subsection \ref{Ch4subsec:Isotropization} and
Appendix \ref{app:pthresh}), $\kappa_\Par$ and $\kappa_\Perp$ are
the coefficients of heat conduction, and $\kappa_m$ is the
coefficient in $q_\Perp$ due to parallel gradients in the strength
of magnetic field~\cite{Snyder1997}. The $\kappa_m$ component of
$q_\Perp$ that arises because of parallel magnetic field gradients
is important for correctly recovering the saturated state for the
mirror instability in the fluid limit, where (in steady state)
$q_{\Par, \Perp} \approx 0$ implies that $T_\Par$ is constant along
the field line, and $T_\Perp$ and magnetic pressure are
anticorrelated.

Given our closure models, the coefficients for the heat fluxes are
given by \ba \label{Ch4eq:kappa_par} \kappa_\Par &=& \frac{8
p_\Par}{\rho} \frac{1}{\sqrt{8 \pi \frac{p_\Par}{\rho}}
k_L + (3\pi-8)\nu_{eff} }, \\
\label{Ch4eq:kappa_perp} \kappa_\Perp &=& \frac{p_\Par}{\rho}
\frac{1}{\sqrt{\frac{\pi}{2}\frac{p_\Par}{\rho}}
k_L + \nu_{eff} }, \\
\label{Ch4eq:kappa_mag} \kappa_m &=&
\left(1-\frac{p_\Perp}{p_\Par}\right) \frac{p_\Perp}{B^2}
\kappa_\Perp, \ea where $k_L$ is the parameter that corresponds to a
typical wavenumber characterizing Landau damping (see ``crude model
of Landau damping" in Section \ref{Ch2sec:LFC}). We consider several values
of $k_L$ to study the effect of Landau damping on different scales.
In particular, we consider $k_L=0.5/\delta z$, $0.25/\delta z$, and
$0.125/\delta z$ which correspond to correctly capturing Landau
damping on scales of $12 \delta z$, $24 \delta z$, $48 \delta z$,
respectively, where $\delta z=L_z/N_z$, $L_z=1$ for all our runs,
and $N_z$ is the number of grid points in the $z$-direction (taken
be $27$ and $54$ for low and high resolution calculations,
respectively). Thus, $k_L=0.25/\delta z$ corresponds to correctly
capturing Landau damping for modes with wavelengths comparable to
the size of the box in the low resolution runs.

The term $\nu_{eff}$ in Eqs. \ref{Ch4eq:kappa_par} and
\ref{Ch4eq:kappa_perp} is an effective collision frequency which is
equal to the real collision frequency $\nu$, as long as $\mu$
conservation is satisfied. However, when the pressure anisotropy is
large enough to drive microinstabilities that break $\mu$ invariance
, and enhance pitch angle scattering, then there is an increase in
the effective collision frequency that decreases the associated
conductivities.  The expressions for $\nu_{eff}$ are given in Eqs.
\ref{Ch4eq:nueff1}, \ref{Ch4eq:nueff2}, and \ref{Ch4eq:nueff3} in
Appendix \ref{app:pthresh}.

Shearing periodic boundary conditions appropriate to the shearing
box are described in \cite{Hawley1995}. Excluding $V_y$, all
variables at the inner $x$- boundary are mapped to sheared ghost
zones at the outer boundary; a similar procedure applies for the
inner ghost zones. $V_y$ has a jump of $(3/2) \Omega L_x$ across the
box while applying the $x$- shearing boundary conditions, to account
for the background shear in $V_y$.

\subsection{Numerical methods}

We have used the shearing box version of the ZEUS MHD code
\cite{Stone1992a,Stone1992b}, and modified it to include kinetic
effects. The ZEUS code is a time explicit, operator split, finite
difference algorithm on a staggered mesh, i.e., scalars and the
diagonal components of second rank tensors are zone centered, while
vectors are located at zone faces, and pseudovectors and offdiagonal
components of second rank tensors are located at the edges. The
location of different variables on the grid is described in detail
in Appendix \ref{app:grid}. Appendix \ref{app:courant} describes how
we choose the time step $\delta t$ to satisfy the Courant condition
(which is modified by pressure anisotropy and heat conduction). We
also require that the choice of $\delta t$ maintain positivity of
$p_\Par$ and $p_\Perp$.

Implementation of the shearing box boundary conditions is described
in \cite{Hawley1995}.  One can either apply boundary conditions on
the components of ${\bf B}$ or the EMFs (whose derivatives give
${\bf B}$). We apply shearing periodic boundary conditions on the
EMFs to preserve the net vertical flux in the box, although applying
boundary conditions directly on ${\bf B}$ gives similar results.

Eqs. \ref{Ch4eq:SB4} and \ref{Ch4eq:SB5} are split into transport
and source steps, analogous to the energy equation in the original
ZEUS MHD. The transport step is advanced conservatively, and the
source step uses centered differences in space. It should be noted
that in Eq. \ref{Ch4eq:SB5} the $\nabla \cdot {\bf q_\Perp}$ term is
not purely diffusive, and it is necessary to carefully treat the
magnetic gradient part of $q_\Perp$ in the transport step for
robustness of the code (Appendix \ref{app:qperp_conservative}).

We have tested the newly added subroutines for evolving anisotropic
pressure and parallel heat conduction.  We tested the anisotropic
conduction routine by initializing a ``hot'' patch in circular
magnetic field lines and assessing the extent to which heat remains
confined along the field.  This is the same test described
in~\cite{Parrish2005}, and we find good agreement with their
results. The method we use for the simulations in this chapter is
the ``asymmetric method," described in Chapter \ref{chap:chap5}
which contains different tests we carried out.
Additional tests include linear (damped and undamped) waves and
instabilities in non-rotating anisotropic plasmas, the Alfv\'en
wave, and the firehose and mirror instabilities (see Appendix
\ref{app:app3}). For mirror simulations we observe the formation of
stationary anticorrelated density and magnetic structures as seen in
the hybrid simulations of \cite{McKean1993}. For firehose we see the
instability with magnetic perturbations developing at small scales
but during saturation the perturbations are at larger scales (as
seen in \cite{Quest1996}); a 2-D test for firehose instability,
where pressure anisotropy is caused by the shearing of plasma, is
presented in Appendix \ref{app:app3}.

Finally, the numerical growth rates of the kinetic MRI were compared
to the analytic results for different pressure anisotropies, $(k_x,
k_z)$, collision frequencies, and angles between the magnetic field
and ${\bf \hat{z}}$; we find good agreement with the results of
\cite{Quataert2002} and \cite{Sharma2003} (described in Chapter
\ref{chap:chap3}). When $k_L = k_\Par$, the growth rate of the
fastest growing mode is within $\sim 3 \%$ of the theoretical
prediction. The simulations with $B_\phi=B_z$ show $\approx$ twice
faster growth as compared to $B_z=0$, as predicted by linear theory.

\subsection{Shearing box and kinetic MHD}

Certain analytic constraints on the properties and energetics of
shearing box simulations have been described in \cite{Hawley1995}.
These constraints serve as a useful check on the numerical
simulations. Here we mention the modifications to these constraints
in KMHD. Conservation of total energy in the shearing box gives \be
\label{Ch4eq:encon} \frac{\partial}{\partial t} \Gamma = \frac{3}{2}
\Omega L_x \int_x dA \left [ \rho V_x \delta V_y - \left ( 1-
\frac{4\pi(p_\Par-p_\Perp)}{B^2} \right) \frac{B_x B_y}{4\pi}
\right], \ee where $\delta V_y = V_y + (3/2) \Omega x$, and $\Gamma$
is the total energy given by,\ \be \label{Ch4eq:energy} \Gamma =
\int d^3x \left [ \rho \left( \frac{V^2}{2} + \phi \right) +
\frac{p_\Par}{2} + p_\Perp + \frac{B^2}{8\pi} \right] \ee where
$\phi=-(3/2) \Omega^2 x^2$ is the tidal effective potential about
$R_0$.  Eq.~(\ref{Ch4eq:encon}) states that the change in the total
energy of the shearing box is due to work done on the box by the
boundaries. Notice that there is an anisotropic pressure
contribution to the work done on the box. Eq. (29) in
\cite{Balbus1998} for conservation of angular momentum in
cylindrical geometry (same as Eq. \ref{Ch1eq:AngularMomentum}) is
also modified because of the anisotropic pressure and is given by
\be \label{Ch4eq:angmom} \frac{\partial}{\partial t} (\rho R V_\phi)
+ \nabla \cdot  \Bigg[ \rho V_\phi {\bf V}R -
\frac{B_\phi}{4\pi}\left(1-\frac{4\pi(p_\Par-p_\Perp)}{B^2} \right)
{\bf B_p}R + \left( p_\Perp + \frac{B_p^2}{8\pi} \right){\bf
\hat{\phi}}R \Bigg] = 0, \ee where ${\bf B_p} = B_R {\bf
\hat{R}}+B_z {\bf \hat{z}}$ is the poloidal field. We can calculate
the level of angular momentum transport (and corresponding heating)
in our simulations by measuring the stress tensor given by \be
W_{xy}= \rho V_x \delta V_y - \frac{B_x B_y}{4\pi} +
\frac{(p_\Par-p_\Perp)}{B^2} B_x B_y \ee Note that the stress tensor
has an additional contribution due to pressure anisotropy. One can
define a dimensionless stress via Shakura and Sunyaev's $\alpha$
parameter by \be \alpha \equiv \frac{W_{xy}}{P_0} = \alpha_{R} +
\alpha_{M} + \alpha_{A} \ee where $\alpha_R$, $\alpha_M$, $\alpha_A$
are the Reynolds, Maxwell and anisotropic stress parameters,
respectively. As in \cite{Hawley1995}, we normalize the stress using
the initial pressure to define an $\alpha$ parameter.

\subsection{Shearing box parameters and initial conditions}
\label{Ch4subsec:IC}

The parameters for our baseline case have been chosen to match the
fiducial run Z4 of \cite{Hawley1995}. The simulation box has a
radial size $L_x=1$, azimuthal size $L_y=2\pi$, and vertical size
$L_z=1$.  The sound speed $V_s=\sqrt{p/\rho}=L_z \Omega$, so that
the vertical size is about a disk scale height (though it is an
unstratified box).  The pressure is assumed to be isotropic
initially, with $p_0 = \rho_0 V_s^2 = 10^{-6}$ and $\rho_0 = 1$. All
of our simulations start with a vertical field with $\beta=8\pi
p_0/B_0^2 =400$. The fastest growing MRI mode for this choice of
parameters is well resolved.  We consider two different numerical
resolutions: $27 \times 59 \times 27$ and $54 \times 118 \times 54$.
Perturbations are introduced as initially uncorrelated velocity
fluctuations. These fluctuations are randomly and uniformly
distributed throughout the box. They have a mean amplitude of
$|\delta V| = 10^{-3} V_s$.

\section{Results}
\begin{table}[hbt]
\begin{center}
\caption{Vertical field simulations with $\beta=400$
\label{Ch4tab:tab1}} \vskip0.05cm
\begin{tabular}{cccccccccc} \hline
Label & $k_L^a$ & $\xi^b$ & $\la \la \frac{B^2}{8\pi p_0} \ra \ra^c$
& $\la \la \frac{V^2}{2 p_0} \ra \ra$ & $\la \la \frac{B_x B_y}{4\pi
p_0} \ra \ra$ & $\la \la \frac{\rho V_x \delta V_y}{p_0} \ra \ra$ &
$\la \la \frac{\Delta p^*}{B^2}\frac{B_xB_y}{p_0} \ra
\ra$ & $ \la \la \frac{4\pi \Delta p}{B^2} \ra \ra$ \\
\hline
$Zl1$ & $\infty$ & $\infty$ & $0.0026$ &$0.094$& $0.0$ & $0.0$ & $0.0$ & $-11.96$ \\
$Zl2$ & $\infty$ & $3.5$ & $0.25$ & $0.28$ & $0.15$ & $0.067$ & $0.14$ & $-0.96$ \\
$Zl3^{\dag}$ & $0.5/\delta z$ & $\infty$ & $-$ & $-$ & $-$ & $-$ & $-$ & $-$ \\
$Zl4$ & $0.5/\delta z$ & $3.5$ & $0.38$ & $0.36$ & $0.23$ & $0.097$ & $0.20$ & $-1.37$ \\
$Zl5$ & $0.5/\delta z$ & $0.5$ & $0.35$ & $0.27$ & $0.197$ & $0.054$ & $0.069$ & $-0.02$ \\
$Zl6$ & $0.25/\delta z$ & $3.5$ & $0.27$ & $0.30$ & $0.16$ & $0.070$ & $0.15$ & $-1.39$ \\
$Zl7$ & $0.125/\delta z$ & $3.5$ & $0.21$ & $0.26$ & $0.124$ & $0.051$ & $0.117$ & $-1.44$ \\
$Zl8$ & $0.5/\delta z$ & $3.5$ & $0.157$ & $0.315$ & $0.094$ & $0.069$ & $0.225$ & $-2.11$ \\
$ZMl$ & $-$ & $-$ & $0.39$ & $0.29$ & $0.22$ & $0.066$ & $-$ & $-$ \\
$Zh1$ & $\infty$ & $\infty$ & $0.0026$ & $0.095$ & $0.0$ & $0.0$ & $0.0$ & $-10.2$ \\
$Zh2$ & $\infty$ & $3.5$ & $0.41$ & $0.32$ & $0.24$ & $0.083$ & $0.18$ & $-1.09$ \\
$Zh3^\dag$ & $0.5/\delta z$ & $\infty$ &$-$& $-$ & $-$ & $-$ & $-$ & $-$ \\
$Zh4$ & $0.5/\delta z$ & $3.5$ & $0.40$ & $0.33$ & $0.22$ & $0.078$ & $0.18$ & $-1.20$ \\
$Zh5$ & $0.5/\delta z$ & $0.5$ & $0.349$ & $0.253$ & $0.186$ & $0.042$ & $0.055$ & $-0.02$ \\
$Zh6$ & $0.25/\delta z$ & $3.5$ & $0.24$ & $0.26$ & $0.13$ & $0.044$ & $0.13$ & $-1.42$ \\
$ZMh$ & $-$ & $-$ & $0.375$ & $0.27$ & $0.204$ & $0.0531$ & $-$ & $-$ \\
\hline
\end{tabular}
\end{center}
Vertical field simulation with initial $\beta=400$. $Z$ indicates
that all simulations start with a vertical field, `$l$', `$h$'
indicate low ($27 \times 59 \times 27$) and high ($54 \times 118
\times 54$) resolution runs respectively. $Zl4$ is the fiducial run.
$Zl1$, $Zh1$  are the runs in CGL limit. $ZMl$ and$ZMh$ are the MHD
runs. \\
$^a$ Wavenumber parameter used in Landau closure for
parallel heat conduction (Eqs. \ref{Ch4eq:qpar1} and
\ref{Ch4eq:qperp1}). \\
$^b$ Imposed limit on pressure anisotropy for pitch angle scattering
due to mirror instability (Eq. \ref{Ch4eq:pitch2}).  Excluding
$Zl1$, $Zh1$, and $Zl8$ all of these calculations also use a
pressure anisotropy limit due to the ion cyclotron instability (Eq.
\ref{Ch4eq:pitch3}). \\
$^c$ $\langle \langle \rangle \rangle$ denotes a time and space
average taken from 5 to 20 orbits. \\
$^*$ $\Delta p=(p_\Par-p_\Perp)$ \\
$^\dag$ These cases run for only $\approx 4$ orbits at which point
the time step becomes very small because regions of large pressure
anisotropy are created (see Section \ref{Ch4sec:pitchangle}).
\end{table}
The important parameters for our simulations are listed in
Table~\ref{Ch4tab:tab1}.  Each simulation is labeled by $Z$ (for the
initial $B_z$ field), and $l$ and $h$ represent low ($27 \times 59
\times 27$) and high ($54 \times 118 \times 54$) resolution runs,
respectively.  We also include low and high resolution MHD runs for
comparison with kinetic calculations (labeled by $ZM$).  Our models
for heat conduction and pressure isotropization have several
parameters: $k_L$, the typical wavenumber for Landau damping used in
the heat flux (Eqs. \ref{Ch4eq:qpar1} and \ref{Ch4eq:qperp1}), and
$\xi$, the parameter that forces the pressure anisotropy to be
limited by $p_\Perp/p_\Par -1 < 2 \xi/\beta_\Perp$ (representing
pitch angle scattering due to small scale mirror modes; Eq.
\ref{Ch4eq:pitch2}).  All calculations except $Zl8$, $Zl1$, and
$Zh1$ also use the ion cyclotron scattering ``hard wall'' from Eq.
\ref{Ch4eq:pitch3}.  In addition to these model parameters,
Table~\ref{Ch4tab:tab1} also lists the results of the simulations,
including the volume and time averaged magnetic and kinetic
energies, and Maxwell, Reynolds, and anisotropic stresses. As Table
~\ref{Ch4tab:tab1} indicates, the results of our simulations depend
quantitatively---though generally not qualitatively---on the
microphysics associated with heat conduction and pressure
isotropization.  Throughout this section we use single brackets
$\langle f \rangle$ to denote a volume average of quantity $f$; we
use double brackets $\langle \langle f \rangle \rangle$ to denote a
volume and time average in the saturated turbulent state, from orbit
5 onwards.
\subsection{Fiducial run}
\label{Ch4sec:fiducial} We have selected run $Zl4$ as our fiducial
model to describe in detail.  This model includes isotropization by
ion cyclotron instabilities and mirror modes, with the former
dominating (for $\xi = 3.5$; see Section \ref{Ch4sec:mirror_scattering})
except at early times.  The conductivity is determined by $k_L =
0.5/\delta z$ which implies that modes with wavelengths $\sim 12
\delta z \sim L_z/2$ are damped at a rate consistent with linear
theory.
\begin{figure}
\begin{center}
\includegraphics[width=4in,height=3in]{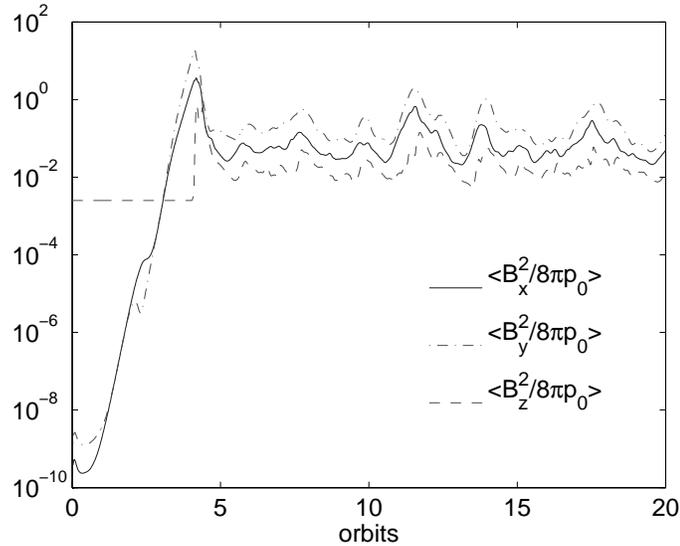}
\caption[Volume-averaged magnetic energy for the run $Zl4$]{Time
evolution of volume-averaged magnetic energy for the fiducial run
$Zl4$.  Time is given in number of orbits. There is a small decrease
in the magnetic energy at $\approx 2$ orbits when the pressure
anisotropy is sufficient to stabilize the fastest growing mode.
However, small-scale kinetic instabilities limit the magnitude of
the pressure anisotropy, allowing the magnetic field to continue to
amplify. As in MHD, there is a channel phase which breaks down into
turbulence at $\approx 4$ orbits. \label{Ch4fig:figure2}}
\end{center}
\end{figure}
\begin{figure}
\begin{center}
\includegraphics[width=5in,height=4in]{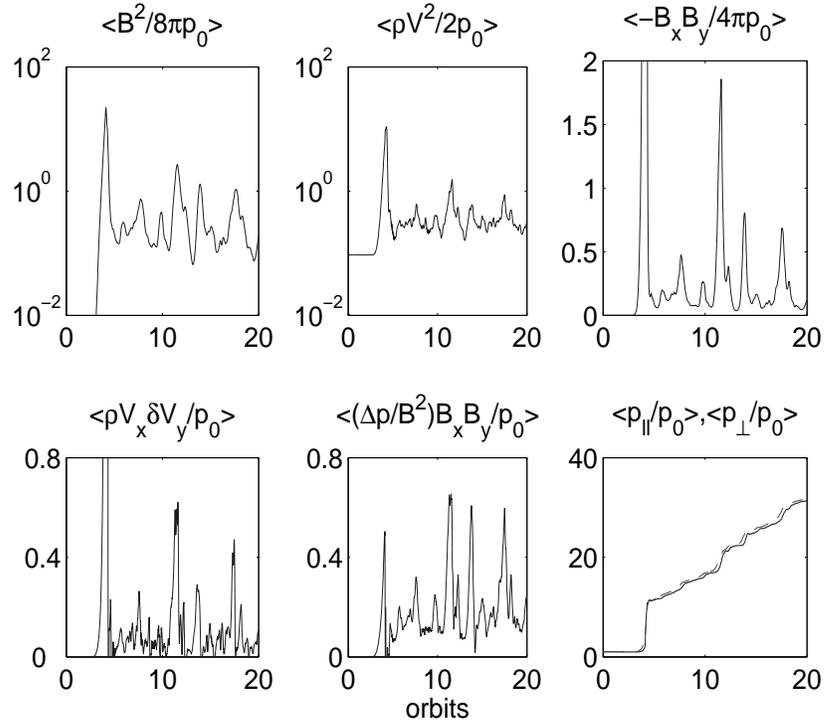}
\caption[Volume averaged quantities for the run $Zl4$]{Time
evolution of volume-averaged magnetic and kinetic energies, Maxwell,
Reynolds, and anisotropic stress, and pressure ($p_\Par$: solid
line, $p_\Perp$: dashed line) for the fiducial model $Zl4$.  Time is
given in orbits and all quantities are normalized to the initial
pressure $p_0$. $\delta V_y=V_y+(3/2)\Omega x $ and $\Delta p =
(p_\Par - p_\Perp)$. \label{Ch4fig:figure3}}
\end{center}
\end{figure}
\begin{figure}
\begin{center}
\includegraphics[width=4in,height=3in]{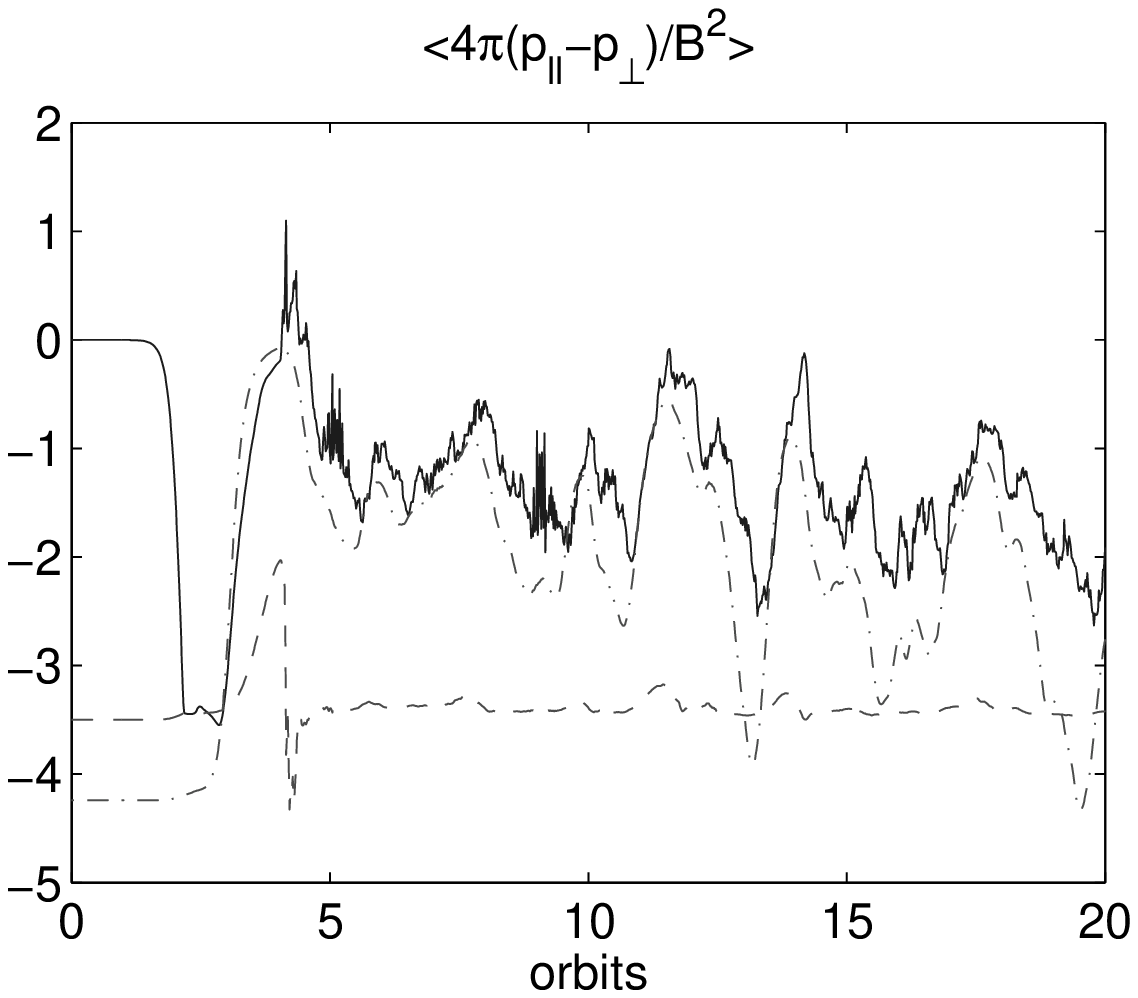}
\caption[Volume averaged pressure anisotropy and anisotropy
thresholds for the run $Zl4$]{Time evolution of volume-averaged
pressure anisotropy ($4\pi(p_\Par-p_\Perp)/B^2$: solid line) for
model $Zl4$. Also plotted are the ``hard wall'' limits on the
pressure anisotropy due to the ion cyclotron (dot dashed line) and
mirror instabilities (dashed line). Ion cyclotron scattering is
generally more efficient in the steady state. The limits on pressure
anisotropy are applied at each grid point while this figure is based
on volume averaged quantities. \label{Ch4fig:figure4}}
\end{center}
\end{figure}

Figures \ref{Ch4fig:figure2}-\ref{Ch4fig:figure4} show the time
evolution of various physical quantities for run $Zl4$.  The early
linear development of the instability is similar to that in MHD,
with the field growing exponentially in time. The key new feature is
the simultaneous exponential growth of pressure anisotropy ($p_\Perp
> p_\Par$) as a result of $\mu$ conservation (up to 2 orbits in Fig.
\ref{Ch4fig:figure4}).  As described in Section \ref{Ch4sec:linmodes}, this
pressure anisotropy tends to stabilize the MRI modes and shut off
the growth of the magnetic field.  Indeed, in simulations that do
not include any isotropization of the pressure tensor, we find that
all MRI modes in the box are stabilized by the pressure anisotropy
and the simulation saturates with the box filled with small
amplitude anisotropic Alfv\'en waves (see
Figure~\ref{Ch4fig:figure6}). This highlights the fact that, unlike
in MHD, the MRI is not an exact nonlinear solution in kinetic
theory. However, the pressure anisotropy required to stabilize all
MRI modes exceeds the pressure anisotropy at which pitch angle
scattering due to mirror and ion cyclotron instabilities become
important.  This takes place at about orbit 2 in run $Zl4$ (see the
small `dip' in the growth of magnetic energies in Figure
\ref{Ch4fig:figure2}), at which point the pressure anisotropy is
significantly reduced and the magnetic field is able to grow to
nonlinear amplitudes.

The nonlinear saturation at orbit $\sim 5$ appears qualitatively
similar to that in MHD, and may occur via analogues of the parasitic
instabilities described by~\cite{Goodman1994}.  The channel solution
is, however, much more extreme in KMHD than MHD (the maximum $B^2$
in Figure \ref{Ch4fig:figure2} is approximately an order of
magnitude larger than in analogous MHD runs). After saturation, the
magnetic and kinetic energies in the saturated state are comparable
in KMHD and MHD (see Table~\ref{Ch4tab:tab1}).  This is essentially
because the pitch angle scattering induced by the kinetic
microinstabilities acts to isotropize the pressure, enforcing a
degree of MHD-like dynamics on the collisionless plasma.

Figure \ref{Ch4fig:figure3} and Table \ref{Ch4tab:tab1} show the
various contributions to the total stress. As in MHD, the Reynolds
stress is significantly smaller than the Maxwell stress.  In kinetic
theory, however, there is an additional component to the stress due
to the anisotropic pressure (Eq. \ref{Ch4eq:angmom}).  In the
saturated state, we find that the Maxwell stress is similar in KMHD
and MHD, but that the anisotropic stress itself is comparable to the
Maxwell stress. Expressed in terms of an $\alpha$ normalized to the
initial pressure, our fiducial run $Zl4$ has $\alpha_M=0.23$,
$\alpha_R=0.097$, and $\alpha_A=0.2$, indicating that stress due to
pressure anisotropy is dynamically important.

Nearly all physical quantities in
Figures~\ref{Ch4fig:figure2}-\ref{Ch4fig:figure4} reach an
approximate statistical steady state.  The exceptions are $p_\Par$
and $p_\Perp$, which increase steadily in time because the momentum
flux on the boundaries does work on the system (Eq.
\ref{Ch4eq:encon}), which is eventually converted to heat in the
plasma by artificial viscosity; there is no cooling for internal
energy to reach a steady state (the same is true in HGB's MHD
simulations). Because of the steadily increasing internal energy and
approximately fixed $B^2$ (although with large fluctuations), the
plasma $\beta$ shows a small secular increase from orbits 5-20 (a
factor of $\approx 3$ increase, though with very large fluctuations
due to the large fluctuations in magnetic energy).

Figure \ref{Ch4fig:figure4} shows the pressure anisotropy thresholds
due to the ion cyclotron and mirror instabilities, in addition to
the volume averaged pressure anisotropy in run $Zl4$. From Eq.
\ref{Ch4eq:pitch3}, the ion cyclotron threshold ($4\pi \Delta
p/B^2$) is expected to scale as $\sqrt{\beta_\Par}$, which is fairly
consistent with the trend in Figure \ref{Ch4fig:figure4}.  The
actual pressure anisotropy in the simulation shows a small increase
in time as well, although less than that of the ion cyclotron
threshold.  These secular changes in $\beta$ and $\Delta p$ are a
consequence of the increasing internal energy in the shearing box,
and are probably not realistic.  In a global disk, we expect
that---except perhaps near the inner and outer boundaries---$\beta$
will not undergo significant secular changes in time. In a small
region of a real disk in statistical equilibrium, the heating would
be balanced by radiation (for thin disks) or by cooler plasma
entering at large $R$ and hotter plasma leaving at small $R$ (in low
luminosity, thick disks).
\begin{figure}
\begin{center}
\includegraphics[width=4in,height=3in]{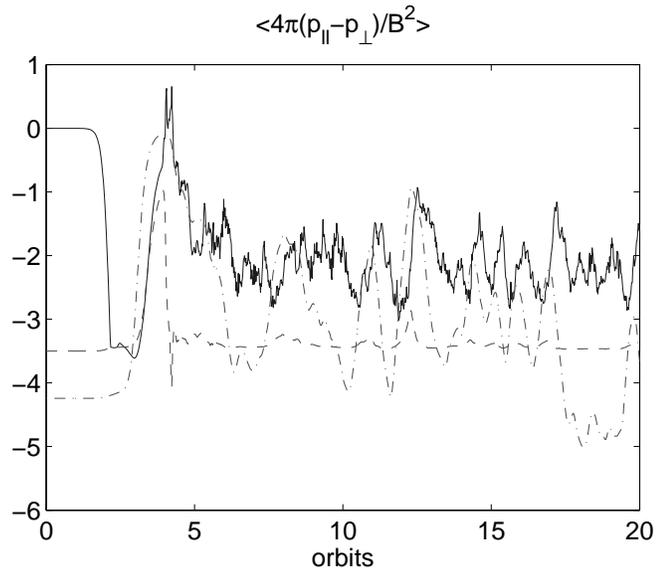}
\caption[Volume averaged pressure anisotropy and anisotropy
thresholds for the run $Zl8$]{Time evolution of volume-averaged
pressure anisotropy ($4\pi(p_\Par-p_\Perp)/B^2$: solid line) for
model $Zl8$. Also plotted are the ``hard wall'' limits on the
pressure anisotropy due to the ion cyclotron (dot dashed line) and
mirror instabilities (dashed line), although the ion cyclotron
scattering limit is not applied in this simulation. The volume
averaged pressure anisotropy saturates at smaller anisotropy than
the mirror threshold at $\xi=3.5$, which is the only limit on
pressure anisotropy used. \label{Ch4fig:figure5}}
\end{center}
\end{figure}

It is interesting to note that in Figure \ref{Ch4fig:figure4}, the
pressure anisotropy ($4 \pi \Delta p/B^2$) is closely tied to the
ion cyclotron threshold at times when $B^2$ is rising (which
corresponds to the channel solution reemerging).  Increasing $B$
leads to a pressure anisotropy with $p_\Perp>p_\Par$ by $\mu$
conservation. At the same time, the ion cyclotron threshold ($\sim
\sqrt{\beta}$) decreases, eventually the pressure limiting threshold
is encountered.  When $B$ is decreasing, however, we do not find the
same tight relationship between the pressure anisotropy and the
imposed threshold.  Figure \ref{Ch4fig:figure4} clearly indicates
that in our fiducial simulation pitch angle scattering is dominated
by the ion cyclotron threshold.  For comparison,
Figure~\ref{Ch4fig:figure5} shows the pressure anisotropy and
thresholds for run $Zl8$ which is identical to the fiducial run,
except that the ion cyclotron threshold is not used and the only
scattering is due to the mirror threshold. In this case, the
saturated pressure anisotropy is somewhat larger than in the
fiducial run, but the pressure anisotropy is not tied to the mirror
threshold.
\begin{table}[hbt]
\begin{center}
\caption{Statistics for Model $Zl4$\label{Ch4tab:tab2}} \vskip0.05cm
\begin{tabular}{cccccc} \hline
Quantity $f$ & $\la \la f \ra \ra$ & $ \la \la \delta f^2 \ra
\ra^{1/2} $ & $(\frac{\tau_{int}}{T})^{1/2} \la \la \delta f^2 \ra
\ra^{1/2}$  & $\mbox{min}(f)$ & $\mbox{max}(f)$ \\
\hline
$\frac{B_x^2}{8\pi p_0}$ & $0.083$ & $0.092$ & $0.016 $ & $0.021$ & $0.662$\\
$\frac{B_y^2}{8\pi p_0}$ & $0.276$ & $0.318$ & $0.048 $ & $0.036$ & $1.987$\\
$\frac{B_z^2}{8\pi p_0}$ & $0.021$ & $0.017$ & $0.0025 $ & $0.0032$ & $0.144$\\
$\frac{\rho V_x^2}{2 p_0}$ & $0.102$ & $0.094$ & $0.014 $ & $0.0184$ & $0.63$\\
$\frac{\rho \delta V_y^2}{2 p_0}$ & $0.125$ & $0.079$ & $0.0127 $ & $0.715$ & $0.0264$\\
$\frac{\rho V_z^2}{2 p_0}$ & $0.037$ & $0.034$ & $0.0032 $ & $0.008$ & $0.348$\\
$\frac{-B_xB_y}{4\pi p_0}$ & $0.229$ & $0.277$ & $0.0434 $ & $0.037$ & $1.856$\\
$\frac{\rho V_x\delta V_y}{p_0}$ & $0.097$ & $0.113$ & $0.0147 $ & $-0.072$ & $0.6211$\\
$\frac{(p_\Par-p_\Perp)}{p_0}\frac{B_xB_y}{B^2}$ & $0.198$ & $0.129$ & $0.0178 $ & $0.017$ & $0.654$\\
$\frac{4\pi(p_\Par-p_\Perp)}{B^2}$ & $-1.366$ & $0.51$ & $0.098$ & $-2.632$ & $-0.083$\\
$\frac{-B_xB_y}{(B^2/2)}$ & $0.5895$ & $0.1043$ & $0.0067$ & $0.3744$ & $0.8611$ \\
$\frac{\rho V_x \delta V_y}{(B^2/8\pi)}$ & $0.3323$ & $0.2725$ & $0.017$ & $-0.5307$ & $1.2704$ \\
$\frac{4\pi (p_\Par-p_\Perp)}{B^2} \frac{B_xB_y}{(B^2/2)}$ &
$0.7356$
& $0.3718$ & $0.0714$ & $0.032$ & $1.807$ \\
$\frac{W_{xy}}{(B^2/8\pi)}$ & $1.6574$ & $0.6598$ & $0.084$ & $0.4364$ & $3.7159$ \\
$\frac{\alpha_R}{\alpha_M}$ & $0.5357$ & $0.3975$ & $0.024$ & $-0.9105$ & $2.084$ \\
$\frac{\alpha_A}{\alpha_M}$ & $1.2287$ & $0.5504$ & $0.119$ & $0.0854$ & $2.7243$ \\
$\frac{\rho}{\rho_0}$ & $0.99935$ & $2.3 \times 10^{-5}$ & $1.1 \times 10^{-5}$ & $0.9993$ & $0.9994$\\
$\frac{p_\Perp B_0}{B p_0}$ & $3.557$ & $1.665$ & $-^a$ & $1.1178$ & $7.929$\\
$\frac{p_\Par B^2 \rho_0^2}{\rho^2 B_0^2 p_0}$ & $3.144 \times 10^3$ & $3.49 \times 10^3$ & $-$ & $585.4$ & $1.993 \times 10^4$\\
\hline
\end{tabular}
\end{center}
$^a$ We calculate the error using the autocorrelation time only for
quantities that saturate to a steady state after 5 orbits. Estimate
for correlation time $\tau_{int}$ is based on the discussion in
\cite{Nevins2005}. $p_\Perp$ and $p_\Par$ show a secular growth with
time, so this way of expressing them as an average and an error is
not applicable.
\end{table}

Table~\ref{Ch4tab:tab2} gives the mean, standard deviation, and
standard error in the mean, for various quantities in the saturated
portion of the fiducial simulation.  The standard errors are
estimated by taking into account the finite correlation time for the
physical quantities in the simulation, as described in Appendix
~\ref{app:error_bars}.  In many cases, the deviations are
significantly larger than the mean.  As in MHD, we find that the
magnetic energy is dominated by the $y$- component, which is about a
factor of $3$ larger than the $x$- component; the vertical component
is smaller yet. The radial and azimuthal kinetic energy fluctuations
are comparable, while the vertical component is smaller. We also
find that, as in MHD, the perturbed kinetic and magnetic energies
are not in exact equipartition: the magnetic energy is consistently
larger. Table~\ref{Ch4tab:tab2} also shows the mean and deviations
for $\langle p_\Perp/B \rangle$ and $\langle p_\Par B^2/\rho^2
\rangle$. Because of pitch angle scattering $\mu= \langle p_\Perp/B
\rangle$ is no longer conserved. $\langle p_\Par B^2/\rho^2 \rangle$
varies because of both, heat conduction and pitch angle scattering.

The pressure anisotropy in our fiducial run saturates at $4\pi
(p_\Perp-p_\Par)/B^2 \approx 1.5$.  By contrast, the threshold for
the mirror instability is $4 \pi (p_\Perp-p_\Par)/B^2=0.5$. This
implies that the model is unstable to generating mirror modes.
However, the mirror modes that can be excited at this level of
anisotropy do not violate $\mu$ conservation and thus do not
contribute to pitch angle scattering (see
Section \ref{Ch4sec:mirror_scattering}).  They can in principle isotropize
the plasma in a volume averaged sense by spatially redistributing
plasma into magnetic wells \cite{Kivelson1996}. This saturation
mechanism is simulated using our kinetic MHD code for a uniform,
anisotropic  plasma (see Appendix \ref{app:mirror}).  It does not
appear to be fully efficient in the saturated state of our turbulent
disk simulations, even at the highest resolution; strong MRI
turbulence dominates over everything else for these parameters.

In the next few sections we compare the fiducial simulation
described above with variations in the pitch angle scattering model
and the parallel conductivity.  A comparison of the total stress in
all of our simulations is shown in Figure~\ref{Ch4fig:figure7}.

\subsection{The double adiabatic limit}
\label{Ch4sec:DA}
\begin{figure}
\begin{center}
\includegraphics[width=5in,height=4in]{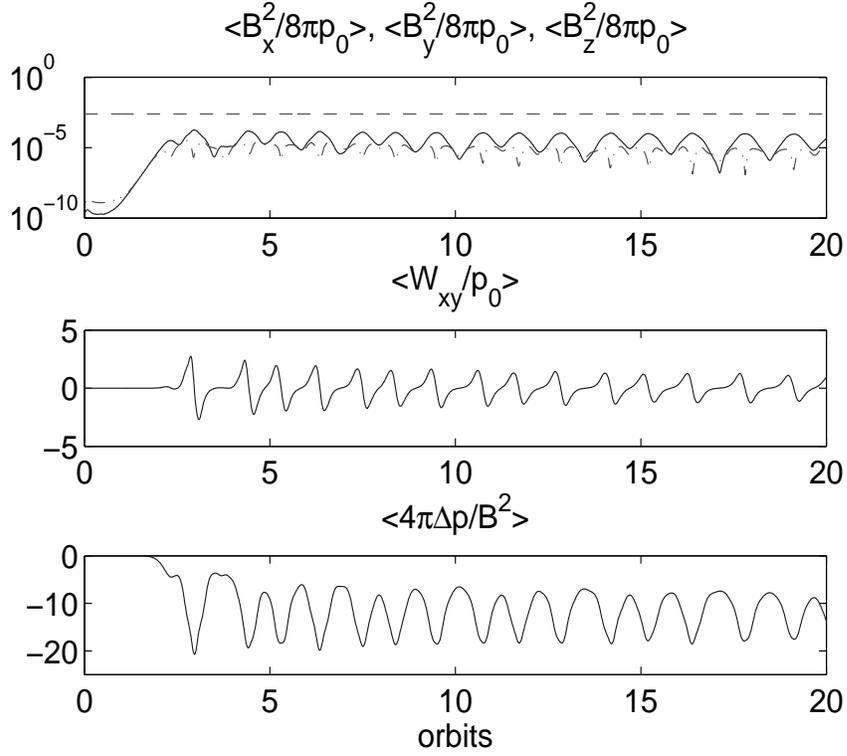}
\caption[Volume averaged quantities for $Zl1$, the CGL run with no
limit on pressure anisotropy]{Time evolution of volume-averaged
magnetic energy (dashed line: $B_z^2/8\pi p_0$, solid line:
$B_x^2/8\pi p_0$, dot dashed line: $B_y^2/8\pi p_0$), total stress
($W_{xy}/p_0$) in units of $10^{-3}$, and pressure anisotropy for
model $Zl1$. Time is given in orbits and all quantities are
normalized to the initial pressure $p_0$. $\delta
V_y=V_y+(3/2)\Omega x$ and $\Delta p = (p_\Par - p_\Perp)$.  In this
calculation there is no heat conduction and no isotropization of the
pressure tensor.  All resolved MRI modes are thus stabilized by
pressure anisotropy and the `saturated' state is linear anisotropic
Alfv\'en waves with no net angular momentum transport.
\label{Ch4fig:figure6}}
\end{center}
\end{figure}
Simulations $Zl1$ and $Zh1$ are simulations in the double adiabatic
limit (no heat conduction), with no limit on the pressure anisotropy
imposed. In this limit both $\mu= \langle p_\Perp/B \rangle$ and
$\langle p_\Par B^2/\rho^2 \rangle$ are conserved.
Figure~\ref{Ch4fig:figure6} shows volume averages of various
quantities as a function of time for the run $Zl1$.  These
calculations are very different from the rest of our results and
show saturation at very low amplitudes ($\delta B^2/B^2 \approx
0.04$). In the saturated state, the box is filled with shear
modified anisotropic Alfv\'en waves and all physical quantities are
oscillating in time. The total stress is also oscillatory with a
vanishing mean, resulting in negligible transport.  In these
calculations, the pressure anisotropy grows to such a large value
that it shuts off the growth of all of the resolved MRI modes in the
box. Table~\ref{Ch4tab:tab1} shows that $\langle \langle 4 \pi
(p_\Par-p_\Perp)/B^2 \rangle \rangle$ saturates at $-11.96$ and
$-10.2$ for the low and high resolution runs, respectively (although
the normalized pressure anisotropy $\langle \langle
(p_\Par-p_\Perp)/p_\Par \rangle \rangle \approx -0.07$ is quite
small).  This is much larger than the anisotropy thresholds for
pitch angle scattering described in Subsection \ref{Ch4subsec:Isotropization}.
As a result, we do not expect these cases to be representative of
the actual physics of collisionless disks.  These cases are of
interest, however, in supporting the predictions of the linear
theory with anisotropic initial conditions considered
in Section \ref{Ch4sec:linmodes}, and in providing a simple test for the
simulations.  They also highlight the central role of pressure
isotropization in collisionless dynamos \cite{Schekochihin2005}.

\subsection{Varying conductivity}
\begin{figure}
\begin{center}
\includegraphics[width=4in,height=3in]{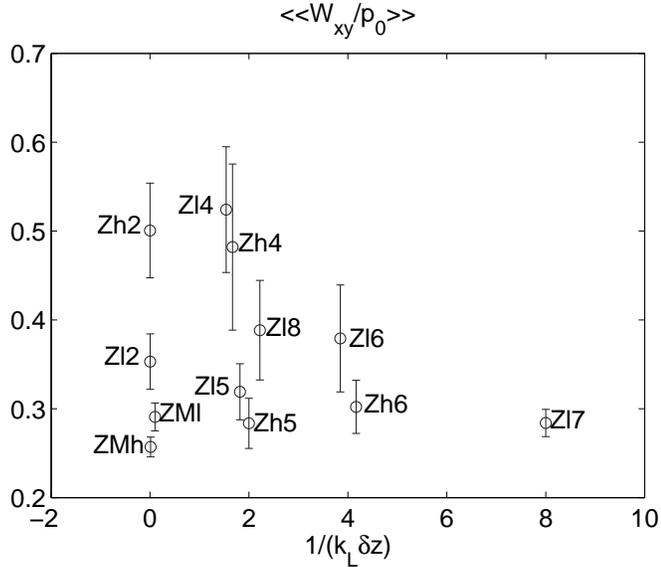}
\caption[Average total stress for different runs]{Space and time
average of the total stress $\langle \langle W_{xy}/p_0 \rangle
\rangle$ versus $1/(k_L \delta z)$ for different runs.  Error bars
shown are based on estimates of the correlation time of the
fluctuations described in \cite{Nevins2005}. \label{Ch4fig:figure7}}
\end{center}
\end{figure}
We have carried out a series of simulations with different
conductivities defined by the parameter $k_L$.  Simulations $Zl2$
and $Zh2$ are in the CGL limit with vanishing parallel heat
conduction, but with the same limits on pressure anisotropy as the
fiducial model. Simulations $Z6$ use $k_L \delta z = 0.25$ while run
$Zl7$ uses $k_L=0.125/\delta z$.  Both of these are smaller than the
value of $k_L \delta z = 0.5$ in the fiducial run, which implies a
larger conductivity. Figure~\ref{Ch4fig:figure7} shows that the
total stress varies by about a factor of 2 depending on the
conductivity and resolution.  Simulations with larger conductivity
tend to have smaller saturation amplitudes and stresses.  This could
be because larger conductivity implies more rapid Landau damping of
slow and fast magnetosonic waves.  In all cases, however, the
anisotropic stress is comparable to the Maxwell stress, as in the
fiducial run. Until a more accurate evaluation is available of the
heat fluxes for modes of all wavelengths in the simulation
simultaneously (either by a more complete evaluation of the nonlocal
heat fluxes, Eqs. \ref{Ch4eq:nonlocal1}-\ref{Ch4eq:nonlocal2}, or
even by a fully kinetic MHD code that directly solves the DKE, Eq.
\ref{Ch4eq:DKE}), it is difficult to ascertain which value of the
conductivity best reflects the true physics of collisionless disks.

\subsection{Different pitch angle scattering models}
\label{Ch4sec:pitchangle} In this section we consider variations in
our model for pitch angle scattering by high frequency waves.  All
of these calculations use $k_L = 0.5/\delta z$.  We note again that
the appropriate pitch angle scattering model remains somewhat
uncertain, primarily because of uncertainties in the nonlinear
saturation of long-wavelength, $\mu$-conserving mirror modes.  The
calculations reported here cover what, we believe, is a plausible
range of models.

Models $Zl5$ and $Zh5$ place a more stringent limit on the allowed
pressure anisotropy, taking $\xi=0.5$ in Eq. \ref{Ch4eq:pitch2}.
This corresponds to the threshold of the mirror instability.  Not
surprisingly, this simulation is the most ``MHD-like'' of our
calculations, with magnetic and kinetic energies, and Maxwell
stresses that are quite similar to those in MHD.  Even with this
stringent limit, however, the anisotropic stress is $\approx 1/3$ of
the Maxwell stress.  It is also interesting to note that although
the dimensionless pressure anisotropy is quite small $\langle
\langle 4 \pi (p_\Par-p_\Perp)/B^2 \rangle \rangle \approx -0.02$,
the dimensionless anisotropic stress $\langle \langle 4 \pi
(p_\Par-p_\Perp)/B^2 \times B_xB_y/p_0 \rangle \rangle \approx
-0.07$ is significantly larger (and larger than Reynolds stress)
because of correlations between the pressure anisotropy and field
strength.
\begin{figure}
\begin{center}
\includegraphics[width=4in,height=3in]{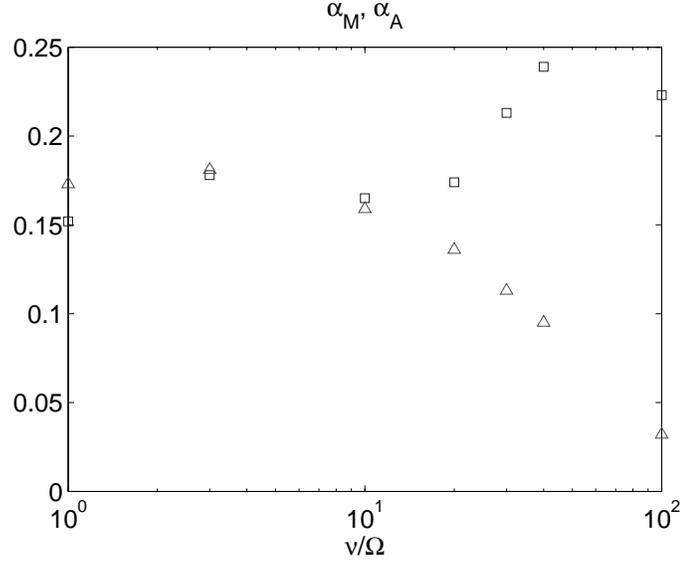}
\caption[Average Maxwell and anisotropic stresses as a function of
collision frequency]{Maxwell ($\alpha_M$: squares) and anisotropic
stress ($\alpha_A$: triangles) plotted against the collision
frequency normalized to rotation frequency ($\nu/\Omega$).
Transition to MHD occurs for $\nu/\Omega \gtrsim 30$ (see
Table~\ref{Ch4tab:tab3}). \label{Ch4fig:figure8}}
\end{center}
\end{figure}
\begin{table}[hbt]
\begin{center}
\caption{Simulations with an explicit collision term
\label{Ch4tab:tab3}} \vskip0.05cm
\begin{tabular}{ccccccc} \hline
$\nu/\Omega$ & $\langle \langle 4\pi \Delta p/B^2 \rangle \rangle$ &
$ \langle \langle - \frac{B_x B_y}{4\pi p_0} \rangle \rangle $ & $
\langle \langle \frac{\rho V_x \delta V_y}{p_0} \rangle \rangle$ &
$\langle \langle \frac{\Delta p^*}{B^2}\frac{B_xB_y}{p_0}
\rangle \rangle$ & $\alpha_A/\alpha_M$ & $\alpha_A/\alpha_A(\nu=0)$ \\
\hline
$0$ & $-1.41$ & $0.18$ & $0.082$ & $0.196$ & $1.09$ & $1$ \\
$1$ & $-1.47$ & $0.152$ & $0.072$ & $0.173$ & $ 1.14$ & $0.88$ \\
$3$ & $-1.43$ & $0.178$ & $0.08 $ & $0.181$ & $1.02$ & $0.92$ \\
$10$ & $-1.35$ & $0.165$ & $0.071 $ & $0.159$ & $0.96$ & $0.81$ \\
$20$ & $-1.24$ & $0.174$ & $0.070 $ & $0.136$ & $0.78$ & $0.69$\\
$30$ & $-1.01$ & $0.213$ & $0.070 $ & $0.113$ & $0.53$ & $0.58$\\
$40$& $-0.87$ & $0.239$ & $0.070 $ & $0.095$ & $0.4$ & $0.48$\\
$100$ & $-0.43$ & $0.223$ & $0.06 $ & $0.032$ & $0.14$ & $0.16$\\
\hline
\end{tabular}
\end{center}
$^*$ $\Delta p=(p_\Par-p_\Perp)$
\end{table}

As a test of how large a collisionality is needed for the results of
our kinetic simulations to rigorously approach the MHD limit, we
have carried out a series of simulations including an explicit
collisionality $\nu$ and varying its magnitude relative to the disk
frequency $\Omega$. Our results are summarized in Table
\ref{Ch4tab:tab3} and Figure~\ref{Ch4fig:figure8}. In these
simulations we start with initial conditions determined by the
saturated turbulent state of our fiducial run $Zl4$, but with an
explicit collision frequency (in addition to the scattering models
described in Section \ref{Ch4sec:anisotropy}). Figure
\ref{Ch4fig:figure8} shows that for $\nu/\Omega \lesssim 20$, the
results are very similar to the collisionless limit.  For larger
collision frequencies the anisotropic stress is reduced and the
simulations quantitatively approach the MHD limit.  These results
are similar to those obtained in Chapter \ref{chap:chap3} (see
Figure \ref{Ch3fig:newfig}), where linear calculations indicate that
the MHD limit for modes with $k \sim \Omega/V_A$ is approached when
$\nu \gtrsim \beta^{3/4} \Omega$.

To consider the opposite limit of low collisionality (because of
pitch angle scattering), run $Zl8$ places a less stringent limit on
the allowed pressure anisotropy, taking $\xi=3.5$ in Eq.
\ref{Ch4eq:pitch2}, and ignoring the limit set by the ion cyclotron
instability in Eq. \ref{Ch4eq:pitch3}. The results of this
calculation are not physical but are useful for further clarifying
the relative importance of the Maxwell and anisotropic stresses as a
function of the pitch angle scattering rate.  In $Zl8$, the
saturated magnetic energy and Maxwell stress are lower than in all
of our other calculations (excluding the double adiabatic models
described in Section \ref{Ch4sec:DA}). Interestingly, however, the total
stress is comparable to that in the other calculations (Figure
\ref{Ch4fig:figure7}) because the anisotropic stress is $\approx
2.4$ times larger than the Maxwell stress (Table \ref{Ch4tab:tab1}).
As discussed briefly in  Section \ref{Ch4sec:fiducial}, the pressure
anisotropy in this simulation is not simply set by the applied
mirror pitch angle scattering threshold (its quite smaller than the
mirror ``hard wall;" see Figure \ref{Ch4fig:figure5}).  It is
possible that resolved mirror modes contribute to decreasing the
volume averaged pressure anisotropy (but see below).

Finally, in models $Z3$ we include parallel heat conduction but do
not limit the pressure anisotropy.  In these calculations, we expect
to be able to resolve the long-wavelength $\mu$-conserving mirror
modes that reduce the pressure anisotropy by forming magnetic wells
\cite{Kivelson1996}.\footnote{At the resolution of $Zl3$, the
fastest growing mirror mode in the computational domain has a linear
growth comparable to that of the MRI.}  In our test problems with
uniform anisotropic plasmas, this is precisely what we find (see
Appendix \ref{app:mirror}). In the shearing box calculations,
however, even at the highest resolutions, we find that the pressure
anisotropy becomes so large that Eqs. \ref{Ch4eq:pitch2}) and
\ref{Ch4eq:pitch3} are violated, so that pitch angle scattering due
to high frequency microinstabilities would become important.  The
resolved mirror modes are thus not able to isotropize the pressure
sufficiently fast at all places in the box.\footnote{In higher
resolution simulations, one can resolve smaller-scale and faster
growing mirror modes, and thus the effects of isotropization by
resolved mirror modes could become increasingly important.  We see
no such indications, however, for the range of resolutions we have
been able to simulate.} However, it is hard to draw any firm
conclusions from these simulations because they stop at around 4
orbits (for both resolutions $Zl3$ and $Zh3$) during the initial
nonlinear transient stage.  At this time the pressure becomes highly
anisotropic and becomes very small at some grid points, and the time
step limit causes $\delta t \rightarrow 0$.

Pitch angle scattering centers are not uniformly distributed in
space and show intermittency. Subsection \ref{Ch2subsec:nu_eff}
gives simple estimates for effective collision frequency and mean
free path assuming a uniform distribution of scattering centers.
Also discussed are simulation results which show that only a very
small fraction of the box undergoes pitch angle scattering (see
Figure \ref{Ch2fig:figure2}). Figure \ref{Ch2fig:figure2} also shows
that pitch angle scattering due to mirror instability dominates
ion-cyclotron instability for $\beta \gtrsim 100$. Intermittency of
pitch angle scattering can be crucial for thermal conduction and
viscous transport in collisionless high-$\beta$ plasmas.

\section{Additional simulations}

Our paper, \cite{Sharma2006}, describes simulations with an initial
$\beta=400$ and an initial vertical field ($B_\phi=0$). The linear
theory predicts that the fastest growing mode for $B_\phi=B_z$ in
the kinetic regime is $\approx$ twice faster than MHD, and at a much
larger scale. The scale separation between the fastest growing
kinetic MHD and MHD modes for $B_\phi=B_z$ is greatest for large
$\beta$ (see \cite{Quataert2002}). In this section we describe
simulations not described in \cite{Sharma2006}---initial conditions
with $B_\phi=B_z$ and only $B_\phi$, and the high $\beta$ regime.
One of the motivations is to see whether a faster growth rate for
$B_\phi=B_z$ in the kinetic regime results in a nonlinear saturation
different from MHD. The fastest growing MRI mode in MHD occurs at a
scale $H/\sqrt{\beta}$, much smaller than the disk height scale
$H=c_s/\Omega$ for large initial $\beta$. Thus, we vary the box size
and resolution to study the effect of these parameters on nonlinear
saturation.

\subsection{High $\beta$ simulations}
\begin{figure}
\begin{center}
\includegraphics[width=4in,height=3in]{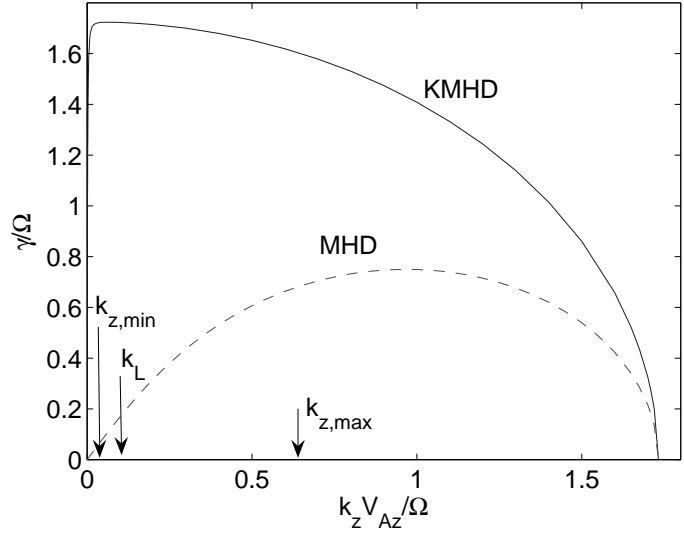}
\caption[MRI growth rates in kinetic and MHD regimes for
$\beta=10^6$ with $k_L=0.5/\delta z$]{The MRI growth rate in kinetic
(using $k_L=0.5/\delta z$) and MHD regimes for $\beta=10^6$. Arrows
$k_{z,min}=2\pi/L_z$ and $k_{z,max} =\pi/\delta z$ mark the minimum
and maximum wavenumbers in the low resolution ($27 \times 59 \times
27$) runs. For a higher resolution ($54 \times 118 \times 54$)
simulation  both $k_L$ and $k_{z,max}$ double as $\delta z$ is
reduced by half. \label{Ch4fig:figure14}}
\end{center}
\end{figure}
Figure \ref{Ch4fig:figure14} shows the growth rate of the MRI in the
kinetic and MHD regimes for $\beta=10^6$---the fastest growing
kinetic MRI is at a much larger length scale. Because of a large
separation of scales between the fastest growing modes in the
kinetic and MHD regimes, it is difficult to resolve both the scales
in a numerical simulation. The figure also marks, by arrows, the
minimum and maximum wavenumber corresponding to the chosen box size
for the low resolution runs ($KYZl$ and $MYZl$); for these runs both
these scales are not resolved. We carry out low ($27 \times 59
\times 27$) and high ($54 \times 118 \times 58$) resolution kinetic
and MHD simulations, with different box sizes. The arrows in Figure
\ref{Ch4fig:figure14} correspond to the low resolution runs with the
smallest boxes that we have considered---increasing the box size
resolves the fastest growing kinetic modes at large scales, while
increasing the number of grid points resolves the fastest MHD modes
at small scales.
\begin{figure}
\begin{center}
\includegraphics[width=2.95in,height=2.5in]{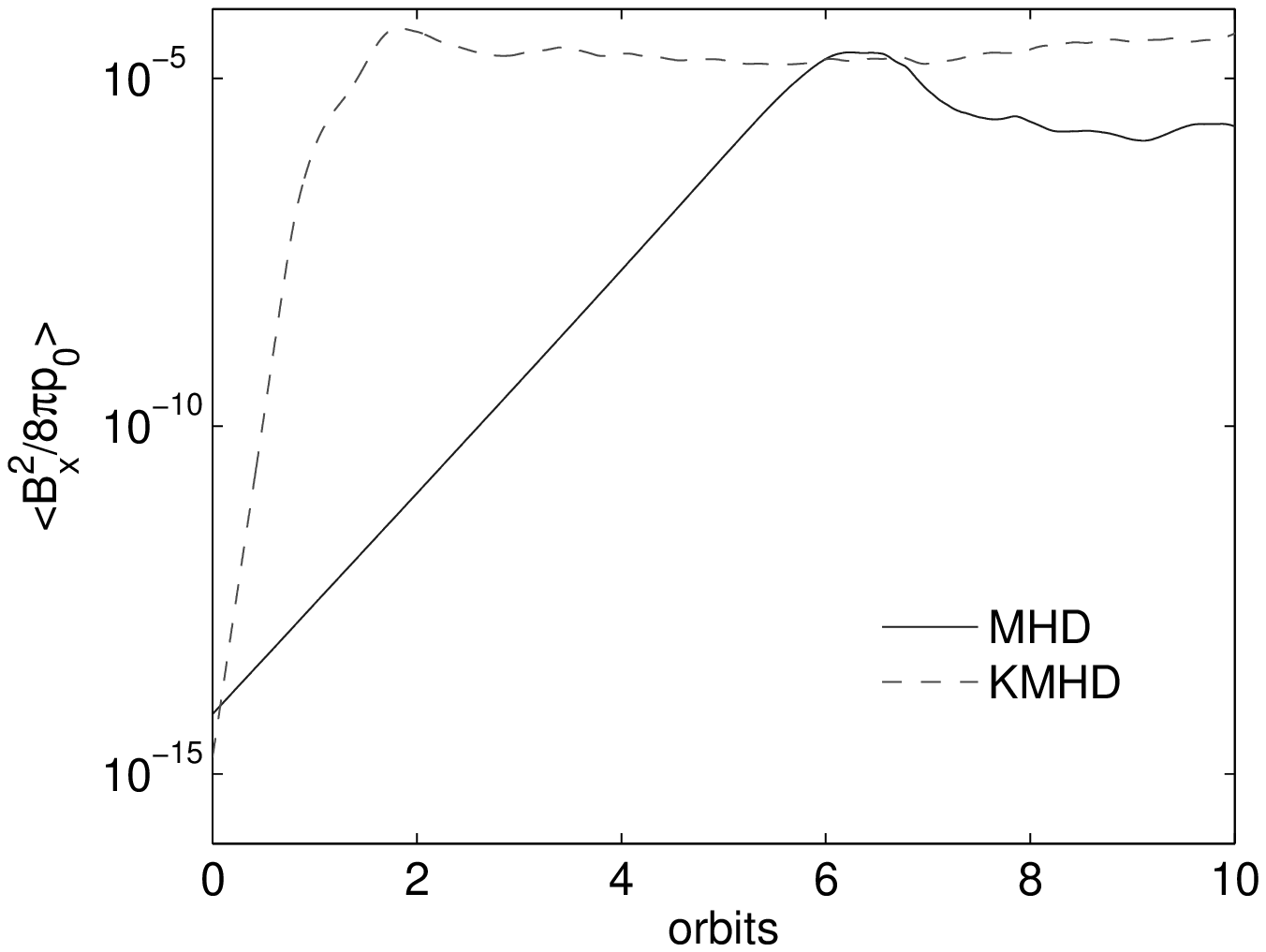}
\includegraphics[width=2.95in,height=2.5in]{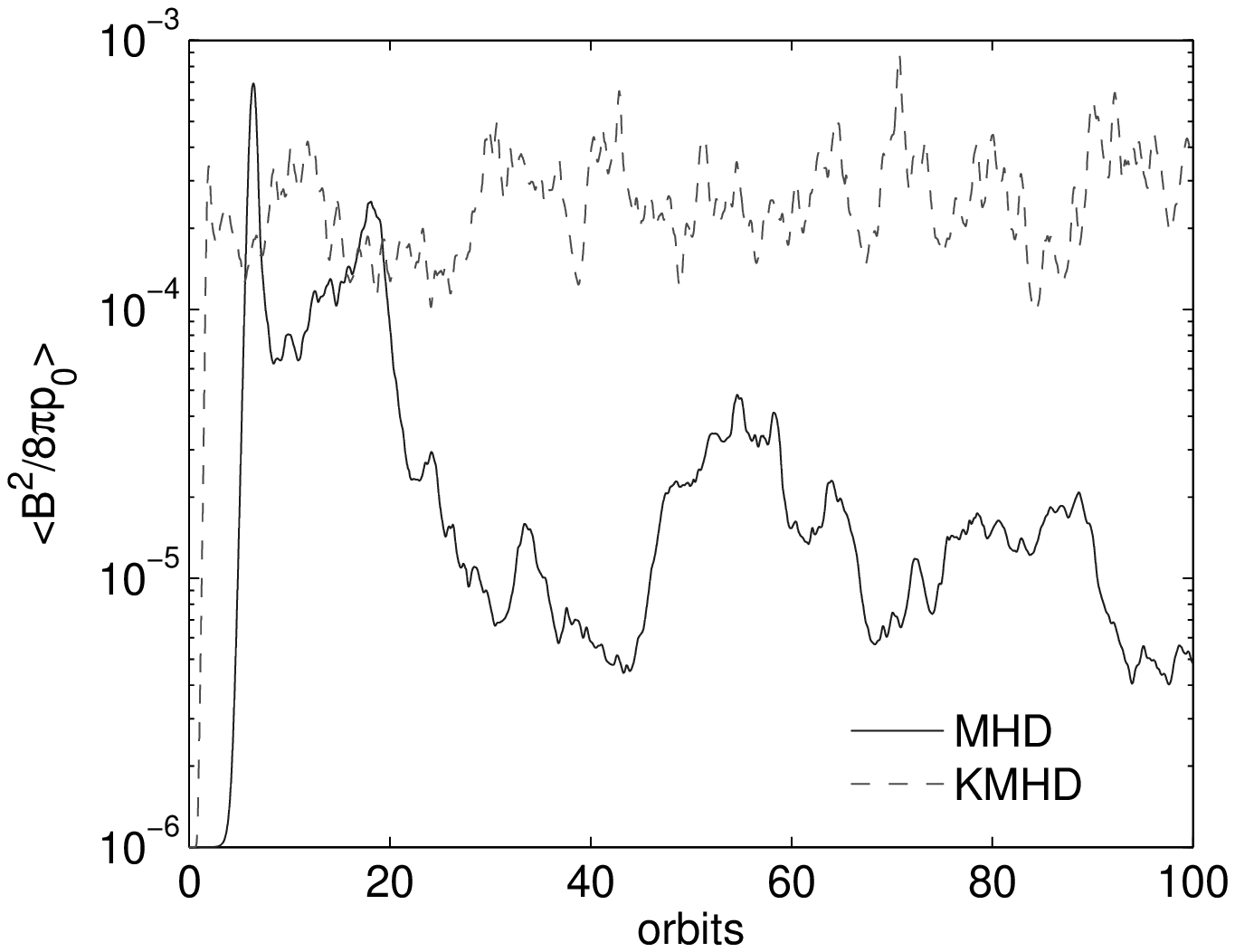}
\caption[Magnetic energy for a linear eigenmode in MHD and
KMHD]{Figure on left shows magnetic energy in the $x$- component,
$\la B_x^2/8\pi p_0 \ra$, plotted as a function of number of orbits
for runs $MYZlin$ and $KYZlin$. The initial disturbance is a linear
eigenmode with an amplitude of $10^{-9}$ and vertical wavenumber
$k_z = 8\pi/L_z = 4 k_{z,min}$. Resolution for both cases is $27
\times 59 \times 27$. As expected, the MRI growth rate is much
faster in kinetic regime than in MHD. The growth rates deduced from
the slope are for MHD: $\gamma/\Omega=0.29$, and for KMHD:
$\gamma/\Omega=1.78$. Figure on right shows the total magnetic
energy for 100 orbits. Saturated magnetic energy in MHD is much
smaller than KMHD at late times. \label{Ch4fig:figure15}}
\end{center}
\end{figure}

Figure \ref{Ch4fig:figure15} shows the magnetic energy in the $x$-
component of the magnetic field, $\la\la B_x^2/8\pi\ra\ra$, for runs
initialized with an MRI eigenmode (runs $KYZlin$ and $MYZlin$ in
Table \ref{Ch4tab:tab4}); the kinetic growth rate is indeed faster
than in MHD, as predicted by linear theory. The MHD growth rate for
$MYZlin$ calculated from the slope of $B_x^2/8\pi$ is
$\gamma/\Omega=0.29$, consistent with linear theory for this
particular mode (Figure \ref{Ch4fig:figure14} shows a similar growth
rate for $k_z=4k_{z,min}$, the mode initialized in $MYZlin$). For
the same run, Table \ref{Ch4tab:tab4} shows that the magnetic energy
and stresses in the saturated state are much smaller than all other
runs unlike the kinetic run $KYZlin$, the presence of a single mode
somehow affects the saturation in MHD! In comparison, similar case
initialized with random noise ($MYZl$) saturates at large amplitude
as shown in Table \ref{Ch4tab:tab4}. It seems that nonlinear
saturation in MHD shows a bifurcation depending on the initial
conditions; somehow the memory of initial conditions is retained
even at late times.

Apart from verifying the linear growth, we also study the
differences between the nonlinear saturation of the kinetic and MHD
simulations; all the runs described in this section use pitch angle
scattering models and conduction parameter ($k_L$) similar to the
fiducial run $Zl4$. We use a range of box sizes, starting from the
smallest boxes ($KYZl$, $KYZh$ and $MYZl$, $MYZh$) to the boxes with
vertical height equal to the disk height scale ($KYZ8l$, $KYZ8h$ and
$MYZ8l$, $MYZ8h$). The nonlinear simulations are done at low
($27\times 59 \times 27$) and high ($54 \times 118 \times 54$)
resolutions. Both MHD and kinetic simulations show that the magnetic
and kinetic energies, and stresses scale with the box size, provided
that the resolution is good enough (see Figure
\ref{Ch4fig:figure18}); this is similar to what was observed by
\cite{Hawley1995} for MHD simulations. The magnetic energy is
$\approx 5$ times smaller for the kinetic simulations, however, the
total stress (dominated by the anisotropic stress for kinetic
simulations) is comparable for kinetic and MHD simulations. Although
the MRI growth in the kinetic regime is double that in MHD, the
kinetic simulations saturate at a smaller magnetic energy compared
to MHD.
\begin{figure}
\begin{center}
\includegraphics[width=2.95in,height=2.5in]{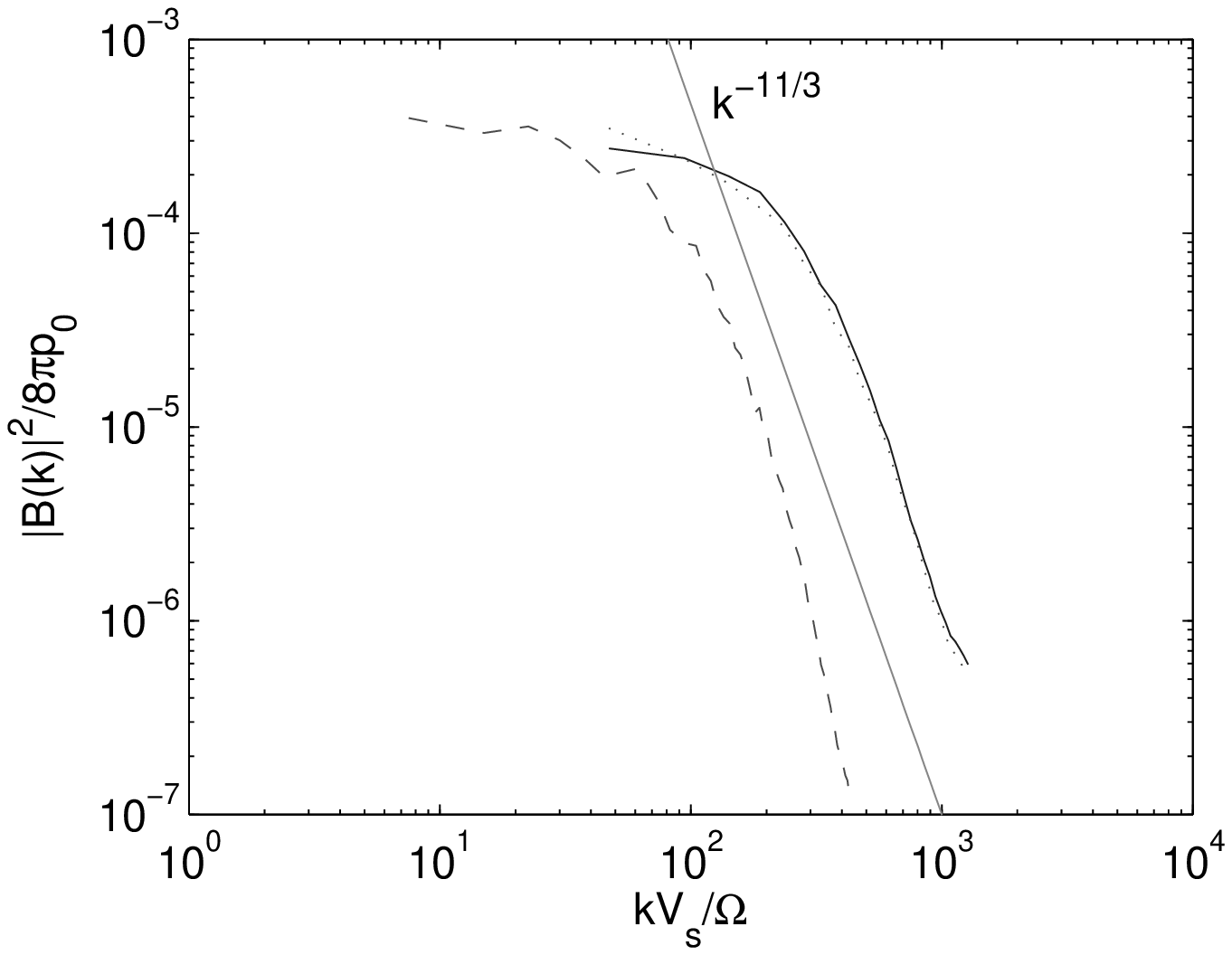}
\includegraphics[width=2.95in,height=2.5in]{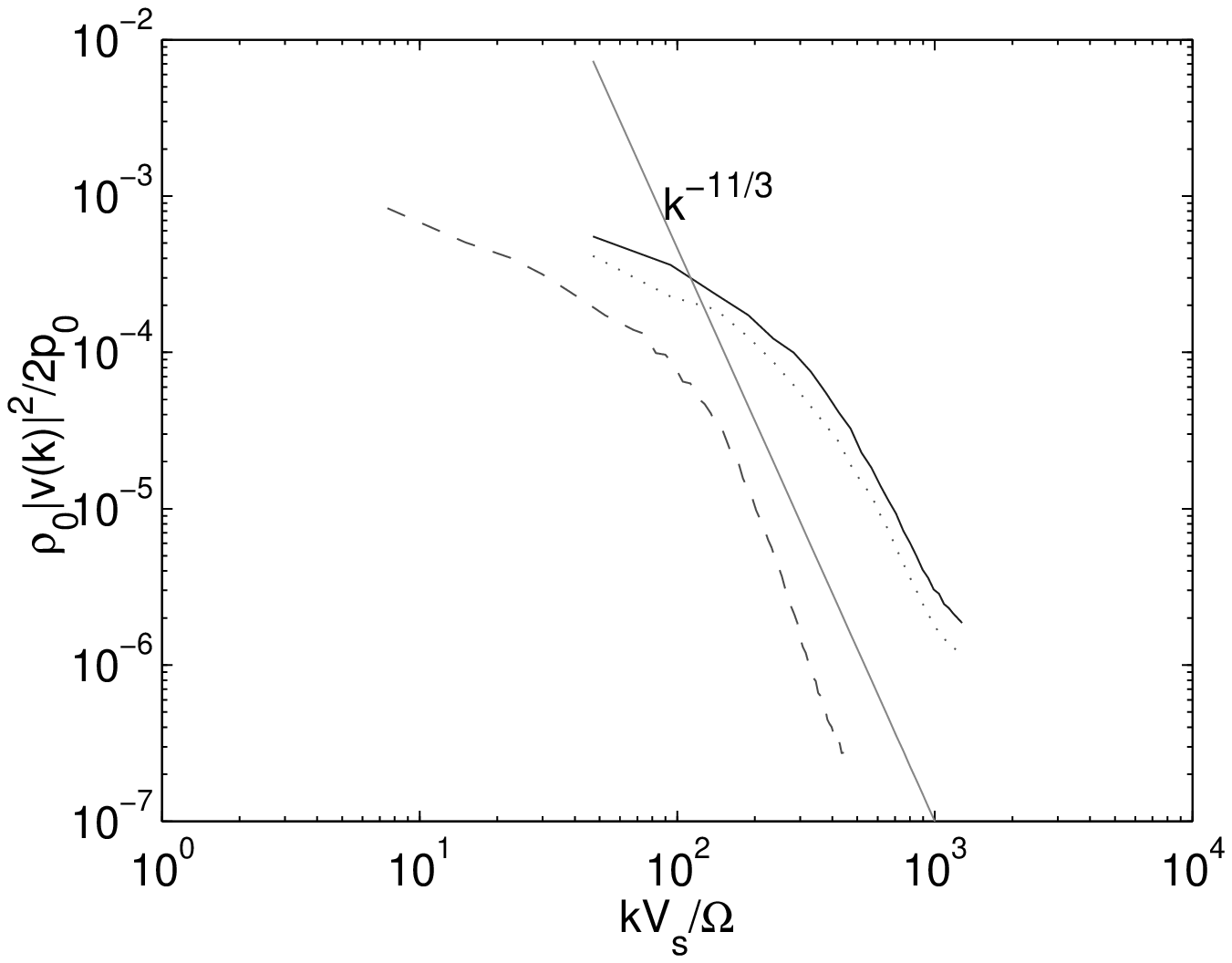}
\includegraphics[width=2.95in,height=2.5in]{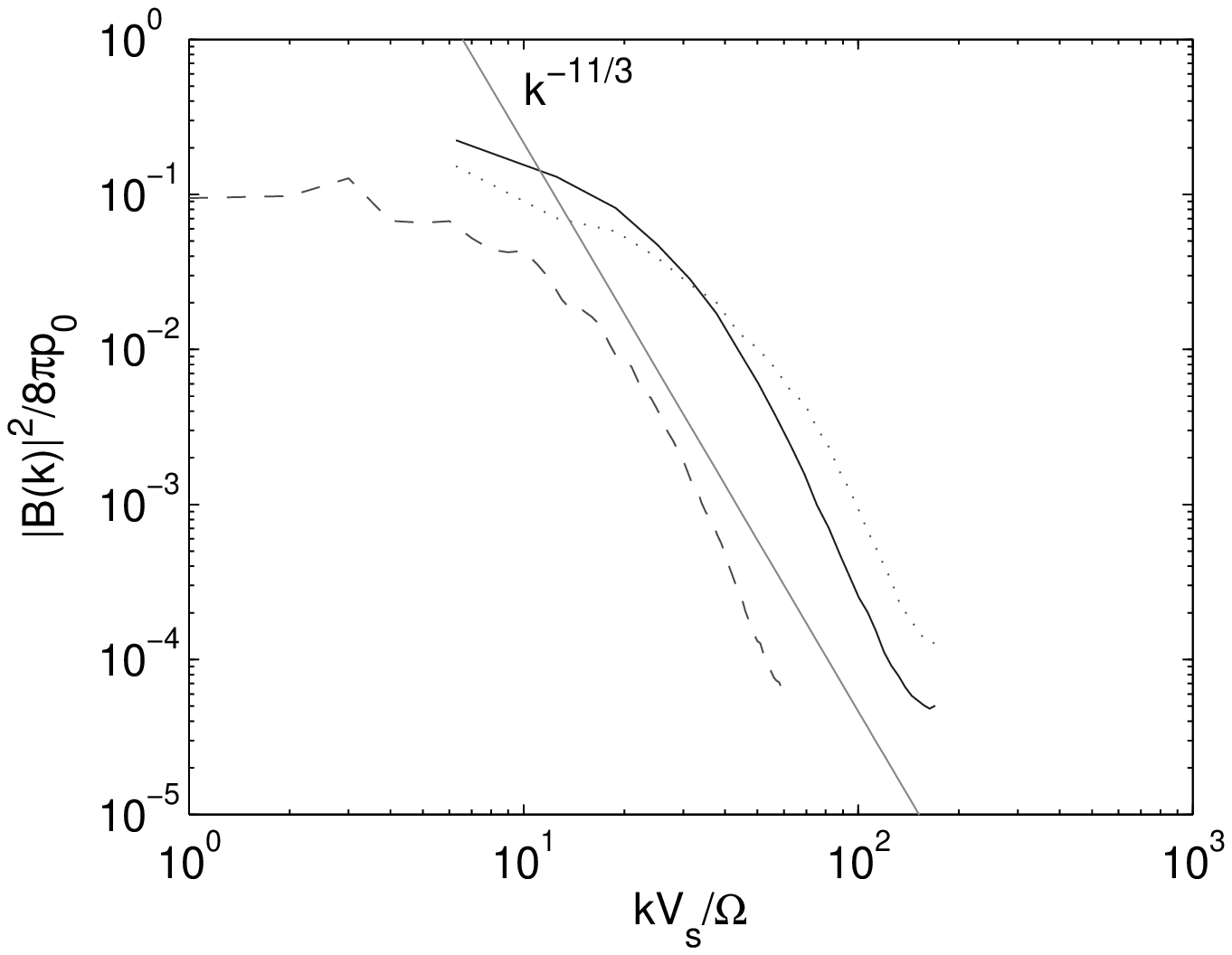}
\includegraphics[width=2.95in,height=2.5in]{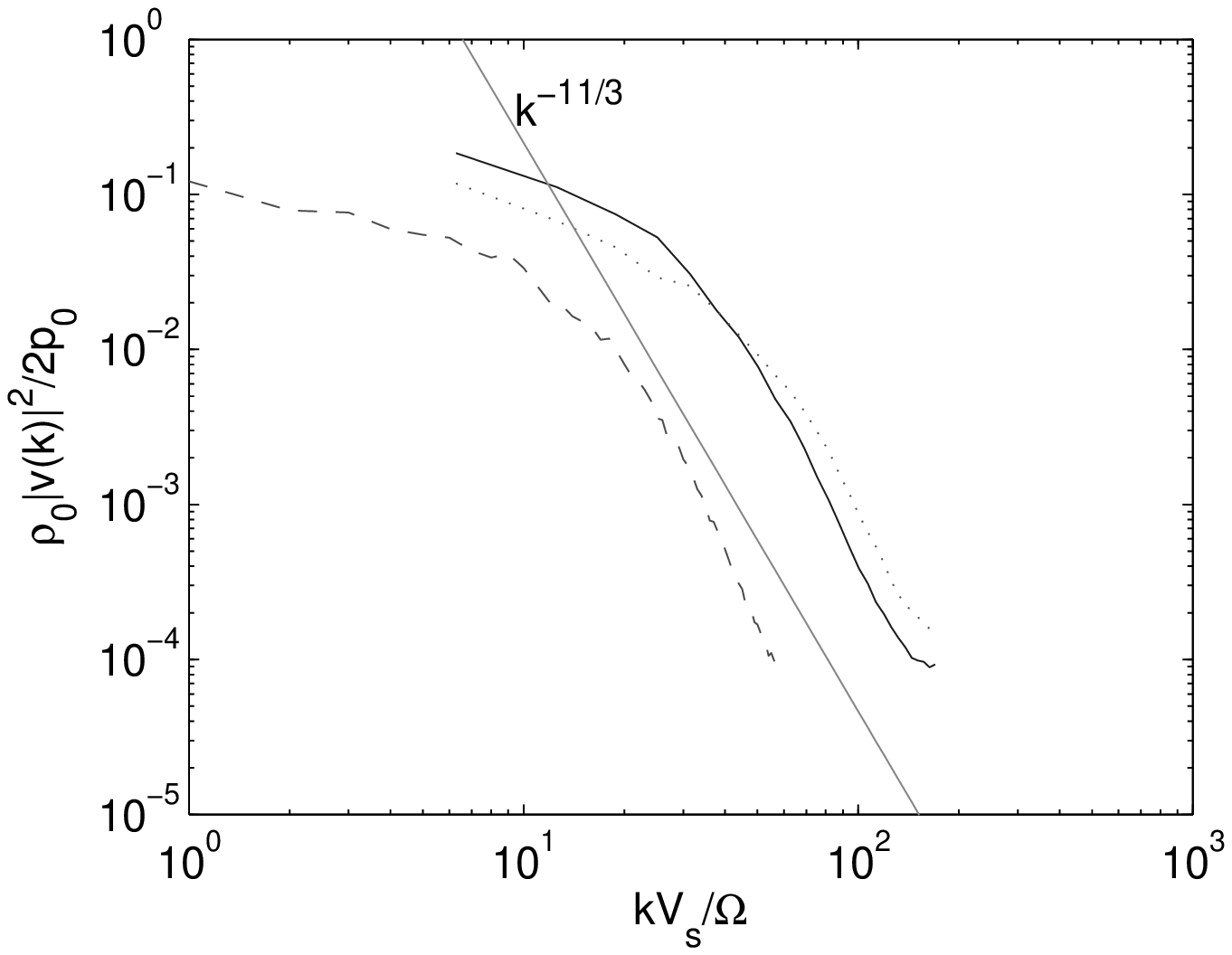}
\caption[Spectra of kinetic and magnetic energies for kinetic runs
$KYZh$ and $Zh4$]{Turbulent magnetic ($|B(k)|^2/8\pi p_0$) and
kinetic energy ($\rho_0 |V(k)|^2/2 p_0$) spectra for kinetic MHD:
the $\beta=10^6$ run $KYZh$ (top), and the $\beta=400$ initial
vertical field case (bottom, run $Zh4$ in \cite{Sharma2006}; see
Table \ref{Ch4tab:tab1}). Spectra with respect to $k_x$ (solid
line), $k_y$ (dashed line) and $k_z$ (dotted line) are shown. Also
shown is the $k^{-11/3}$ Kolmogorov spectrum. The magnetic and
kinetic energies for $KYZh$ are much smaller than for $Zh4$. For top
figures, spectra with respect to $k_y$ are steeper, because for
$\beta=10^6$ cases shear in velocity $V_y$ dominates the
fluctuations, causing the eddies to be elongated in the $y$-
direction, with a steeper spectrum. The spectra are averaged in the
other two directions in $k$- space; e.g., for a spectrum with
respect to $k_x$, $|{\bf B}|^2(k_x)= \int dk_y dk_z {\bf
B}(k_x,k_y,k_z) {\bf B^*}(k_x,k_y,k_z)$. \label{Ch4fig:figure16}}
\end{center}
\end{figure}
\begin{figure}
\begin{center}
\includegraphics[width=2.95in,height=2.5in]{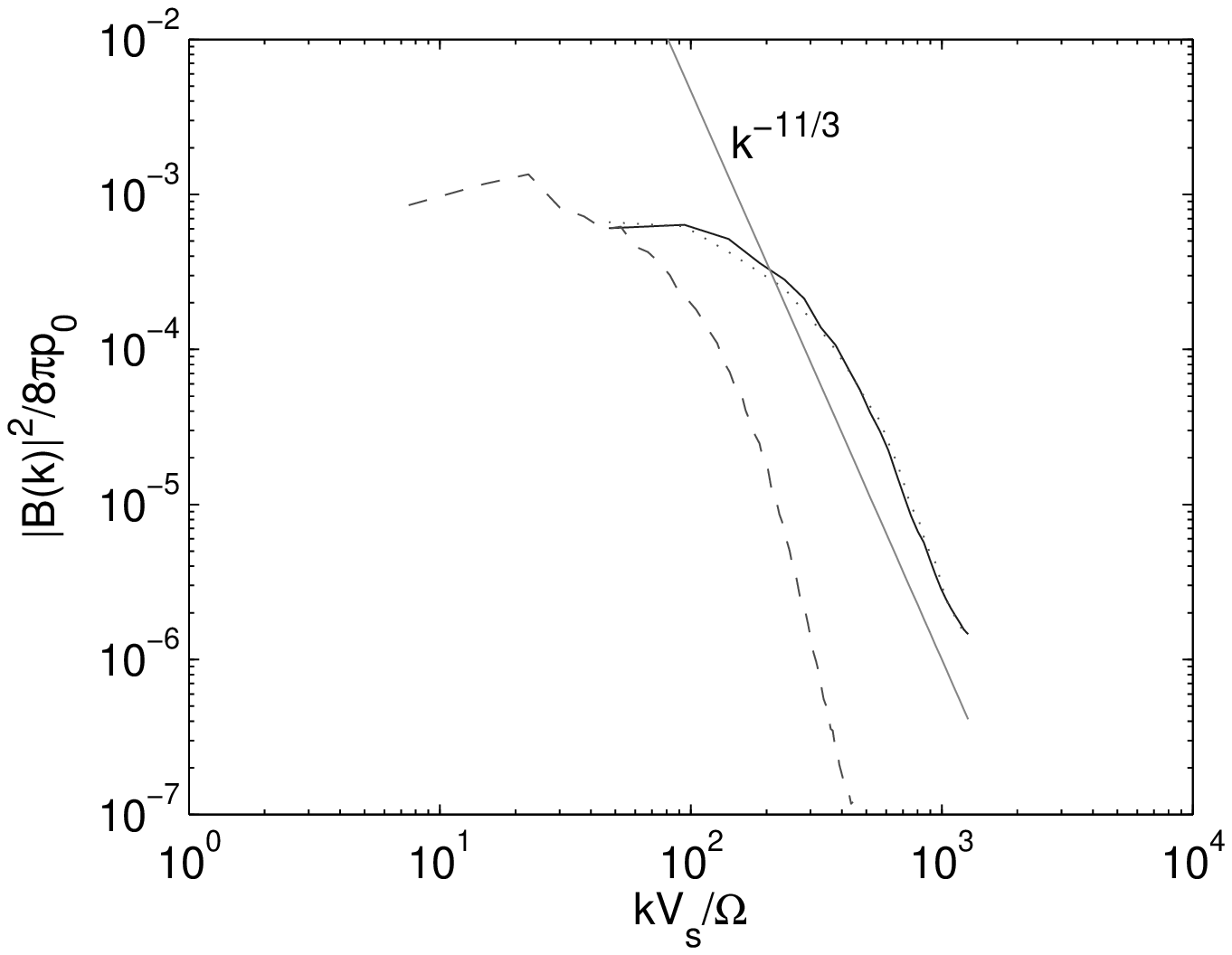}
\includegraphics[width=2.95in,height=2.5in]{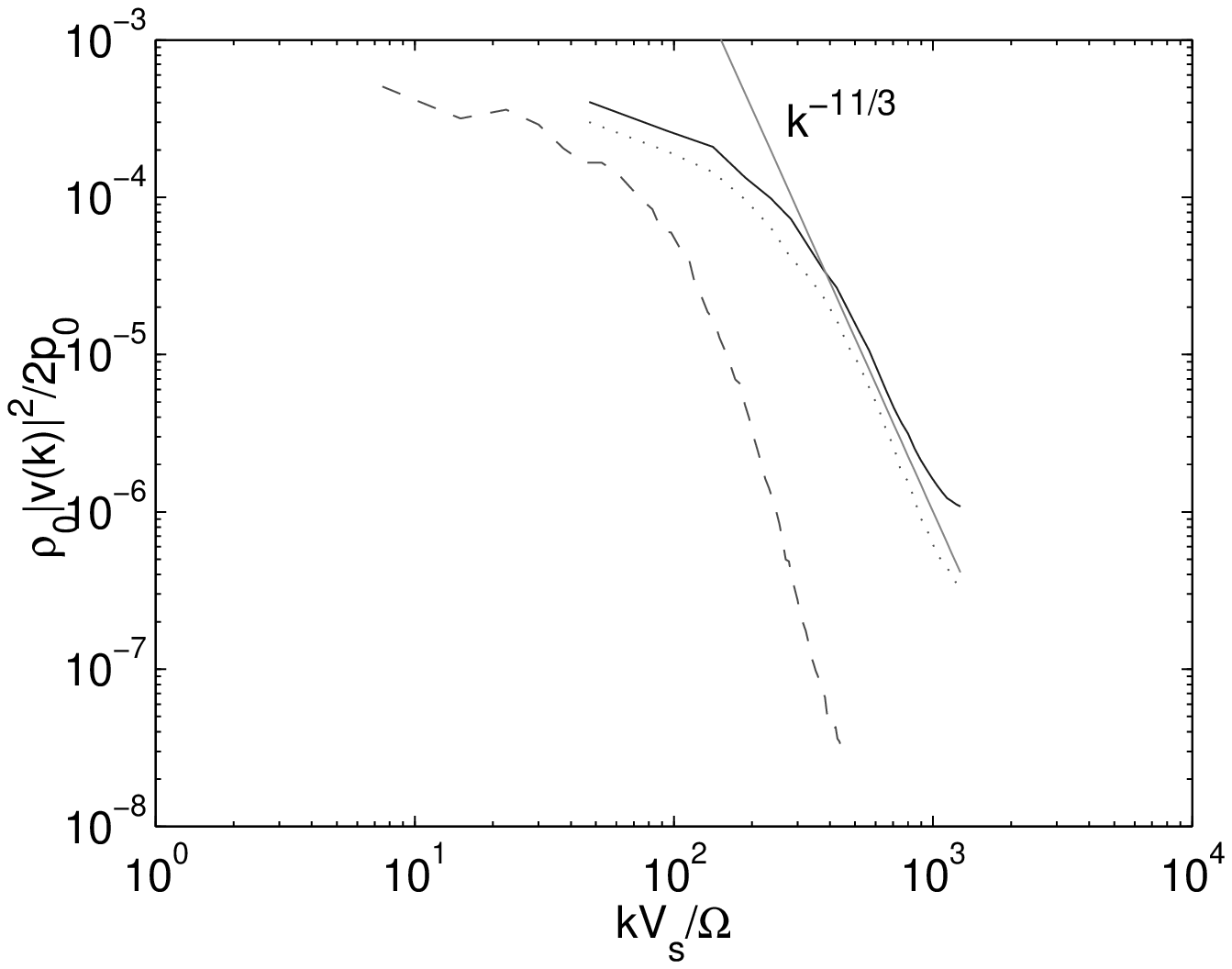}
\includegraphics[width=2.95in,height=2.5in]{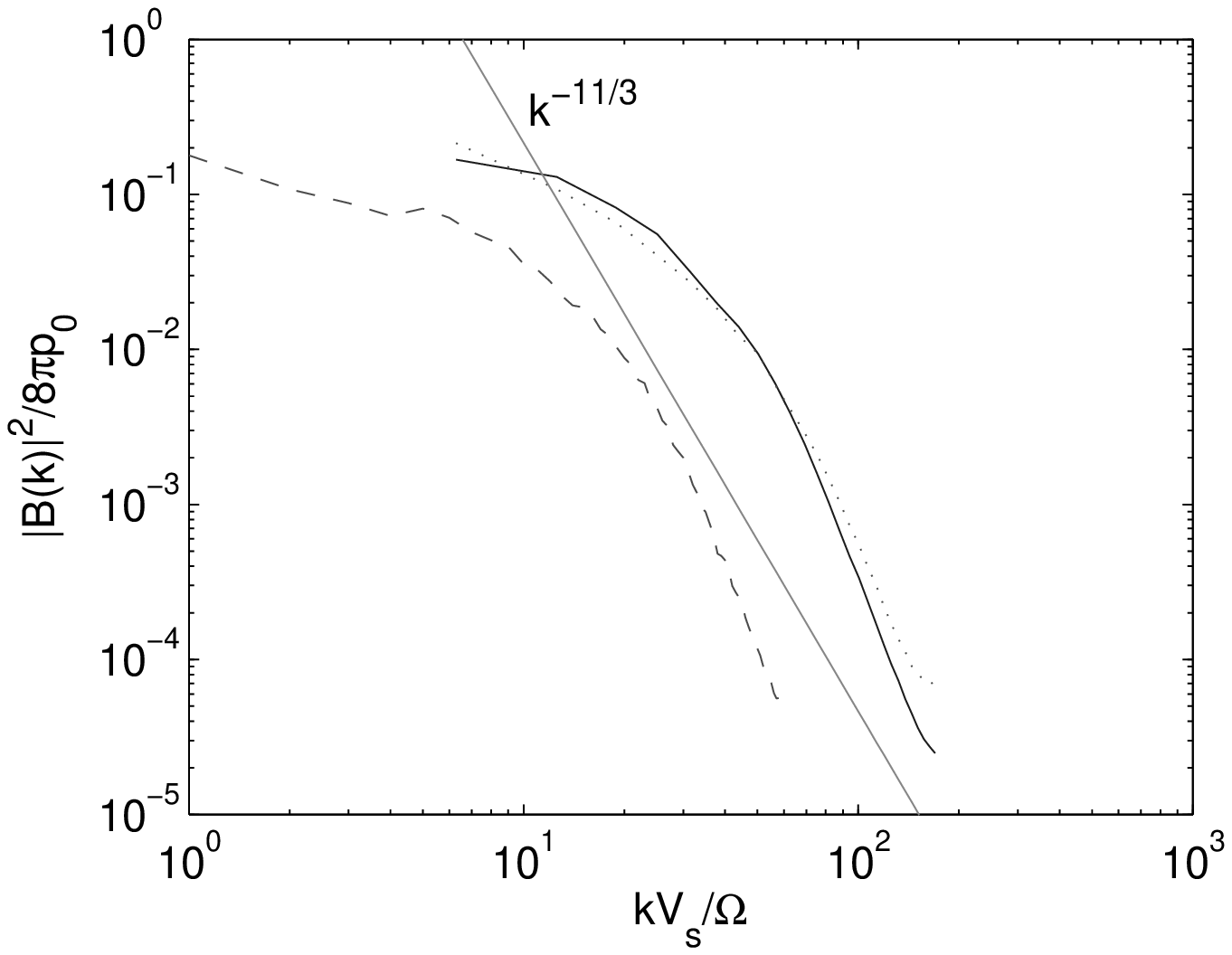}
\includegraphics[width=2.95in,height=2.5in]{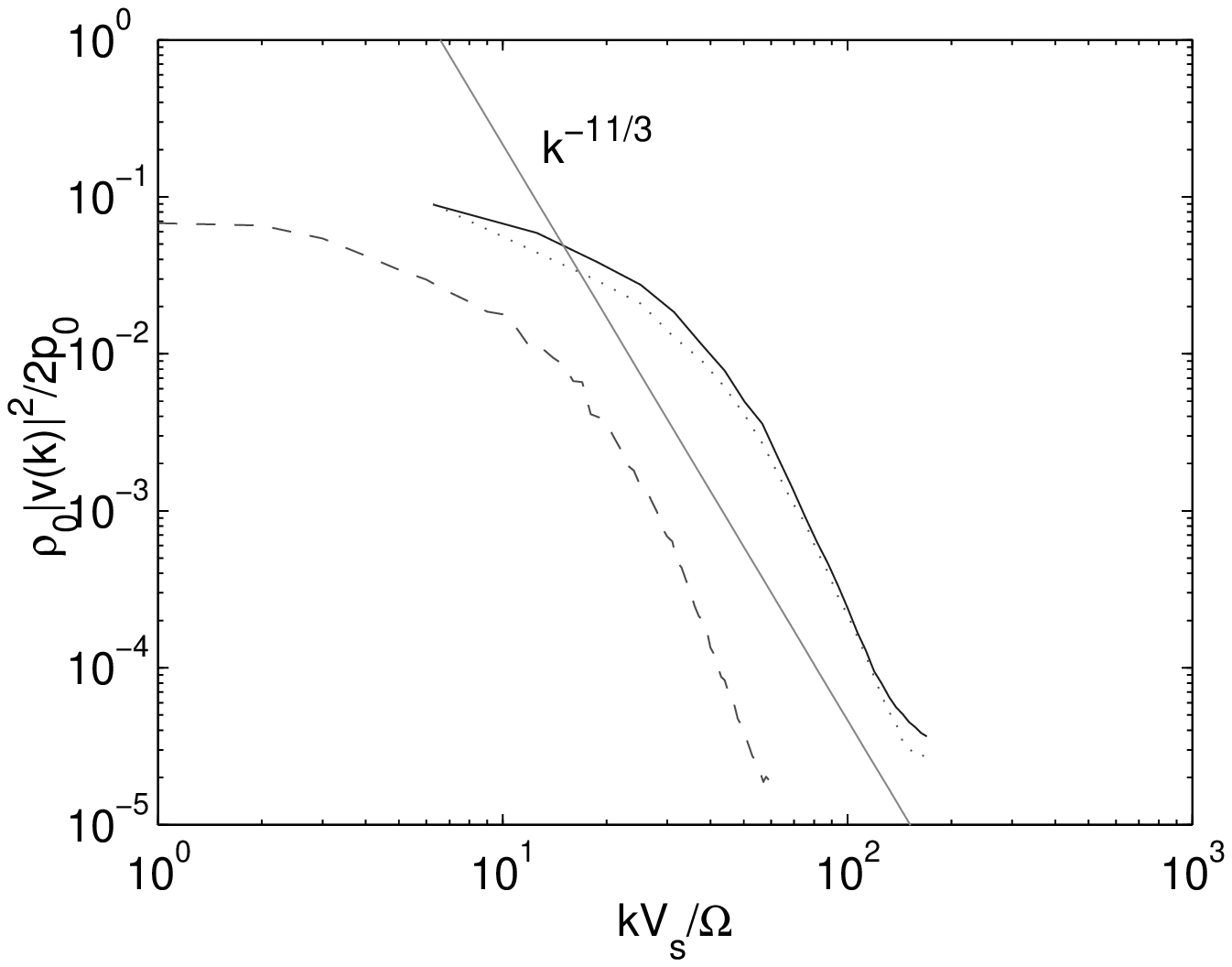}
\caption[Spectra of kinetic and magnetic energies for MHD runs
$MYZh$ and $ZMh$]{Turbulent magnetic ($|B(k)|^2/8\pi p_0$) and
kinetic energy ($\rho_0 |V(k)|^2/2 p_0$) spectra for MHD: the
$\beta=10^6$ run $MYZh$ (top), and the $\beta=400$ initial vertical
field case (bottom, run $ZMh$ in \cite{Sharma2006}; see Table
\ref{Ch4tab:tab1}). Spectra with respect to $k_x$ (solid line),
$k_y$ (dashed line) and $k_z$ (dotted line) are shown. Also shown is
the $k^{-11/3}$ Kolmogorov spectrum. The magnetic and kinetic
energies for $MYZh$ are much smaller than for $MZh$. For top
figures, spectra with respect to $k_y$ are steeper, because for
$\beta=10^6$ cases shear in velocity $V_y$ dominates the
fluctuations, causing the eddies to be elongated in the $y$-
direction, with a steeper spectrum. The spectra are averaged in the
other two directions in $k$- space; e.g., for a spectrum with
respect to, $|{\bf B}|^2(k_x)= \int dk_y dk_z {\bf B}(k_x,k_y,k_z)
{\bf B^*}(k_x,k_y,k_z)$. \label{Ch4fig:figure17}}
\end{center}
\end{figure}

Figure \ref{Ch4fig:figure16} shows the spectra of magnetic and
kinetic energies for runs $KYZh$ and $Zh4$; a $k^{-11/3}$ Kolmogorov
spectrum is a good fit for the kinetic and magnetic energies. The
spectra as a function of $k_y$ look slightly steeper for high
$\beta$ simulations. This may be because fluctuation energy is small
compared to the energy in the radial shear of $V_\phi$, which
elongates the eddies in the azimuthal direction. The spectra for MHD
$B_\phi=B_z$ runs are similar (see Figure \ref{Ch4fig:figure17}) to
the kinetic runs. Although the spectra are similar to the Kolmogorov
spectrum for isotropic, homogeneous turbulence, MRI turbulence is
anisotropic with non-zero correlations between radial and azimuthal
fields, resulting in sustained Maxwell and Reynolds stresses.

\begin{sidewaystable}[hbt]
\begin{center}
\caption{$B_\phi=B_z$, $\beta=10^6$ simulations\label{Ch4tab:tab4}}
\vskip0.05cm
\begin{tabular}{cccccccccc} \hline
Label & $L_x$ & $L_y$ & $L_z$ & $\la \la \frac{B^2}{8\pi p_0} \ra
\ra$ & $\la \la \frac{V^2}{2 p_0} \ra \ra$ & $\la \la \frac{B_x
B_y}{4\pi p_0} \ra \ra$ & $\la \la \frac{\rho V_x \delta V_y}{p_0}
\ra \ra$ & $\la \la \frac{\Delta p^*}{B^2}\frac{B_xB_y}{p_0} \ra
\ra$ & $ \la \la \alpha \ra \ra$ \\
\hline $KYZl$ & $0.133$ & $0.837$ & $0.133$ & $2.51 \times 10^{-4}$
& $0.0021$& $1.12 \times 10^{-4}$ &
$1.04 \times 10^{-4}$ & $4.59 \times 10^{-4}$ & $7.75 \times 10^{-4}$ \\
$KYZlin$ & $0.133$ & $0.837$ & $0.133$ &
$2.84 \times 10^{-4}$ &$0.0021$ & $1.24 \times 10^{-4}$ & $1.11 \times 10^{-4}$ & $5.00 \times 10^{-4}$ & $7.34 \times 10^{-4} $ \\
$KYZh$ & $0.133$ & $0.837$ & $0.133$ & $3.68 \times 10^{-4}$ &
$0.0021$ &
$1.63 \times 10^{-4}$ & $1.26 \times 10^{-4}$ & $5.53 \times 10^{-4}$ & $8.42 \times 10^{-4} $ \\
$KYZ2l$ & $0.267$ & $1.675$ & $0.267$ & $4.36 \times 10^{-4}$
&$0.0073$&
$1.84 \times 10^{-4}$ & $1.84 \times 10^{-4}$ & $8.00 \times 10^{-4}$ & $0.0012$ \\
$KYZ2h$ & $0.267$ & $1.675$ & $0.267$ & $5.17 \times 10^{-4}$ &
$0.0074$ &
$2.25 \times 10^{-4}$ & $2.05 \times 10^{-4}$ & $8.54 \times 10^{-4}$ & $0.0013$ \\
$KYZ4l$ & $0.533$ & $3.350$ & $0.533$ & $4.14 \times 10^{-4}$
&$0.0273$&
$1.52 \times 10^{-4}$ & $2.85 \times 10^{-4}$ & $8.91 \times 10^{-4}$ & $0.0013$ \\
$KYZ4h$ & $0.533$ & $3.350$ & $0.533$ & $0.001$ & $0.0277$ &
$4.5 \times 10^{-4}$ & $5.16 \times 10^{-4}$ & $0.0016$ & $0.0026$ \\
$KYZ8l$ & $1.0$ & $6.283$ & $1.0$ & $8.63 \times 10^{-5}$ &$0.1024$&
$2.63 \times 10^{-5}$ & $0.0012$ & $0.0010$ & $0.0022$ \\
$KYZ8h$ & $1.0$ & $6.283$ & $1.0$ &
$0.0023$ & $0.0948$ & $9.62 \times 10^{-4}$ & $0.0013$ & $0.0030$ & $0.0053$ \\
$MYZl$ & $0.133$ & $0.837$ & $0.133$ &
$8.21 \times 10^{-4}$ & $0.0019$ & $3.56 \times 10^{-4}$ & $7.05 \times 10^{-4}$ & $-$ & $0.0011$ \\
$MYZlin$ & $0.133$ & $0.837$ & $0.133$ & $1.40 \times 10^{-5}$ &
$0.0015$ &
$2.66 \times 10^{-6}$ & $1.69 \times 10^{-6}$ & $-$ & $4.35 \times 10^{-6}$ \\
$MYZh$ & $0.133$ & $0.837$ & $0.133$ & $7.57 \times 10^{-4}$ &
$0.0019$ &
$3.43 \times 10^{-4}$ & $6.96 \times 10^{-5}$ & $-$ & $0.0010$ \\
$MYZ2l$ & $0.267$ & $1.675$ & $0.267$ & $0.0020$ & $0.0071$ &
$8.35 \times 10^{-4}$ & $1.96 \times 10^{-4}$ & $-$ & $0.0010$ \\
$MYZ2h$ & $0.267$ & $1.675$ & $0.267$ & $0.0019$ & $0.0072$ &
$8.68 \times 10^{-4}$ & $2.10 \times 10^{-4}$ & $-$ & $0.0011$ \\
$MYZ4l$ & $0.533$ & $3.350$ & $0.533$ & $0.0088$ &$0.0297$ &
$0.0036$ & $0.0012$ & $-$ & $0.0048$ \\
$MYZ4h$ & $0.533$ & $3.350$ & $0.533$ & $0.0052$ & $0.0282$ &
$0.0023$ & $7.88 \times 10^{-4}$ & $-$ & $0.0031$ \\
$MYZ8l$ & $1.0$ & $6.283$ & $1.0$ & $0.0012$ &$0.0959$&
$1.56 \times 10^{-4}$ & $2.32 \times 10^{-4}$ & $-$ & $3.88 \times 10^{-4}$ \\
$MYZ8h$ & $1.0$ & $6.283$ & $1.0$ & $0.0111$ & $0.0967$ &
$0.0047$ & $0.0023$ & $-$ & $0.007$ \\
\hline
\end{tabular}
\end{center}
`$YZ$' represents both $Y$ an $Z$ fields. `$l$' and `$h$' stand for
low ($27 \times 59 \times 27$) and high resolution ($54 \times 118
\times 54$) runs. `$lin$' stands for an initial linear eigenmode
with $k_z=8\pi/L_z$. `$K$' and `$M$' stand for kinetic and MHD
respectively. \\
$^* \Delta p = (p_\parallel-p_\perp)$
\end{sidewaystable}

We have carried out vertical field simulation with $\beta=10^6$ to
compare with high $\beta$ $B_\phi=B_z$ simulations. For an initial
vertical field, the growth rate for the fastest growing mode is the
same in MHD and kinetic regimes (see Figure \ref{Ch3fig:Fig2}). The
parameters for these simulations are similar to the $B_\phi=B_z$
simulations; the volume and time averaged quantities are listed in
Table \ref{Ch4tab:tab5}. Although the growth rates for $B_\phi=B_z$
cases are larger than for a vertical field, Figure
\ref{Ch4fig:figure18} shows that the saturation energies and
stresses are $\approx$ 1-2 times larger for the vertical field
cases. Similarly, MHD simulations show slightly larger stresses and
energies for pure vertical field cases. Also, as in $B_\phi=B_z$
simulations, pure vertical field simulations show that the saturated
magnetic energy is $\approx 3-5$ times smaller for the kinetic
regime compared to MHD, whereas, the total stress (dominated by the
anisotropic stress in the kinetic regime) is comparable. This
demonstrates that $B_\phi=B_z$ simulations are not very different
from the pure vertical field simulations in the kinetic regime.
Although the magnetic energy in MHD regime is larger, anisotropic
stress results in a comparable total stress in the kinetic and MHD
regimes.

\begin{figure}
\begin{center}
\includegraphics[width=2.95in,height=2.5in]{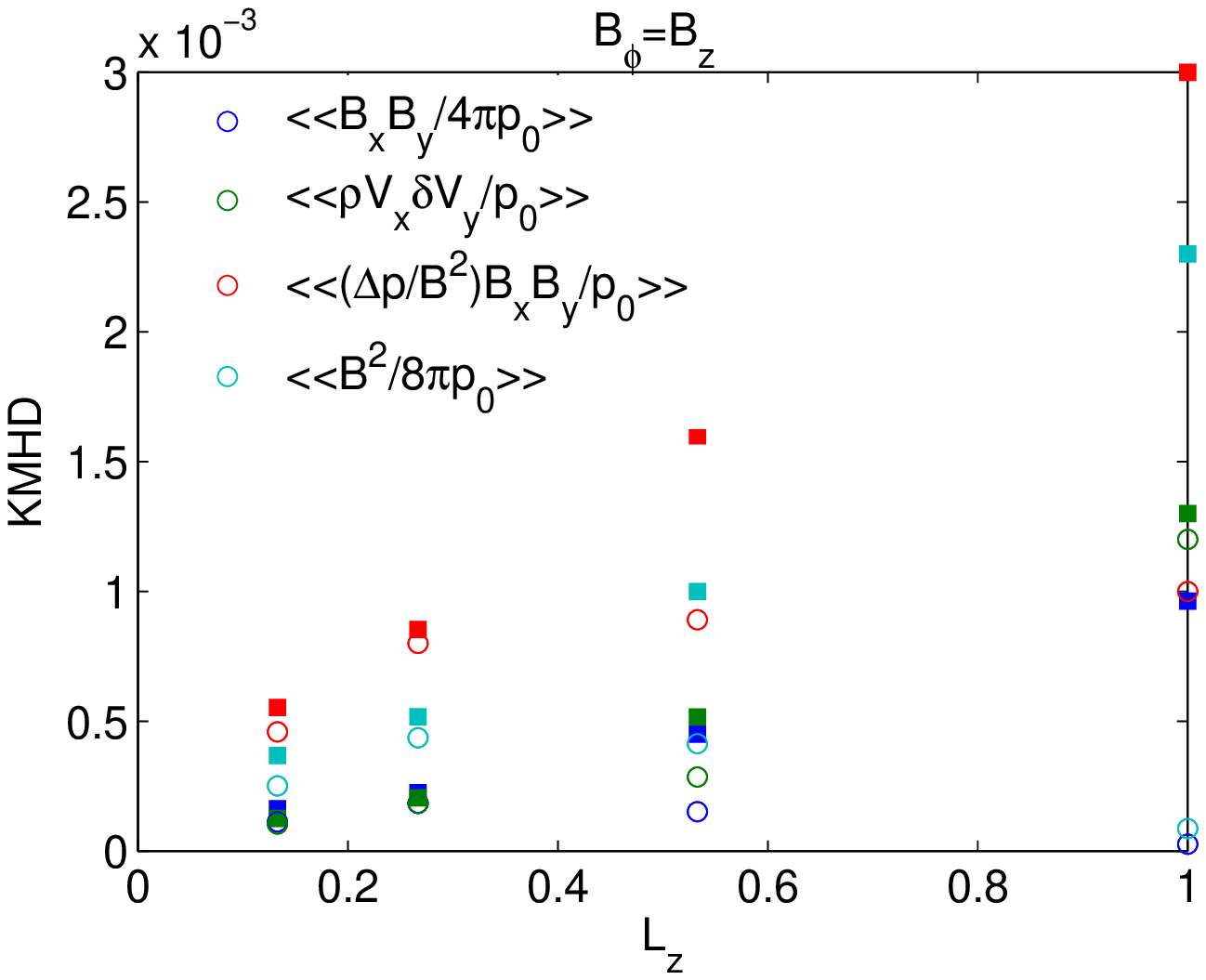}
\includegraphics[width=2.95in,height=2.5in]{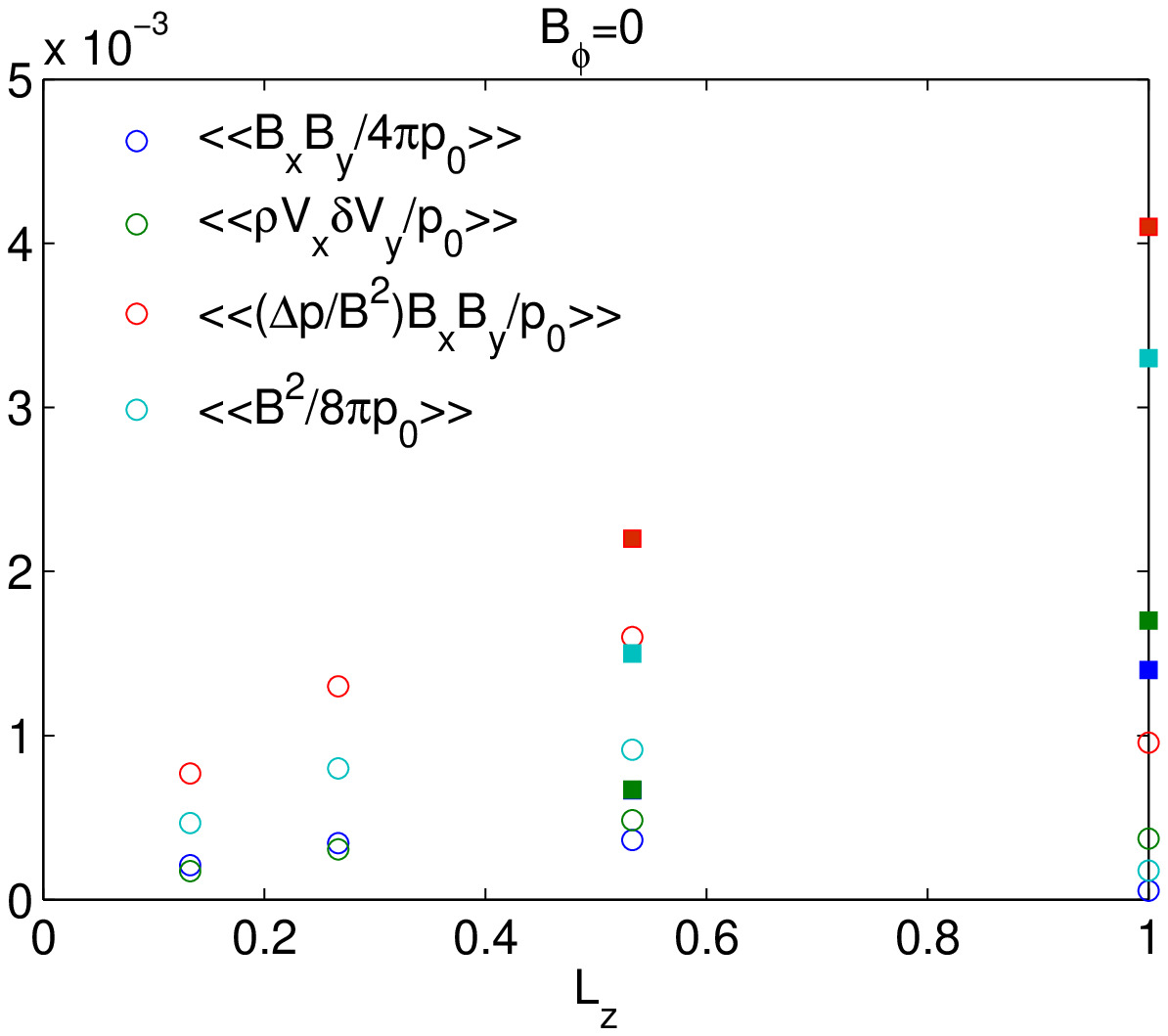}
\includegraphics[width=2.95in,height=2.5in]{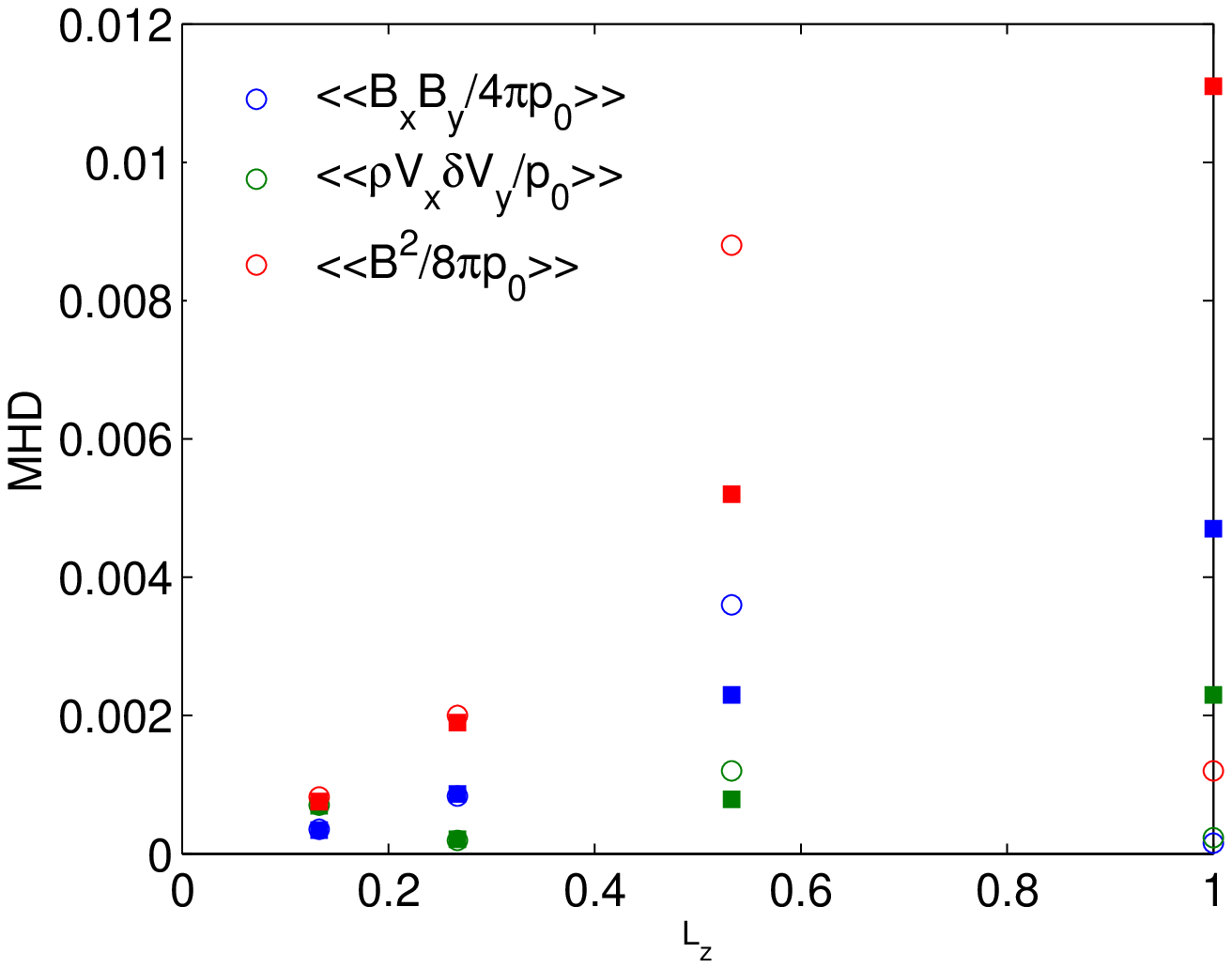}
\includegraphics[width=2.95in,height=2.5in]{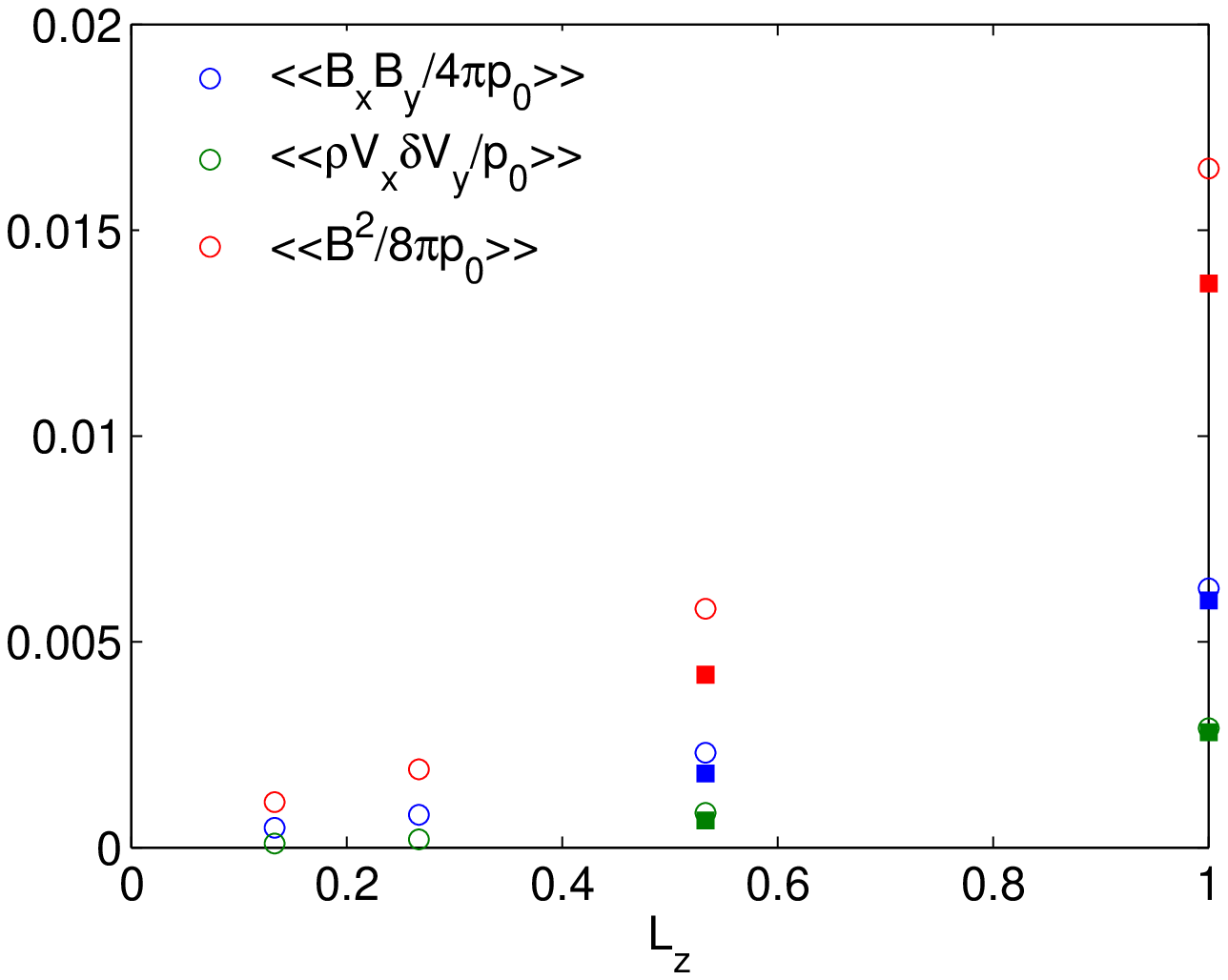}
\caption[Convergence studies for $\beta=10^6$ MRI simulations with
$B_\phi=B_z$ and only $B_z$]{The top plots shows the Maxwell,
Reynolds, and anisotropic stresses, and magnetic energy for
$\beta=10^6$ runs in the kinetic regime; the left one with
$B_\phi=B_z$ and the right one with only $B_z$. The bottom plots
show the Maxwell and Reynolds stresses, and magnetic energy for MHD
runs; the left one with $B_\phi=B_z$ and the right one with only
$B_z$. Open circles represent low resolution runs ($27\times 59
\times 27$), while filled squares represent high resolution runs
($54 \times 118 \times 54$). The magnetic energy in the saturated
state is $\approx 3-5$ times larger in the MHD regime, while the
total stress is comparable in the two regimes. The stresses and
magnetic energy increases with the box size, except for the low
resolution kinetic and $B_\phi=B_z$ MHD runs with the vertical box
size equal to the box height scale (runs labeled by `$8l$').
\label{Ch4fig:figure18}}
\end{center}
\end{figure}

\begin{sidewaystable}[hbt]
\begin{center}
\caption{Only $B_z$, $\beta=10^6$ simulations\label{Ch4tab:tab5}}
\vskip0.05cm
\begin{tabular}{cccccccccc}
\hline
Label & $L_x$ & $L_y$ & $L_z$ & $\la \la \frac{B^2}{8\pi p_0} \ra
\ra$ & $\la \la \frac{V^2}{2 p_0} \ra \ra$ & $\la \la \frac{B_x
B_y}{4\pi p_0} \ra \ra$ & $\la \la \frac{\rho V_x \delta V_y}{p_0}
\ra \ra$ & $\la \la \frac{\Delta p^*}{B^2}\frac{B_xB_y}{p_0} \ra
\ra$ & $ \la \la \alpha \ra \ra$ \\
\hline $KZl$ & $0.133$ & $0.837$ & $0.133$ & $4.66 \times 10^{-4}$ &
$0.0023$& $2.09 \times 10^{-4}$ & $1.74 \times 10^{-4}$ & $7.68
\times 10^{-4}$ &
$0.0012$ \\
$KZ2l$ & $0.267$ & $1.675$ & $0.267$ & $7.99 \times 10^{-4}$ &
$0.0077$& $3.45 \times 10^{-4}$ & $3.07 \times 10^{-4}$ & $0.0013$ &
$0.0019$ \\
$KZ4l$ & $0.533$ & $3.350$ & $0.533$ & $9.14 \times 10^{-4}$ &
$0.0278$& $3.63 \times 10^{-4}$ & $4.84 \times 10^{-4}$ & $0.0016$ &
$0.0024$ \\
$KZ4h$ & $0.533$ & $3.350$ & $0.533$ & $0.0015$ & $0.0282$& $6.66
\times 10^{-4}$ & $6.71 \times 10^{-4}$ & $0.0022$ &
$0.0035$ \\
$KZ8l$ & $1.0$ & $6.283$ & $1.0$ & $1.77 \times 10^{-4}$ & $0.0992$&
$5.35 \times 10^{-5}$ & $3.72 \times 10^{-4}$ & $9.57 \times
10^{-4}$ &
$0.0014$ \\
$KZ8h$ & $1.0$ & $6.283$ & $1.0$ & $0.0033$ &
$0.0946$& $0.0014$ & $0.0017$ & $0.0041$ & $0.0072$ \\

$MZl$ & $0.133$ & $0.837$ & $0.133$ & $0.0011$ & $0.0020$ &
$4.79 \times 10^{-4}$ & $9.41 \times 10^{-5}$ & $-$ & $5.73 \times 10^{-4}$ \\
$MZ2l$ & $0.267$ & $1.675$ & $0.267$ & $0.0019$ & $0.0072$ &
$7.93 \times 10^{-4}$ & $1.96 \times 10^{-4}$ & $-$ & $9.89 \times 10^{-4}$ \\
$MZ4l$ & $0.533$ & $3.350$ & $0.533$ & $0.0058$ & $0.0283$ &
$0.0023$ & $8.44 \times 10^{-4}$ & $-$ & $0.0031$ \\
$MZ4h$ & $0.533$ & $3.350$ & $0.533$ & $0.0042$ & $0.0276$ &
$0.0018$ & $6.59 \times 10^{-4}$ & $-$ & $0.0025$ \\

$MZ8l$ & $1.0$ & $6.283$ & $1.0$ & $0.0165$ & $0.0978$ &
$0.0063$ & $0.0029$ & $-$ & $0.0092$ \\

$MZ8h$ & $1.0$ & $6.283$ & $1.0$ & $0.0137$ & $0.0986$ &
$0.006$ & $0.0028$ & $-$ & $0.0089$ \\
\hline
\end{tabular}
\end{center}
`$Z$' represents a vertical field. `$l$' and `$h$' stand for low
($27 \times 59 \times 27$) and high resolution ($54 \times 118
\times 54$) runs. `$K$' and `$M$' stand for kinetic and MHD
respectively. \\
$^* \Delta p = (p_\parallel-p_\perp)$
\end{sidewaystable}

\subsection{Runs with $\beta=400$}
\begin{figure}
\begin{center}
\includegraphics[width=2.95in,height=2.5in]{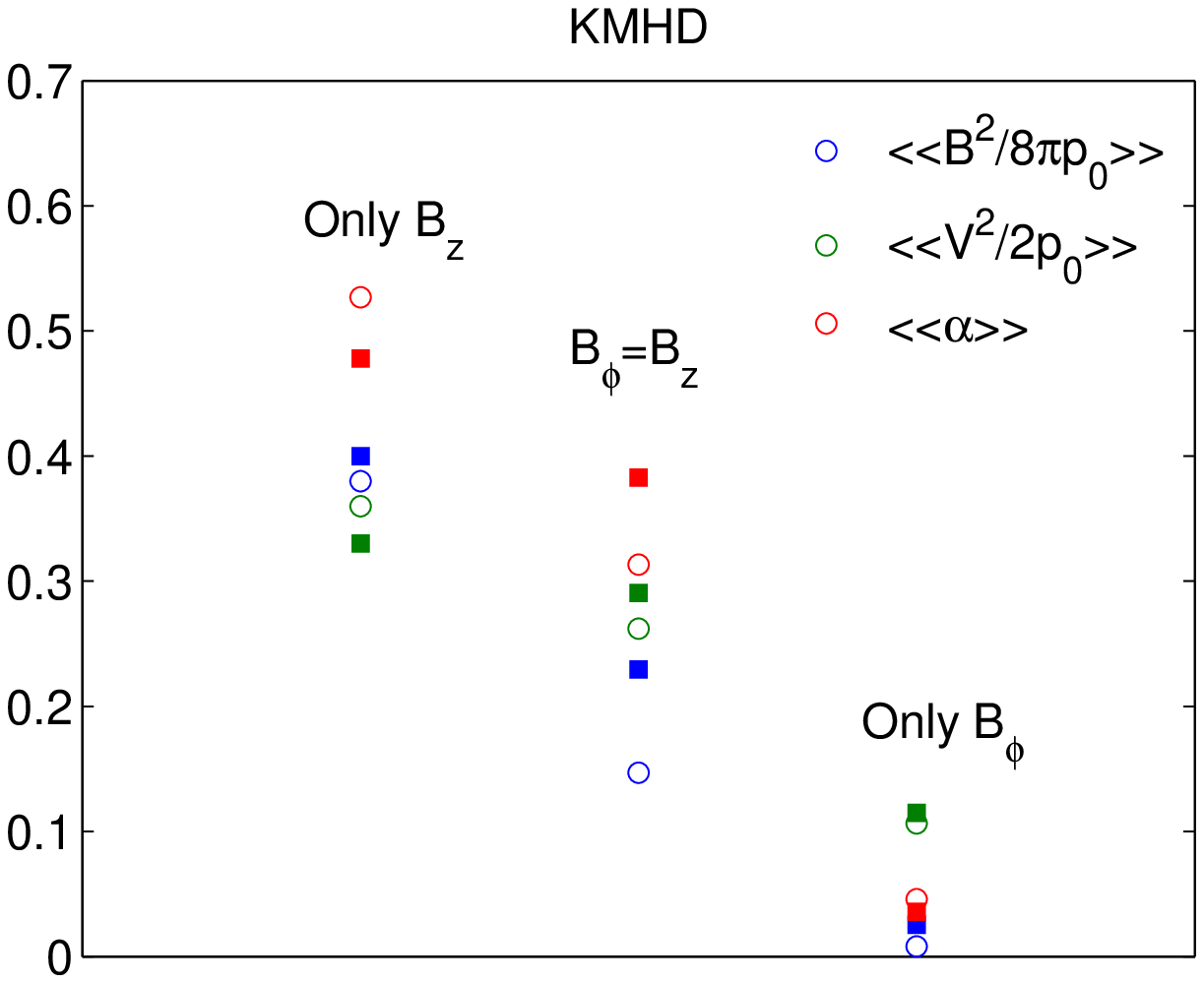}
\includegraphics[width=2.95in,height=2.5in]{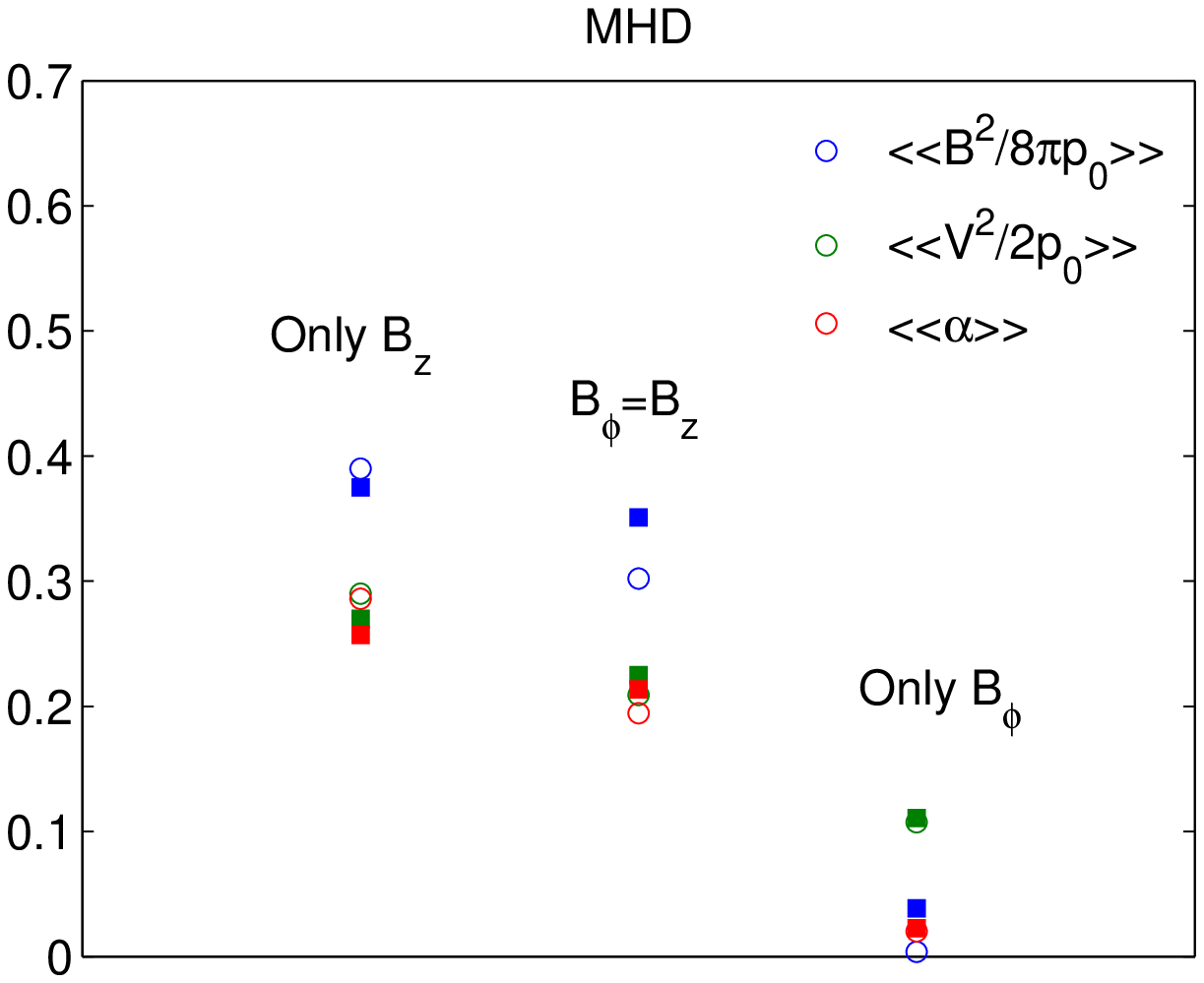}
\caption[Magnetic and kinetic energies, and total stress for
$\beta=400$ MRI simulations with different field orientations]{The
magnetic and kinetic energies, and the total stress for kinetic
(left) and MHD (right) simulations for $\beta=10^6$. For both cases,
energies and stresses are the largest for vertical field
simulations, followed by the $B_\phi=B_z$ runs, and the pure
azimuthal field runs. The total stress is $\approx$ twice larger for
the kinetic runs, whereas the magnetic energy is comparable with MHD
(smaller for the case of $B_\phi=B_z$). The kinetic and MHD
simulations with an azimuthal field give similar results. The
fluctuating kinetic energy is small compared to the energy in the
velocity shear for azimuthal field simulations; this is the reason
kinetic energy is larger than other quantities for azimuthal
simulations. \label{Ch4fig:figure19}}
\end{center}
\end{figure}

We also carried runs with $\beta=400$ to compare different field
geometries. Figure \ref{Ch4fig:figure19} shows that kinetic and
magnetic energies, and stresses are largest for the pure vertical
field cases (similar to $\beta=10^6$ simulations), followed by
$B_\phi=B_z$, and azimuthal field cases, for both MHD and kinetic
regime. Another point to be taken from Figure \ref{Ch4fig:figure19}
is that in the kinetic regime, unlike MHD, the total stress is
larger than the magnetic energy. For azimuthal field simulations in
both kinetic and MHD regimes, the fluctuation energy is smaller than
the energy in the shear flow; smaller fluctuations correspond to
lower level of turbulence and transport. For $B_\phi=B_z$
simulations, the magnetic energy is ($\approx$ twice) larger in the
MHD than in kinetic regime; reminiscent of $\beta=10^6$ results
where magnetic energy in MHD is even larger. Comparing simulations
where the initial $\beta=10^6$ with simulations where the initial
$\beta=400$ suggest that magnetic and kinetic energies and stresses
increase as we reduce the initial $\beta$. This behavior is not
fully understood but is similar to that observed in MHD simulations
with a net flux (see Figure 8 in \cite{Hawley1996}). MHD simulations
with no net flux result in a saturated $\beta$ independent of the
initial $\beta$ \cite{Hawley1996,Sano2004}, we expect the same to be
true in the kinetic regime.

\begin{sidewaystable}[hbt]
\begin{center}
\caption{$\beta=400$ simulations with different field orientations
\label{Ch4tab:tab6}} \vskip0.05cm
\begin{tabular}{cccccccccc}
\hline Label & $L_x$ & $L_y$ & $L_z$ & $\la \la \frac{B^2}{8\pi p_0}
\ra \ra$ & $\la \la \frac{V^2}{2 p_0} \ra \ra$ & $\la \la \frac{B_x
B_y}{4\pi p_0} \ra \ra$ & $\la \la \frac{\rho V_x \delta V_y}{p_0}
\ra \ra$ & $\la \la \frac{\Delta p^*}{B^2}\frac{B_xB_y}{p_0} \ra
\ra$ & $ \la \la \alpha \ra \ra$ \\
\hline
$Zl4^{\dagger}$ & $1.0$ & $6.283$ &
$1.0$ & $0.38$ & $0.36$ & $0.23$ & $0.097$ & $0.20$ & $0.527$ \\
$KYZl400$ & $1.0$ & $6.283$ & $1.0$ &
$0.147$ & $0.262$& $0.0838$ & $0.0537$ & $0.1757$ & $0.3132$ \\
$KYl400$ & $1.0$ & $6.283$ & $1.0$ & $0.008$ &
$0.1063$& $0.032$ & $0.0032$ & $0.0106$ & $0.0169$ \\
$Zh4^{\dagger}$ & $1.0$ & $6.283$ & $1.0$ & $0.40$ &
$0.33$ & $0.22$ & $0.078$ & $0.18$ & $0.478$ \\
$KYZh400$ & $1.0$ & $6.283$ & $1.0$ & $0.2294$ &
$0.2904$& $0.1211$ & $0.0571$ & $0.2046$ & $0.3828$ \\
$KYh400$ & $1.0$ & $6.283$ & $1.0$ & $0.0253$ &
$0.1148$& $0.0108$ & $0.0067$ & $0.0183$ & $0.0358$ \\
$MZl^{\dagger}$ & $1.0$ & $6.283$ & $1.0$ & $0.39$
& $0.29$ & $0.22$ & $0.066$ & $-$ & $0.286$ \\
$MYZl400$ & $1.0$ & $6.283$ & $1.0$ & $0.302$ & $0.209$ &
$0.1595$ & $0.0350$ & $-$ & $0.1945$ \\
$MYl400$ & $1.0$ & $6.283$ & $1.0$ & $0.0372$ & $0.1073$ &
$0.015$ & $0.0051$ & $-$ & $0.0201$ \\
$MZh^{\dagger}$ & $1.0$ & $6.283$ & $1.0$ & $0.375$
& $0.27$ & $0.204$ & $0.0531$ & $-$ & $0.257$ \\
$MYZh400$ & $1.0$ & $6.283$ & $1.0$ & $0.351$ & $0.225$ &
$0.1793$ & $0.0342$ & $-$ & $0.2135$ \\
$MYh400$ & $1.0$ & $6.283$ & $1.0$ & $0.0385$ & $0.1107$ &
$0.017$ & $0.0057$ & $-$ & $0.0227$ \\
\hline
\end{tabular}
\end{center}
'$Z$' and $Y$ represent vertical and azimuthal initial field. `$l$'
and `$h$' stand for low and high resolution runs. `$K$' and `$M$'
stand for kinetic and MHD respectively.\\
$^\dagger$ These runs are from \cite{Sharma2006}; see Table
\ref{Ch4tab:tab1} \\
$^* \Delta p = (p_ \parallel-p_\perp)$
\end{sidewaystable}

\section{Summary and Discussion}

In this chapter we have described our local shearing box simulations
of the magnetorotational instability in a collisionless plasma
\cite{Sharma2006}. We are motivated by the application to hot
radiatively inefficient flows which are believed to be present in
many low-luminosity accreting systems (see Section \ref{Ch1sec:RIAFs}).  Our
method for simulating the dynamics of a collisionless plasma is
fluid-based, and relies on evolving a pressure tensor with closure
models for the heat flux along magnetic field lines. These heat flux
models can be thought of as approximating the collisionless (Landau)
damping of linear modes in the simulation.

By adiabatic invariance, a slow increase (decrease) in the magnetic
field strength tends to give rise to a pressure anisotropy with
$p_\Perp > p_\Par$ ($p_\Par > p_\Perp$), where the sign of
anisotropy is defined by the local magnetic field.  Such a pressure
anisotropy can, however, give rise to small scale kinetic
instabilities (firehose, mirror, and ion cyclotron) which act to
isotropize the pressure tensor, effectively providing an enhanced
rate of pitch angle scattering (``collisions'').  We have included
the effects of this isotropization via a subgrid model which
restricts the allowed magnitude of the pressure anisotropy (see
Section \ref{Ch4sec:anisotropy}).

We find that the nonlinear evolution of the MRI in a collisionless
plasma is qualitatively similar to that in MHD, with comparable
saturation magnetic field strengths and magnetic stresses.  The
primary new effect in kinetic theory is the existence of angular
momentum transport due to the anisotropic pressure stress (Eq.
\ref{Ch4eq:angmom}).  For the allowed pressure anisotropies
estimated in Section \ref{Ch4sec:anisotropy}, the anisotropic stress is
dynamically important and is as large as the Maxwell stress (Table
\ref{Ch4tab:tab1}). The high $\beta$ $B_\phi=B_z$ simulations,
although showing the expected faster growth rate than in MHD, show a
smaller magnetic energy (factor of $\sim 5$) in the kinetic regime,
but the total stress is comparable to MHD. Although the MRI in
kinetic and MHD regimes is different linearly (with the fastest
growing mode in the kinetic regime twice faster than in MHD), they
are qualitatively similar in the nonlinear regime.

The precise rate of transport in the present simulations is
difficult to quantify accurately and depends---at the factor of
$\sim 2$ level---on some of the uncertain microphysics in our
kinetic analysis (e.g., the rate of heat conduction along magnetic
field lines and the exact threshold for pitch angle scattering by
small-scale instabilities; see Figure \ref{Ch4fig:figure7}).  For
better results, it would be interesting to extend these calculations
with a more accurate evaluation of the actual nonlocal heat fluxes,
Eqs. \ref{Ch4eq:nonlocal1}-\ref{Ch4eq:nonlocal2}, or even to
directly solve the drift kinetic equation, Eq. \ref{Ch4eq:DKE}, for
the particle distribution function. Further kinetic studies in the
local shearing box, including studies of the small-scale
instabilities that limit pressure anisotropy, would be helpful in
developing appropriate fluid closures for global simulations.

It is interesting to note that two-temperature RIAFs can only be
maintained below a critical luminosity $\sim \alpha^2 L_{\rm EDD}$
\cite{Rees1982}.  Thus enhanced transport in kinetic theory due to
the anisotropic pressure stress would extend upward in luminosity
the range of systems to which RIAFs could be applicable.  This is
important for understanding, e.g., state transitions in X-ray
binaries \cite{Esin1997}.

In addition to angular momentum transport by anisotropic pressure
stresses, Landau damping of long-wavelength modes can be important
in heating collisionless accretion flows.  Because the version of
ZEUS MHD code we use is non-conservative, we cannot carry out a
rigorous calculation of heating by different mechanisms such as
Landau damping and reconnection.  Following the total
energy-conserving scheme of ~\cite{Turner2003}, however, we estimate
that the energy dissipated by collisionless damping (present in the
form of work done by anisotropic stress) is comparable to or larger
than that due to numerical magnetic energy loss (which is the major
source of heating in MHD simulations), which represents both energy
lost due to reconnection and the energy cascading beyond the scales
at the resolution limit. One caveat to studying energetics in a
local shearing box is that in local simulations, the pressure
increases in time due to heating, while $B^2 \sim {\rm constant}$.
Thus $\beta$ increases in time and the turbulence becomes more and
more incompressible. This will artificially decrease the importance
of compressible channels of heating. Clearly it is of significant
interest to better understand heating and energy dissipation in
RIAFs, particularly for the electrons. The one fluid simulations
provide some indications of electron heating in RIAFs;
electrons will also be anisotropic because of magnetic energy
fluctuations. The pressure anisotropy in electrons is also limited
due to microinstabilities, e.g., the electron whistler instability
considered by \cite{Gary1996}. The heating rate of electrons due to
anisotropic stress ($d\ln p/dt$) is comparable to that of ions
because pressure anisotropy is comparable for electrons and ions.
For RIAFs, it may mean that electrons cannot be kept too cool
compared to the ion; but systematic 2-fluid simulations that account
for all sources of heating are needed to draw firm conclusions.

In all of our calculations, we have assumed that the dominant source
of pitch angle scattering is high frequency microinstabilities
generated during the growth and nonlinear evolution of the MRI. We
cannot, however, rule out that there are other sources of high
frequency waves that pitch angle scatter and effectively decrease
the mean free path of particles relative to that calculated here
(e.g., shocks and reconnection).  As shown in Table
\ref{Ch4tab:tab3} and Figure \ref{Ch4fig:figure8}, this would
decrease the magnitude of the anisotropic stress; we find that for
$\nu \gtrsim 30 \, \Omega$, the results of our kinetic simulations
quantitatively approach the MHD limit. In this context it is
important to note that the incompressible part of the MHD cascade
launched by the MRI is expected to be highly anisotropic with
$k_\Perp \gg k_\Par$ \cite{Goldreich1995}.  As a result, there is
very little power in high frequency waves that could break $\mu$
conservation. It is also interesting to note that satellites have
observed that the pressure anisotropy in the solar wind near 1 AU is
approximately marginally stable to the firehose instability
\cite{Kasper2002}, consistent with our assumption that
microinstabilities dominate the isotropization of the plasma. There
is evidence for pressure anisotropy in other collisionless
plasmas, e.g., the solar wind \cite{Marsch1982,Kasper2002} and
magnetosphere \cite{Tsurutani1982,Gary1993}.

In this chapter we have focused on kinetic modifications to angular
momentum transport via anisotropic pressure stresses and parallel
heat conduction.  In addition, kinetic effects substantially modify
the stability of thermally stratified low collisionality plasmas
such as those expected in RIAFs. Balbus \cite{Balbus2000} showed
that in the presence of anisotropic heat conduction, thermally
stratified plasmas are unstable when the temperature decreases
outwards, rather than when the entropy decreases outwards (the usual
Schwarzschild criterion).  This has been called the magnetothermal
instability (MTI). Parrish and Stone \cite{Parrish2005} show that in
non-rotating atmospheres the MTI leads to magnetic field
amplification and efficient heat transport.  In future global
simulations of RIAFs, it will be interesting to explore the combined
dynamics of the MTI, the MRI, and angular momentum transport via
anisotropic pressure stresses. Apart from affecting the local
dynamics and energetics, collisionless effects can affect the global
structure of hot, collisionless disks.

\chapter{Anisotropic conduction with large temperature gradients}
\label{chap:chap5}
A natural step, after local studies of the MRI in the collisionless
regime, is to investigate the effects of collisionless plasma
processes on the global structure of collisionless disks in
radiatively inefficient accretion flows (RIAFs). Instead of
including both anisotropic pressure and anisotropic conduction, as
in the local studies described in Chapters \ref{chap:chap3} and
\ref{chap:chap4}, we began by looking at just the effects of
anisotropic thermal conduction. Anisotropic conduction is important
for global disk structure because  an anisotropically conducting
plasma is convectively stable if the temperature increases outwards
($dT/dr \geq 0$; see \cite{Balbus1998,Balbus2000}). Whereas, convective
stability in collisional fluids require the entropy
($s=p/\rho^\gamma$) to increase outwards. Local, 2-D, vertically
stratified MHD simulations of Parrish and Stone \cite{Parrish2005}
have confirmed that the convective instability in plasma with
anisotropic thermal conduction, christened the magnetothermal
instability (MTI), is driven by temperature gradients. If convection
is important, as in hydrodynamic disks
\cite{Stone1999,Quataert2000}, anisotropic conduction can modify the
global structure (and hence the luminosity) of RIAFs.

\begin{figure}
\begin{center}
\includegraphics[width=4in,height=3in]{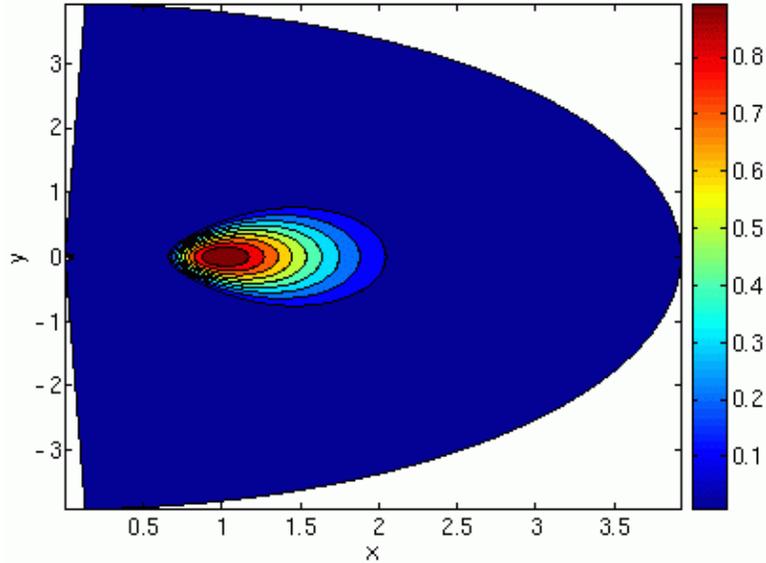}
\caption[Initial density for a typical global MHD simulations]{The
initial density for a typical global MHD disk simulation (e.g.,
\cite{Stone2001,Hawley2002}). A high density, constant angular
momentum torus is surrounded by a non-rotating, low density corona.
Temperature (and density) jumps by $\sim 100$ at the torus-corona
interface. Magnetic field vanishes in the corona, while it is along
the density contours in the torus ($\beta \sim 100$).
\label{Ch5fig:fig1}}
\end{center}
\end{figure}

Our aim was to include anisotropic thermal conduction in global, 2-D
MHD simulations of RIAFs, and to see if the structure of turbulent,
quasi-steady disk changes. We began by adding an anisotropic
conduction routine, based on centered differencing, to the global ZEUS
MHD code used by Stone and Pringle \cite{Stone2001}. The initial
condition for most global MHD disk simulations
\cite{Hawley2002,Stone2001} is a constant angular momentum, high
density torus surrounded by a low density, non-rotating corona (see
Figure \ref{Ch5fig:fig1} for a typical example). Pressure balance at
the torus-corona interface requires a big jump in temperature across
it (ratio of temperatures = inverse ratio of densities $\sim 100$).

The implementation of anisotropic thermal conduction in presence of
large temperature gradients was far from trivial. The simulations
with anisotropic conduction and MHD-disk initial conditions (with a
large temperature gradient; see Figure \ref{Ch5fig:fig1}) did not
run for long, eventually becoming numerically unstable, even though
we were using a Courant stable time step.
We found that, unlike isotropic conduction, the centered differencing of
anisotropic conduction allowed for heat to flow
from higher to lower temperatures.
The heat flow in the ``wrong" direction can lead to negative
temperatures in regions with large temperature gradients. An
implementation of anisotropic thermal conduction that does not give
rise to negative temperatures required considerable time and effort.
Thus, global MHD disk simulations with anisotropic thermal conduction have
been left for the future.

Anisotropic diffusion, in which the rate of diffusion of some quantity is
faster in some directions than others, occurs in many different physical
systems and applications.  Examples include diffusion in geological
formations, thermal properties of structural materials and crystals, image
processing~\cite{Caselles1998,Mrazek2001}, biological systems,
and plasma physics.
Diffusion Tensor Magnetic Resonance Imaging makes use of anisotropic
diffusion to distinguish different types of tissue as a medical
diagnostic~\cite{Basser2002}.  In plasma physics, the collision operator gives
rise to anisotropic diffusion in velocity space, as does the quasilinear
operator describing the interaction of particles with waves~\cite{Stix1992}. In
magnetized plasmas, thermal conduction can be much more rapid along a field
line than across it; this will be the main application in mind for
this chapter.

In this chapter we show that anisotropic thermal conduction based on
centered differences is not always consistent with the second law of
thermodynamics. Test problems that result in negative temperature
with centered ``asymmetric" and ``symmetric" differencing are
presented. This happens because heat can flow from lower to higher
temperature in regions with large temperature gradient.
Temperature gradients in anisotropic heat fluxes need to be limited
to ensure that temperature extrema are not accentuated. We tried
several different approaches, and eventually developed slope-limited
methods that successfully avoid the negative temperature problem, by
using limiters analogous to those used in numerical solution of
hyperbolic equations \cite{Leveque2002}.
Perpendicular numerical diffusion ($\chi_{\perp,num}$) scales as
$\sim \chi_\parallel \Delta x^2$ in case of the least diffusive
slope limited schemes. The limited methods are more diffusive than
the ``symmetric" method, but comparable to the ``asymmetric" method.
Also, like the ``asymmetric" method, the limited methods lack the
desirable property of ``symmetric" method that the perpendicular
numerical diffusion ($\chi_{\perp,num}$) is independent of the parallel
conduction $\chi_\parallel$. The main advantage of slope limited methods
is that they do not give rise to
negative temperatures in presence of large temperature gradients.
Thus, limited methods will be useful to simulate hot, dilute
astrophysical plasmas where conduction is anisotropic and
temperature gradients are enormous, e.g., disk-corona boundary,
energetic reconnection events, and collisionless shocks.

\section{Introduction}
\label{Ch5sec:introduction}

When the plasma collision frequency, $\nu$~($\propto nT^{-3/2}$, $n$
is the number density and $T$ is the temperature), is small compared
to the cyclotron frequency $\Omega_{c}=qB/mc$, key transport
quantities like stress and thermal conduction become anisotropic
with respect to the magnetic field direction (the ratio of parallel
to perpendicular transport coefficients is $\sim (\Omega_c/\nu)^2$);
heat and momentum transport parallel to the field is much larger
than in the cross-field direction~\cite{Braginskii1965}. In a plasma
with comparable electron and proton temperatures, heat transport is
dominated by electrons, which are faster than ions by the ratio
$\sqrt{m_i/m_e}$, and momentum transport is dominated by the
protons. Anisotropic plasmas are abundant in nature~(e.g., solar
corona, solar wind, magnetosphere, and radiatively inefficient
accretion flows) as well as high temperature laboratory devices like
tokamaks. In order to simulate dilute, anisotropic plasmas, accurate
and robust numerical methods are needed.

Numerical methods based on finite differences~\cite{Gunter2005} and
higher order finite elements~\cite{Sovinec2004} have been useful in
simulating highly anisotropic conduction~($\chi_\parallel/\chi_\perp
\sim 10^9$, where $\chi_\parallel$ and $\chi_\perp$ are parallel and
perpendicular conduction coefficients) in laboratory plasmas.
``Symmetric" differencing introduced in~\cite{Gunter2005} is
particularly simple and has some desirable properties---perpendicular
numerical diffusion independent of $\chi_\parallel$, and self
adjointness of the numerical heat flux operator. The scheme based on
asymmetric centered differences, with components of the heat flux
vector located at the cell faces, have been used to study convection
in anisotropically conducting plasmas~\cite{Parrish2005} and for
local simulations of collisionless accretion
disks~\cite{Sharma2006}. Anisotropic thermal conduction plays a
crucial role in the convective stability of dilute plasmas; Parrish
and Stone \cite{Parrish2005} have confirmed the prediction that
convection in stratified anisotropic plasmas is governed, not by the
entropy gradient~(the classic Schwarzchild criterion, $ds/dr>0$ for
convective stability of fluids), but by the temperature
gradient~($dT/dr>0$ for convective stability of plasmas with
anisotropic conduction; see \cite{Balbus2000,Balbus2001}).

An important fact that has not been discussed before (to our
knowledge) is that the methods based on centered differences can
give rise to heat fluxes inconsistent with the second law of
thermodynamics, i.e., heat can flow from lower to higher
temperatures! Temperature extrema can be accentuated unphysically,
and negative temperatures can arise if centered differencing is used. We
show, using simple numerical test problems, that both symmetric and
asymmetric centered methods can give rise to negative temperatures
at some grid points. Negative temperature results in numerical instability 
because the sound speed becomes imaginary. 

We show that the symmetric and asymmetric methods can be modified so
that the temperature extrema are not accentuated. The components of
anisotropic heat flux, e.g., $q_x$, consist of two contributions:
the normal term, $q_{xx} = -n\chi b_x^2 \partial T/\partial x$, and
the transverse term, $q_{xy} = -n\chi b_xb_y \partial T/\partial y$.
The normal term for the asymmetric method, like isotropic
conduction, is from higher to lower temperatures, but the transverse
term can be of any sign. The transverse term needs to be ``limited"
to ensure that temperature extrema are not accentuated. We use slope
limiters, analogous to those used in second order methods for
hyperbolic problems \cite{VanLeer1979,Leveque2002}, to limit the
transverse heat fluxes. However, for the symmetric method where
primary heat fluxes are located at cell corners, $q_{xx,i+1/2,j}$
need not be the same sign as $\partial T/\partial x|_{i+1/2,j}$.
Thus, both the normal and transverse terms have to be limited for
the symmetric method.
Methods based on the entropy-like function ($\dot{s}^* \equiv -{\bf q \cdot \nabla} T
\geq 0$; see Appendix \ref{app:app5} to see how this is different from the entropy function), 
which limit the transverse component of the heat flux, are also discussed.

Limiting introduces numerical diffusion in the perpendicular
direction, and the desirable property of the symmetric method that
perpendicular pollution is independent of $\chi_\parallel$ no longer
holds. The ratio of perpendicular numerical diffusion and the
physical parallel conductivity with a Monotonized Central (MC; see
\cite{Leveque2002} for discussion of slope limiters) limiter is
$\chi_{\perp, num} / \chi_\parallel \sim 10^{-3}$ for a modest
number of grid points ($\sim 100$ in each direction). This clearly
is not adequate for simulating laboratory plasmas which require
$\chi_\parallel/\chi_\perp \sim 10^9$, as perpendicular numerical
diffusion will swamp the true perpendicular diffusion. For
laboratory plasmas, the temperature profile is relatively smooth and
the negative temperature problem does not arise, so symmetric
differencing \cite{Gunter2005} or higher order finite elements
\cite{Sovinec2004} will be adequate.

However, astrophysical plasmas can have sharp gradients in
temperature~(e.g., the transition region of the sun separating the
hot corona and the much cooler chromosphere, the disk-corona
interface in accretion flows), and centered differencing can give
rise to negative temperatures. Thus, symmetric and asymmetric
centered methods cannot be used (the sound speed becomes imaginary
with negative temperature and can give rise to spurious
instabilities). The slope limited methods will introduce somewhat
larger  perpendicular numerical
diffusion~($\chi_{\perp,num}/\chi_\parallel \sim 10^{-3}$) but will
always ensure the correct direction of heat fluxes, and hence the
positivity of the temperature. Even a modest anisotropy in
conduction~($\chi_\parallel/\chi_\perp \sim 10^3$) should be useful
to study the qualitatively new effects of anisotropic conduction on
dilute astrophysical plasmas, but the positivity condition on
temperature is a must for
robust numerical simulations~(ruling out the use of centered
differencing for plasmas with large temperature gradients). Figure 3
in~\cite{Parrish2005} shows that the linear growth rate of the
magnetothermal instability~(the convective instability of stratified
anisotropic plasmas discussed in~\cite{Balbus1998,Balbus2000}) is
not much different for $\chi_\perp/\chi_\parallel \rightarrow 0$ and
$\chi_\perp/\chi_\parallel \lesssim 0.1$, and a numerical method that
gives rise to slightly larger (compared to the symmetric method, but still
$\chi_{\perp,num}/\chi_\parallel<0.1$) pollution of perpendicular
conduction looks acceptable. We have tested our slope-limited
methods on the magnetothermal instability and get results similar to
\cite{Parrish2005}, both linearly and nonlinearly.

The chapter is organized as follows. We begin with the equation for
anisotropic conduction and its numerical implementation using
asymmetric and symmetric centered differencing. We present simple
2-D test problems for which asymmetric and symmetric centered
differencing give rise to negative temperatures. The slope limited
methods for anisotropic heat conduction are introduced, followed by
the limiting of the symmetric method based on the entropy-like condition.
We discuss some mathematical properties of the slope limited
methods. We present further test problems comparing different
methods and study their convergence properties. In the end we
conclude and discuss the applications of the methods that we have
developed.

\section{Anisotropic thermal conduction}
Anisotropic thermal conduction can be important in a magnetized
plasmas if the mean free path (much larger than the gyroradius) is
comparable to the dynamical length scales. In such cases, a
divergence of anisotropic heat flux has to be added to the energy
equation. Such a term can modify the characteristic structure of the
MHD equations and can be evolved separately by using operator
splitting, as done in \cite{Parrish2005}. In operator splitting, MHD
evolution operator and anisotropic thermal conduction are applied
alternately, and their numerical implementations are independent.
The equation for the evolution of internal energy due to anisotropic
conduction is \ba \label{Ch5eq:anisotropic_conduction}
\frac{\partial e}{\partial t} &=& - {\bf \nabla \cdot q}, \\
 {\bf q} &=& - {\bf \hat{b}} n (\chi_\Par-\chi_\perp)
\nabla_\parallel T - n \chi_\perp {\bf \nabla} T \ea where $e$ is
the internal energy per unit volume, ${\bf q}$ is the heat flux,
$\chi_\Par$ and $\chi_\perp$ are the coefficients of parallel and
perpendicular conduction with respect to the local field
direction~(with dimensions $L^2T^{-1}$), $n$ is the number density,
$T=(\gamma-1)e/n$ is the temperature, $\gamma=5/3$ is the ratio of
the specific heats for an ideal gas, ${\bf \hat{b}}$ is the unit
vector along the field line, and $\nabla_\parallel={\bf
\hat{b}}\cdot {\bf \nabla}$ represents the derivative along the
direction of the magnetic field. In the test problems that we
present, $\gamma=2$ is chosen to avoid factors of $2/3$ and $5/3$;
qualitative features are independent of $\gamma$.
\begin{figure}
\begin{center}
\includegraphics[width=3in,height=3in]{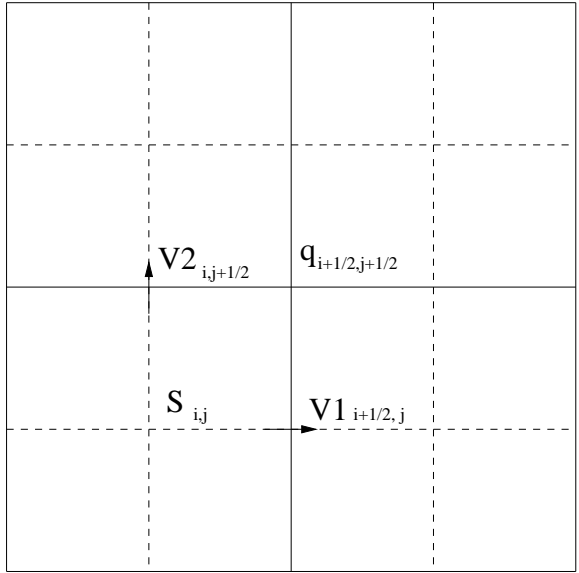}
\caption[Staggered grid with vectors at cell faces and scalars at
centers]{A staggered grid with scalars $S_{i,j}$ (such as $n$, $e$,
and $T$) located at cell centers. The components of vectors, e.g.,
${\bf \hat{b}}$ and ${\bf q}$ are located at cell faces. However,
for the symmetric centered scheme the primary heat fluxes are
located at cell corners \cite{Gunter2005}, and the face centered
flux is obtained by interpolation.\label{Ch5fig:fig2}}
\end{center}
\end{figure}

On a staggered grid, scalars like $n$, $e$, and $T$ are located at
the cell centers whereas the components of vectors like ${\bf
\hat{b}}$ and ${\bf q}$ are located at cell faces, as shown in
Figure \ref{Ch5fig:fig2}. The face centered components of vectors
naturally represent the flux of scalars out of a cell. Notice
however, as we describe later, that in G\"{u}nter et al.'s symmetric
method \cite{Gunter2005}, primary heat fluxes are located at cell
corners which are averaged to get the face centered heat fluxes.

All the schemes presented here are conservative and fully explicit.
It should be possible to take longer time steps with an implicit
generalization of these schemes, but the construction of a fast
implicit scheme for anisotropic conduction is non-trivial. In two
dimensions the internal energy density is updated as follows, \be
\label{Ch5eq:e_evolve} e^{n+1}_{i,j} = e^{n}_{i,j} - \Delta t \left[
\frac{q^n_{x,i+1/2,j}-q^n_{x,i-1/2,j}}{\Delta x} +
\frac{q^n_{y,i,j+1/2}-q^n_{y,i,j-1/2}}{\Delta y} \right], \ee where
the time step, $\Delta t$, satisfies the stability condition (ignoring 
density variations) 
\be
\label{Ch5eq:TimeStep} \Delta t \leq \frac{\mbox{min}[\Delta x^2,
\Delta y^2]}{4(\chi_\parallel + \chi_\perp)}, \ee $\Delta x $ and
$\Delta y$ are grid sizes in the two directions. The generalization
to three dimensions is straightforward.

The methods we discuss differ in the way heat fluxes are calculated
at the faces. In rest of the section we discuss the methods based on
asymmetric and symmetric centered differencing, as discussed in
\cite{Gunter2005}. The asymmetric method was used by
\cite{Parrish2005} and \cite{Sharma2006} for simulations of hot,
dilute, anisotropic astrophysical plasmas. We show in Section
\ref{Ch5sec:Negative} that both symmetric and asymmetric methods can
give rise to negative temperatures in regions with large temperature
gradients. From here on  $\chi$ will represent parallel conduction
coefficient in cases where an explicit perpendicular diffusion is
not considered~(i.e., the only perpendicular diffusion is due to
numerical effects).

\subsection{Centered asymmetric scheme}
\begin{figure}
\begin{center}
\psfrag{A}{\large{$(n\chi)_{-1/2}$}}
\psfrag{B}{\large{$(n\chi)_{1/2}$}} \psfrag{C}{\large{$T_0$}}
\psfrag{D}{\large{$T_{1}$}} \psfrag{E}{\large{$T_{-1}$}}
\includegraphics[width=3in,height=3in]{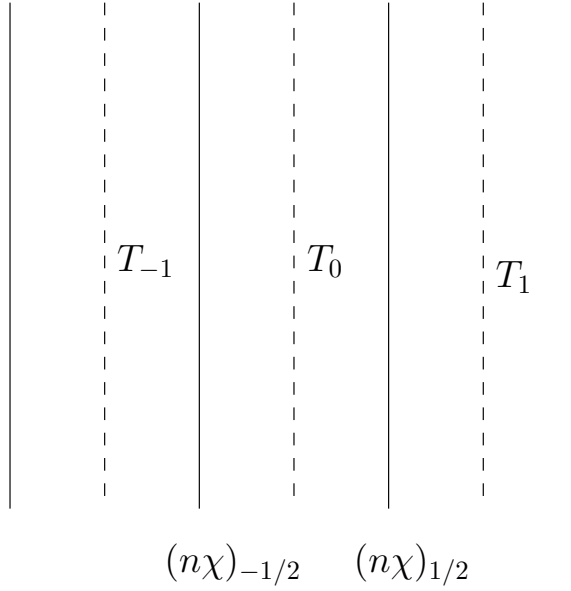}
\caption[Harmonic averaging for $\ol{n \chi}$]{This figure provides
a motivation for using a harmonic average for $\ol{n}\ol{\chi}$.
Consider a 1-D case with the temperatures and $n\chi$'s as shown in
the figure. Given $T_{-1}$ and $T_1$ and the $n\chi$'s at the faces,
we want to calculate an average $\ol{n}\ol{\chi}$ between cells $-1$
and $1$. Assumption of a constant heat flux gives,
$q_{-1/2}=q_{1/2}=\ol{q}$, i.e., $-(n\chi)_{-1/2}
(T_0-T_{-1})/\Delta x = -(n\chi)_{1/2} (T_1-T_0)/\Delta x =
-\ol{n}\ol{\chi} (T_1-T_{-1})/ 2 \Delta x$. This immediately gives a
harmonic mean, which is weighted towards the smaller of the two
arguments, for the interpolation $\ol{n}\ol{\chi}$.
\label{Ch5fig:fig2.1}}
\end{center}
\end{figure}

The heat flux in the $x$- direction (in 2-D), using the asymmetric
method is given by \be \label{Ch5eq:q1_asymmetric} q_{x,i+1/2,j}=-
\overline{n}\overline{\chi}  b_x \left[ b_x \frac{\partial
T}{\partial x} + \overline{b_y} \overline{\frac{\partial T}{\partial
y}} \right], \ee where overline represents the variables
interpolated to the face at $(i+1/2,j)$. The variables without an
overline are naturally located at the face. The interpolated
quantities at the face are given by simple arithmetic averaging, \ba
\label{Ch5eq:asymmetric_bavg} \overline{b_y} &=&
(b_{y,i,j-1/2}+b_{y,i+1,j-1/2}
+ b_{y,i,j+1/2}+b_{y,i+1,j+1/2})/4, \\
\label{Ch5eq:asymmetric_Tavg} \overline{\partial T/\partial y} &=&
(T_{i,j+1}+T_{i+1,j+1}-T_{i,j-1}-T_{i+1,j-1})/4 \Delta y. \ea

We use a harmonic mean to interpolate the product of number density
and conductivity, \be \label{Ch5eq:asymmetric_navg}
\frac{2}{\ol{n}\ol{\chi}}=\frac{1}{(n\chi)_{i,j}}+\frac{1}{(n\chi)_{i+1,j}};
\ee this is second order accurate for smooth regions, but
$\ol{n}\ol{\chi}$ becomes proportional to the minimum of the two
$n\chi$'s on either side of the face when the two differ
significantly.
Figure \ref{Ch5fig:fig2.1}
gives the motivation for the use of a harmonic average. Harmonic
averaging is also necessary for the method to be stable with the
present time step given in Eq. \ref{Ch5eq:TimeStep}. Instead, if we
use a simple mean, the stable time step condition becomes severe by
a factor $ \sim \mbox{max}[n_{i+1,j},n_{i,j}]/2\mbox{min}
[n_{i+1,j},n_{i,j}]$, which can result in unacceptably small time
steps for initial conditions with large density contrast.
Physically, this is because the heat capacity is very small in a low
density region, so a small amount of heat flow into that region
causes very fast changes in the temperature.

Analogous expressions can be written for the heat fluxes in other
directions. This method is used in astrophysical MHD simulations of
\cite{Parrish2005} and \cite{Sharma2006}, who include anisotropic
conduction in a cartesian geometry.

\subsection{Centered symmetric scheme}
The notion of symmetric differencing was introduced in
\cite{Gunter2005}, where primary heat fluxes are located at the cell
corners, with \be \label{Ch5eq:q1_symmetric} q_{x,i+1/2,j+1/2} =
-\overline{n}\overline{\chi} \overline{b_x} \left [ \overline{b_x}
\overline{\frac{\partial T}{\partial x}} + \overline{b_y}
\overline{\frac{\partial T}{\partial y}}   \right ], \ee where
overline represents the interpolation of variables at the corner
given by a simple  arithmetic average \ba
\label{Ch5eq:symmetric_bxavg}
\overline{b_x} &=& (b_{x,i+1/2,j}+b_{x,i+1/2,j+1})/2, \\
\label{Ch5eq:symmetric_byavg}
\overline{b_y} &=& (b_{y,i,j+1/2}+b_{y,i+1,j+1/2})/2, \\
\label{Ch5eq:symmetric_Tavg} \overline{\partial T/\partial x} &=&
(T_{i+1,j}+T_{i+1,j+1}-T_{i,j}-T_{i,j+1})/2 \Delta x, \\
\overline{\partial T/\partial y} &=&
(T_{i,j+1}+T_{i+1,j+1}-T_{i,j}-T_{i+1,j})/2 \Delta y. \ea As before
(and for the same reasons), a harmonic average is used for the
number density \be \label{Ch5eq:symmetric_navg}
\frac{4}{\ol{n}\ol{\chi}}= \frac{1}{(n\chi)_{i,j}} +
\frac{1}{(n\chi)_{i+1,j}} +
\frac{1}{(n\chi)_{i,j+1}}+\frac{1}{(n\chi)_{i+1,j+1}}. \ee This is
different from \cite{Gunter2005} who use an arithmetic average for
$n$ and $\chi$.
Analogous expression can be written for $q_{y,i+1/2,j+1/2}$.

The heat fluxes located at the cell faces, $q_{x,i+1/2,j}$ and
$q_{y,i,j+1/2}$, to be used in Eq.~(\ref{Ch5eq:e_evolve}) are given
by an arithmetic average, \ba
q_{x,i+1/2,j} &=& (q_{x,i+1/2,j+1/2}+q_{x,i+1/2,j-1/2})/2, \\
q_{y,i,j+1/2} &=& (q_{y,i+1/2,j+1/2}+q_{y,i-1/2,j+1/2})/2.\ea As
demonstrated in \cite{Gunter2005}, the symmetric heat flux satisfies
the self adjointness property (equivalent to $\dot{s}^* \equiv -{\bf
q \cdot \nabla} T \geq 0$ at cell corners) and has the desirable
property that the perpendicular numerical diffusion
($\chi_{\perp,num}$) is independent of $\chi_\parallel/\chi_\perp$.
But, as we show later, both symmetric and asymmetric schemes do not
satisfy the very important local property that heat must flow from
higher to lower temperatures; the violation of this at temperature
minima can result in negative temperature in regions with large
temperature gradients.

As mentioned earlier, the heat flux in the $x$- direction, $q_x$,
consists of two terms, the normal term $q_{xx}=-n \chi b_x^2
\partial T/\partial x$ and the transverse term $q_{xy}=-n \chi b_x
b_y \partial T/\partial y$. The asymmetric scheme uses a 2 point
stencil to calculate the normal gradient and a 6 point stencil to
calculate the transverse gradient, as compared to the symmetric
method that uses a 6 point stencil for both (hence the name
symmetric). This makes the symmetric method less sensitive to the
orientation of coordinate system with respect to the field lines.
\begin{figure}
\begin{center}
\includegraphics[width=2.95in,height=2.5in]{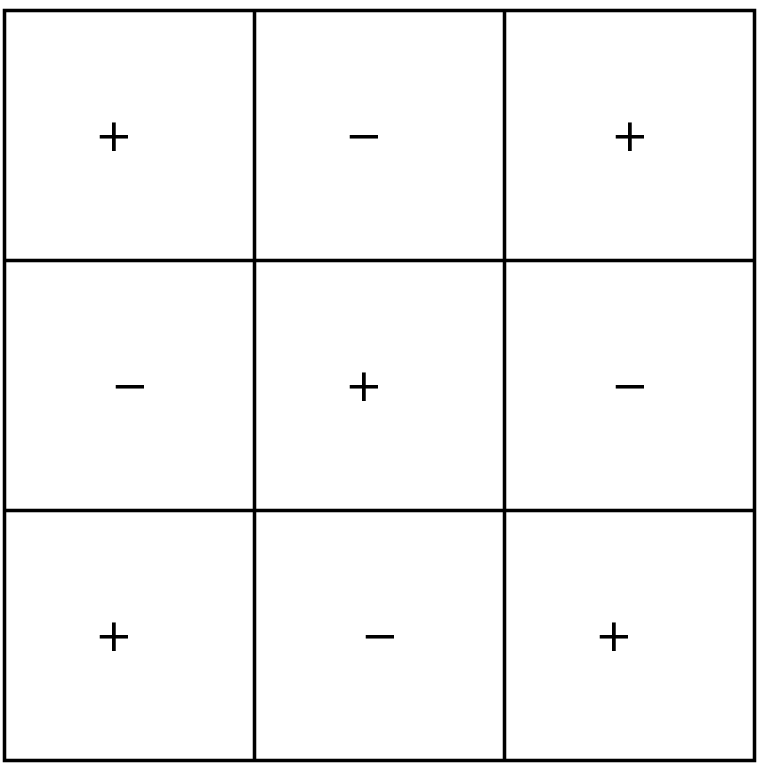}
\end{center}
\caption[Symmetric method's inability to diffuse a chess-board
pattern]{The symmetric method is unable to diffuse a temperature
distributed in a chess-board pattern. The plus ($+$) and minus ($-$)
symbols denote two unequal temperatures. Temperature gradients at
the cell corners vanish, resulting in a vanishing heat flux
independent of the magnetic orientation, e.g., average of $\partial
T/\partial x|_{i+1/2,j}=(T_+ - T_-)/\Delta x$ and $\partial
T/\partial x|_{i+1/2,j+1}=(T_- - T_+)/\Delta x$ to calculate
$\partial T/\partial x|_{i+1/2,j+1/2}= \partial T/\partial
x|_{i+1/2,j} + \partial T/\partial x|_{i+1/2,j+1}$ vanishes, similarly
$\partial T/\partial y|_{i+1/2,j+1/2}=0$. \label{Ch5fig:fig3}}
\end{figure}

A problem with the symmetric method which is immediately apparent is
its inability to diffuse away a chess-board temperature pattern, as
$\overline{\partial T/\partial x}$ and $\overline{\partial
T/\partial y}$, located at the cell corners, vanish for this initial
condition (see Figure \ref{Ch5fig:fig3}). All heat fluxes evaluated
with the symmetric method vanish and the temperature pattern is
stationary in time. This problem is alleviated if the perpendicular
diffusion coefficient, $\chi_\perp$, is large enough to diffuse the
temperature gradients at small scales.

\section{Negative temperature with centered differencing}
\label{Ch5sec:Negative} In this section we present two simple test
problems that demonstrate that negative temperatures can arise
because of centered differencing, for both asymmetric and symmetric
methods.

\subsection{Asymmetric method}

\begin{figure}
\begin{center}
\includegraphics[width=3in,height=3in]{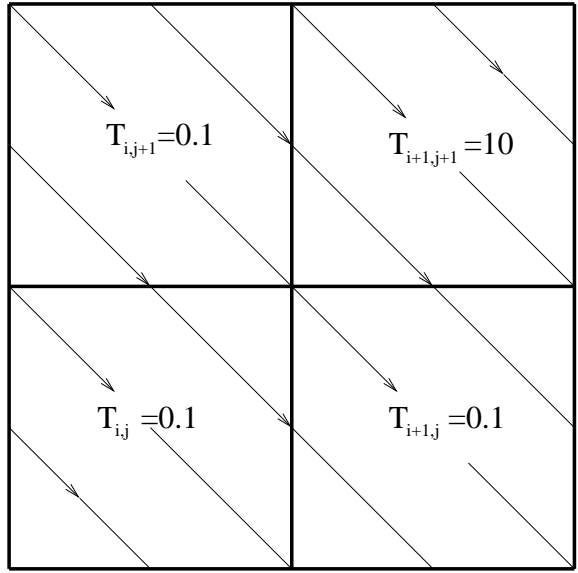}
\includegraphics[width=5in,height=4in]{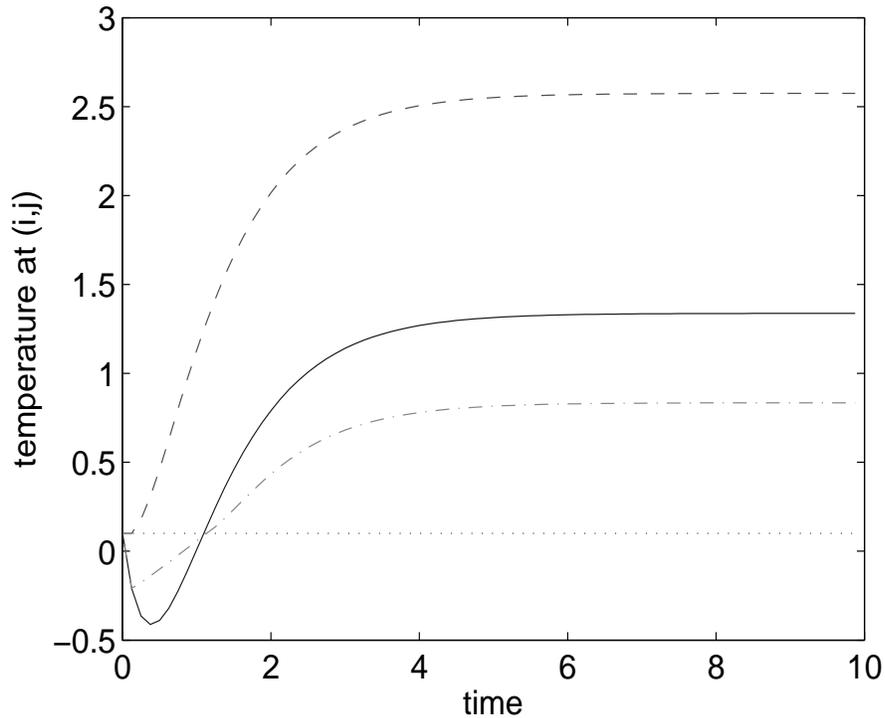}
\end{center}
\caption[Test problem that shows negative temperature with
asymmetric method]{Test problem that shows the asymmetric method can
give rise to negative temperature. Magnetic field lines are along
the diagonal with $b_x=-b_y=1/\sqrt{2}$. With the asymmetric method,
heat flows out of the grid located at southwest corner, resulting in
a negative temperature $T_{i,j}$. However, at late times the
temperature becomes positive again. The temperature at $(i,j)$ is
shown for different methods: asymmetric (solid line), symmetric
(dotted line), asymmetric and symmetric with slope limiters (dashed
line; both give the same result), and symmetric with entropy
limiting (dot dashed line).\label{Ch5fig:fig4}}
\end{figure}

Consider a $2 \times 2$  grid with a hot zone ($T=10$) in the first
quadrant and cold temperature ($T=0.1$) in the rest, as shown in
Figure \ref{Ch5fig:fig4}. Magnetic field is uniform over the box
with $b_x=-b_y=1/\sqrt{2}$. Number density is a constant equal to
unity. Reflecting boundary conditions are used. Using the asymmetric
scheme for heat fluxes out of the grid point $(i,j)$ (the third
quadrant) gives, $q_{x,i-1/2,j}=q_{y,i,j-1/2}=0$, and $q_{x,i+1/2,j}
= q_{y,i,j+1/2} = (9.9/8) n \chi/\Delta x$ (where $\Delta x=\Delta
y$ is assumed). Thus, heat flows out of the grid point $(i,j)$,
already a temperature minimum. This gives rise to temperature
becoming negative. Figure \ref{Ch5fig:fig4} shows the temperature in
the third quadrant with time for different methods. The asymmetric
method gives negative temperature ($T_{i,j}<0$) for first few time
steps, which eventually becomes positive. All other methods (except
the one based on entropy limiting) give positive temperatures at all
times for this problem. Methods based on limiting
temperature gradients will be discussed later. This test
demonstrates that the asymmetric method may not be suitable for
cases with large temperature gradients because negative temperatures
result in numerical instabilities.

\subsection{Symmetric method}
\label{Ch5subsec:symmfail}
\begin{figure}
\begin{center}
\includegraphics[width=3in,height=3in]{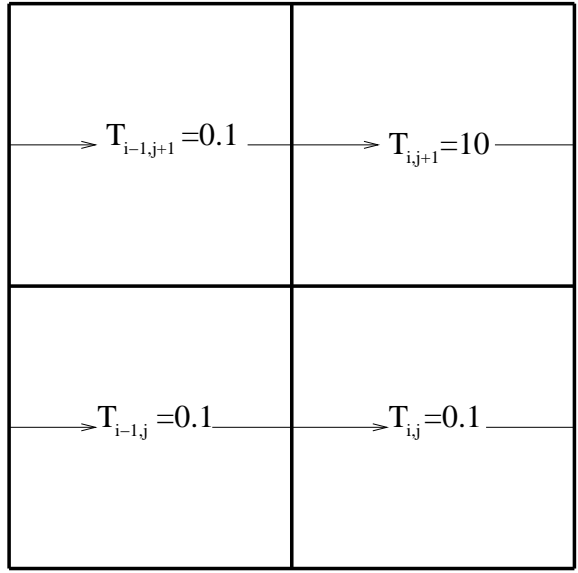}
\includegraphics[width=5in,height=4in]{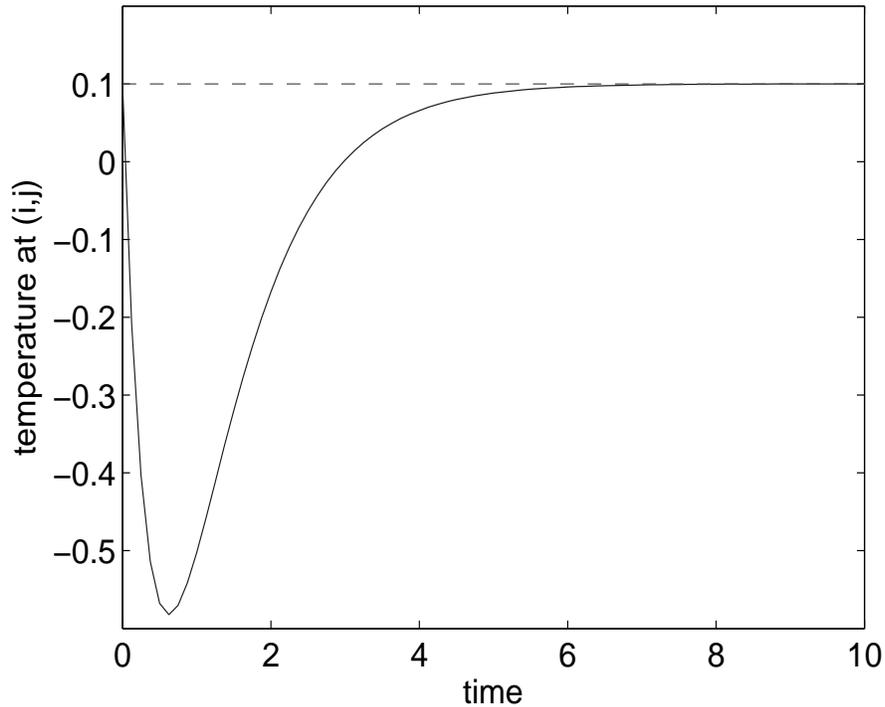}
\caption[Test problem that shows negative temperature with symmetric
method]{The result of the test problem for which the symmetric
method gives negative temperature at $(i,j)$. Magnetic field is
along the $x$- direction, $b_x=1$ and $b_y=0$. With this initial
condition, all heat fluxes into $(i,j)$ should vanish and the
temperature $T_{i,j}$ should not evolve. All methods except the
symmetric method (asymmetric, and slope and entropy limited methods)
give a constant temperature $T_{i,j}=0.1$ at all times. But with the
symmetric method, the temperature at $(i,j)$ becomes negative due to
the heat flux out of the corner at $(i-1/2,j+1/2)$. The temperature
$T_{i,j}$ eventually becomes equal to the initial value of $0.1$.
\label{Ch5fig:fig5}}
\end{center}
\end{figure}
The symmetric method does not give negative temperature with the
test problem of the previous section. In fact, the symmetric method
gives the correct result for temperature with no numerical diffusion
in the perpendicular direction (zero heat flux out of the grid point
$(i,j)$, see Figure \ref{Ch5fig:fig4}). Other methods resulted in a
temperature increase at $(i,j)$ because of perpendicular numerical
diffusion. Here we consider a case where the symmetric method gives
negative temperature.

As before, consider a $2 \times 2$ grid with a hot zone ($T=10$) in
the first quadrant and cold temperature ($T=0.1$) in the rest; the
only difference from the previous test problem is that the magnetic
field lines are along the $x$- axis, $b_x=1$ and $b_y=0$ (see Figure
\ref{Ch5fig:fig5}). Reflective boundary conditions are used, as
before. Since there is no temperature gradient along the field lines
for the grid point $(i,j)$, we do not expect the temperature there to
change. While all other methods give a stationary temperature in
time, the symmetric method results in a heat flux out of the grid
$(i,j)$ through the corner at $(i-1/2,j+1/2)$.  With the initial
condition as shown in Figure \ref{Ch5fig:fig5}, the only
non-vanishing symmetric heat flux out of $(i,j)$ is,
$q_{x,i-1/2,j+1/2}=- (9.9/2) n\chi \Delta x$. The only non-vanishing
face-centered heat flux entering the box through a face is $q_{x,i-1/2,j}=- (9.9/4)
n\chi \Delta x<0$; i.e., heat flows out of $(i,j)$ which is already
a temperature minimum. This results in the temperature becoming
negative at $(i,j)$, although at late times it becomes equal to the
initial temperature at $(i,j)$. This simple test shows that the
symmetric method can give negative temperatures (and associated
numerical problems) in presence of large temperature gradients.

\section{Slope limited fluxes}

The heat flux $q_x$ is composed of two terms, the normal $q_{xx} =
-n \chi b_x^2 \partial T/\partial x$ term, and the transverse
$q_{xy}=-n\chi b_xb_y \partial T/\partial y$ term. For the
asymmetric method, the discrete form of the term $q_{xx} = -n \chi
b_x^2 \partial T/\partial x$ is of the same sign as the $x$-
component of isotropic heat flux ($-n \chi
\partial T/\partial x$), and hence it guarantees that heat flows from
higher to lower temperatures. However, $q_{xy}=-n\chi b_xb_y
\partial T/\partial y$ can have an arbitrary sign, and can give rise
to heat flowing in the ``wrong" direction. We use slope limiters,
analogous to those used for linear reconstruction of variables in
numerical simulation of hyperbolic systems
\cite{VanLeer1979,Leveque2002}, to ``limit" the transverse terms.
Both asymmetric and symmetric methods can be modified with slope
limiters. The slope limited heat fluxes ensure that temperature
extrema are not accentuated. Thus, unlike the symmetric and
asymmetric methods, slope limited methods can never give negative
temperatures.

\subsection{Limiting the asymmetric method}

Since the normal heat flux term, $q_{xx}$, is naturally located at
the face, no interpolation for $\partial T/\partial x$ is required
for its evaluation. However, an interpolation at the $x$- face is
required to evaluate $\overline{\partial T/\partial y}$ used in
$q_{xy}$ (the term with overlines in Eq. \ref{Ch5eq:q1_asymmetric}).
The arithmetic average used in Eq. \ref{Ch5eq:asymmetric_Tavg} for
$\ol{\partial T/\partial y}$ to calculate $q_{xy}$ was found to
result in heat flowing from lower to higher temperatures (see Figure
\ref{Ch5fig:fig4}). To remedy this problem we have used slope
limiters to interpolate temperature gradients in the transverse heat
fluxes.

Slope limiters are widely used in numerical simulations of
hyperbolic equations (e.g., computational gas dynamics; see
\cite{VanLeer1979,Leveque2002}). Given the initial values for
variables at grid centers, slope limiters (e.g., minmod, van Leer,
and Monotonized Central (MC)) are used to calculate the slopes of
conservative piecewise linear reconstructions in each grid cell.
Limiters use the variable values in the nearest grid cells to come
up with slopes which ensure that no new extrema are created for
conserved variables, a property of hyperbolic equations. We use
slope limiters to interpolate temperature gradients in transverse
heat flux terms. Analogous to hyperbolic problems where limiters
prevent new unphysical extrema, limiters prevent amplification of
temperature extrema; this may result in negative temperatures.

The slope limited asymmetric heat flux in the $x$- direction is
still given by Eq. \ref{Ch5eq:q1_asymmetric}, with the same
$\partial T/\partial x$ as in the asymmetric method, but a slope
limited interpolation for the transverse temperature gradient,
$\overline{\partial T/\partial y}$ is needed, \be \label{Ch5eq:slope_asymm} \left .
\overline{\frac{\partial T}{\partial y}} \right |_{i+1/2,j} = L
\left \{ L \left [\left . \frac{\partial T}{\partial y} \right
|_{i,j-1/2}, \left . \frac{\partial T}{\partial y} \right
|_{i,j+1/2} \right ], \right . \left . L \left [ \left .
\frac{\partial T}{\partial y} \right |_{i+1,j-1/2}, \left .
\frac{\partial T}{\partial y} \right |_{i+1,j+1/2} \right ] \right
\}, \ee where $L$ is a slope limiter like minmod, van Leer, or
Monotonized Central (MC) limiter \cite{Leveque2002}; e.g., the van
Leer limiter is \ba \label{Ch5eq:VanLeer} \nonumber L(a,b) &=&
\frac{2ab}{a+b}~\mbox{if}~ab>0, \\
&=& 0~\mbox{otherwise.} \ea Slope limiters weights the interpolation
towards the argument smallest in magnitude, and returns a zero if
the two arguments are of opposite signs. An analogous expression for
the transverse temperature gradient at the $y$- face,
$\overline{\partial T/\partial x}$, is used to evaluate the heat
flux $q_y$. Averaging similar to the asymmetric method is used for
all other interpolations (Eqs. \ref{Ch5eq:asymmetric_bavg} and
\ref{Ch5eq:asymmetric_navg}).

\subsection{Limiting the symmetric method}

In the symmetric method, primary heat fluxes in both directions are 
located at the cell corners (see Eq.
\ref{Ch5eq:q1_symmetric}). Temperature gradients in both directions have
to be interpolated at the corners. Thus, to ensure that temperature extrema are
not amplified with the symmetric method, both $\overline{\partial
T/\partial x}$ and $\overline{\partial T/\partial y}$ need to be
limited.

\ignore{The limited expression for $q_{xx}$ at the corners is given
by \ba \label{Ch5eq:qxx_symm_lim_up} q_{xx,i+1/2,j+1/2} &=&
-\overline{n}\overline{\chi} \overline{b_x}^2 L2 \left [ \left .
\frac{\partial T}{\partial x} \right |_{i+1/2,j}, \left .
\frac{\partial T}{\partial x}
\right |_{i+1/2,j+1} \right ], \\
\label{Ch5eq:qxx_symm_lim_down} q_{xx,i+1/2,j-1/2} &=&
-\overline{n}\overline{\chi} \overline{b_x}^2 L2 \left [ \left .
\frac{\partial T}{\partial x} \right |_{i+1/2,j}, \left .
\frac{\partial T}{\partial x} \right |_{i+1/2,j-1} \right ], \ea
with $q_{xx,i+1/2,j} = (q_{xx,i+1/2,j+1/2} + q_{xx,i+1/2,j-1/2})/2$;
the interpolated quantities (indicated with an overline) are given
by Eqs. \ref{Ch5eq:symmetric_bxavg}, \ref{Ch5eq:symmetric_byavg},
and \ref{Ch5eq:symmetric_navg}. The limiter $L2$, very different
from standard slope limiters (e.g., minmod, van Leer, and MC), is
defined as
\ba \nonumber
L2(a,b) &=& (a+b)/2, \mbox{ if } \alpha a < (a+b)/2 < a/\alpha \mbox{ or }
 \alpha a > (a+b)/2 > a/\alpha \mbox{, otherwise, } \\
 \nonumber       &=&  \alpha a, \mbox{ if sgn(a) $\neq$ sgn(b),} \\
        &=&  a/\alpha, \mbox{ if sgn(a) $=$ sgn(b),}
\ea where $0<\alpha<1$ is a parameter; this reduces to a simple
averaging if temperature is smooth. We choose $\alpha=3/4$; results
are not very sensitive to the exact value of $\alpha$.}

The face-centered $q_{xx,i+1/2,j}$ is calculated by averaging
$q_{xx}$ from the adjacent corners, which are given by the following
slope-limited expressions:
\ba \label{Ch5eq:qxx_symm_lim_up} q^N_{xx,i+1/2,j+1/2} &=&
-\overline{n}\overline{\chi} \overline{b_x^2} L2 \left [ \left .
\frac{\partial T}{\partial x} \right |_{i+1/2,j}, \left .
\frac{\partial T}{\partial x}
\right |_{i+1/2,j+1} \right ], \\
\label{Ch5eq:qxx_symm_lim_down} q^S_{xx,i+1/2,j-1/2} &=&
-\overline{n}\overline{\chi} \overline{b_x^2} L2 \left [ \left .
\frac{\partial T}{\partial x} \right |_{i+1/2,j}, \left .
\frac{\partial T}{\partial x} \right |_{i+1/2,j-1} \right ], \ea
where $S$ and $N$ superscripts indicate the south-biased corner heat flux 
or the north-biased heat flux. The face centered heat flux used in Eq.
\ref{Ch5eq:e_evolve} is
$q_{xx,i+1/2,j} = (q^N_{xx,i+1/2,j+1/2} + q^S_{xx,i+1/2,j-1/2})/2$;
the interpolated quantities (indicated with an overline) are the
same as in Eq. \ref{Ch5eq:q1_symmetric}. The limiter $L2$, which is
somewhat different from standard slope limiters, is defined as
\ba \nonumber L2(a,b) &=& (a+b)/2, \mbox{ if } \min(\alpha a, a /
\alpha)
        < (a+b)/2 < \max(\alpha a, a /\alpha), \\
\nonumber
        &=&  \min(\alpha a, a / \alpha), \mbox{ if } (a+b)/2 \leq
             \min(\alpha a, a / \alpha), \\
        &=&  \max(\alpha a, a / \alpha), \mbox{ if } (a+b)/2 \geq
             \max(\alpha a, a / \alpha),
\ea
where $0<\alpha<1$ is a parameter; this reduces to a simple
averaging if the temperature is smooth while restricting the
interpolated temperature ($\ol{\partial T/\partial x}$) to not 
differ too much from $\partial T /
\partial x |_{i+1/2,j}$ (and be of the same sign). 
We choose $\alpha=3/4$; results are not
very sensitive to the exact value of $\alpha$. 
The $L2$ limiter is not symmetric with respect to its arguments (and thus
the definition of $q^{S}_{xx,i+1/2,j+1/2}$ is slightly different than the
definition of $q^{N}_{xx,i+1/2,j+1/2}$). It ensures that
$q_{xx,i+1/2,j \pm 1/2}$ is of the same sign as $-\partial
T/\partial x |_{i+1/2,j}$; i.e., the interpolated normal heat flux
flows from higher to lower temperatures. If we use a standard slope
limiter (e.g., minmod, van Leer, or MC) in Eqs.
\ref{Ch5eq:qxx_symm_lim_up} and \ref{Ch5eq:qxx_symm_lim_down}, for
the chessboard pattern shown in Figure \ref{Ch5fig:fig3}, all heat
fluxes vanish as with the symmetric method. However, the $L2$
limiter gives, $$ q_{xx,i+1/2,j+1/2}=q_{xx,i+1/2,j-1/2}=
-\ol{n}\ol{\chi} \ol{b_x}^2 \alpha \left . \frac{\partial
T}{\partial x} \right |_{i+1/2,j},
$$
a heat flux from higher to lower temperatures which can diffuse the
chessboard pattern.

The transverse temperature gradient is limited in a way similar to
the asymmetric method. The temperature gradient $\ol{\partial
T/\partial y}$ to be used in Eq. \ref{Ch5eq:q1_symmetric} is given
by \be \label{Ch5eq:slope_symm} \left . \ol{\frac{\partial T}{\partial y}} \right
|_{i+1/2,j+1/2} = L \left [\left . \frac{\partial T}{\partial y}
\right |_{i+1,j+1/2} , \left . \frac{\partial T}{\partial y} \right
|_{i,j+1/2} \right ], \ee with
$q_{xy,i+1/2,j}=L(q_{xy,i+1/2,j+1/2},q_{xy,i+1/2,j-1/2})$, where $L$
is a standard slope limiter.

\section{Limiting using the entropy-like source function}
\label{Ch5sec:ent_limiting} If the entropy-like source function, which we
define as $\dot{s}^*=-{\bf q \cdot \nabla} T$ (see Appendix \ref{app:app5} 
to see how this is different from the entropy function), is positive at all
spatial locations, heat is guaranteed to flow from higher to lower
temperatures. For the symmetric method, $\dot{s}^*$ evaluated at the
cell corners is positive definite, but need not be positive definite if
evaluated at the cell faces, and thus allows the heat to flow across
faces from lower to higher temperatures. This can cause temperature
to decrease at a minimum; temperature can also become negative if
temperature gradients are large (see Figure \ref{Ch5fig:fig5}).
Thus, $\dot{s}^* \geq 0$ satisfied at
all corners on the grid is not sufficient for the heat to flow from
higher to lower temperatures. We use the following entropy-like
condition, applied at all face-pairs, to limit the transverse heat
flux terms ($q_{xy}$ and $q_{yx}$)
 \be \label{Ch5eq:ent_lim} \dot{s}^* = -
q_{x,i+1/2,j} \left . \frac{\partial T}{\partial x} \right
|_{i+1/2,j} - q_{y,i,j+1/2} \left . \frac{\partial T}{\partial y}
\right |_{i,j+1/2} \geq 0. \ee

The limiter $L2$ is used to calculate the normal gradients $q_{xx}$
and $q_{yy}$ at the faces, as in the slope limited symmetric method.
The use of $L2$ ensures that $-q_{xx,i+1/2,j}\partial T/\partial x
|_{i+1/2,j} \geq 0$, and only the transverse terms $q_{xy}$ and
$q_{yx}$ need to be reduced
to satisfy Eq. \ref{Ch5eq:ent_lim}. That is, if on evaluating
$\dot{s}^*$ the entropy-like condition (Eq. \ref{Ch5eq:ent_lim}) is
violated, the transverse terms are reduced to make $\dot{s}^*$ vanish.
The attractive feature of the entropy limited symmetric method is that
it reduces to the symmetric method (least diffusive of all the methods;
see Figure \ref{Ch5fig:fig8})
when Eq. \ref{Ch5eq:ent_lim} is satisfied, and the limiting of transverse
terms may help with the amplification of temperatures at extrema.

The problem with entropy limiting is that the temperature extrema can
still be amplified (see Figures \ref{Ch5fig:fig4} and \ref{Ch5fig:fig7}).
For example, when $\partial T/\partial x|_{i+1/2,j} =
\partial T/\partial y|_{i,j+1/2}=0$, Eq. \ref{Ch5eq:ent_lim}  is satisfied for
arbitrary heat fluxes $q_{x,i+1/2,j}$ and $q_{y,i,j+1/2}$. In such a
case, transverse heat fluxes $q_{xy}$ and $q_{yx}$ can cause heat to
flow across the zero temperature gradient and result in a new
temperature extremum, which may even be a negative. However, this
unphysical behavior can only occur for one time step, after which
$\grad T \neq 0$ and Eq. \ref{Ch5eq:ent_lim} becomes a useful limit
again. The result is that the overshoots are not as pronounced as in
the asymmetric and symmetric methods, as shown in Figures
\ref{Ch5fig:fig6} and \ref{Ch5fig:fig7}.
With entropy limiting, unlike the symmetric and asymmetric methods,
the spurious temperature oscillations (reminiscent of unphysical
oscillations near discontinuities in hyperbolic systems) are damped
(see Figure \ref{Ch5fig:fig7}).
Although temperature minimum can be accentuated by the entropy
limited method, early on one can choose sufficiently small time
steps to ensure that temperature does not become negative; this is
equivalent to saying that entropy limited method will not give
negative temperatures at late times (see Figure \ref{Ch5fig:fig7}
and Tables \ref{Ch5tab:tab1}-\ref{Ch5tab:tab4}). This trick will not
work for the centered symmetric and asymmetric methods where
temperatures can be negative even at late times (see Figure
\ref{Ch5fig:fig7}).

To guarantee that temperature extrema are not amplified, in addition
to entropy limiting at all points, one should also use slope
limiting of transverse temperature gradients at extrema.
This results in a method that does not amplify the extrema, but is
more diffusive compared to just entropy limiting (see Figure
\ref{Ch5fig:fig8}). Because of simplicity of slope limited methods
and their desirable mathematical properties (discussed in the next
section), they are preferred over the cumbersome entropy limited
methods.

\section{Mathematical properties}
In this section we prove that the slope limited fluxes do not
amplify the temperature extrema. Also discussed are global and local
properties related to the entropy-like condition, $\dot{s}^* = - {\bf
q \cdot \nabla} T \geq 0$.

\subsection{Behavior at temperature extrema}
Slope limiting of both asymmetric and symmetric methods guarantees
that the temperature extrema are not amplified further, i.e., the
maximum temperature does not increase and the minimum does not
decrease. This ensures that the temperature is always positive and
numerical problems because of imaginary sound speed do not arise.
The normal heat flux in the asymmetric method ($=-\ol{n}\ol{\chi}
b_x^2 \partial T/\partial x$) and the L2 limited normal heat flux
term in the symmetric method (Eqs. \ref{Ch5eq:qxx_symm_lim_up} and
\ref{Ch5eq:qxx_symm_lim_down}) allows the heat to flow only from
higher to lower temperatures. Thus, the terms responsible for
unphysical behavior at temperature extrema are the transverse heat
fluxes $q_{xy}$ and $q_{yx}$. Slope limiters ensure that the
transverse heat terms vanish at extrema and heat flows down the
temperature gradient at those grid points.

The operator $L(L(a,b),L(c,d))$, where $L$ is a slope limiter like
minmod, van Leer, or MC, is symmetric with respect to all its
arguments, and hence can be written as $L(a,b,c,d)$. For the slope
limiters considered here (minmod, van Leer, and MC), $L(a,b,c,d)$ vanishes
unless all four arguments $a,b,c,d$ have the same sign.

At a local temperature extremum (say at $(i,j)$), the $x$- (and $y$-)
face-centered slopes 
$\partial T/\partial y|_{i,j+1/2}$ and $\partial T/\partial
y|_{i,j-1/2}$ (and $\partial T/\partial x|_{i+1/2,j}$ and $\partial
T/\partial x|_{i-1/2,j}$) are of opposite signs or at least one of 
them is zero. This ensures that the slope limited transverse temperature gradients
($\overline{\partial T/\partial y}$ and $\overline{\partial
T/\partial x}$) vanish (from Eqs. \ref{Ch5eq:slope_asymm} and
\ref{Ch5eq:slope_symm}). The heat fluxes become $q_{x,i \pm 1/2, j}
= -\ol{n}\ol{\chi} \ol{b_x}^2 \partial T/\partial x|_{i \pm 1/2, j}$
and $q_{y,i, j\pm 1/2} = -\ol{n}\ol{\chi} \ol{b_y}^2 \partial
T/\partial y|_{i, j \pm 1/2}$
at the temperature extrema, which are always down the temperature
gradient. This ensures that
temperature never becomes negative, unlike the methods based on
centered differencing.

\subsection{The entropy-like condition, $\dot{s}^* = -{\bf q \cdot \nabla} T \geq 0$}

If the number density, $n$, remains constant in time, then
multiplying Eq. \ref{Ch5eq:anisotropic_conduction} with $T$ and
integrating over all space gives \be \label{Ch5eq:selfadjointness}
\frac{1}{(\gamma-1)} \frac{\partial}{\partial t} \int n T^2 dV  = -
\int T {\bf \nabla \cdot q} dV = \int {\bf q \cdot \nabla} T dV = -
\int n \chi |\nabla_\parallel T|^2 dV \le 0, \ee assuming that the
surface contributions vanish. This analytic constraint implies that
temperature fluctuations cannot increase in time (on an average).

G\"{u}nter et al. \cite{Gunter2005} have shown that the symmetric method 
is self-adjoint and satisfies the entropy-like condition, 
Eq. \ref{Ch5eq:selfadjointness}. The
local entropy-like source function $\dot{s}^*=-{\bf q \cdot \grad} T$
evaluated at the corner $(i+1/2,j+1/2)$ for the symmetric method is  \be
\dot{s}^*_{i+1/2,j+1/2} =  -q_{x,i+1/2,j+1/2} \left.
\ol{\frac{\partial T}{\partial x}} \right |_{i+1/2,j+1/2} -
q_{y,i+1/2,j+1/2} \left. \ol{\frac{\partial T}{\partial y}} \right
|_{i+1/2,j+1/2}. \ee Using the form for symmetric heat fluxes (Eq.
\ref{Ch5eq:q1_symmetric}), the entropy-like function becomes, \ba
\nonumber \dot{s}^* &=& \ol{n}\ol{\chi} \ol{b_x}^2 \ol{\frac{\partial
T}{\partial x}}^2 + \ol{n}\ol{\chi} \ol{b_y}^2 \ol{\frac{\partial
T}{\partial y}}^2 + 2 \ol{n}\ol{\chi} \ol{b_x}~\ol{b_y}
\ol{\frac{\partial T}{\partial x}}~\ol {\frac{\partial T}{\partial
y}}, \\ &=& \ol{n}\ol{\chi} \left [ \ol{b_x} \ol{\frac{\partial
T}{\partial x}} +  \ol{b_y} \ol{\frac{\partial T}{\partial y}}
\right ]^2 \geq 0. \ea Thus, ${\bf q \cdot \grad} T \leq 0$, and
integration over the whole space implies Eq. \ref{Ch5eq:selfadjointness}.
Although the entropy-like condition is satisfied by the symmetric method
at the corners (both locally and globally), this condition is not sufficient 
to guarantee local positivity of temperature at cell centers, as we 
demonstrate in Subsection \ref{Ch5subsec:symmfail}.
Also notice that the
modification of the symmetric method to satisfy entropy-like condition at
face pairs (see Section \ref{Ch5sec:ent_limiting}) does not cure the
problem of negative temperature. Thus, a method which satisfy the
entropy-like condition ($\dot{s}^* = -{\bf q \cdot \nabla} T \geq 0$) does
not necessarily satisfy the condition that temperature extrema
should not be amplified.

With an appropriate interpolation, the asymmetric method and the slope
limited asymmetric methods
can be shown to satisfy the global entropy-like condition, $\dot{S}^* = -\int
{\bf q \cdot \grad} T dV/V \geq 0$. Consider \be \dot{S}^* =
\frac{-1}{N_x N_y} \sum_{i,j} \left [ q_{x,i+1/2,j} \left .
\frac{\partial T}{\partial x} \right |_{i+1/2,j} + q_{y,i,j+1/2}
\left . \frac{\partial T}{\partial y} \right |_{i,j+1/2}  \right ],
\ee where $N_x$ and $N_y$ are the number of grid points in each
direction. Substituting the form of asymmetric heat fluxes,
\ba \nonumber \label{Ch5eq:Sdot}  \dot{S}^* &=& \frac{1}{N_x N_y}
\sum_{i,j} \left [ \left ( \ol{n} \ol{\chi} b_x^2 \left |
\frac{\partial T}{\partial x} \right |^2 \right )_{i+1/2,j} + \left
( \ol{n} \ol{\chi} b_y^2 \left | \frac{\partial T}{\partial
y} \right |^2 \right)_{i,j+1/2} \right . \\
&+& \left . \left ( \ol{n \chi b_x b_y \frac{\partial T}{\partial
y}} \right )_{i+1/2,j} \left . \frac{\partial T}{\partial x} \right
|_{i+1/2,j} +  \left ( \ol{n \chi b_x b_y \frac{\partial T}{\partial
x}} \right )_{i,j+1/2} \left . \frac{\partial T}{\partial y} \right
|_{i,j+1/2} \right ], \ea where overlines represent appropriate
interpolations.
We define \ba G_{x,i+1/2,j} &=& \sqrt{ \left( \ol{n}\ol{\chi}
\right)}_{i+1/2,j} b_{x,i+1/2,j} \left . \frac{\partial
T}{\partial x} \right |_{i+1/2,j}, \\
G_{y,i,j+1/2} &=& \sqrt{ \left( \ol{n}\ol{\chi} \right)}_{i,j+1/2}
b_{y,i,j+1/2} \left . \frac{\partial
T}{\partial y} \right |_{i,j+1/2}, \\
\ol{G}_{y,i+1/2,j} &=& \ol{ \sqrt{n \chi}
b_y \left . \frac{\partial
T}{\partial y} \right | }_{i+1/2,j}, \\
\ol{G}_{x,i,j+1/2} &=& \ol{ \sqrt{n \chi}
b_x \left . \frac{\partial T}{\partial x} \right |
}_{i,j+1/2}.
 \ea
In terms of $G$'s, Eq. \ref{Ch5eq:Sdot} can be written as \be \dot{S}^*
= \frac{1}{N_x N_y} \sum_{i,j} \left [  G_{x,i+1/2,j}^2 +
G_{y,i,j+1/2}^2 + G_{x,i+1/2,j} \ol{G}_{y,i+1/2,j} +
\ol{G}_{y,i,j+1/2} G_{y,i,j+1/2} \right ]. \ee

A lower bound on $\dot{S}^*$ is obtained by assuming the cross terms
to be negative, i.e., \be \dot{S}^* \geq  \frac{1}{N_x N_y} \sum_{i,j}
\left [ G_{x,i+1/2,j}^2 + G_{y,i,j+1/2}^2 - \left | G_{x,i+1/2,j}
 \ol{G}_{y,i+1/2,j} \right | - \left |
\ol{G}_{y,i,j+1/2} G_{y,i,j+1/2} \right | \right ]. \ee Now define
$\ol{G}_{y,i+1/2,j}$ and $\ol{G}_{x,i,j+1/2}$ as follows (the
following interpolation is necessary for the proof to hold): \ba
\ol{G}_{y,i+1/2,j} &=& L ( G_{y,i,j+1/2}, G_{y,i,j-1/2},
G_{y,i+1,j+1/2}, G_{y,i+1,j-1/2} ),  \\
\ol{G}_{x,i,j+1/2} &=& L ( G_{x,i+1/2,j}, G_{x,i-1/2,j},
G_{x,i+1/1,j+1}, G_{y,i-1/2,j+1} ), \ea where $L$ is an arithmetic
average (as in centered asymmetric method) or a slope limiter
(e.g., minmod, van Leer, or MC) which satisfy the property that $
|L(a,b,c,d)| \leq (|a|+|b|+|c|+|d|)/4$, to put a lower bound on
$\dot{S}^*$. Thus, \ba \nonumber \dot{S}^* &\ge& \frac{1}{N_xN_y}
\sum_{i,j} G_{x,i+1/2,j}^2 + G_{y,i,j+1/2}^2  - \frac{1}{4} \left [
\left |G_{x,i+1/2,j} G_{y,i,j+1/2} \right | \right . \\ \nonumber
&+& \left |G_{x,i+1/2,j} G_{y,i,j-1/2} \right |  + \left
|G_{x,i+1/2,j} G_{y,i+1,j+1/2} \right | + \left |G_{x,i+1/2,j}
G_{y,i+1,j-1/2} \right |
\\  \nonumber &+& \left |G_{y,i,j+1/2} G_{x,i+1/2,j}\right | + \left |
G_{y,i,j+1/2} G_{x,i-1/2,j} \right | + \left |G_{y,i,j+1/2} G_{x,i+1/2,j+1}
\right |  \\
&+& \left . \left |G_{y,i,j+1/2} G_{x,i-1/2,j+1} \right | \right ].
\ea Shifting the dummy indices and combining various terms give, \ba
\nonumber \dot{S}^* &\ge& \frac{1}{N_xN_y} \sum_{i,j} G_{x,i+1/2,j}^2
+ G_{y,i,j+1/2}^2 -\frac{1}{2} \left [  \left |G_{x,i+1/2,j}
G_{y,i,j+1/2} \right |  \right . \\
&+& \nonumber \left . \left |G_{x,i+1/2,j} G_{y,i,j-1/2} \right | + 
\left | G_{x,i+1/2,j} G_{y,i+1,j+1/2} \right |
+ \left |G_{x,i+1/2,j} G_{y,i+1,j-1/2} \right | \right ] \\
\nonumber &=& \frac{1}{4N_xN_y} \sum_{i,j} \left [ \left (
G_{x,i+1/2,j}-G_{y,i,j+1/2} \right )^2 + \left (
G_{x,i+1/2,j}-G_{y,i,j-1/2} \right )^2 \right . \\
&+& \left . \left ( G_{x,i+1/2,j}-G_{y,i+1,j+1/2} \right )^2  +
\left ( G_{x,i+1/2,j}-G_{y,i+1,j-1/2} \right )^2 \right ] \geq 0.
\ea 

Thus, an appropriate interpolation (for the asymmetric and the
slope limited asymmetric methods) can result in a
scheme that satisfies the global entropy-like condition just as it does 
for the non-limited symmetric method. A variation of this proof can be
used to prove the global true entropy condition $\dot{S} \geq 0$ by multiplying 
Eq. \ref{Ch5eq:anisotropic_conduction} with $1/T$ instead of $T$ (see 
Appendix \ref{app:app5}), although the form of limiting would need to be modified 
slightly. It is useful to know that introducing a limiter to the asymmetric method
does not break the global entropy-like condition, if the right combination of 
quantities is limited in the interpolation.
However, it is
important to remember that the entropy-like (or entropy) condition does not guarantee
a local heat flow in the correct direction, and hence temperature can
still become negative. Thus, to get a robust method for anisotropic diffusion, 
it is necessary that heat flows in the correct direction  at temperature 
extrema.

\section{Further tests}
\begin{tiny}
\begin{table}[hbt]
\begin{center}
\caption{Diffusion in circular field lines: $50 \times 50$ grid
\label{Ch5tab:tab1} } \vskip0.05cm
\begin{tabular}{ccccccc} \hline
Method   & L1 error & L2 error  & L$\infty$ error & T$_{\rm max}$ &
T$_{\rm min}$ &
$\chi_{\perp,num}/\chi_\parallel$ \\
\hline
asymmetric & 0.0324 & 0.0459 & 0.0995 & 10.0926 & 9.9744 & 0.0077 \\
asymmetric minmod & 0.0471 & 0.0627 & 0.1195 & 10.0410 & 10 & 0.0486 \\
asymmetric MC & 0.0358 & 0.509 & 0.1051 & 10.0708 & 10 & 0.0127 \\
asymmetric van Leer & 0.0426 & 0.0574 & 0.1194 & 10.0519 & 10 & 0.0238 \\
symmetric & 0.0114 & 0.0252 & 0.1425 & 10.2190 & 9.9544 & 0.00028 \\
symmetric entropy & 0.03332 & 0.0477 & 0.0997 & 10.0754 & 10 & 0.0088 \\
symmetric entropy extrema & 0.0341 & 0.0487 & 0.1010 & 10.0751 & 10 & 0.0101 \\
symmetric minmod & 0.0475 & 0.0629 & 0.1322 & 10.0406 & 10 & 0.0490 \\
symmetric MC & 0.0289 & 0.0453 & 0.0872 & 10.0888 & 10 & 0.0072 \\
symmetric van Leer & 0.0438 & 0.0585 & 0.1228 & 10.0519 & 10 & 0.0238 \\
\hline
\end{tabular}
\end{center}
\end{table}
\begin{table}[hbt]
\begin{center}
\caption{Diffusion in circular field lines: $100 \times 100$ grid
\label{Ch5tab:tab2} } \vskip0.05cm
\begin{tabular}{ccccccc} \hline
Method   & L1 error & L2 error  & L$\infty$ error & T$_{\rm max}$ &
T$_{\rm min}$ &
$\chi_{\perp,num}/\chi_\parallel$ \\
\hline
asymmetric & 0.0256 & 0.0372 & 0.0962 & 10.1240 & 9.9859 & 0.0030 \\
asymmetric minmod & 0.0468 & 0.0616 & 0.1267 & 10.0439 & 10 & 0.0306 \\
asymmetric MC & 0.0261 & 0.0405 & 0.0907 & 10.1029 & 10 & 0.0040 \\
asymmetric van Leer & 0.0358 & 0.0502 & 0.1002 & 10.0741 & 10 & 0.0971 \\
symmetric & 0.0079 & 0.0173 & 0.1206 & 10.2276 & 9.9499 & 0.000041 \\
symmetric entropy & 0.0285 & 0.0420 & 0.0881 & 10.0961 & 10 & 0.0042 \\
symmetric entropy extrema & 0.0291 & 0.0425 & 0.0933 & 10.0941 & 10 & 0.0041 \\
symmetric minmod & 0.0471 & 0.0618 & 0.1275 & 10.0433 & 10 & 0.0305 \\
symmetric MC & 0.0123 & 0.0252 & 0.1133 & 10.1406 & 10 & 0.00084 \\
symmetric van Leer & 0.0374 & 0.0514 & 0.1038 & 10.0697 & 10 & 0.0104 \\
\hline
\end{tabular}
\end{center}
\end{table}
\begin{table}[hbt]
\begin{center}
\caption{Diffusion in circular field lines: $200 \times 200$ grid
\label{Ch5tab:tab3} } \vskip0.05cm
\begin{tabular}{ccccccc} \hline
Method   & L1 error & L2 error  & L$\infty$ error & T$_{\rm max}$ &
T$_{\rm min}$ &
$\chi_{\perp,num}/\chi_\parallel$ \\
\hline
asymmetric & 0.0165 & 0.0281 & 0.0949 & 10.1565 & 9.9878 & 0.0012 \\
asymmetric minmod & 0.0441 & 0.0585 & 0.1214 & 10.0511 & 10 & 0.0191 \\
asymmetric MC & 0.0161 & 0.0289 & 0.0930 & 10.1397 & 10 & 0.0015 \\
asymmetric van Leer & 0.0264 & 0.0407 & 0.0928 & 10.1006 & 10 & 0.0035 \\
symmetric & 0.0052 & 0.0132 & 0.1125 & 10.2216 & 9.9509 & $1.90 \times 10^{-5}$  \\
symmetric entropy & 0.0256 & 0.0385 & 0.0959 & 10.1103 & 10 & 0.0032 \\
symmetric entropy extrema & 0.0260 & 0.0391 & 0.0954 & 10.1074 & 10 & 0.0032 \\
symmetric minmod & 0.0444 & 0.0588 & 0.1219 & 10.0503 & 10 & 0.0192 \\
symmetric MC & 0.0053 & 0.0160 & 0.0895 & 10.1676 & 10 & 0.0002 \\
symmetric van Leer & 0.0281 & 0.0426 & 0.0901 & 10.0952 & 10 & 0.0038 \\
\hline
\end{tabular}
\end{center}
\end{table}
\begin{table}[hbt]
\begin{center}
\caption{Diffusion in circular field lines: $400 \times 400$ grid
\label{Ch5tab:tab4} } \vskip0.05cm
\begin{tabular}{ccccccc} \hline
Method   & L1 error & L2 error  & L$\infty$ error & T$_{\rm max}$ &
T$_{\rm min}$ &
$\chi_{\perp,num}/\chi_\parallel$ \\
\hline
asymmetric & 0.0118 & 0.0234 & 0.0866 & 10.1810 & 9.9898 & $5.9 \times 10^{-4}$ \\
asymmetric minmod & 0.0399 & 0.0539 & 0.1120 & 10.0629 & 10 & 0.0115 \\
asymmetric MC & 0.0102 & 0.0230 & 0.0894 & 10.1708 & 10 &  $6.8 \times 10^{-4}$ \\
asymmetric van Leer & 0.0167 & 0.0290 & 0.1000 & 10.1321 & 10 & 0.0013 \\
symmetric & 0.0033 & 0.0104 & 0.1112 & 10.2196 & 9.9504 & $8.37 \times 10^{-6}$  \\
symmetric entropy & 0.0252 & 0.0384 & 0.0969 & 10.1144 & 10 & 0.0027 \\
symmetric entropy extrema & 0.0253 & 0.0383 & 0.0958 & 10.1135 & 10 & 0.0026 \\
symmetric minmod & 0.0401 & 0.0541 & 0.1124 & 10.0622 & 10 & 0.0116 \\
symmetric MC & 0.0032 & 0.0122 & 0.0896 & 10.1698 & 10 & $6.5 \times 10^{-5} $ \\
symmetric van Leer & 0.0182 & 0.0307 & 0.1026 & 10.1260 & 10 & 0.0013 \\
\hline
\end{tabular}
\end{center}
\end{table}
\end{tiny}
We use test problems discussed in \cite{Parrish2005} and
\cite{Sovinec2004} to compare different methods. 
The first test problem (taken from \cite{Parrish2005}) initializes a
hot patch in circular field lines; ideally the hot patch should
diffuse only along the field lines, but perpendicular numerical
diffusion can cause some cross-field diffusion. There is a
discontinuity in the initial temperature of the hot patch and the
background temperature. If the temperature jump is large temperature
can become negative on using centered differencing (asymmetric and
symmetric methods). The second test problem includes a source term
and an explicit perpendicular perpendicular diffusion coefficient
($\chi_\perp$). The steady state temperature gives a measure of the
perpendicular numerical diffusion, $\chi_{\perp,num}$.

\subsection{Circular diffusion of hot patch}
\begin{figure}
\begin{center}
\includegraphics[width=2.95in,height=2.5in]{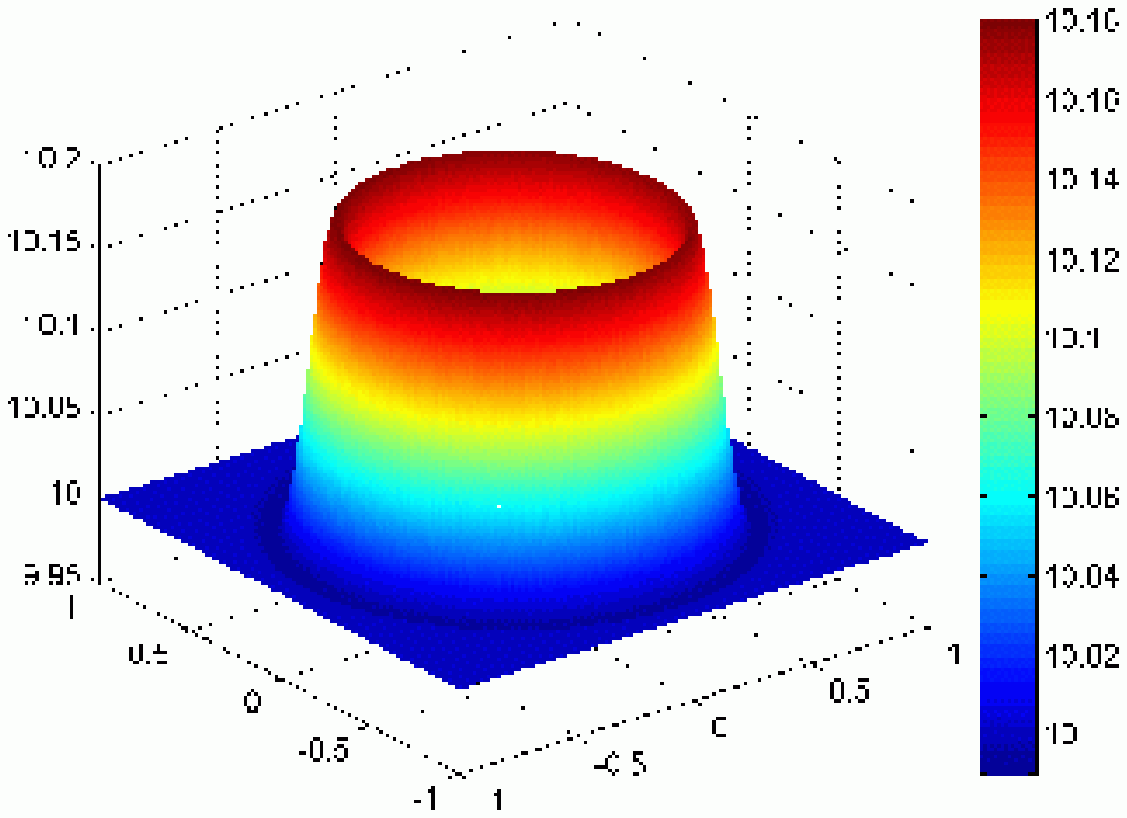}
\includegraphics[width=2.95in,height=2.5in]{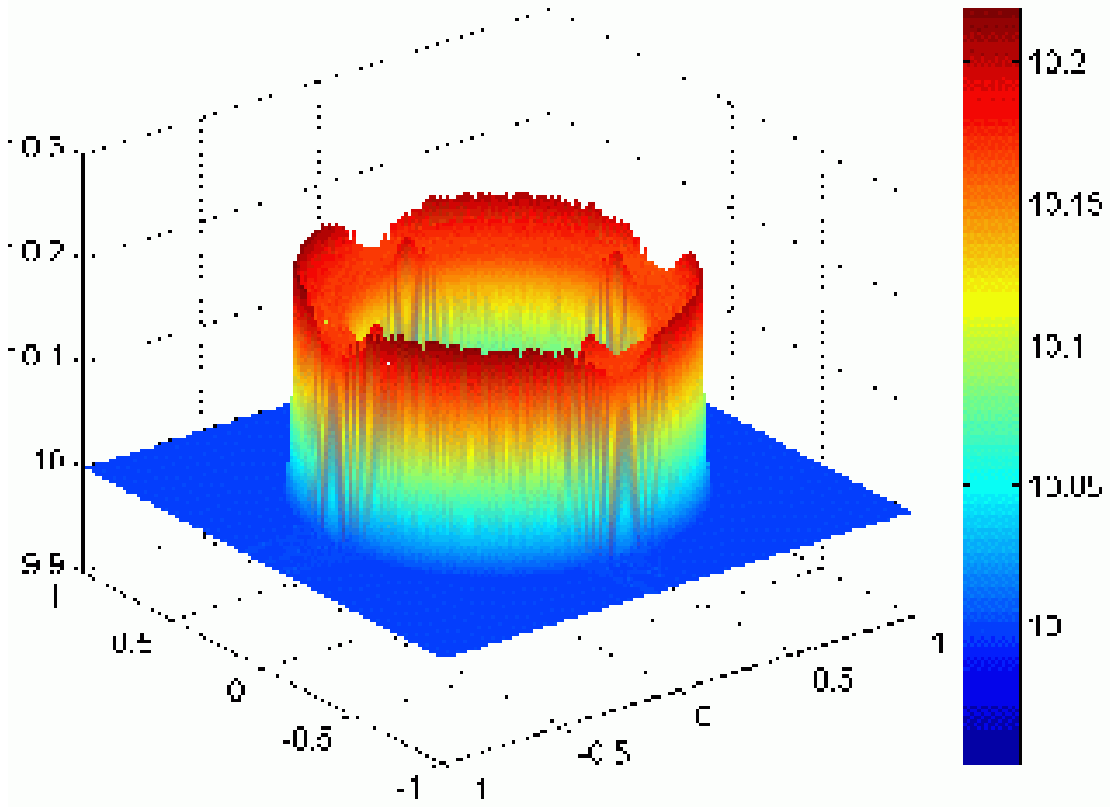}
\includegraphics[width=2.95in,height=2.5in]{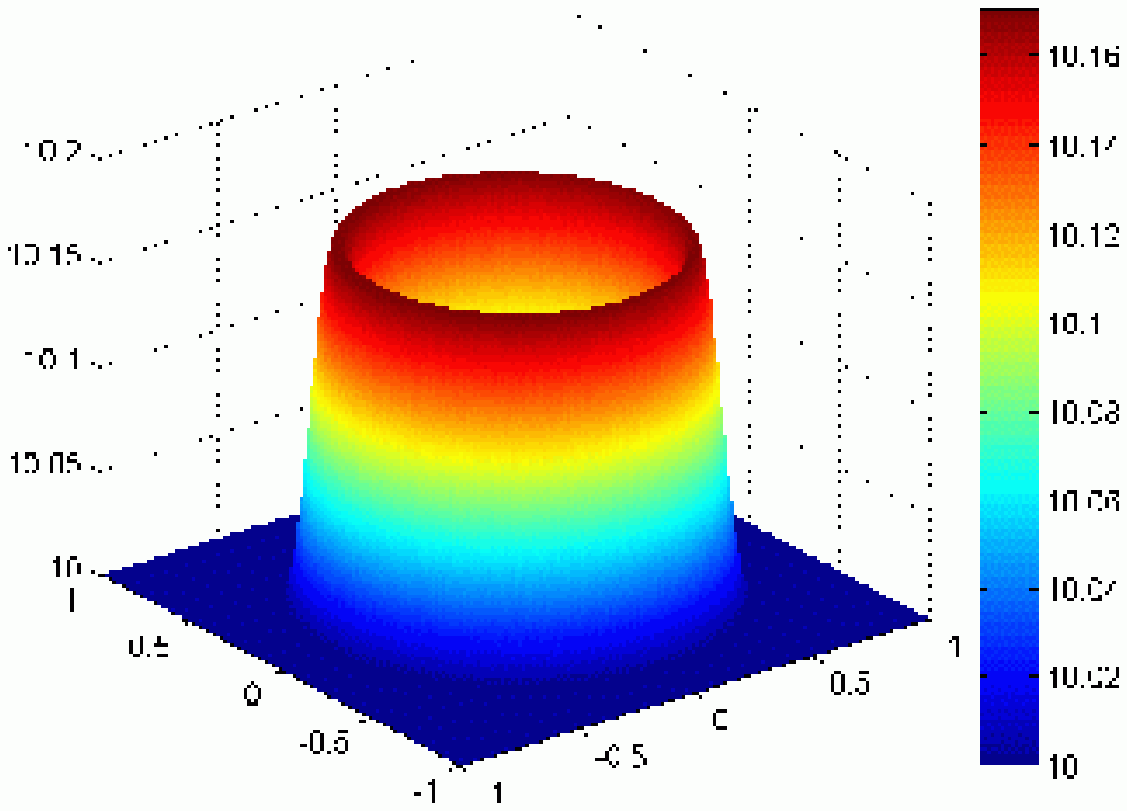}
\includegraphics[width=2.95in,height=2.5in]{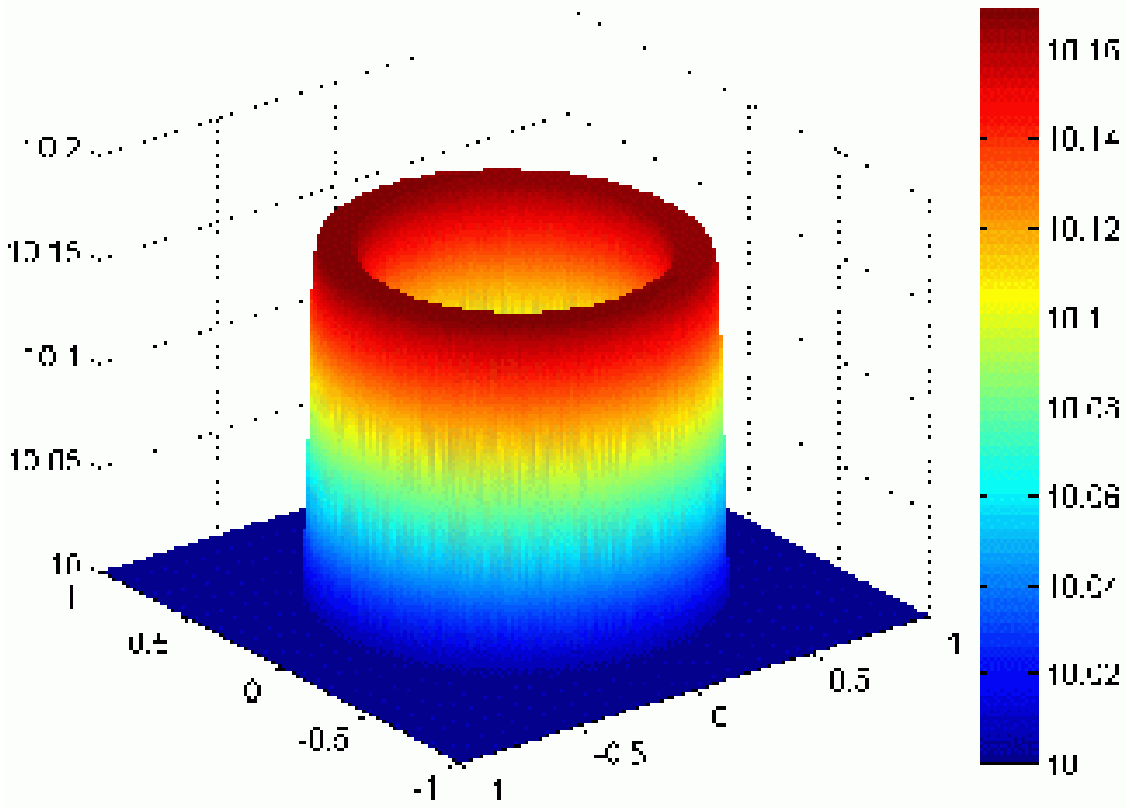}
\includegraphics[width=2.95in,height=2.5in]{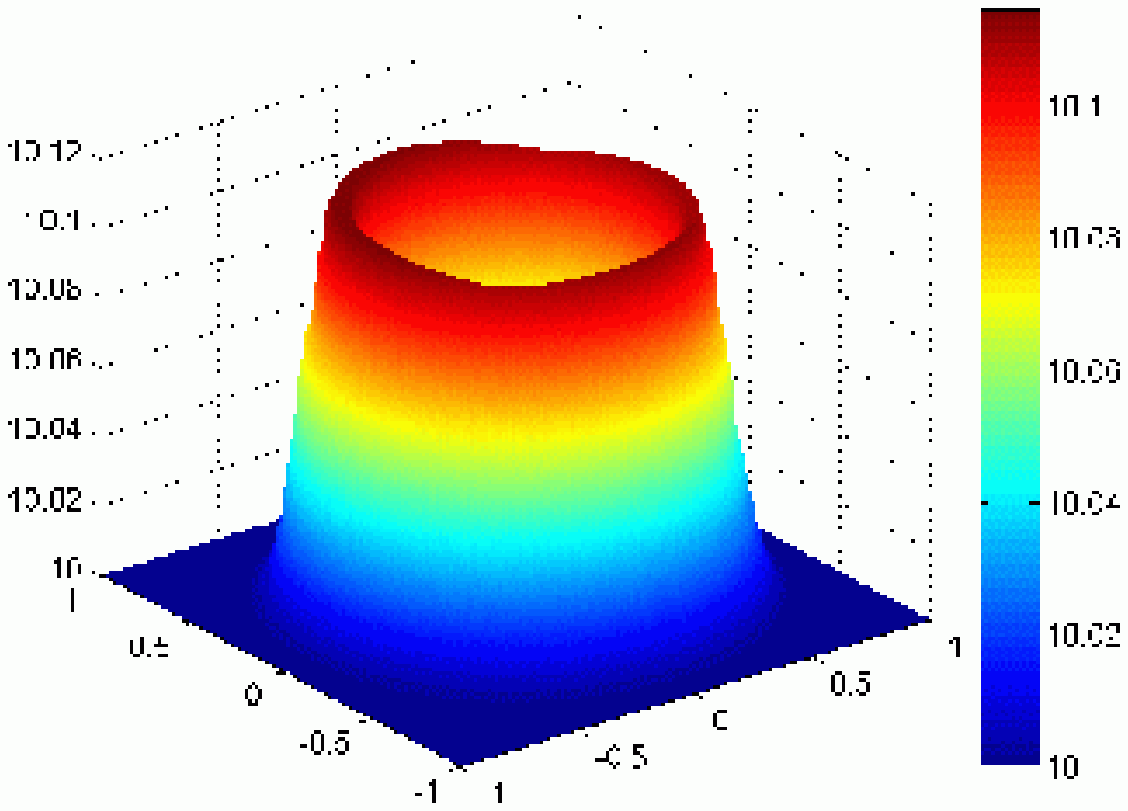}
\includegraphics[width=2.95in,height=2.5in]{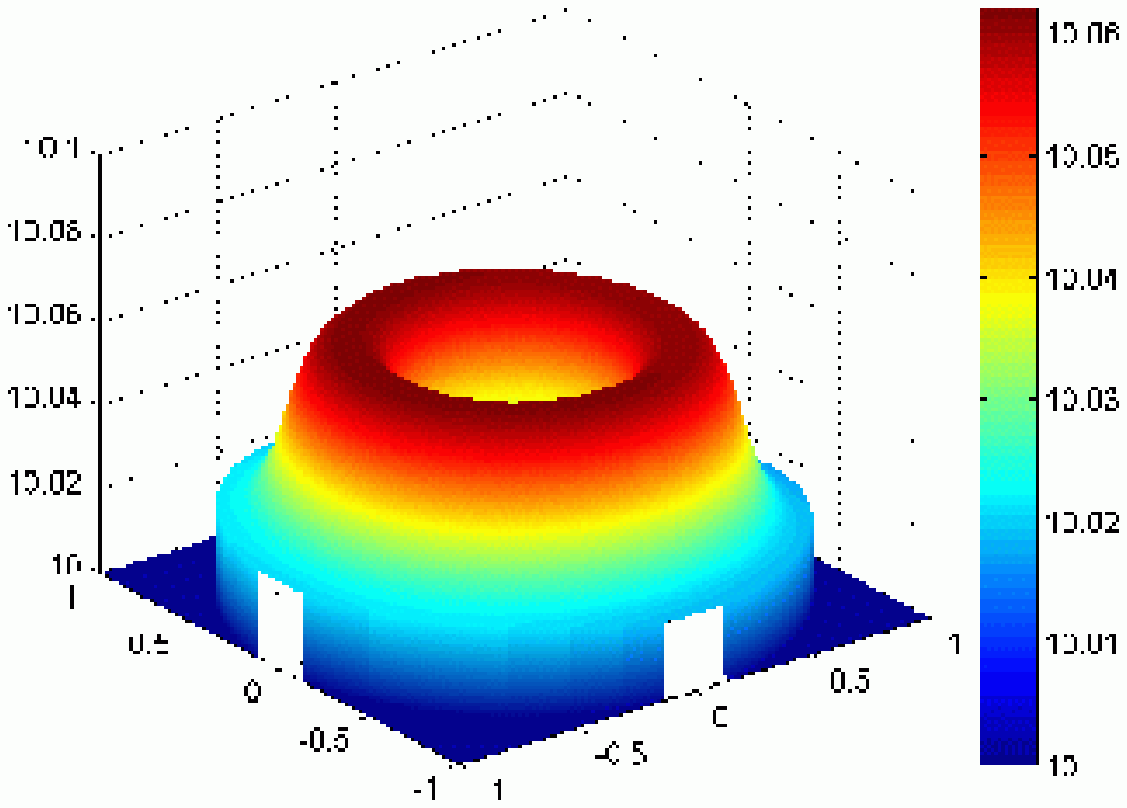}
\caption[Temperature at late times for the ring diffusion test
problem using different methods]{The temperature at $t=200$ for
different methods initialized with the ring diffusion problem on a
400 $\times$ 400 grid. Shown from left to right and top to bottom
are the temperatures for: asymmetric, symmetric, asymmetric-MC,
symmetric-MC, entropy limited symmetric, and minmod methods. Both
the asymmetric and symmetric methods give temperatures below 10 (the
initial minimum temperature). The result with a minmod limiter is
very diffusive. The slope limited symmetric method is less diffusive
than the slope limited asymmetric method. Entropy limited method
does not show non-monotonic behavior at late times, but is diffusive
compared to the better slope limited methods.\label{Ch5fig:fig6}}
\end{center}
\end{figure}

The circular diffusion test problem
was proposed in \cite{Parrish2005}. A hot patch surrounded by a
cooler background is initialized in circular field lines; the
temperature drops discontinuously across the patch boundary. At late
times, we expect the temperature to become uniform (and higher) in a
ring along the magnetic field lines. The computational domain is a
$[-1,1]\times[-1,1]$ cartesian box, with reflective boundary
conditions. The initial temperature distribution is given by \ba
\label{Ch5eq:ring_problem} \nonumber T &=& 12 \hspace{0.25 in}
\mbox{if} \hspace{0.1 in} 0.5<r<0.7 \hspace{0.1 in}
\mbox{and} \hspace {0.1 in}  \frac{11}{12}\pi<\theta<\frac{13}{12}\pi, \\
&=& 10 \hspace{0.25 in} \mbox{otherwise}, \ea where
$r=\sqrt{x^2+y^2}$ and $\tan\theta=y/x$. A set of circular field
lines centered at the origin is initialized. The parallel conduction
coefficient $\chi$ is chosen to be 0.01; there is no explicit
perpendicular diffusion. We evolve the anisotropic conduction
equation (\ref{Ch5eq:e_evolve}) till time = 200, using different
methods that we have discussed. By this time we expect the
temperature to be almost uniform along the circular ring
$0.5<r<0.7$. In steady state (at late times), energy conservation
implies that the the ring temperature should be 10.1667, while the
temperature outside the ring should be maintained at 10.

Figure \ref{Ch5fig:fig6} shows the temperature distribution for
different methods at time=200. All methods result in a higher
temperature in the annulus $r \in [0.5,0.7]$. The slope limited
schemes show larger perpendicular diffusion (Tables
\ref{Ch5tab:tab1}-\ref{Ch5tab:tab4} and Figure \ref{Ch5fig:fig9})
compared to the symmetric and asymmetric schemes. The perpendicular
numerical diffusion ($\chi_{\perp,num}$) scales with the parallel
diffusion coefficient $\chi$ for all methods. However, for Sovinec's
test problem (discussed in the next subsection) where temperature is
always smooth, and an explicit $\chi_\perp$ is present,
perpendicular numerical diffusion for the symmetric method does not
scale with $\chi_\parallel$.

\begin{figure}
\begin{center}
\includegraphics[width=4in,height=3in]{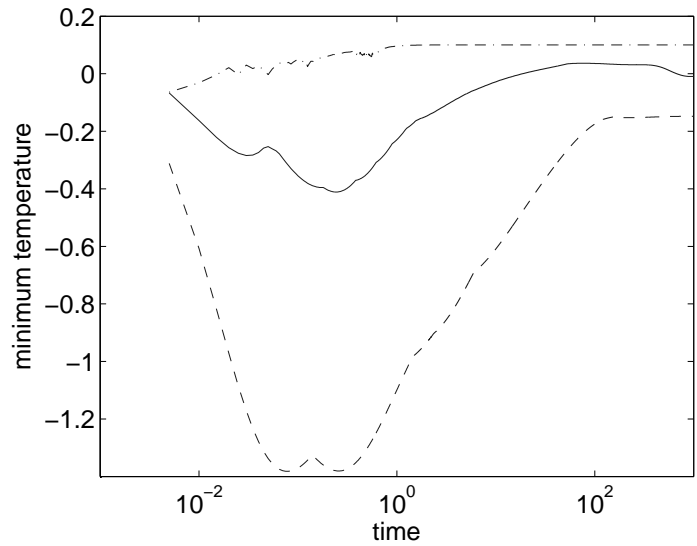}
\caption[Minimum temperature in the box for the ring diffusion
problem using asymmetric, symmetric, and entropy-limited
methods]{The minimum temperature over the whole box for symmetric
(dashed line), asymmetric (solid line), and entropy limited
symmetric (dot dashed line) methods in presence of circular field
lines. Initially the temperature of the hot patch is 10 and the
background is at 0.1. Both asymmetric and symmetric result in
negative temperature, even at late times. The nonmonotonic behavior
with the entropy limited method is considerably less pronounced; the
minimum temperature quickly becomes equal to the initial minimum
$0.1$. The limited heat fluxes keep the minimum at 0.1, as expected
physically.\label{Ch5fig:fig7}}
\end{center}
\end{figure}

The minmod limiter is much more diffusive than van Leer and MC
limiters. Both symmetric and asymmetric methods give a minimum
temperature below the initial minimum of 10, even at late times (see
Tables \ref{Ch5tab:tab1}-\ref{Ch5tab:tab4}). At late times the
symmetric method gives a temperature profile full of non-monotonic
oscillations (Figure \ref{Ch5fig:fig6}). Although, the slope limited
fluxes are more diffusive than the symmetric and asymmetric methods,
they never show undershoots below 10. Although the entropy limited
symmetric method gives temperature undershoots at early times, the
minimum temperature is still $10$ at late times (see Tables
\ref{Ch5tab:tab1}-\ref{Ch5tab:tab4} and Figure \ref{Ch5fig:fig7}).
Entropy limiting combined with a slope limiter at the extrema
behaves similar to the slope limiter based schemes.

Strictly speaking, a hot ring surrounded by a cold background is not
a steady solution for the ring diffusion problem. Temperature in the
ring will diffuse in the perpendicular direction (because of
perpendicular numerical diffusion, although very slowly) until the
whole box is at a constant temperature. A rough estimate for time
averaged perpendicular numerical diffusion $\la \chi_{\perp,num}
\ra$ follows from Eq. \ref{Ch5eq:anisotropic_conduction}, \be
\label{Ch5eq:chiperp_num} \la \chi_{\perp,num} \ra = \frac{ \int
(T_f - T_i) dV} {\int dt \left ( \int \grad^2 T dV \right )}, \ee
where the space integral is taken over the hot ring $0.5<r<0.7$, and
$T_i$ and $T_f$ are the initial and final temperature distributions
in the ring. Figure \ref{Ch5fig:fig9} plots the numerical
perpendicular diffusion (using Eq. \ref{Ch5eq:chiperp_num}) for the
runs in Tables \ref{Ch5tab:tab1}-\ref{Ch5tab:tab4}. The estimates
for perpendicular diffusion agree roughly with the more accurate
calculations using Sovinec's test problem described in the next
subsection (compare Figures \ref{Ch5fig:fig8} and
\ref{Ch5fig:fig9}). Table \ref{Ch5tab:tab6} lists the convergence of
$\la \chi_{\perp,num} \ra$ for the ring diffusion problem using
different methods; as with Sovinec's test, the symmetric method is
the least diffusive.

To study the very long time behavior of different methods (in
particular to check whether the symmetric and asymmetric methods
give negative temperatures even at very late times) we initialize
the same problem with the hot patch at 10 and the cooler background
at 0.1. Figure \ref{Ch5fig:fig7} shows the minimum temperature with
time for the symmetric, asymmetric, and entropy limited symmetric
methods; slope limited methods give the correct result for the
minimum temperature ($T_{\rm min}=0.1$) at all times. With a large
temperature contrast, both symmetric and asymmetric methods give
negative values for the temperature minimum at all times.
Such points where temperature becomes negative, when coupled with
MHD equations, can give numerical instability because of an
imaginary sound speed.
The minimum temperature with the entropy limited symmetric method
shows small undershoots at early times which are damped quickly and
the minimum temperature is 
equal to the initial minimum ($0.1$) after time=1.

\subsection{Convergence studies: measuring $\chi_{\perp,num}$}
\begin{figure}
\begin{center}
\includegraphics[width=4 in, height=3 in]{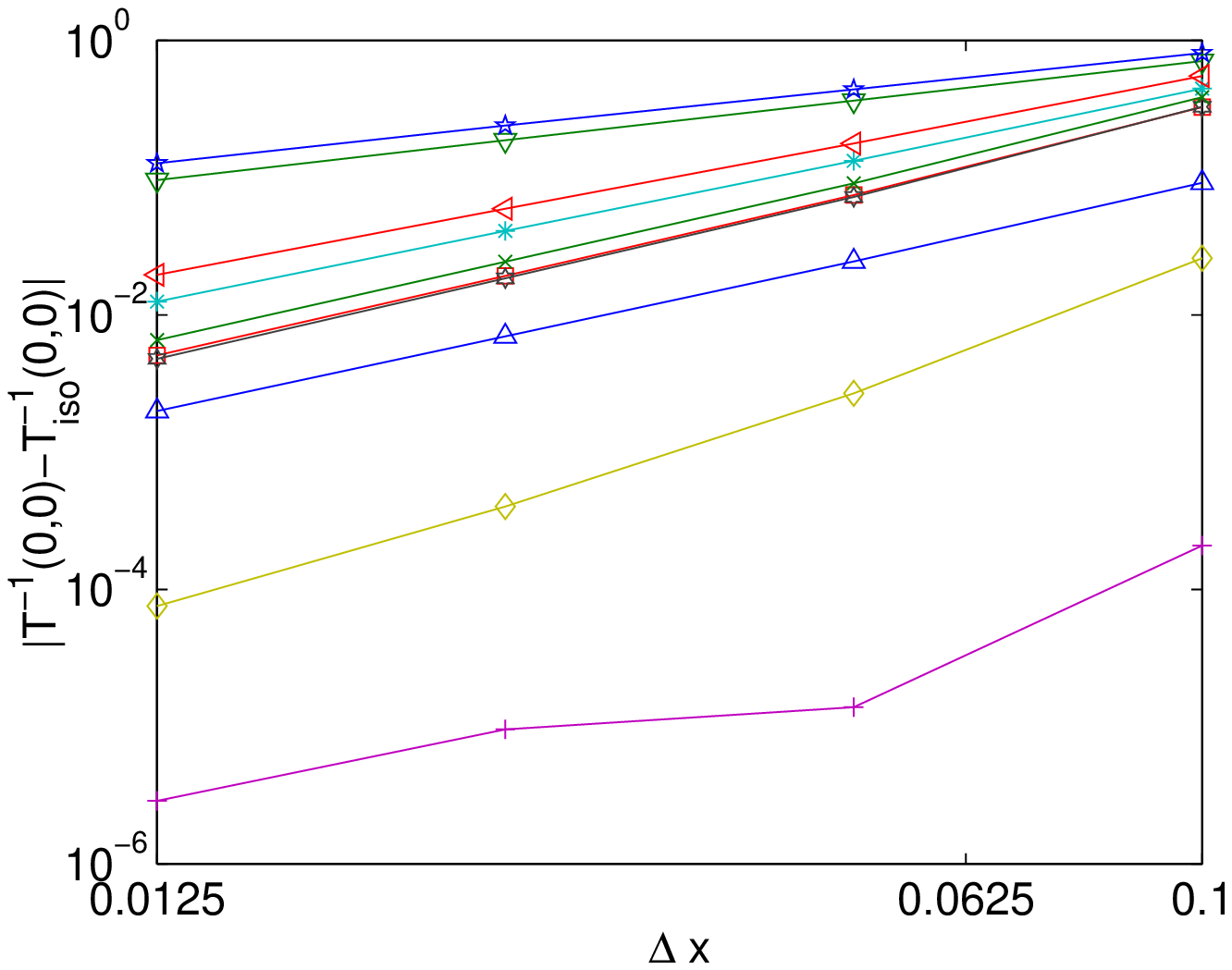}
\includegraphics[width=4 in, height=3 in]{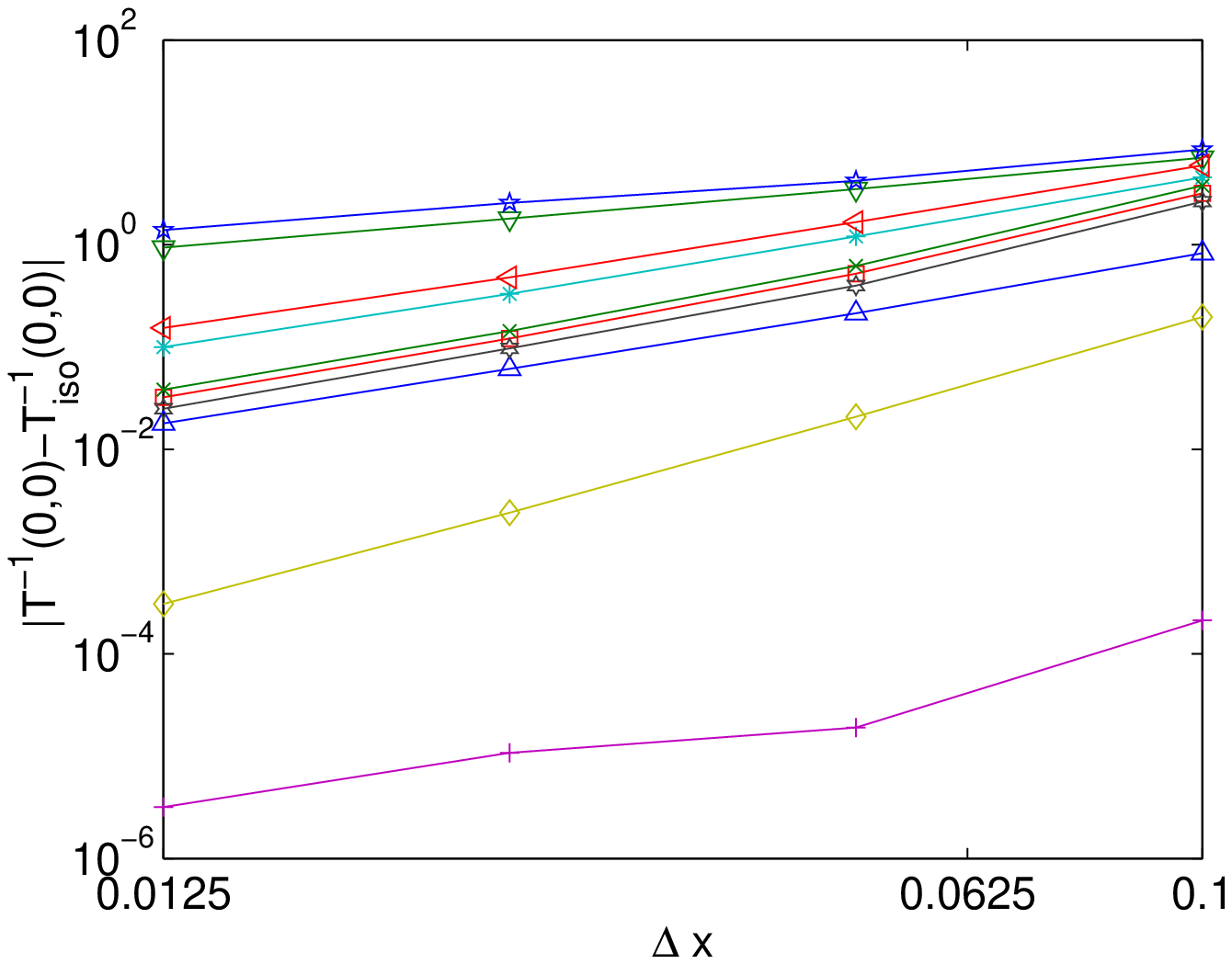}
\caption[Numerical perpendicular diffusion for
$\chi_\parallel/\chi_\perp=$10, 100 using different methods]{A
measure of perpendicular numerical diffusion $\chi_{\perp,num} = 
(T^{-1}(0,0)-T^{-1}_iso)$ for
$\chi_\parallel/\chi_\perp=10$ (top curve) and for
$\chi_\parallel/\chi_\perp=100$ (bottom curve), using different
methods for heat conduction. The different schemes are: 
asymmetric ($\triangle$), asymmetric with minmod ($\triangledown$), 
asymmetric with MC ($\square$), asymmetric
with van Leer ($\ast$), symmetric ($+$), symmetric with entropy
limiting ($\diamond$), symmetric with entropy and extrema limiting
($\triangleright$), symmetric with minmod ($\star$), symmetric with
MC ($\times$), and symmetric with van Leer limiter
($\triangleleft$). The numerical diffusion scales with
$\chi_\parallel$ for all methods except the symmetric differencing
\cite{Gunter2005}. The slope limited methods using the van Leer and
MC limiters show a second order convergence of the L1 error, like
the methods based on centered differencing. Limiting both symmetric
and asymmetric methods give similar results, but the desirable property
of the symmetric method, that the error is independent of
$\chi_\parallel/\chi_\perp$, no longer holds.\label{Ch5fig:fig8}}
\end{center}
\end{figure}

We have use the steady state test problem described in
\cite{Sovinec2004} to measure the perpendicular numerical diffusion
coefficient, $\chi_\perp$. The computational domain is a unit square
$[-0.5,0.5]\times[-0.5,0.5]$, with vanishing temperature at the
boundaries.
The source term $Q=2\pi^2 \cos(\pi x) \cos(\pi y)$ that drives the
lowest eigenmode of the temperature distribution is added to the
anisotropic diffusion equation, Eq.
\ref{Ch5eq:anisotropic_conduction}; the anisotropic diffusion
equation with a source term possesses a steady state solution. The
equation that we evolve is \be
\label{Ch5eq:anisotropic_conduction_source} \frac{\partial
e}{\partial t} = - {\bf \nabla \cdot q} + Q ;\ee a forward in time
centered in space (FTCS) differencing is used to add the source
term.

The magnetic field is derived from the flux function of the form
$\psi \sim \cos(\pi x) \cos(\pi y)$; this results in circular field
lines centered at the origin. The temperature eigenmode driven by
the source function $Q$ is constant along the field lines. The
steady state solution for the temperature is $T(x,y)=\chi_\perp^{-1}
\cos(\pi x) \cos(\pi y)$, independent of $\chi_\parallel$. The
perpendicular diffusion coefficient, $\chi_\perp$, is chosen to be
unity, and $T^{-1}(0,0)$ provides a measure of total perpendicular
diffusion, the sum of $\chi_\perp$ (the explicit perpendicular
diffusion) and $\chi_{\perp, num}$ (the perpendicular numerical
diffusion).

Figure \ref{Ch5fig:fig8} shows the perpendicular numerical
diffusivity $\chi_{\perp,num}= |T^{-1}(0,0)-T^{-1}_{iso}(0,0)|$ for
$\chi_\parallel/\chi_\perp=10$, $100$ using different methods
(where $T^{-1}_{iso}(0,0)$ is the
temperature at the origin when $\chi_\parallel=\chi_\perp$ is used
for the same resolution). G\"{u}nter et al. \cite{Gunter2005} and Sovinec et
al. \cite{Sovinec2004} use $\chi_{\perp,num}=|T^{-1}(0,0)-1|$ to measure 
perpendicular numerical diffusion; this is not precise and exaggerates the 
error for the symmetric method.

The perpendicular perpendicular diffusion ($\chi_{\perp,num}$) for all
methods except the symmetric method increases linearly with
$\chi_\parallel/\chi_\perp$. This property has been emphasized by
\cite{Gunter2005} to motivate the use of symmetric differencing for
fusion applications
which require the error (perpendicular numerical diffusion) to be
small for $\chi_\parallel/\chi_\perp \sim 10^9$. Higher order finite
elements, which maintain such high anisotropy, have also been used
for fusion applications \cite{Sovinec2004}.

The slope limited methods (with a reasonable resolution) are not
suitable for the applications which require
$\chi_\parallel/\chi_\perp \gg 10^4$; this rules out the fusion
applications mentioned in \cite{Gunter2005,Sovinec2004}. However,
only the slope limited methods give physically appropriate behavior
at temperature extrema, thereby avoiding negative temperatures in
presence of sharp temperature gradients. The slope limited method
with an MC limiter appears to be the most accurate method which does
not result in the amplification of temperature extrema.
\begin{figure}
\begin{center}
\includegraphics[width=4in,height=3in]{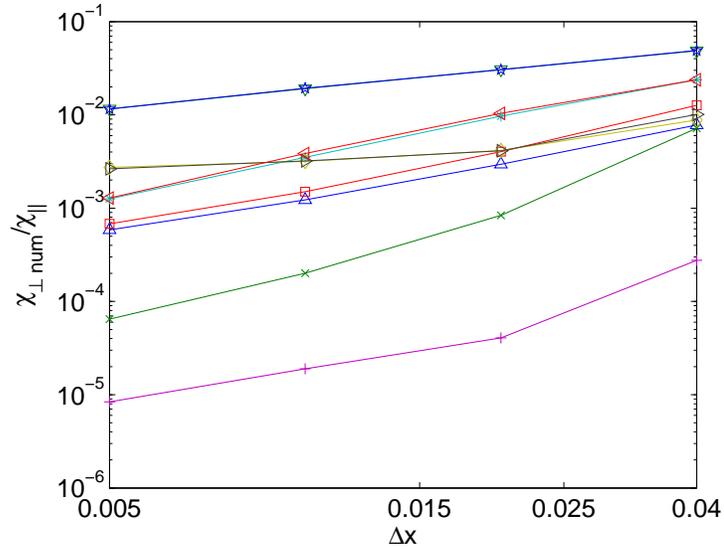}
\caption[Convergence of $\chi_{\perp,num}/\chi_\parallel$ for the
ring diffusion problem]{Convergence of
$\chi_{\perp,num}/\chi_\parallel$ as number of grid points is
increased for the ring diffusion problem. The numerical
perpendicular diffusion, $\chi_\perp$, is calculated numerically, by
measuring the heat diffusing out of the circular ring. The different
schemes are: asymmetric ($\triangle$), asymmetric with minmod
($\triangledown$), asymmetric with MC ($\square$), asymmetric with
van Leer ($\ast$), symmetric ($+$), symmetric with entropy limiting
($\diamond$), symmetric with entropy and extrema limiting
($\triangleright$), symmetric with minmod ($\star$), symmetric with
MC ($\times$), and symmetric with van Leer limiter
($\triangleleft$). The numerical diffusion linearly scales with
$\chi_\parallel$ for all methods, even with symmetric differencing
for this problem. The slope limited methods using the van Leer and
MC limiters show a second order convergence of L1 error, like the
methods based on centered differences. The slopes for asymptotic
convergence are listed in
Table~\ref{Ch5tab:tab6}.\label{Ch5fig:fig9}}
\end{center}
\end{figure}

\begin{table}[hbt]
\begin{center}
\caption{Asymptotic slopes for convergence of error
$\chi_{\perp,num} = |T^{-1}(0,0)-T^{-1}_{iso}(0,0)|$ \label{Ch5tab:tab5}} \vskip0.05cm
\begin{tabular}{ccc}
\hline
Method & $\chi_\parallel/\chi_\perp=10$ & $\chi_\parallel/\chi_\perp=100$ \\
\hline
asymmetric & 1.802  & 1.770 \\
asymmetric minmod & 0.9674 & 0.9406 \\
asymmetric MC & 1.9185 & 1.9076 \\
asymmetric van Leer & 1.706 & 1.728 \\
symmetric & 1.726 & 1.762 \\
symmetric entropy & 2.407 & 2.966 \\
symmetric entropy extrema & 1.949 & 1.953 \\
symmetric minmod & 0.9155 & 0.8761 \\
symmetric MC & 1.896 & 1.9049 \\
symmetric van Leer & 1.6041 & 1.6440 \\
\hline
\end{tabular}
\end{center}
\end{table}

\begin{table}[hbt]
\begin{center}
\caption{Asymptotic slopes for convergence of $\chi_{\perp, num}$ in
the ring diffusion test \label{Ch5tab:tab6}} \vskip0.05cm
\begin{tabular}{cc}
\hline
Method & slope \\
\hline
asymmetric & 1.066  \\
asymmetric minmod & 0.741 \\
asymmetric MC & 1.142 \\
asymmetric van Leer & 1.479 \\
symmetric & 1.181 \\
symmetric entropy & 0.220 \\
symmetric entropy extrema & 0.282 \\
symmetric minmod & 0.735 \\
symmetric MC & 1.636 \\
symmetric van Leer & 1.587 \\
\hline
\end{tabular}
\end{center}
\end{table}

The error (perpendicular numerical diffusion,
$\chi_{\perp,num}=|T^{-1}(0,0)-T^{-1}_{iso}(0,0)|$) for all methods, 
except the one which uses a minmod limiter, shows a second order 
convergence (see Table \ref{Ch5tab:tab5}). Figures \ref{Ch5fig:fig8} and
\ref{Ch5fig:fig9} show that the perpendicular numerical diffusivity
with a van Leer (or an MC) slope limiter is $\sim 10^{-3}$ for
$\approx 100$ grid points in each direction. This anisotropy is more
than sufficient to study qualitatively new effects of anisotropic
conduction on dilute astrophysical plasmas
\cite{Balbus1998,Balbus2000,Parrish2005,Sharma2006}. Among the
various limiters discussed, MC is the least diffusive, followed by
the van Leer limiter, and minmod is the most diffusive of all.

\section{Conclusions}
It is shown that simple centered differencing of anisotropic
conduction can result in negative temperatures in presence of large
temperature gradients. We have presented simple test problems where
asymmetric and symmetric methods give rise to heat flowing from
lower to higher temperatures, leading to negative temperatures at
some grid points. Negative temperature results in numerical
instabilities, as the sound speed becomes imaginary. Numerical
schemes based on slope limiters are proposed to solve this problem.

The methods developed here will be useful in numerical studies of
hot, dilute, anisotropic astrophysical plasmas
\cite{Parrish2005,Sharma2006}, where large temperature gradients may
arise. Anisotropic conduction can play a crucial role in determining
the global structure
of hot, nonradiative accretion flows (e.g.,
\cite{Balbus2001,Sharma2006,Menou2005}). Therefore, it will be
useful to extend ideal MHD codes used in previous global numerical
studies (e.g., \cite{Stone2001}) to include anisotropic conduction.
Because of the huge temperature gradients that may occur in global
disk simulations with a hot, dilute corona and a cold, dense disk,
slope limited methods, which guarantee the positivity of
temperature, must be used.

Although the slope and entropy limited methods in the present form
are not suitable for fusion applications that require accurate
resolution of perpendicular diffusion for huge anisotropy
($\chi_\parallel/\chi_\perp \sim 10^9$), they are appropriate for
astrophysical applications with large temperature gradients. A
relatively small anisotropy of thermal conduction is sufficient to
study the effects of anisotropic conduction.
The primary advantage of the limited methods is their robustness in
presence of large temperature gradients. Apart from the simulations
of dilute astrophysical plasmas with large temperature gradients
(e.g., solar corona, magnetosphere, and magnetized collisionless
shocks), our methods may find a use in diverse fields where
anisotropic diffusion is important, e.g., image processing,
biological transport, and geological systems.

Chapters \ref{chap:chap3} and \ref{chap:chap4} explored local
(linear and nonlinear) properties of the MRI in the collisionless
regime, but global calculations are required to study the relative
roles of conduction, convection, and outflows, which determine the
radial profile of different quantities, e.g., density, temperature,
and radiation. Anisotropic conduction (and pressure) is crucial to
understand the structure of hot, thick, collisionless RIAFs (see
Section \ref{Ch1sec:RIAFs}), and the slope limited methods are the only
option for robust nonlinear simulations because large temperature
gradients (e.g., the disk corona interface) arise naturally in
global disk simulations.

\chapter{Conclusions}
\label{chap:chap6}
The main goal of the thesis was to study plasma kinetic
processes operating in radiatively inefficient accretion flows (RIAFs)
around compact objects, such as the supermassive black hole in the
Galactic center and other nearby galactic centers (see
\ref{Ch1sec:RIAFs} for details). Global MHD simulations of hot,
thick accretion disks show that very little of the gas initially
accreted from the outer regions actually makes it to the last stable
orbit; most of the matter is lost as magnetized outflows
\cite{Stone2001,Hawley2001,Hawley2002,Igumenshchev2003,Pen2003,Proga2003b}.
Although the reduction of the net mass accretion rate is part of the
reason for the low luminosity, it is required by most models that
the electrons radiate much less efficiently than the standard 10\%
efficiency for such low observed luminosities (\cite{Quataert2003},
see \ref{Ch1sec:RIAFs}). Some models, e.g., ADAFs, ascribe the low
luminosity to low electron temperature compared to ions. To
understand whether electrons can be maintained much cooler than
ions, one needs to understand the conversion of gravitational energy
into internal energy of electrons and ions.

We began by looking into the MRI in the collisionless regime and
studied the transition from collisionless to collisional regimes as
the collision frequency is increased (see Chapter \ref{chap:chap3}).
We show the equivalence of the drift kinetic equation and its
moments closed with a Landau fluid closure for parallel heat flux,
in both collisional and collisionless regimes. Unlike MHD, where
energy is dissipated (resistively and viscously) only at small
scales, the collisionless plasmas have damped modes at all scales
which can heat electrons and ions differently (see Figure
\ref{Ch3fig:compare}). The linear studies were followed by 3-D local
unstratified shearing box simulations of magnetized collisionless
plasmas, using the kinetic MHD (KMHD) formalism closed with a local form of
Landau fluid closure for parallel heat flux (see Chapter
\ref{chap:chap4}). Although, both linear studies and nonlinear
simulations were carried out in a one fluid plasma with $T_i \gg
T_e$, we can roughly estimate the heating rate for both electrons
and ions. It is important to investigate what collisionless effects
can do to the structure of RIAFs; especially to consider anisotropic
thermal conduction, since it has important implications for the
convective stability of plasmas \cite{Balbus2000,Balbus2001}. While
implementing anisotropic conduction we discovered that the centered
finite differencing of anisotropic conduction can give negative
temperature in regions with large temperature gradients. To tackle
this problem we developed a method where the transverse temperature
gradient is obtained, not by simple averaging, but by using slope
limiters. The method based on slope limiters guarantees the
positivity of temperature (see Chapter \ref{chap:chap5}).

\section{Summary}

To assess the importance of plasma kinetic effects in RIAFs we began
with the study of collisionless MRI in the linear regime. The effect
of collisions was introduced through a BGK collision operator. We
use 3+1 Landau fluid closure for parallel thermal fluxes, which is
equivalent to a Pad\'e approximation for the fully kinetic plasma
response. The Landau closure gives a good approximation to linear
collisionless effects like Landau/Barnes damping.

We verify the equivalence of a fully kinetic analysis and the one
based on Landau closure by considering the modes of a magnetized
Keplerian disk, in both high and low collisionality regimes. Heating
in a collisionless disk can occur at all scales due to
Landau/Barnes damping of the fast and slow modes; whereas, in MHD
resistive and viscous heating at small scales is the only source of
heating. Since collisionless damping is a resonant phenomenon, it
can heat electrons or ions preferentially ($T_p \gg T_e$ is required
by some RIAF models; see \ref{Ch1sec:RIAFs}). The fastest growing MRI is
twice as fast in the collisionless regime as compared to MHD. More
importantly, it occurs at much larger length scales compared to MHD.
Fast growth at large scales can in principle result in a different
nonlinear saturation (for magnetic energy and stress) compared to
MHD (though our nonlinear simulations to date find that in practice
the final nonlinear spectra are similar). The MRI transitions from 
the collisionless to the Braginskii regime (when the mean free path
becomes short compared to the wavelength, $\nu \gtrsim \Omega
\sqrt{\beta}$), and then to the MHD regime (when the parallel
viscous damping becomes negligible, $\nu \gtrsim \Omega \beta$), as
the collision frequency is increased.

Balbus and Islam (see \cite{Balbus2004,Islam2005}) have studied
collisionless effects on the MRI by adding Braginskii anisotropic
stress to the MHD equations, and verified our results; they
emphasize the importance of anisotropic stress and call it the
``magnetoviscous" instability because the instability occurs at long
wavelengths even for an arbitrarily small field strength.

The linear studies were followed by local shearing box simulations
of magnetized collisionless disks. The ZEUS MHD code was modified to
include the kinetic MHD terms: anisotropic pressure in the equation
of motion, and equations evolving $p_\parallel$ and $p_\perp$ closed
by a local Landau fluid closure for heat flux along the field lines.
Adiabatic invariant ($\mu=p_\perp/B$) is conserved for collisionless
plasmas at length scales much larger than the Larmor radius and time
scales much larger than the gyroperiod. Pressure anisotropy
($p_\perp>p_\parallel$) is created naturally as magnetic field is
amplified by the MRI. Small scale instabilities---mirrror,
ion-cyclotron, and firehose---are excited even at at small pressure
anisotropies ($\Delta p/p \gtrsim $ few$/\beta$). Although,
mirror and firehose instabilites are correctly captured in Landau
MHD, we have to include a subgrid model for pressure isotropization
due to these and ion-cyclotron instabilities because at large
pressure anisotropies the fastest growing instabilities occur at the
gyroradius scale and violate adiabatic invariance.

The result of pressure anisotropy is that there is a qualitatively
new mechanism to transport angular momentum, the anisotropic stress.
Apart from appearing in the equation of motion, anisotropic stress
also appears in the internal energy equation, resulting in heating.
The anisotropic stress is as important as the Maxwell stress, and
depends only weakly on $k_L$ (the parameter in the local Landau heat
fluxes) and the pitch angle scattering model.

Pitch angle scattering due to microinstabilities limit the pressure
anisotropy and results in  MHD-like behavior---the reason MHD often
provides a good approximation for large scale dynamics of
astrophysical systems. What MHD does not tell us is how the energy
released from accretion is dissipated---whether it goes into
electrons or ions? A fully kinetic simulation with huge resolution
can address the issue of plasma heating; but insights can be gained
from fluid treatments like kinetic MHD (e.g., anisotropic stress can
heat both electrons and ions).

The kinetic MHD simulations also show that the kinetic and magnetic
energies are peaked at large scales (as in MHD). The simulations
with $B_\phi=B_z$ initially, confirm that the linear growth rate in
the kinetic regime is twice faster than in MHD; but the nonlinear
saturation is not very different in the two regimes. In fact,
somewhat counter-intuitively, the saturated magnetic energy for
$B_\phi=B_z$ simulations is smaller compared to simulations with
only a vertical field with the same $\beta$. Anisotropic stress can
be larger than the Maxwell stress for $\beta>$ a few $100$. To sum
up, the nonlinear saturation of the MRI is quite similar for MHD and
kinetic regimes.

Along with the local studies, it is crucial to understand the global
structure of hot collisionless accretion flows. Global MHD
simulations have shown that very little of the mass initially
accreted from the outer regions actually accreted on to the black
hole; most of it is lost in outflows. Anisotropic thermal conduction
can be crucial for the structure of hot collisionless accretion
flows; collisionless plasmas with long mean free path can transport
heat very efficiently along the field lines. When we used finite
differencing to implement anisotropic conduction in a global
simulation, we discovered that the temperature became negative at
the torus-corona interface. This led us to investigate numerical
algorithms for anisotropic thermal conduction in presence of large
temperature gradients. We devised simple test problems that
demonstrated that existing algorithms (both symmetric and asymmetric
differencing) can result in heat flux out of a cold region, causing
temperature to become negative in regions with high temperature
gradient. This problem was solved by using slope limiters to obtain
the transverse temperature gradient, instead of using a simple
arithmetic average. The limiter-based methods are slightly more
diffusive across field lines than the asymmetric method, but still
show second order convergence. Although the symmetric method has
very small numerical diffusion, it gives rise to high frequency
non-monotonic temperature fluctuations with large temperature
gradients.

\section{Future directions}

There are several directions for future work, for both local and
global studies. Till now we have only done single fluid simulations,
assuming the electrons to be cold. We can extended these simulations
to include electrons to study comparative heating of electrons and
ions. The original ZEUS code did not conserve energy (up to 90\% of
energy released from accretion was lost numerically), but energy
conservation can be restored to a large extent by adding the energy
lost while updating velocities and magnetic fields into heating of
the plasma \cite{Turner2003}, or by switching to codes using
conservative algorithms (such as the recently developed ATHENA code
\cite{Gardiner2005}). In the absence of explicit resistivity and
viscosity, the sources in the internal energy equation are: energy
lost in updating magnetic fields (mimics magnetic dissipation),
energy lost when updating velocity (represents viscous losses), the
$-p {\bf \nabla \cdot V}$ heating, and the work done by anisotropic
stress.

The energy-conserving one fluid simulations show that the work done
by anisotropic stress is comparable to (or even larger than) the
energy lost in magnetic field or velocity update; this means that
the physical anisotropic heating is not negligible compared to
resistive or viscous heating. This has important implications for
local two-fluid simulations. The electron pressure will also be
anisotropic ($T_{\perp, e}>T_{\parallel, e}$) because of adiabatic
invariance, and the anisotropy will be limited by pitch angle
scattering due to electron whistler instability with $\Delta p/p
\sim $ (a~few)$/\beta$ (see \cite{Kennel1966,Gary1996}). This means
that the heating rate due to anisotropic stress, $(1/e)de/dt$, is
comparable for electrons and ions, and is comparable to resistive or
viscous heating. Thus, local two-fluid simulations which conserve
energy can shed some light on electron/ion heating and whether
$T_p/T_e \gg 1$ is possible. This approach where both electrons and
ions are heated because of the energy released from accretion is
different from an approach where one looks for collisionless heat
transport from hot ions to cold electrons (e.g.,
\cite{Begelman1988}).

Another area of progress is to implement more accurate non-local
closures for thermal conduction in nonlinear simulations (see
Chapter \ref{chap:chap2}); till now we have used a crude, local
approximation with a parameter $k_L$ that exaggerates damping for
scales smaller than $2\pi/k_L$ and reduces damping for larger
scales. A local approximation may be fine if pitch-angle
scattering due to microinstabilities reduces the effective mean free
path to be comparable to the fastest growing MRI mode, which reduces 
the sensitivity to
the parameter $k_L$. However, pitch angle scattering due to
microinstabilities is not uniform. This intermittency may lead to a larger 
effective mean free path than simple estimates at first suggest.
It is important to understand
the role of intermittent scattering structures in imposing MHD-like
dynamics in collisionless plasmas.

It's important to realize that astrophysical plasmas are very
different from fusion plasmas; magnetic fields are strong in fusion
devices, with only small perturbation from the equilibrium
condition, but in astrophysical plasmas with subthermal fields,
strong shear flows can mix the fields and magnetic fields can be
chaotic. Chaotic fields reduce thermal conduction as the effective
mean free path is reduced to the field correlation length
\cite{Chandran1998}, this may mean that results do not sensitively
depend on thermal conduction.

Another approach, which is computationally more challenging but
feasible for some problems, is to evolve the drift kinetic equation
(DKE, Eq. \ref{Ch2eq:DKE}) to evolve the distribution function in a
5-D phase space and to use its moments for $p_\parallel$ and
$p_\perp$ to close the kinetic MHD moment hierarchy. Many
hydrodynamic codes are based on Riemann solvers; given a
discontinuity at grid boundaries, Riemann solvers divide the
discontinuity into wave families of the system and give the
evolution of the variables due to flux through the boundaries
\cite{Leveque2002}. The number of modes of the drift kinetic
equation is huge, and it is impossible to solve the Riemann problem
exactly. One approach to solve hyperbolic equations that does not
require the solution of Riemann problems is based on central methods
(alternatively, they can be related to a simple, approximate Riemann
solver; see \cite{Nessyahu1990,Kurganov2000}); central methods have
also been applied to MHD simulations \cite{Balbas2006}. The DKE
simulations do not require the closure approximation, but like in
KMHD with Landau closures, subgrid models for pitch angle scattering
to microinstabilities will be required. It is also possible to carry
out full Vlasov or particle-in-cell (PIC) simulations where a
subgrid model for microinstabilities is not required, but for such
simulations to be applicable to RIAFs they will need to resolve both
the large MRI scale and the Larmor radius scale (8 orders of
magnitudes smaller than the disk height scale).

Recent global MHD simulations were responsible for understanding
that only a small fraction of gas accreted in outer region actually
make it to the black hole, most of it is lost in outflows. A small
accretion rate is one reason for small radiative luminosities of
RIAFs. An important direction for future research is to include
kinetic MHD effects like anisotropic conduction in global
simulations. Since plasma in RIAFs is hot and collisionless,
anisotropic thermal conduction is rapid. This can be important in
determining the structure of RIAFs. The structure of the
self-similar solution for a RIAF changes dramatically if a saturated
form of thermal conduction (due to free streaming of particles) is
included \cite{Menou2005}. Another reason that anisotropic
conduction can be important is because the convective stability
criterion for anisotropic plasmas is that temperature decreases
outwards, $dT/dr<0$, instead of the usual Schwarzschild condition of
entropy increasing outwards, $ds/dr<0$ \cite{Balbus1998,Balbus2000}.
The effect of thermal conduction is subtle because the MRI may
generate chaotic fields and suppress thermal conduction and impose
more MHD-like behavior, instead of giving a state which is stable to
the magnetothermal instability. Thus, it will important to know
whether anisotropic thermal conduction will be a small effect due to
its suppression because of MHD turbulence, or it will alter the
structure of RIAFs.

\appendix
\chapter{Accretion models}
\label{app:app1}
\section{Efficiency of black hole accretion}
\label{app:BHefficiency} Black holes are different from neutron
stars and white dwarfs as they do not have a surface. Although there
is no surface, black holes are characterized by an event horizon, a
region from which nothing, not even light, can escape. For a
non-rotating (Schwarzschild) black hole, Newtonian arguments (speed
of light = escape velocity at the event horizon) can be used to
calculate the Schwarzschild radius, $r_g=2GM_*/c^2$, radius of the
event horizon for a black hole of mass $M_*$.

To calculate accretion efficiency one needs to know the form of the
effective potential. In Newtonian theory, the energy equation for a
mass with specific angular momentum $l$ is \be \frac{1}{2} \left
(\frac{dr}{dt} \right )^2 + \Phi_{\rm eff}(r) = E, \ee where $E=$
constant is the total energy per mass, and $\Phi_{\rm
eff}(r)=l^2/2r^2-GM_*/r$ is the effective potential. Newtonian
approximation is not valid for a black hole, and a full general
relativistic treatment is required. However, Paczynski and Wiita
\cite{Paczynski1980} introduced a pseudo-Newtonian potential for a
non-rotating black hole, $\Phi_{\rm PW}=GM_*/(r-r_g)$, which gives a
good approximation for the effective potential of a non-rotating
black hole. Using the Paczynski-Wiita potential, the energy equation
becomes \be \frac{1}{2} \left ( \frac{dr}{dt} \right )^2 +
\frac{l^2}{2 r^2} - \frac{GM_*}{r - r_g} = E \ee
\begin{figure}
\begin{center}
\includegraphics[width=4in,height=3in]{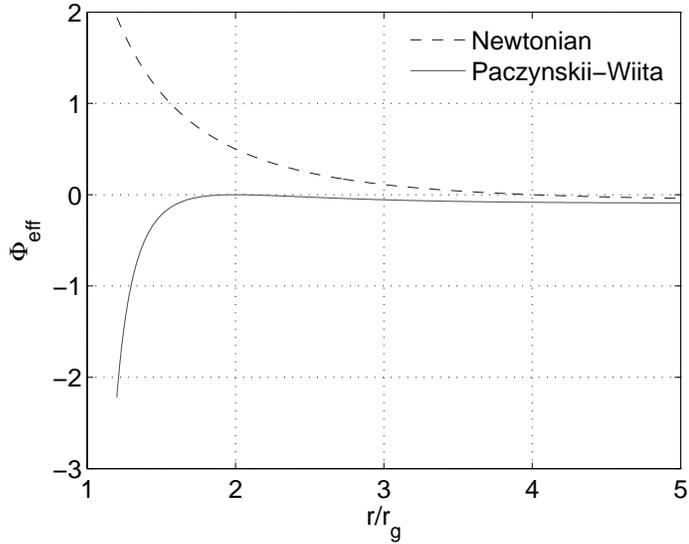}
\caption[Newtonian and Paczynski-Wiita potential for
$l=4GM_*/c$]{Comparison of the Newtonian and Paczynski-Wiita
potential for $l=4GM_*/c$, corresponding to a marginally bound orbit
in general relativity. Notice that $\Phi \rightarrow 0$ as $r
\rightarrow \infty$. \label{Ch1fig:PseudoNewtonian}}
\end{center}
\end{figure}

The Paczynski-Wiita potential is useful because the effective
potential, as in the case of general relativistic potential, has a
minimum and a maximum if the specific angular momentum
$l>2\sqrt{3}GM_*/c$ \cite{Schutz1985}. In comparison, the Newtonian
effective potential has a single minimum (corresponding to the
circular Keplerian orbit) for any non-zero angular momentum. The
general relativistic consequences are: 1) for any given angular
momentum, particles with sufficiently high energy can overcome the
centrifugal barrier and fall in, and 2) particles with low (not zero
as in the Newtonian case) angular momentum are captured by the hole
\cite{Frank2002}. The Paczynski-Wiita potential obtains the correct
general relativistic result for marginally stable (corresponding to
$r=3r_g$ and $l=2\sqrt{3}GM_*/c$, within which all orbits are
unstable), and the marginally bound orbit (with $r=2r_g$ and
$l=4GM_*/c$, particles with $E>0$ can fall directly on to the hole
for specific angular momentum smaller than this). Figure
\ref{Ch1fig:PseudoNewtonian} shows the Newtonian, and the
Paczynski-Wiita potential for a marginally bound orbit, with
$l=4GM_*/c$.

The presence of a last stable orbit has important consequences for
accretion efficiency; beyond this, matter plunges in the black hole
with no time to radiate. Thus, for a Schwarzschild black hole,
matter radiates half the released gravitational energy (and retains
the other half as the kinetic energy) till the last stable orbit
($3r_g$). This gives a radiative efficiency of
$\eta=(GM_*/6r_g)/c^2=1/12$. The relativistic relativistic result of
$6\%$ is not too far off. For a rotating Kerr hole the last stable
orbit moves further in, resulting in a larger efficiency; a
maximally rotating black hole has an efficiency of $42.3\%$ (see
\cite{Misner1973} for detailed introduction to spinning black
holes).

\section{Bondi accretion}
\label{app:Bondi} Bondi accretion \cite{Bondi1952}, a model for
steady, spherical accretion of matter with vanishing angular
momentum (e.g., a star accreting from a stationary gas cloud), is
commonly used to estimate the accretion rate $\dot{M}$ from the
measurement of ambient density and temperature. The following
presentation is based on \cite{Frank2002}.

We will solve the spherically symmetric, hydrodynamic equations in
steady state using spherical polar coordinates ($r,\theta,\phi$)
with origin at the center of the star. The fluid variables are
independent of $\theta$ and $\phi$, and the gas has only a radial
velocity component $V_r=V$. The equation of continuity \be
\label{Ch1eq:Continuity} \frac{1}{r^2}\frac{d}{dr} \left ( r^2 \rho
V \right) = 0, \ee gives a constant inward flux of matter
$\dot{M}=-4\pi r^2 \rho V =$ constant; for accretion $V<0$, as
matter falls in. The Euler equation becomes \be \label{Ch1eq:Euler}
V \frac{dV}{dr} + \frac{1}{\rho} \frac{dp}{dr} + \frac{GM_*}{r^2} =
0. \ee A polytropic equation of state is used, $p=K\rho^\gamma$,
with $1<\gamma<5/3$.
\begin{figure}
\begin{center}
\includegraphics[width=4in,height=3in]{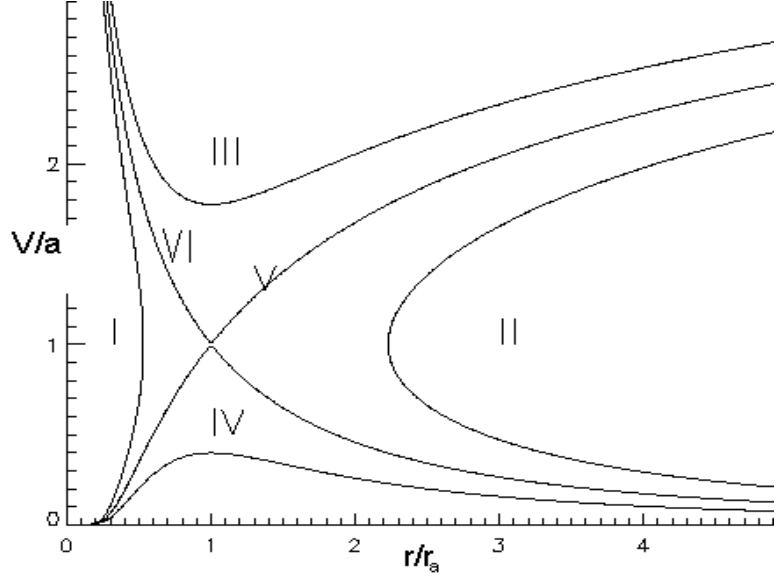}
\caption[Mach number with radius for Bondi spherical accretion]{Mach
number (the ratio of fluid velocity and sound speed) as a function
of radius for spherical accretion with different inner and outer
boundary conditions. Solution $VI$ corresponds to accretion and $V$
to a spherical wind. Taken from Alan Hood's lecture notes, 
http://www-solar.mcs.st-and.ac.uk/\~{}alan/sun\_course/.
\label{Ch1fig:SphericalAccretion}}
\end{center}
\end{figure}

Using $dP/dr = a^2 d\rho/dr$, where $a=\sqrt{\gamma p/\rho}$ is the
sound speed, the continuity and Euler equations can be combined to
give \be \label{Ch1eq:Rearrangement} \frac{1}{2} \left ( 1 -
\frac{a^2}{V^2} \right ) \frac{d}{dr} V^2 = - \frac{GM_*}{r^2} \left
[ 1 - \left ( \frac{2 a^2 r}{GM_*} \right) \right ]. \ee This form
is useful to draw inferences about steady, spherically symmetric
accretion. At large distances ($r \gg GM_*/2a^2$), the right side of
Eq. \ref{Ch1eq:Rearrangement} is positive, and $dV^2/dr<0$ at large
distances where gas is expected to be at rest. This implies that the
gas is subsonic ($V^2<a^2$) for $r\gg GM_*/2a^2$; this is reasonable
because far from the star, the gas has a non-zero temperature. One
need to specify the inner boundary condition in addition to the
outer (ambient) boundary conditions to uniquely specify the
solution; we choose $V^2>a^2$ for small $r$ for accretion (see
Figure \ref{Ch1fig:SphericalAccretion}).

At a radius $r_a=GM_*/2a^2$, either $V^2=a^2$ or $d/dr(V^2)=0$;
latter is true for accretion solution with a supersonic flow for
$r<r_a$. The sonic point condition, $r_a=GM_*/2a^2$ leads to the
relation between  the accretion rate $\dot{M}$ and the ambient
conditions. The integral form of Eq. \ref{Ch1eq:Euler} is the
Bernoulli integral: \be \label{Ch1eq:Bernoulli} \frac{V^2}{2} +
\frac{a^2}{\gamma-1} - \frac{GM}{r} = {\rm Be,~a~constant}. \ee The
boundary condition at $r \rightarrow \infty$, and the sonic point
condition $a^2(r_a)=GM_*/2r_a$ combine to give
$a(r_a)=a(\infty)\sqrt{2/(5-3\gamma)}$, which leads to the constant
accretion rate in terms of sonic point variables, $\dot{M}=4\pi
r_a^2 \rho(r_a) a(r_a)$. Since $a \propto \rho^{\gamma-1}$, \be
\rho(r_a) = \rho(\infty) \left [ \frac{a(r_a)}{a(\infty)} \right
]^{2/(\gamma-1)}; \ee this combined with $\dot{M}$ in terms of sonic
point variables gives the required expression for $\dot{M}$ in terms
of conditions at infinity: \be \label{Ch1eq:Mdot} \dot{M} = \pi G^2
M_*^2 \frac{\rho(\infty)}{a^3(\infty)} \left [ \frac{2}{5-3\gamma}
\right ]^{(5-3\gamma)/2(\gamma-1)}. \ee The dependence of $\dot{M}$
on $\gamma$ is weak. For $\gamma=1.4$, Eq. \ref{Ch1eq:Mdot} gives
\be \label{Ch1eq:MdotNum} \dot{M} \cong 1.4 \times 10^{11} \left (
\frac{M}{M_\odot} \right ) \left ( \frac{\rho(\infty)}{10^{-24}~{\rm
g}~{\rm cm}^{-3}} \right ) \left ( \frac{a(\infty)}{10~{\rm km}~{\rm
s}^{-1}} \right)^{-3}~{\rm g}~{\rm s}^{-1}. \ee

For $r \ll r_a$, matter falls freely, $v^2 \cong 2 GM_*/r$; the
continuity equation gives $\rho \cong \rho(r_a) (r_a/r)^{3/2}$ for
$r<r_a$. One can define an effective accretion radius, beyond which
the thermal energy of the gas is larger than the gravitational
binding energy. The ratio of the thermal and gravitational binding
energy is $(ma^2(r)/2)/(GM_*m/r) \sim r/r_{acc}$, for $r \gtrsim
r_{acc}$, since $a(r) \sim a(\infty)$ for $r>r_{acc} \equiv
2GM_*/a^2(\infty)$ \cite{Frank2002}. Hence, for $r \gg r_{acc}$ the
gravitational pull of the star has negligible effect on the gas. In
terms of the Bondi radius, an approximation for the mass accretion
rate is given by $\dot{M} \sim \pi r_{acc}^2 a(\infty)\rho(\infty)$.

\chapter{Linear closure for high and low collisionality}
\label{app:app2}
\section{Closure for high collisionality: $|\zeta| \gg 1$}

For $|\zeta| \gg 1$, $Z\left(\zeta\right) \approx
-1/\zeta-1/2\zeta^3 -3/4\zeta^5$, $R \approx -1/2 \zeta^2-
3/4\zeta^4$, $1+2\zeta^2 R\approx -3/2 \zeta^2 - 15/4\zeta^4$, $Z-2
\zeta R \approx 1/\zeta^3+3/\zeta^5$.
Equation~(\ref{Ch3eq:closure_perp}) then becomes \be \frac{\delta
n}{n_0}- \frac{\delta p_{\Perp}}{p_0}=- \frac{\delta B}{B_0}\left(1+
\frac{1}{2 \zeta^2}\right)-\frac{\zeta_2}{\zeta}
\left(1+\frac{1}{2\zeta^2}\right)\left(\frac{\delta T}{T_0}
-\frac{\delta B}{B_0}\right). \ee Assuming $|\zeta_1/\zeta_2| \ll
1$~(a high collisionality limit $\omega \ll \nu$) and using the
binomial expansion we get \be \frac{\delta n}{n_0} -\frac{\delta
p_{\Perp}}{p_0}=-\left\{1-\frac{\zeta_1}
{\zeta_2}+\frac{1}{\zeta_2^2}\left(\frac{1}{2}+\zeta_1^2\right)-\frac{\zeta_1}
{\zeta_2^3}\left(\frac{1}{2}+\zeta_1^2\right)\right\}
\left(\frac{\zeta_1}{\zeta_2}\frac{\delta B} {B_0} + \frac{\delta
T}{T_0}\right). \label{Ch3eq:bino_perp} \ee To the lowest nonvanishing
order one gets \be \frac{\delta
n}{n_0}\frac{\zeta_1}{\zeta_2}-\frac{\delta p_{\Perp}}{p_0}
\left(\frac {1}{3} +\frac{2}{3}\frac{\zeta_1}{\zeta_2}\right)+
\frac{\delta p_{\Par}}{p_0}\left(\frac{1}{3}
-\frac{\zeta_1}{3\zeta_2}\right)=
-\frac{\zeta_1}{\zeta_2}\frac{\delta B}{B_0}. \ee Expanding
equation~(\ref{Ch3eq:closure_par}) gives \ba \nonumber &&-\frac{\delta
n}{n_0}\left(\frac{3}{2 \zeta^2}+\frac{15}{4 \zeta^4}\right)+
\frac{\delta p_{\Par}}{p_0}\left(\frac{1}{2 \zeta^2}+\frac{3}{4
\zeta^4}\right)=
-\frac{\delta B}{B_0}\left(\frac{1}{\zeta^2}+\frac{3}{\zeta^4}\right) \\
&& + \left(\frac{ \delta B}{B_0} -\frac{\delta n}{n_0}+\frac{\delta
T}{2 T_0}\right)\zeta_2 \left(\frac {1}{\zeta^3}+
\frac{3}{\zeta^5}\right). \ea Again using the binomial expansion for
$|\zeta_1/\zeta_2| \ll 1$ we get \ba \nonumber &&\frac{\delta
n}{n_0}\left(-\frac{3}{2} \frac{\zeta_1}{\zeta_2} +\frac{9}{2}
\left(\frac{\zeta_1}{\zeta_2}\right)^2+\frac{3}{4 \zeta_2^2}\right)
+ \frac{\delta
p_{\Par}}{p_0}\left(\frac{1}{3}-\frac{1}{2}\frac{\zeta_1}{\zeta_2} +
\frac {1}{2}\left(\frac{\zeta_1}{\zeta_2}\right)^2+\frac{1}{4
\zeta_2^2}\right)
\\
&&+\frac{\delta
p_{\Perp}}{p_0}\left(-\frac{1}{3}+\frac{\zeta_1}{\zeta_2}-2\left(\frac
{\zeta_1}{\zeta_2}\right)^2 -\frac{1}{\zeta_2^2}\right) =
\frac{\delta B}{B_0} \left(-
\frac{\zeta_1}{\zeta_2}+3\left(\frac{\zeta_1}{\zeta_2}\right)^2\right).
\label{Ch3eq:bino_par} \ea The lowest order solution is \be -\frac{3
\zeta_1}{2 \zeta_2} \frac{\delta n}{n_0}
+\left(\frac{1}{3}-\frac{\zeta_1} {2\zeta_2}\right)\frac{\delta
p_{\Par}}{p_0}+\left(-\frac{1}{3} + \frac{\zeta_1}
{\zeta_2}\right)\frac{\delta p_{\Perp}}{p_0}= -
\frac{\zeta_1}{\zeta_2} \frac {\delta B}{B_0}. \ee We shall expand
the parallel and perpendicular pressure perturbations as $\delta
p_{\Perp}=\delta^0p_{\Perp}+\zeta_1/\zeta_2\delta^1p_{\Perp}+
(\zeta_1/\zeta_2)^2 \delta^2p_{\Perp} + ... $ and $\delta
p_{\Par}=\delta^0p_{\Par}+\zeta_1/\zeta_2\delta^1p_{\Par}+
(\zeta_1/\zeta_2)^2\delta^2p_{\Par} + ...$ From
equations~(\ref{Ch3eq:bino_perp}) and~(\ref{Ch3eq:bino_par}) one gets $
\delta^0 p_{\Par}/p_0=\delta^0 p_{\Perp}/p_0=5 \delta n/3 n_0$ for
the lowest order, and $ (\delta p_{\Perp}-\delta^1
p_{\Par})/p_0=3\delta B/B_0- 2 \delta n/n_0$. To the next order we
can expand the solution as \ba &&\frac{\delta p_{\Par}}{p_0}=\frac{5
\delta n}{3n_0}+\frac{\zeta_1} {\zeta_2}\frac{\delta^1
p_{\Par}}{p_0}+\left(\frac{\zeta_1}{\zeta_2}\right)^2 \frac
{\delta^2 p_{\Par}}{p_0}, \\
&& \frac{\delta p_{\Perp}}{p_0}= \frac{5 \delta n}{3
n_0}+\frac{\zeta_1}{\zeta_2}\left(\frac{\delta^1p_{\Par}}
{p_0}+3\frac{\delta B}{B_0}-2 \frac{\delta n}{n_0}\right) +
\left(\frac{\zeta_1} {\zeta_2}\right)^2 \frac {\delta^2
p_{\Perp}}{p_0}. \ea To the next order in $\zeta_1/\zeta_2$ in
equation~(\ref{Ch3eq:bino_perp}) one gets \be -\frac{1}{2
\zeta_1^2}\frac{\delta n}{n_0} + \frac{1}{2} \frac{\delta^1
p_{\Par}}{p_0}+ \frac{1}{3}\left(\frac{\delta^2 p_{\Par}}{p_0}-
\frac{\delta^2 p_{\Perp}}{p_0}\right)=0. \label{Ch3eq:first_perp} \ee
To the next order in equation~(\ref{Ch3eq:bino_par}) we get \be
\left(2+\frac{1}{3 \zeta_1^2}\right)\frac{\delta n}{n_0} -
\frac{\delta^1 p_{\Par}} {p_0}+ \frac{1}{3}\left(\frac{\delta^2
p_{\Par}}{p_0}-\frac{\delta^2 p_{\perp}} {p_0}\right)= \frac{3
\delta B}{B_0}. \label{Ch3eq:first_par} \ee
Equations~(\ref{Ch3eq:high_closure_perp})
and~(\ref{Ch3eq:high_closure_par}) follow from
equations~(\ref{Ch3eq:first_perp}) and~(\ref{Ch3eq:first_par}).

\section{Closure for low collisionality: $|\zeta| \ll 1$}
This regime is useful for low collisionality $\nu \ll k_{\Par} c_0$
and high $\beta$, where the MRI is low frequency as compared to the
sound wave frequency. Using the asymptotic expansion for $|\zeta|
\ll 1$, $Z\left(\zeta\right) \approx
i\sqrt{\pi}\left(1-\zeta^2\right)-2 \zeta$ and $R\left(\zeta\right)
\approx 1+i\sqrt{\pi} - 2 \zeta^2$, we simplify
equation~(\ref{Ch3eq:closure_perp}) to get \be \frac{\delta n}{n_0} -
\frac{\delta p_{\Perp}}{p_0}=\frac{\delta B}{B_0} \zeta \left(i
\sqrt{\pi} - 2 \zeta\right) + \left(\frac{\delta T}{T_0}- \frac
{\delta B}{B_0}\right)\zeta_2\left(i\sqrt{\pi}-2 \zeta\right). \ee
The lowest order term in $\zeta$ gives $\delta p_{\Perp}/p_0=\delta
n/n_0$. Let $\delta p_{\Perp}/p_0 \approx \delta n/n_0+\zeta
\delta^1 p_{\Perp}/p_0$. To the next order one gets \be \zeta
\frac{\delta^1 p_{\Perp}}{p_0}=-i\sqrt{\pi}\zeta\frac{\delta B}{B_0}
+ i \sqrt{\pi} \zeta_2 \frac{\delta B}{B_0}=-i \sqrt{\pi} \zeta_1
\frac{ \delta B}{B_0}. \ee Therefore to second order in $\zeta$,
$\delta p_{\Perp}/p_0 \approx \delta n/n_0 - i \sqrt{\pi} \zeta_1
\delta B/B_0 +\zeta^2 \delta^2 p_{\Perp}/p_o$. On using the
asymptotic formula for $Z$ and $R$ in
equation~(\ref{Ch3eq:closure_par}), one gets \be \frac{\delta n}{n_0}-
\left(1+i\sqrt{\pi} \zeta\right) \frac{\delta p_{\Par}}{p_0}=-i
\sqrt{\pi} \zeta \frac{\delta B}{B_0} -\zeta_2 \left(i\sqrt{\pi} -4
\zeta\right)\left(\frac{\delta n}{n_0}-\frac{\delta T_{\Par}}{2T_0}
-\frac{\delta B}{B_0}\right). \ee To the lowest order one gets
$\delta p_{\Par}/{p_0} = {\delta n}/{n_0}$, so let ${\delta
p_{\Par}}/{p_0} \approx {\delta n}/{n_0} + \zeta {\delta^1
p_{\Par}}/{p_0}$. To the next order, \be \zeta \frac{\delta^1
p_{\Par}}{p_0}=-i\sqrt{\pi} \zeta_1 \frac{\delta n}{n_0} + i
\sqrt{\pi} \zeta_1 \frac{\delta B}{B_0}. \ee Therefore through
second order $\delta p_{\Par}/p_0 \approx \delta n/n_0 + i\sqrt{\pi}
\zeta_1 \left(\delta B/B_0-\delta n/n_0\right)+ \zeta^2\delta^2
p_{\Par}/p_0$. The comparison of the terms of the order $\zeta^2$ in
equation~(\ref{Ch3eq:closure_perp}) give \be \zeta^2 \frac{\delta^2
p_{\Perp}}{p_0}= 2 \zeta_1 \zeta \frac{\delta B}{B_0} -
\frac{\pi}{3} \zeta_1 \zeta_2\left(\frac{\delta B}{B_0}+\frac
{\delta n}{n_0}\right), \label{Ch3eq:second_ord_perp} \ee and the terms
of the order $\zeta^2$ in equation~(\ref{Ch3eq:closure_par}) give \be
\zeta^2 \frac{\delta^2 p_{\Par}}{p_0}=\left(4 \zeta_1 \zeta_2 - \pi
\zeta_1^2 -\frac{7\pi}{6}\zeta_1 \zeta_2 \right)\frac{\delta n}{n_0}
+ \left( \sqrt{\pi} \zeta_1 \zeta -\frac{\pi}{6} \zeta_1 \zeta_2 - 2
\zeta^2 - 4 \zeta_2 \zeta \right)\frac{\delta B}{B_0}.
\label{Ch3eq:second_ord_par} \ee From
equations~(\ref{Ch3eq:second_ord_perp}) and~(\ref{Ch3eq:second_ord_par})
the asymptotic expansion in equations~(\ref{Ch3eq:low_closure_perp})
and~(\ref{Ch3eq:low_closure_par}) follow.

\chapter{Kinetic MHD simulations: modifications to ZEUS}
\label{app:app3}
\section{Grid and variables} \label{app:grid}
\begin{figure}
\begin{center}
\psfrag{A}{\large{-($V_x$, $B_x$, $q_{\parallel x}$, $q_{\perp
x}$)$_{i,j,k}$}} \psfrag{B}{\large{$P_{xy_{i,j,k}}$}}
\psfrag{C}{\large{$E_{y_{i,j,k}}$}}
\psfrag{D}{\large{$P_{xz_{i,j,k+1}}$}}
\psfrag{E}{\large{$E_{z_{i,j,k}}$}} \psfrag{F}{\large{($V_z$, $B_z$,
$q_{\parallel z}$, $q_{\perp z}$)$_{i,j,k+1}$}}
\psfrag{G}{\large{-($V_y$, $B_y$, $q_{\parallel y}$, $q_{\perp
y}$)$_{i,j,k}$}} \psfrag{H}{\large{$E_{x_{i,j,k}}$}}
\psfrag{I}{\large{($d$, $p_\parallel$, $p_\perp$)$_{i,j,k}$}}
\psfrag{J}{\large{$P_{yz_{i,j,k+1}} $}} \psfrag{X}{\large{$x$}}
\psfrag{Y}{\large{$y$}} \psfrag{Z}{\large{$z$}}
\includegraphics[width=5in,height=5in]{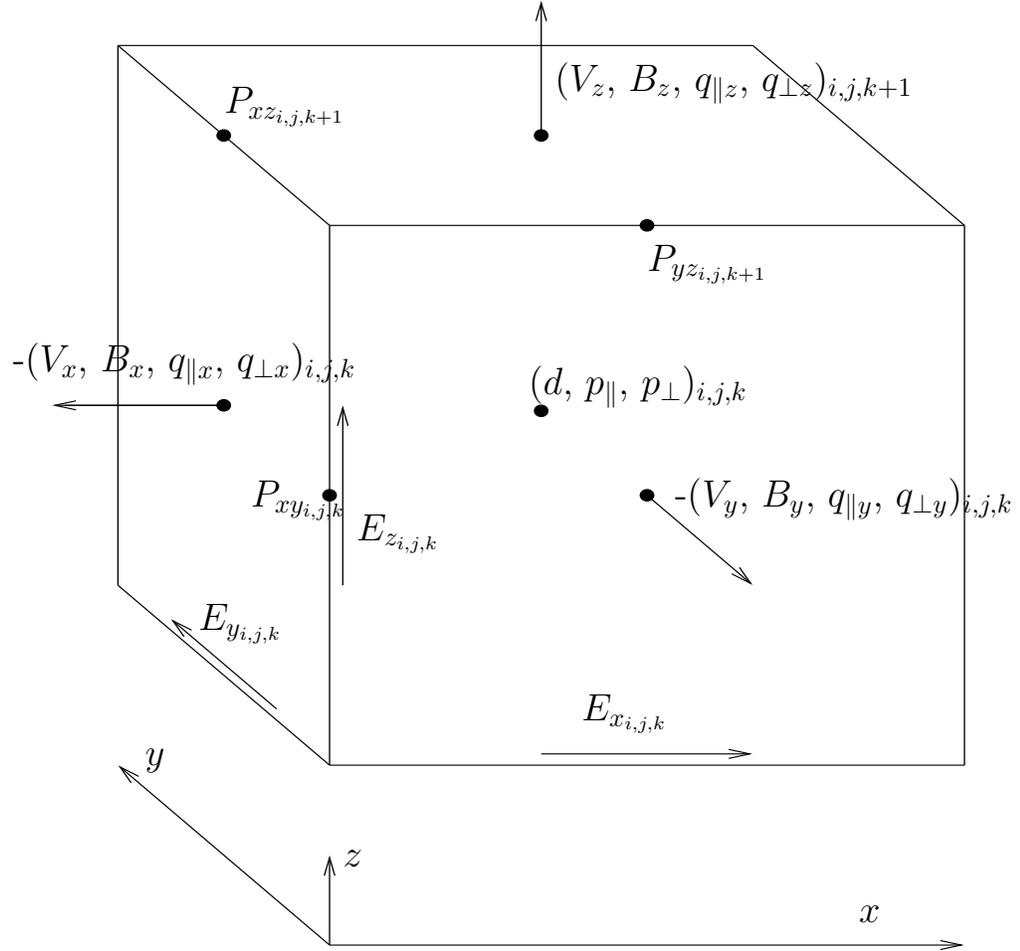}
\caption[Location of different variables on a 3-D staggered
grid]{Location of different variables on a 3-D staggered grid.
Vectors ${\bf V}$, ${\bf B}$, and ${\bf q_{\Par,\Perp}}$ are located
at the face centers. Density ($\rho$) and diagonal components of the
pressure tensor ($p_\Perp$, $p_\Par$) are located at the zone
centers. EMF's ($E_x$, $E_y$, $E_z$), and off diagonal components of
the pressure tensor ($P_{xy}$, $P_{xz}$, $P_{yz}$) are located on
appropriate edges. \label{Ch4fig:figure9}}
\end{center}
\end{figure}
Figure \ref{Ch4fig:figure9} shows the location of variables on the
grid. Scalars and diagonal components of second rank tensors
($\rho$, $p_\Par$, and $p_\Perp$) are zone centered. Vectors,
representing fluxes out of the box, are located at the cell faces
(${\bf V}$, ${\bf B}$, and ${\bf q_{\Par, \Perp}}$). The inductive
electric field ({\bf E}) is located at cell edges such that the
contribution of each edge in calculating $\oint {\bf E} \cdot {\bf
dl}$ over the whole box cancels, and $\nabla \cdot {\bf B}=0$ is
satisfied to machine precision. The off diagonal part of the
pressure tensor in Cartesian coordinates is related to ${\bf \Pi} =
{\bf \hat{b}\hat{b}}(p_\Par-p_\Perp)$. This is a symmetric tensor
whose components $P{xy}$, $P{xz}$, and $P{yz}$ are located such that
the finite difference formulae for the evolution of velocities due
to off diagonal components of stress are given by \ba &&
{Vx_{i,j,k}}^{n+1}={Vx_{i,j,k}}^n - \frac{\delta t}{\delta
y}(P{xy}_{i,j+1,k}^n-P{xy}_{i,j,k}^n) - \frac{\delta t}{\delta
z}(P{xz}^n_{i,j,k+1}-P{xz}_{i,j,k}^n), \\ &&
{Vy_{i,j,k}}^{n+1}={Vy_{i,j,k}}^n - \frac{\delta t}{\delta
x}(P{xy}_{i+1,j,k}^n-P{xy}_{i,j,k}^n) - \frac{\delta t}{\delta
z}(P{yz}^n_{i,j,k+1}-P{yz}_{i,j,k}^n), \\ &&
{Vz_{i,j,k}}^{n+1}={Vz_{i,j,k}}^n - \frac{\delta t}{\delta
x}(P{xz}_{i+1,j,k}^n-P{xz}_{i,j,k}^n) - \frac{\delta t}{\delta
y}(P{yz}^n_{i,j+1,k}-P{yz}_{i,j,k}^n).  \ea

\subsection{Determination of $\delta t$: Stability and positivity}
\label{app:courant} A time explicit algorithm must limit the time
step in order to satisfy the Courant-Friedrichs-Levy (CFL) stability
condition. Physically, $\delta t$ must be smaller than the time it
takes any signal (via fluid or wave motion) to cross one grid zone.
There is also a limit imposed on $\delta t$ for numerical stability
of the diffusive steps. Additionally, since there are quantities
which must be positive definite ($\rho$, $p_\Par$, $p_\Perp$), we
also require $\delta t$ to satisfy positivity. We adopt the
following procedure to choose $\delta t$: \ba && \delta t_{adv} =
\frac{\mbox{min}\{\delta x,\delta y,\delta z\}}{
(|V|+|V_A|+|V_s|+|\Omega L_x|)}, \\ && \delta t_\Par =
\frac{\mbox{min}\{\delta x^2,\delta y^2,\delta z^2\}}
{2 \kappa_\Par}, \\
&& \delta t_\Perp = \frac{\mbox{min}\{\delta x^2,\delta y^2,\delta
z^2\}}{2 \kappa_\Perp}, \ea where $V_A=B/\sqrt{4\pi}$ is the
Alfv\'en speed, and $V_s=\mbox{max}\{ \sqrt{3 p_\Par/\rho}, \sqrt{2
p_\Perp/\rho} \}$ is the maximum sound speed, taking the anisotropy
into account. $\delta t_{adv}$, $\delta t_\Par$, and and $\delta
t_\Perp$ correspond to limits on the time step for stability to
advection, and parallel and perpendicular heat conduction,
respectively.

The source steps for $p_\Par$ and $p_\Perp$ are given by \ba &&
\frac{p_\Par^{n+1}-p_\Par^n}{\delta t} = \left ( -\nabla \cdot {\bf
q_\Par} - 2 p_\Par {\bf \hat{b}} \cdot \nabla {\bf V} \cdot {\bf
\hat{b}} + 2 q_\Perp \nabla \cdot \hat{b} \right)^n = A1, \\ &&
\frac{p_\Perp^{n+1}-p_\Perp^n}{\delta t} = \left ( -\nabla \cdot
{\bf q_{\Perp T} } - p_\Perp \nabla \cdot {\bf V} + p_\Perp {\bf
\hat{b}} \cdot \nabla {\bf V} \cdot {\bf \hat{b}} - q_\Perp \nabla
\cdot \hat{b} \right)^n = A2, \ea where ${\bf q_{\Perp T}
}=-\kappa_\Perp \nabla_\Par T_\Perp$ denotes the temperature
gradient part of ${\bf q_\Perp}$. For positivity of $p_\Par^{n+1}$
and $p_\Perp^{n+1}$ we require that the following conditions are
satisfied: whenever $A1$ and $A2$ are negative, $\delta t_{pos} =
\mbox{min} \{ -p_\Par^n/A1, - p_\Perp^n/A2 \}$; if $A1>0$, $A2<0$,
then $\delta t_{pos} = -p_\Perp^n/A2$; if $A1<0$, $A2>0$, then
$\delta t_{pos} = -p_\Par^n/A1$.  Thus, our final constraint on the
timestep $\delta t$ is given by \be \delta t = C_0 \times \mbox{min}
\left \{ 1/[\mbox{max} \{ \delta t_{adv}^{-2} + \delta t_\Par^{-2} +
\delta t_\Perp^{-2} \}]^{1/2}, \mbox{min} \{\delta t_{pos} \} \right
\} \ee where the $\mbox{max}$ and $\mbox{min}$ are taken over all
zones in the box and $C_0$ is a safety factor (Courant Number) which
we take to be $0.5$.

\section{Implementation of the pressure anisotropy ``hard wall''}
\label{app:pthresh} If the pressure anisotropy is larger than the
constraints given in \S2 by equations
(\ref{Ch4eq:pitch1})-(\ref{Ch4eq:pitch3}),  then microinstabilities will
turn on that will enhance the pitch-angle scattering rate and
quickly reduce the pressure anisotropy to near marginal stability.
Because this is a numerically stiff problem, we use an implicit
approach, following the treatment of \cite{Birn2001}. Whenever
equation~(\ref{Ch4eq:pitch1}) is violated, we use the following
prescription for pitch angle scattering: \ba \label{Ch4eq:fire1} &&
p_\Par^{n+1} = p_\Par^n - \frac{2}{3}\nu_p \delta t \left(
\frac{p_\Par^{n+1}}{2}
-p_\Perp^{n+1} - \frac{B^2}{4\pi} \right), \\
\label{Ch4eq:fire2} && p_\Perp^{n+1} = p_\Perp^n + \frac{1}{3}\nu_p
\delta t \left( \frac{p_\Par^{n+1}}{2} -p_\Perp^{n+1} -
\frac{B^2}{4\pi} \right), \ea where $\nu_p$ is a very large ($\gg
1/\delta t$) rate at which marginal stability is approached.  This
implicit implementation (which can be solved by inverting a $2
\times 2$ matrix) with large $\nu_p$ ensures that each time step the
pressure anisotropy will drop to be very near marginal stability for
the firehose instability to break $\mu$ invariance. Given this pitch
angle scattering, the collisionality parameter $\nu_{eff}$ in the
thermal conductivity (Eqs. \ref{Ch4eq:kappa_par}-\ref{Ch4eq:kappa_mag})
is obtained by comparing equations (\ref{Ch4eq:fire1}) and
(\ref{Ch4eq:fire2}) with equations (\ref{Ch4eq:SB4}) and (\ref{Ch4eq:SB5}):
\be \label{Ch4eq:nueff1} \nu_{eff}= \mbox{max} \left\{ \nu_p
\frac{\left( \frac{p_\Par^{n+1}}{2} -p_\Perp^{n+1}-\frac{B^2}{4\pi}
\right) } {\left( p_\Par^{n+1}-p_\Perp^{n+1} \right)}, \nu \right\}.
\ee The effective pitch angle scattering rate $\nu_{eff}$ is
independent of $\nu_p$ (and much smaller than $\nu_p$) in the limit
of large $\nu_p$, and is by definition just large enough to balance
other terms in equations (\ref{Ch4eq:SB4}-\ref{Ch4eq:SB5}) that are trying
to increase the pressure anisotropy beyond marginal stability.

The prescriptions for pitch angle scattering due to mirror modes and
ion cyclotron waves are similar.  For mirror modes we use \ba &&
p_\Par^{n+1} = p_\Par^n - \frac{2}{3}\nu_p \delta t \left(
p_\Par^{n+1} -p_\Perp^{n+1} + 2 \xi
\frac{p_\Par^{n+1}}{\beta_\Perp^{n+1}} \right),
\\ && p_\Perp^{n+1} = p_\Perp^n + \frac{1}{3}\nu_p \delta t \left(
p_\Par^{n+1} -p_\Perp^{n+1} + 2 \xi
\frac{p_\Par^{n+1}}{\beta_\Perp^{n+1}} \right) \ea to limit the
pressure anisotropy ($\xi=3.5$ for our fiducial run $Zl4$) and
$\nu_{eff}$ is given by \be \label{Ch4eq:nueff2} \nu_{eff}= \mbox{max}
\left\{ \nu_p \frac{\left( p_\Par^{n+1} -p_\Perp^{n+1}+ 2 \xi
\frac{p_\Par^{n+1}}{\beta_\Perp^{n+1}} \right) } {\left(
p_\Par^{n+1}-p_\Perp^{n+1} \right)}, \nu \right\}. \ee For ion
cyclotron pitch angle scattering we use \ba && p_\Par^{n+1} =
p_\Par^n - \frac{2}{3}\nu_p \delta t \left( p_\Par^{n+1}
-p_\Perp^{n+1} + S \frac{p_\Par^{n+1}}{\sqrt{\beta_\Par^{n+1}}}  \right), \\
&& p_\Perp^{n+1} = p_\Perp^n + \frac{1}{3}\nu_p \delta t \left(
p_\Par^{n+1} -p_\Perp^{n+1} + S
\frac{p_\Par^{n+1}}{\sqrt{\beta_\Par^{n+1}}}  \right), \ea and
$\nu_{eff}$ is given by \be \label{Ch4eq:nueff3} \nu_{eff}= \mbox{max}
\left\{ \nu_p \frac{\left( p_\Par^{n+1} -p_\Perp^{n+1}+ S
\frac{p_\Par^{n+1}}{\sqrt{ \beta_\Par^{n+1}} } \right) } {\left(
p_\Par^{n+1}-p_\Perp^{n+1} \right)}, \nu \right\}. \ee

\subsection{Implementation of the advective part of $\nabla \cdot {\bf q_\Perp}$}
\label{app:qperp_conservative} The flux of $p_\Perp$, ${\bf q_\Perp}
= q_\Perp {\bf \hat{b}}$, is given by \be q_\Perp = - \kappa_\Perp
\nabla_\Par \left( \frac{p_\Perp}{\rho} \right) + \left[
\frac{(p_\Par - p_\Perp)}{\rho \left(
\sqrt{\frac{\pi}{2}\frac{p_\Par}{\rho}} k_L + \nu_{eff} \right) }
\frac{{\bf B} \cdot \nabla B}{B^2} \right] p_\Perp = - \kappa_\Perp
\nabla_\Par \left( \frac{p_\Perp}{\rho} \right) + V_{mag} p_\Perp
\ee where the quantity in square brackets can be thought of as an
advection speed due to parallel magnetic gradients.  Because of this
term, ${\bf q_\Perp}$ is not a purely diffusive operator, but also
has an advective part characterized by the velocity $V_{mag}$. If
one treats the advective part via a simple central difference
method, it does not preserve monotonicity.  Instead, to treat the
advective part of ${\bf q_\Perp}$ properly, we include the advective
part in the transport step. After including the advective heat flux
in the transport step, it takes the form \be \frac{\partial
p_\Perp}{\partial t} + \nabla \cdot \left[ ({\bf V} +V_{mag}{\bf
\hat{b}} ) p_\Perp \right] = 0.  \ee Thus, for updating $p_\Perp$ in
the transport step we calculate fluxes on the cell faces using ${\bf
V} + V_{mag} {\bf \hat{b}}$ instead of just ${\bf V}$. The transport
step is then directionally split in the three directions. The
procedure for monotonicity preserving schemes for calculating fluxes
is described in \cite{Stone1992a}.

\section{Numerical tests}

The kinetic modifications to the ZEUS MHD code have been tested for
the ability to capture the collisionless effects.

\subsection{Tests for anisotropic conduction}
The kinetic MHD code used for the shearing box simulations of the
collisionless MRI uses the asymmetric method for anisotropic thermal
conduction \cite{Sharma2006}. Although, the asymmetric method can
result in negative temperature, its fine to use it for local
simulations as there are no sharp temperature gradients (see Chapter
\ref{chap:chap5}). Anisotropic conduction tests have been discussed
extensively in Chapter \ref{chap:chap5}.

\subsection{Collisionless damping of fast mode in 1-D}
\begin{figure}
\begin{center}
\includegraphics[width=2.95in,height=2.5in]{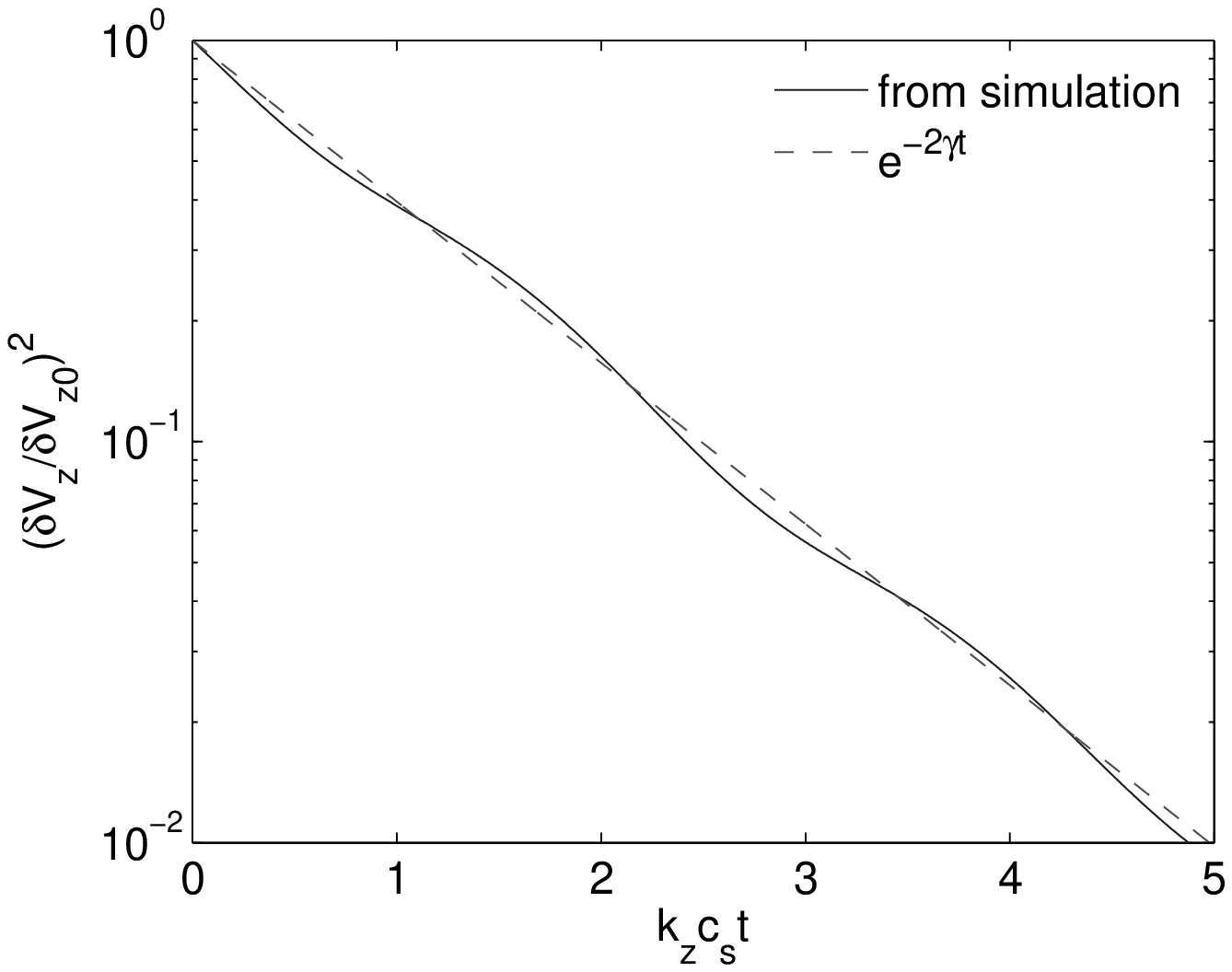}
\includegraphics[width=2.95in,height=2.5in]{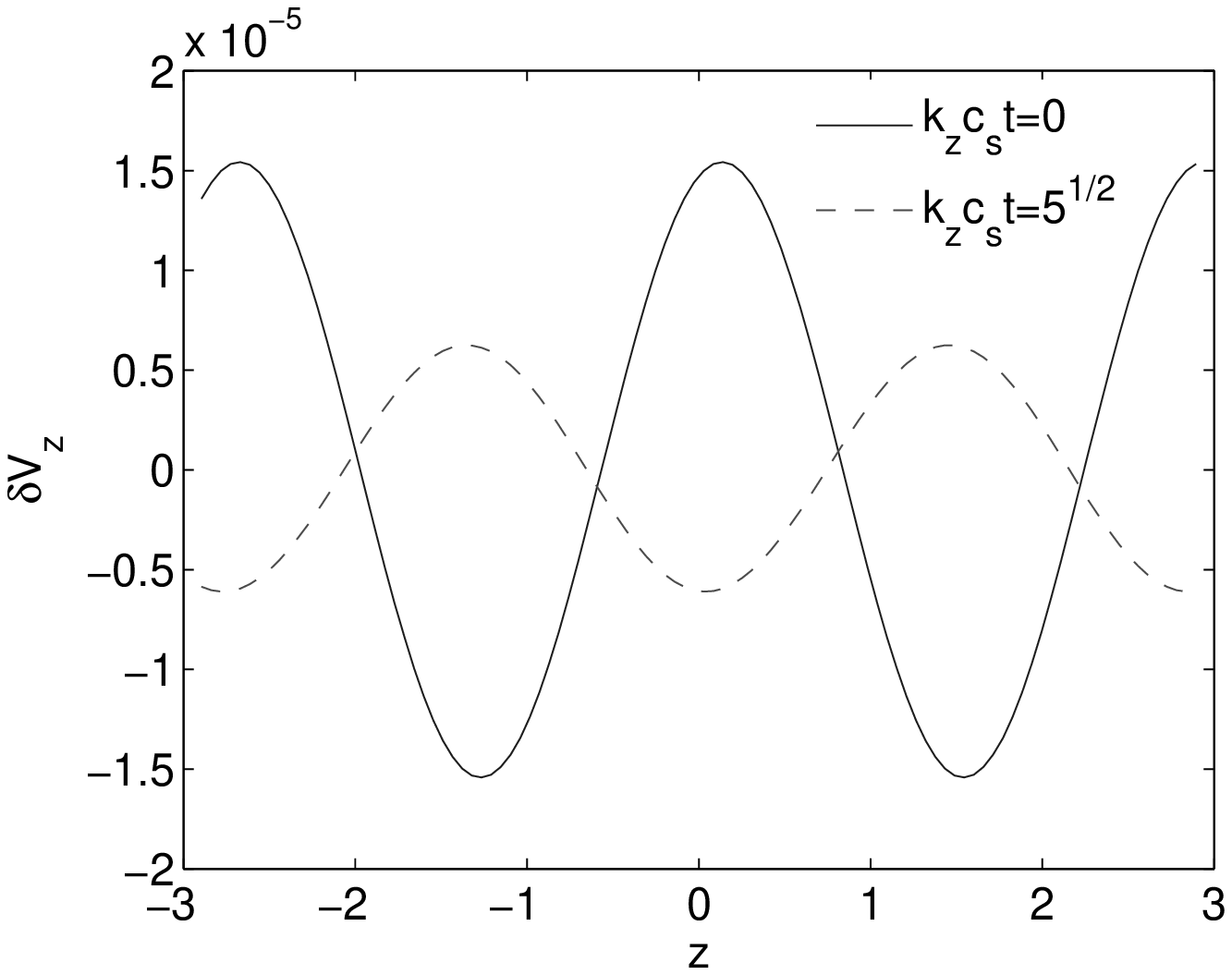}
\includegraphics[width=2.95in,height=2.5in]{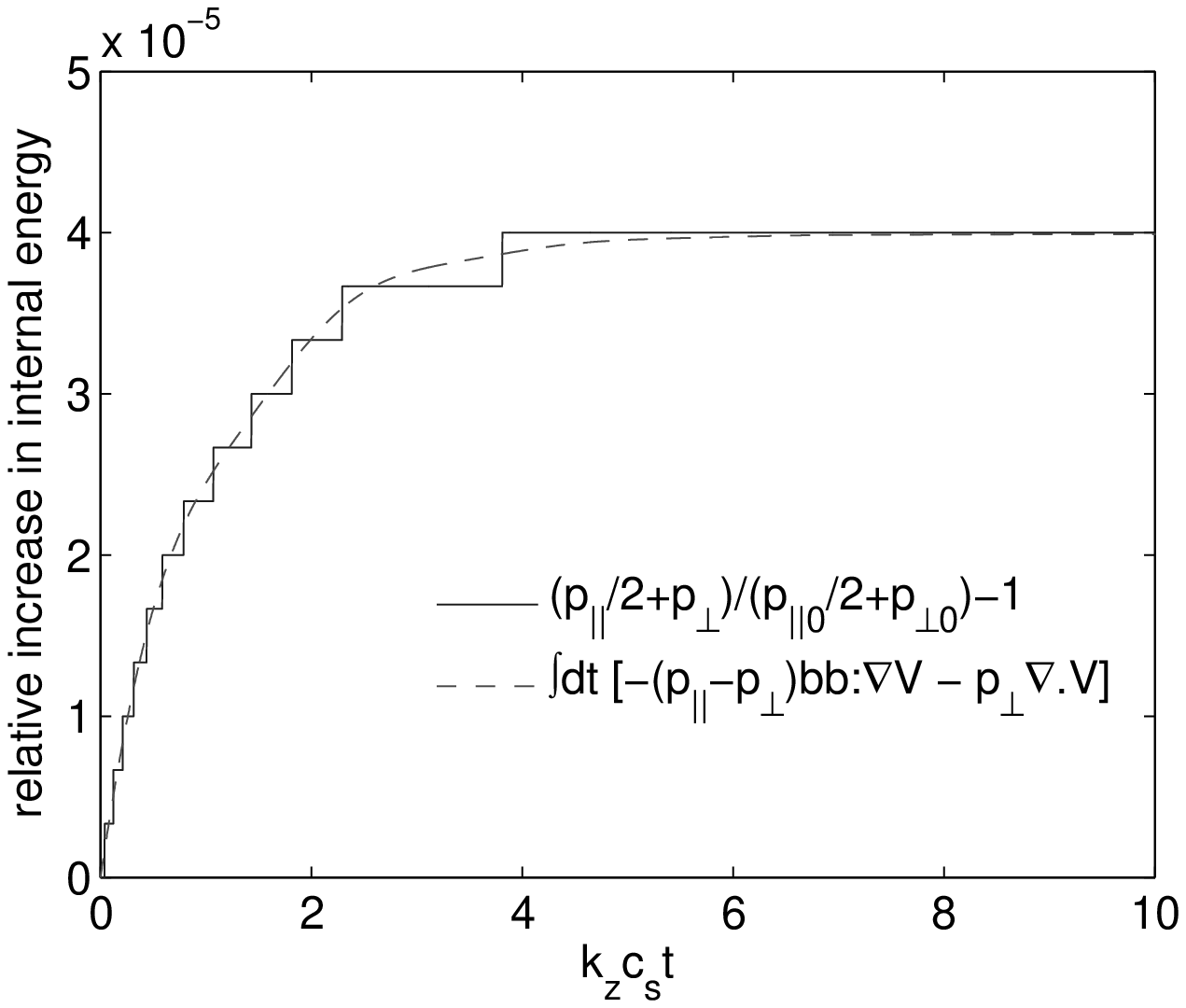}
\includegraphics[width=2.95in,height=2.5in]{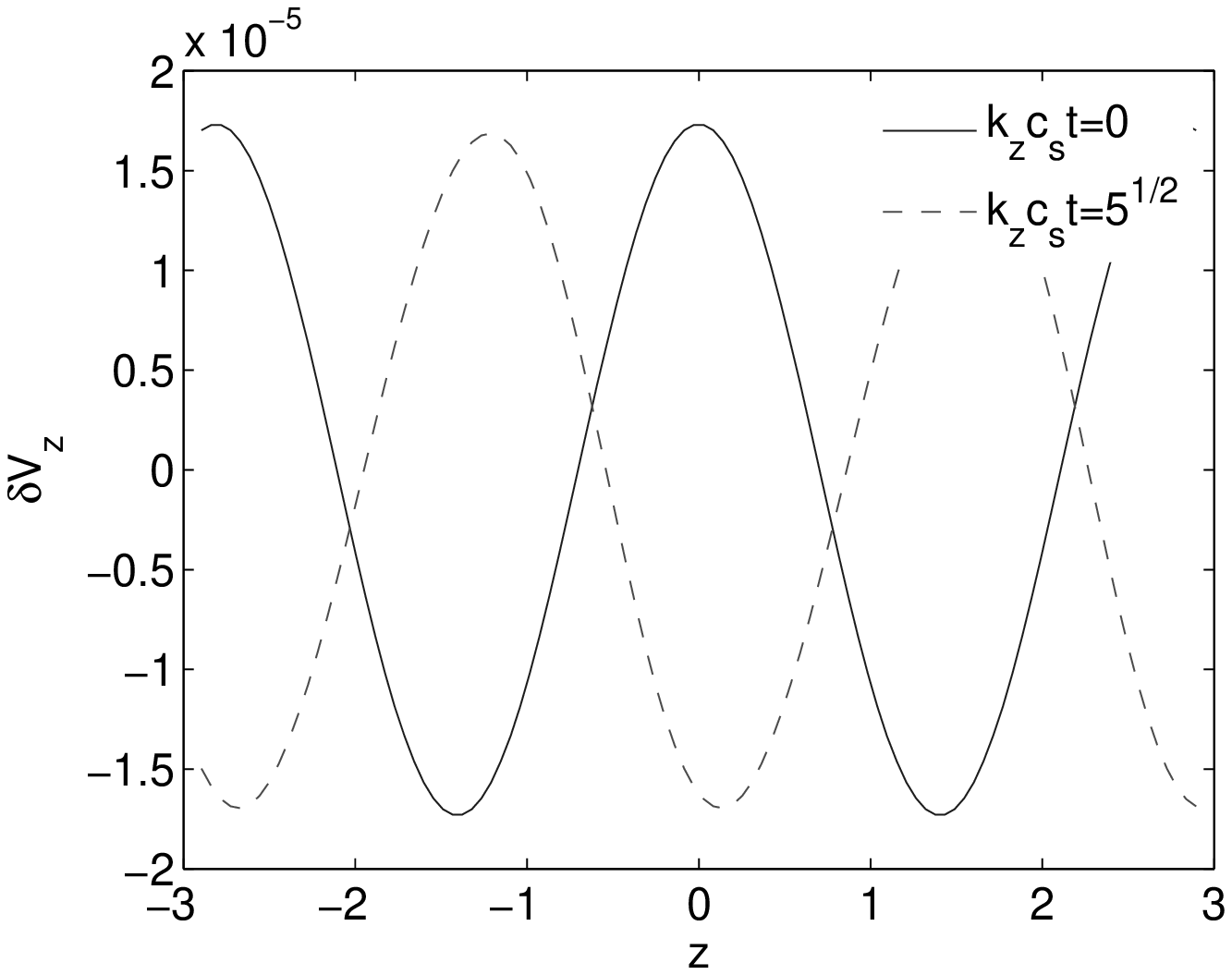}
\caption[Collisionless damping of a fast mode in 1-D]{Figure on top
left shows damping of kinetic energy in time; solid line is from the
simulation and the dashed line is the result from eigenmode analysis
in MATHEMATICA. Top right figure shows the initial eigenmode (solid
line) and the damped eigenmode (dashed line) at a later time. Bottom
left figure shows the increase in internal energy (solid line) and
the result expected from the heating term (dashed line). Bottom
right figure shows the initial fast mode eigenmode (solid line) and
the eigenmode at a late time (dashed line) with the CGL equation of
state, and as expected, there is no damping.
\label{Ch4fig:figure10}}
\end{center}
\end{figure}
We initialize a fast wave eigenmode traveling along the field lines
to verify that the Landau closure reproduces the correct damping
rate. We choose the following parameters: $\rho_0=1.0$,
$p_{\parallel 0}=p_{\perp 0}=10^{-6}$, $k V_{A 0}=10^{-3}$, and
$\beta=10$. A periodic box with the size of two wavelengths is used.
Since we initialize a parallel propagating fast mode, there is no
magnetic perturbation, and the initial eigenmode is given by \ba
\delta \rho &=& A \cos(kz), \\
\delta p_\parallel &=& A 10^{-6} \left (3 p \cos(kz) + 1.36 \sin(kz)
\right ),\\
\delta p_\perp &=& A 10^{-6} \cos(kz), \\
\delta V_z &=& A \left ( 0.0015 \cos(kz) + 0.00046 \sin(kz) \right
), \ea where A=0.01 is the amplitude. Figure \ref{Ch4fig:figure10}
shows the results from the ZEUS code modified to include kinetic
effects. Simulation recovers the correct phase speed and damping
rate. Velocity perturbations are damped and the energy goes to
internal energy. Figure also shows that a fast mode eigenmode in the
CGL limit shows no damping.

Since the magnetic perturbation vanishes for this case, there is no
$-\mu \nabla_\parallel B$ Barnes damping. This leaves only the
parallel $eE_\parallel$ Landau damping. With $eE_\parallel=ik\delta
p_\parallel (m_e/m_i)/n$, assuming cold electrons, Landau damping is
hidden in pressure terms in the equation of motion and the internal
energy equation.

\subsection{Mirror instability in 1-D}
\label{app:mirror} 
\begin{figure}
\begin{center}
\includegraphics[width=2.95in,height=2.5in]{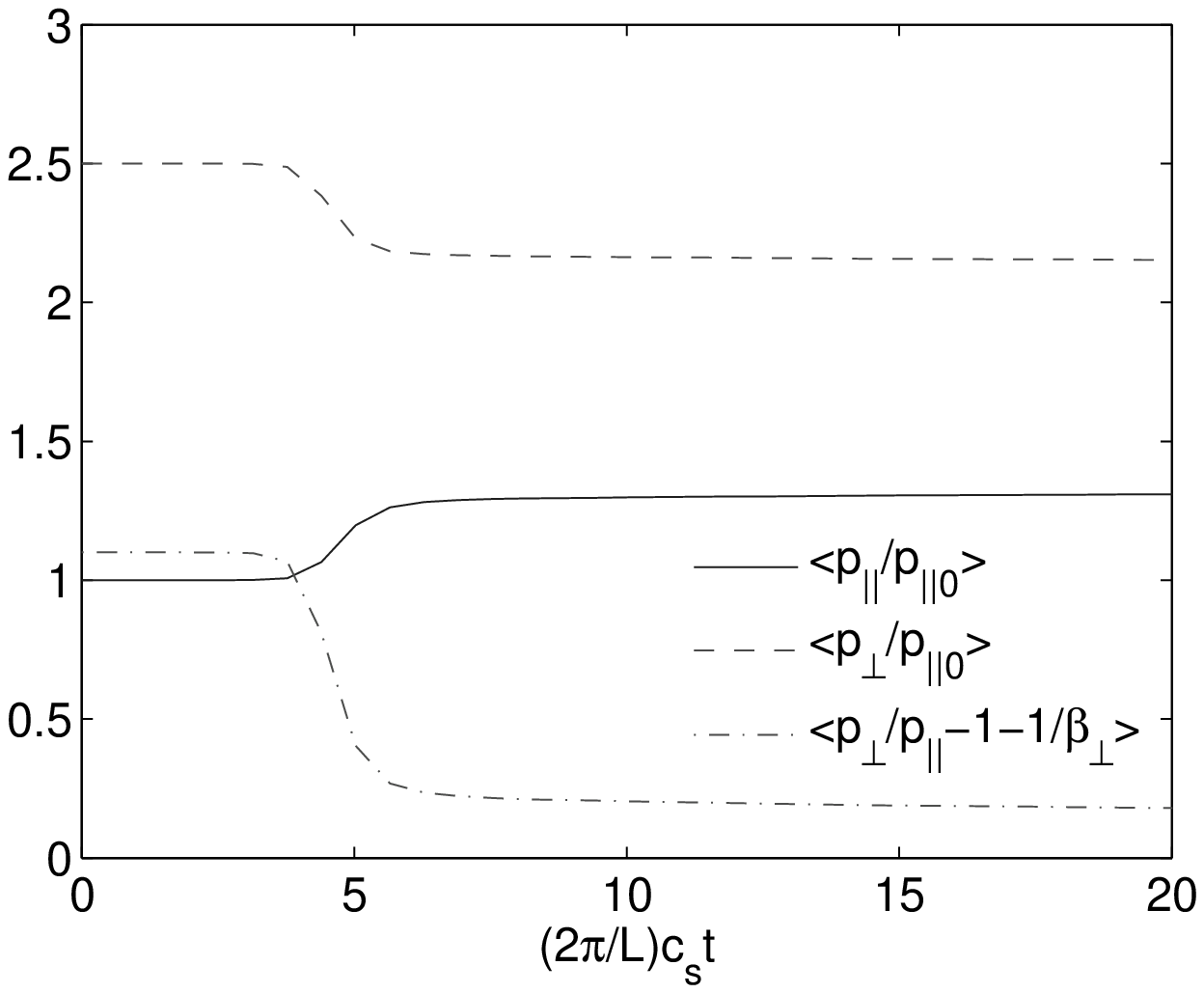}
\includegraphics[width=2.95in,height=2.5in]{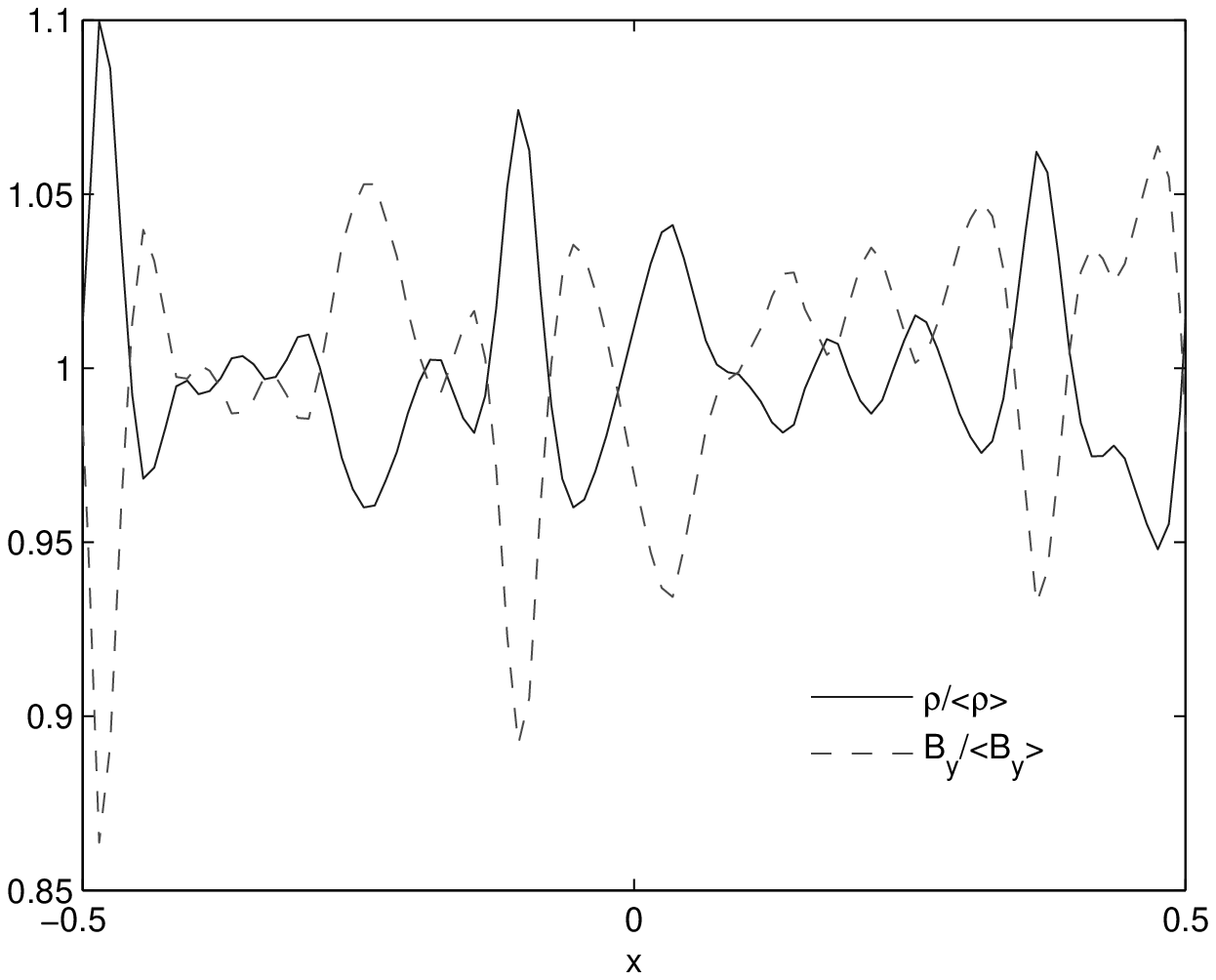}
\caption[Kinetic MHD simulations of the mirror instability in
1-D]{Figure on left shows normalized parallel and perpendicular
pressure (solid and dashed lines respectively), and difference from
marginal stability, $p_\perp/p_\parallel -1 -1/\beta_\perp$
(dot-dashed line), with time. Pressure anisotropy is reduced towards
marginal stability. Right figure shows anticorrelated density (solid
line) and magnetic field strength (dashed line), normalized to their
mean value, in saturated state ($c_{s \parallel}t/2\pi=2$).
\label{Ch4fig:figure11}}
\end{center}
\end{figure}

The mirror instability criterion in the CGL limit
is $p_\perp/6p_\parallel -1 -1/\beta_\perp>0$ as compared to the
criterion in the kinetic regime, $p_\perp/p_\parallel
-1-1/\beta_\perp>0$ \cite{Kulsrud1983,Snyder1997}. We test Landau
closure by initializing an anisotropic pressure
($p_\perp/p_\parallel=2.5$, $\beta_\parallel=1$) which is unstable
according to the kinetic criterion but stable by the CGL criterion.
Landau closure with parameter $k_L=12\pi/L$ (gives correct kinetic
behavior for 6 wavelengths in the box, for larger wavenumber growth
rate is faster than the kinetic result, see Figure
\ref{Ch2fig:Crude}). Figure \ref{Ch4fig:figure11} shows the results
of nonlinear simulations initialized with a small amplitude random
white noise. Pressure anisotropy is reduced to marginality with
time. Particles are trapped in low magnetic field regions due to the
mirror force, and density and magnetic field strengths are
anticorrelated. Growth rate increases linearly with the resolution,
as $\gamma \propto k_\parallel$.

For small pressure anisotropy, adiabatic invariance is obeyed and
plasma rearranges itself in the form form of mirrors and becomes
marginally stable. These 1-D results are consistent with previous
fluid \cite{Baumgartel2001} and kinetic \cite{McKean1993} studies.
Similar 1-D tests for the firehose instability results have shown
results consistent with the previous kinetic simulations
\cite{Quest1996}. Here too, the growth rate is proportional to the
grid resolution (as $\gamma \propto k_\parallel$). Fastest growing
firehose mode is the one with a parallel wavenumber. The transverse
magnetic field disturbances grow until the plasma becomes marginally
stable to the firehose instability. A 2-D test for firehose
instability is discussed in the next section.

\subsection{Shear generated pressure anisotropy: Firehose instability in 2-D}
We have also devised a test problem where the magnetic field
strength decreases because of the shear in the box; a decreasing
field strength causes pressure to become anisotropic
($p_\parallel>p_\perp$). The firehose instability is excited when
the pressure anisotropy increases beyond the firehose instability
threshold ($p_\parallel/p_\perp -1 - 2/\beta_\perp>0$). The shearing
rate is small so that the firehose instability locks the pressure
anisotropy to the marginal value.
\begin{figure}
\begin{center}
\includegraphics[width=2.95in,height=2.5in]{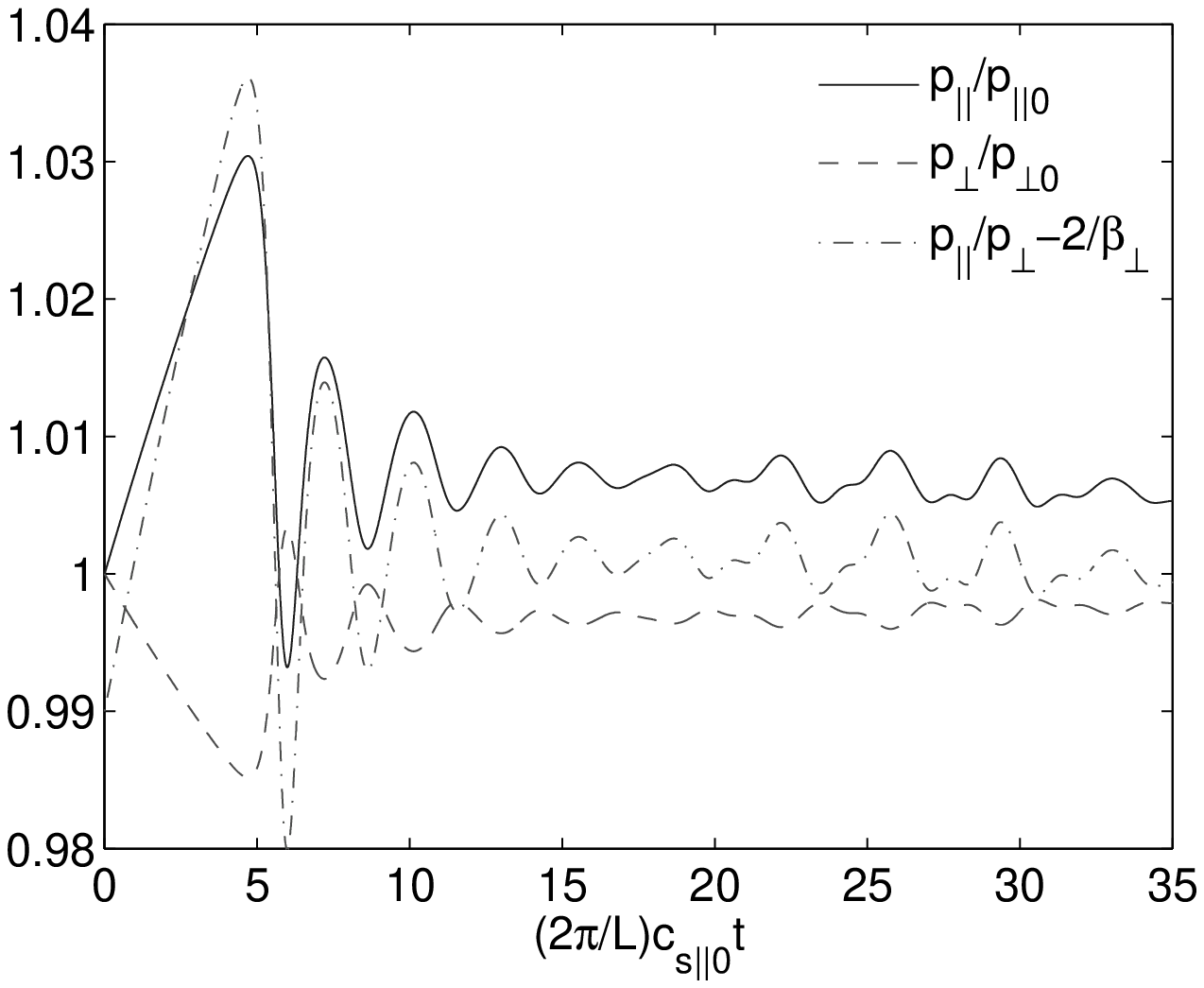}
\includegraphics[width=2.95in,height=2.5in]{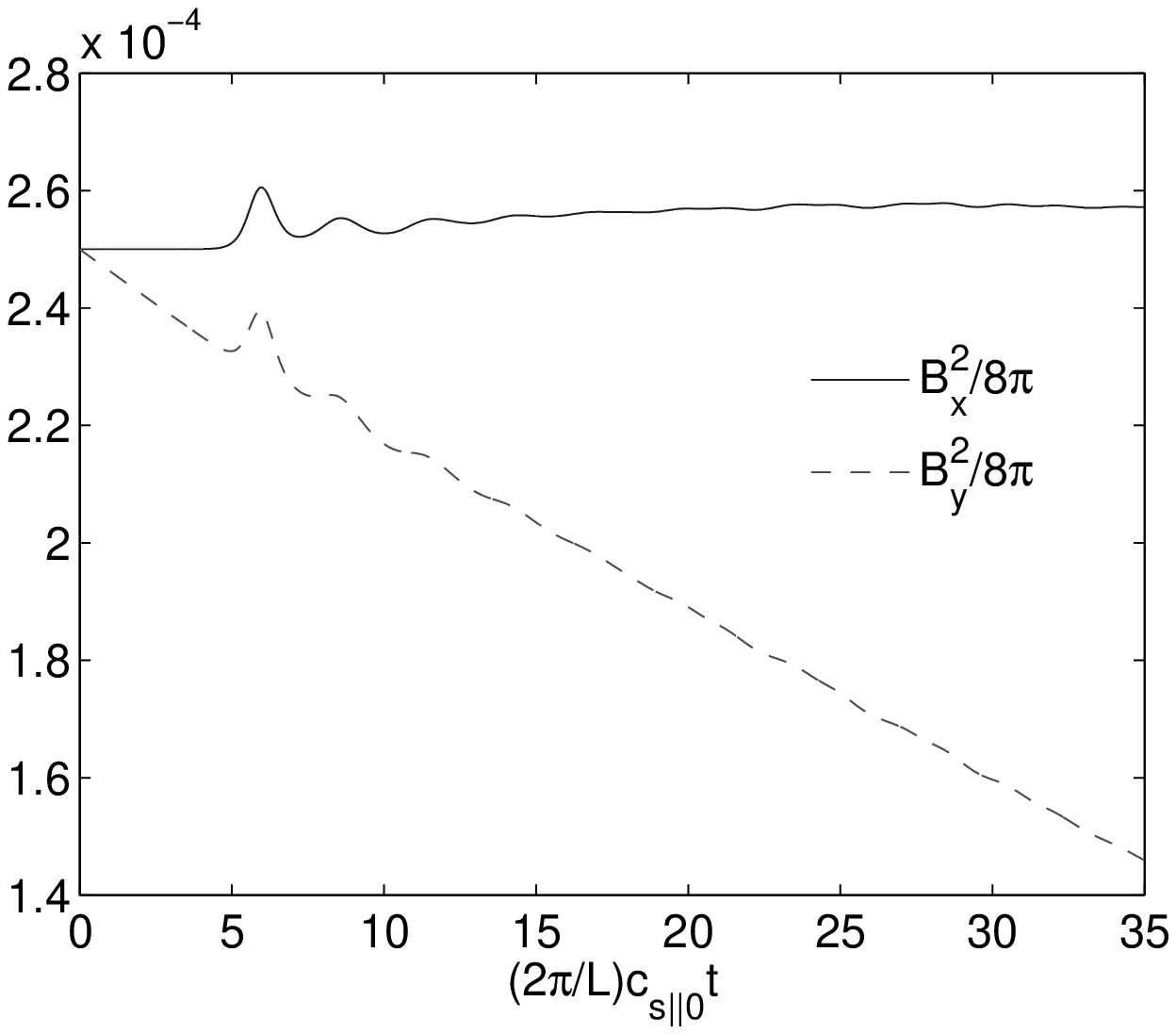}
\caption[Firehose instability in 2-D, with pressure anisotropy
created by the shear in the box]{Figure on the left shows normalized
parallel (solid line) and perpendicular (dashed line) pressure, and
the firehose marginal stability criterion (dot-dashed line),
$p_\parallel/p_\perp-2/\beta_\perp$. At early times pressure
anisotropy is caused by the shear, but as the firehose instability
sets in, pressure anisotropy saturates at the marginal state. Right
figure shows magnetic field strengths, $B_x^2/8\pi$ (solid line) and
$B_y^2/8\pi$ (dashed line). Magnetic strength in the $y-$ direction
is reduced by the shear, but at late times there is a bump in field
strengths showing the firehose instability. \label{Ch4fig:figure12}}
\end{center}
\end{figure}

We use a $50 \times 50$ 2-D box with $L_x=L_y=1$, $p_{\parallel
0}=p_{\perp 0} = 0.1$, $\beta=200$ with $B_{x0}=B_{y0}$,
$V_y(-L_x/2)-V_y(L_x/2)=(3/2)\Omega L_x$ with $\Omega=0.01$, and the
collision frequency $\nu=0.1$. The Landau parameter is
$k_L=0.5/\delta x$; firehose instability is insensitive to the
parallel thermal conduction and the CGL equations give the correct
instability threshold \cite{Kulsrud1983,Snyder1997}. Parameters are
chosen such that the shearing rate is the smallest followed by the
collision frequency and the sound crossing frequency, $\Omega \ll
\nu \ll (2\pi/L) c_{s \parallel 0}$. In this ordering, plasma is
effectively collisionless, and pressure anisotropy is driven slowly
by the shear, so that the firehose instability saturates in the
marginal state.
\begin{figure}
\begin{center}
\includegraphics[width=2.95in,height=2.5in]{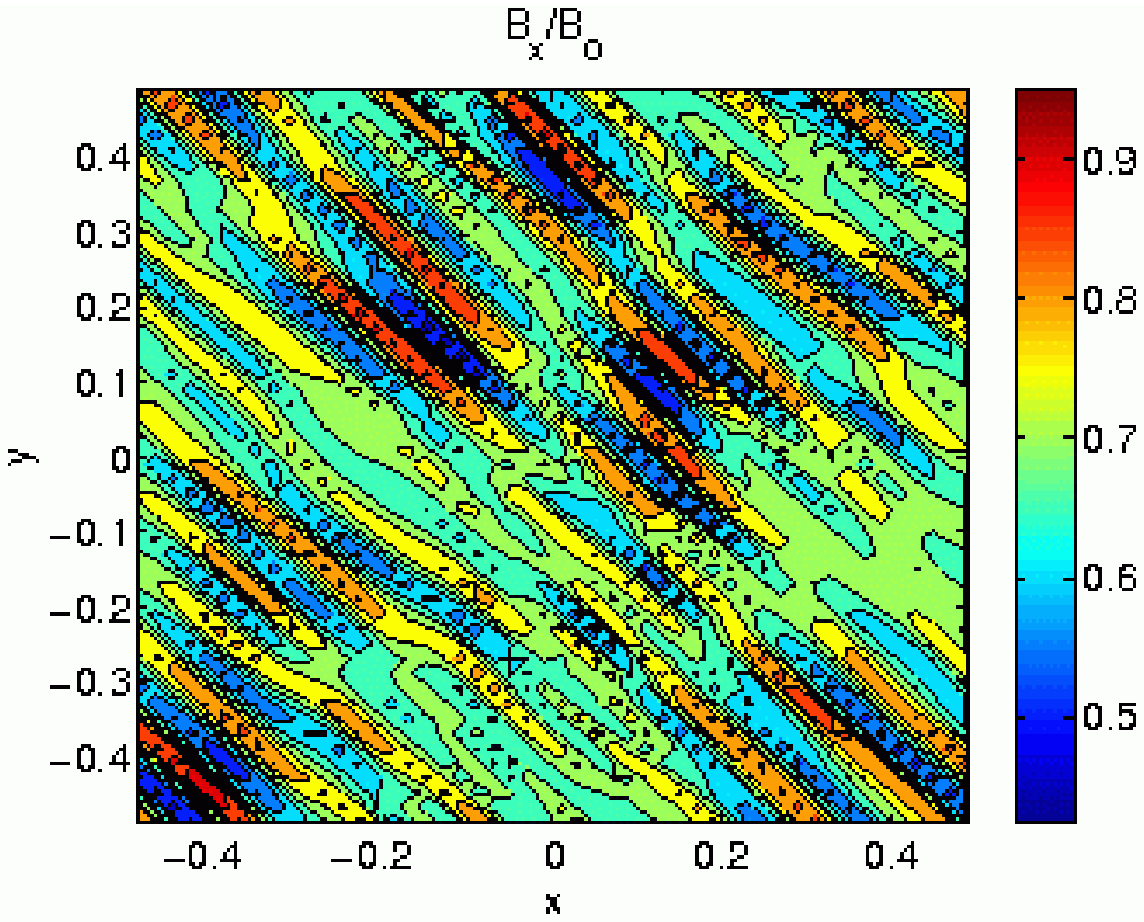}
\includegraphics[width=2.95in,height=2.5in]{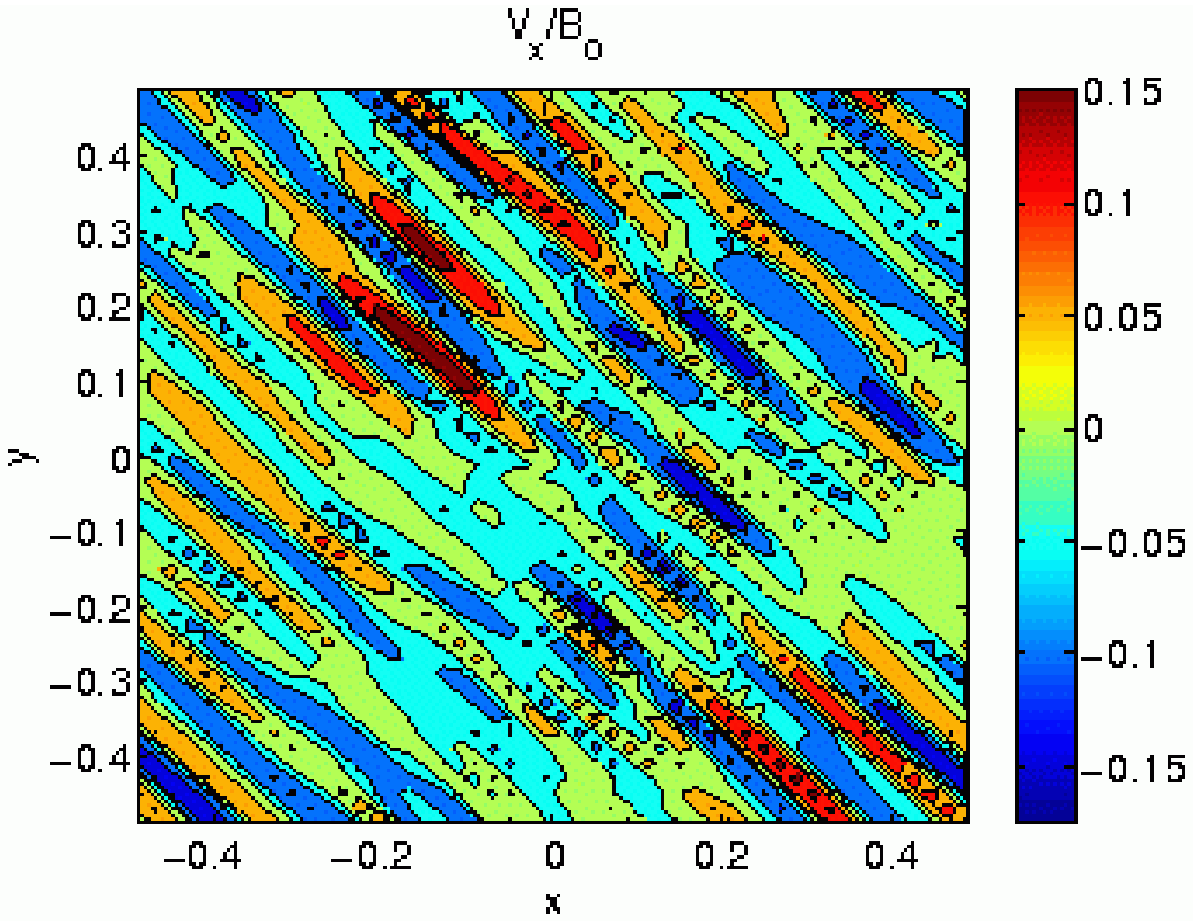}
\includegraphics[width=2.95in,height=2.5in]{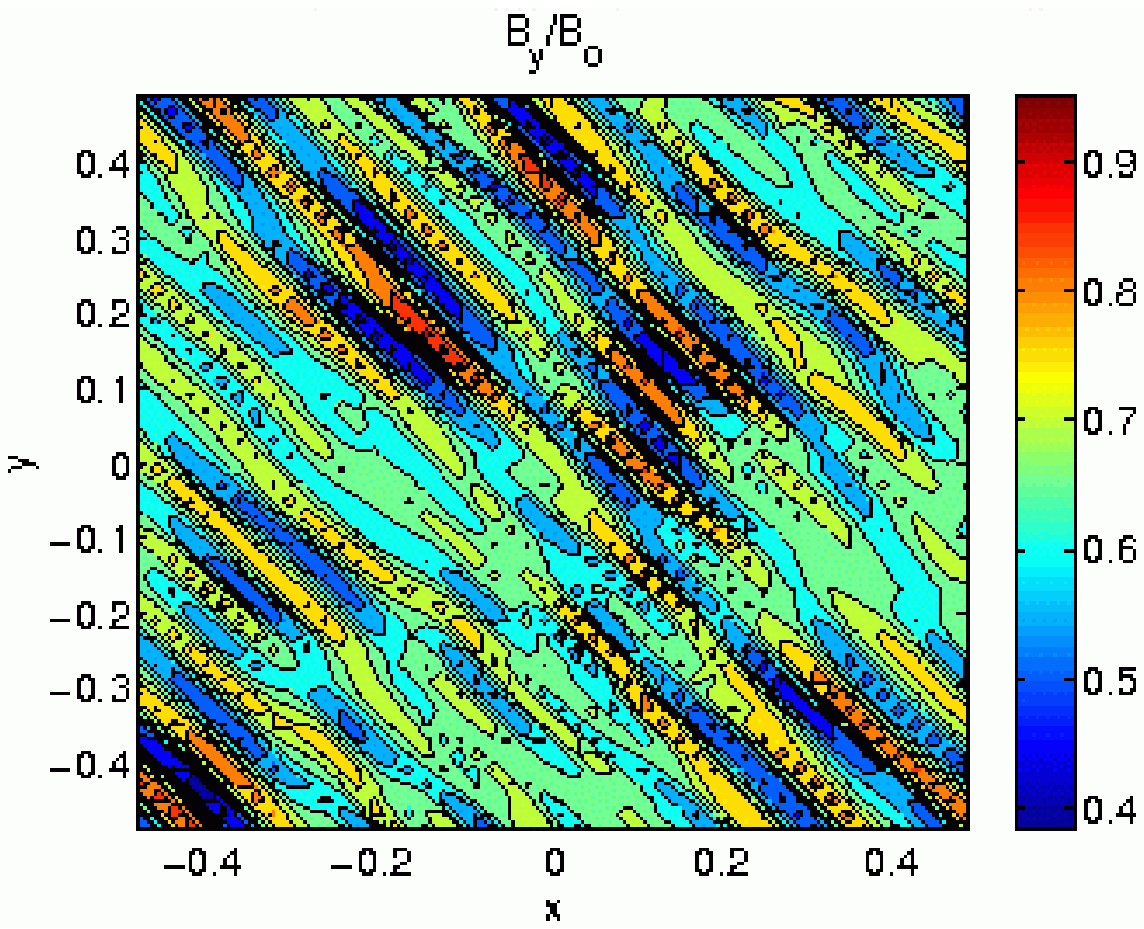}
\includegraphics[width=2.95in,height=2.5in]{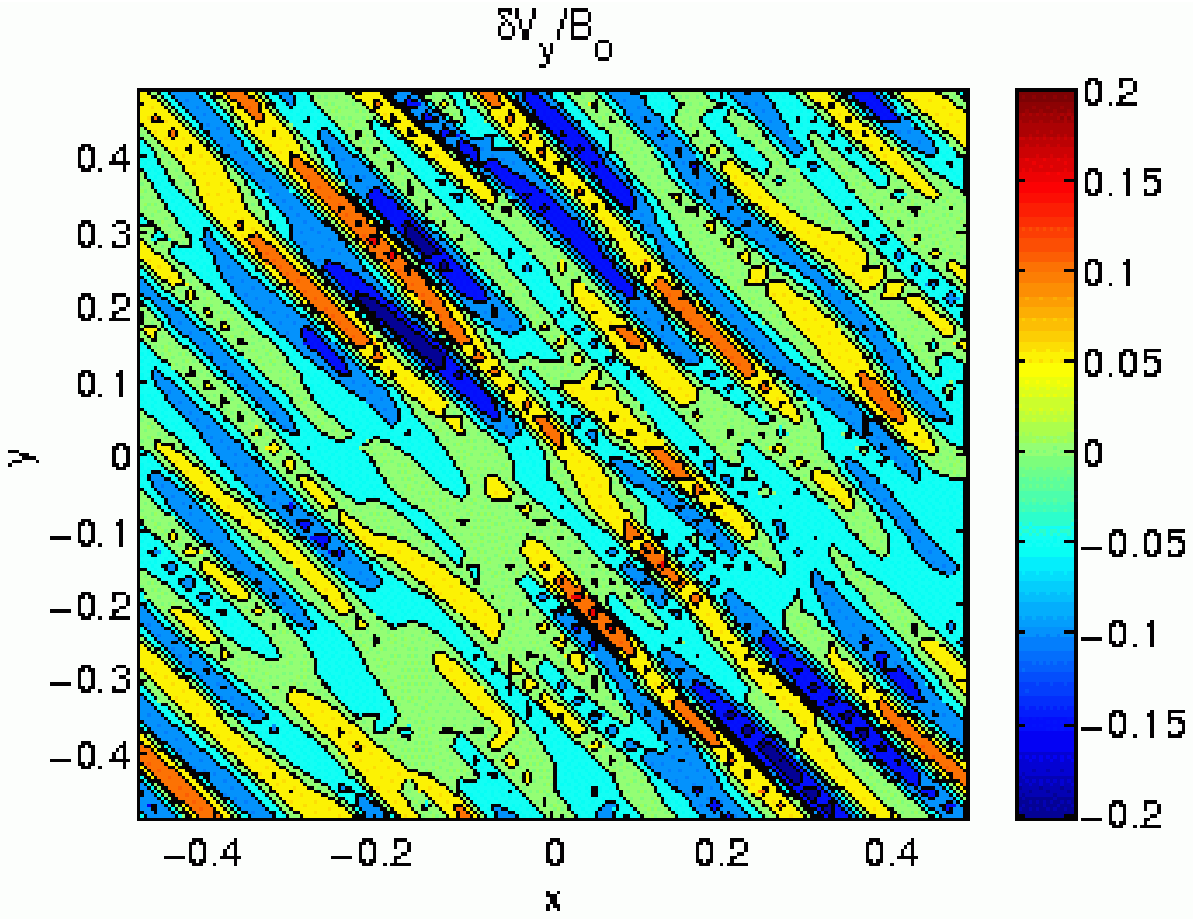}
\caption[Plots of $B_x$, $V_x$, $B_y$, and $\delta V_y$ for a
firehose unstable plasma in a 2-D shearing box]{The 2-D plots of
$B_x$, $V_x$, $B_y$, and $\delta V_y$, normalized to the initial
field strength, at $2\pi c_{s\parallel 0}t \approx 8$. The initially
random perturbations give rise to the firehose instability
propagating primarily along the field lines.
\label{Ch4fig:figure13}}
\end{center}
\end{figure}

Figures \ref{Ch4fig:figure12} and \ref{Ch4fig:figure13} show the
results: at early times $p_\parallel \propto B^{-2}$ increases and
$p_\perp \propto B$ decreases as magnetic field decreases. When
pressure anisotropy crosses the firehose threshold, the instability
reduces the pressure anisotropy to the marginal state by increasing
the transverse (to the mean magnetic field) magnetic perturbations.

\chapter{Error analysis}
\label{app:app4}
\label{app:error_bars}
The standard errors in the time averages
reported in Table~\ref{Ch4tab:tab2} and in
Figure~\ref{Ch4fig:figure7} are estimated by taking into account the
finite correlation time for the physical quantities in the
simulation, using techniques recommended by \cite{Nevins2005}. Given a
finite time series, we want to calculate the ensemble average and the
uncertainty around the ensemble average. The standard deviation  
of the time series does not represent the error (uncertainty) because
the data in the time series are correlated.

For a time series with non-zero correlation time, the standard error 
for the time average $\langle x \rangle =
\int dt \, x(t) / T$ of a signal $x(t)$ is given by $\sigma_{\langle
x \rangle} = \sqrt{{\rm Var}(x)/N_{eff}}$, where ${\rm Var}(x) =
\int dt \, (x(t)-\langle x \rangle )^2 / T$ is the variance of $x$,
$N_{eff} = T/(2 \tau_{int})$ is the effective number of independent
measurements, $T=15$ orbits is the averaging time for the simulations
described in Chapter \ref{chap:chap4}, and $\tau_{int}$
is an estimate of the integrated autocorrelation time. There are
significant subtleties in determining the integrated autocorrelation
time from data.  To deal with this, we use a windowing technique as
recommended by \cite{Nevins2005}, using $\tau_{int} = \int_0^T d\tau
\, C(\tau) W(\tau/\tau_w) $, where $C(\tau)$ is the 2-time
correlation function from the data, $W(\tau/\tau_w)$ is a smooth
window function that cuts off the integral at $\tau \sim \tau_w$,
and $\tau_w \sim \sqrt{T \tau_{int}}$ (this gives results
insensitive to the choice of window width for $\tau_{int} \ll T$).
If windowing is not used, i.e., $W=1$, then the integral for 
$\tau_{int}$ vanishes; therefore, an appropriate windowing function 
is necessary. The two-time autocorrelation function is defined as
\be
C(\tau) = \frac{1}{T} \int_0^{T-\tau} dt \tilde{x}(t) \tilde{x}(t+\tau),
\ee
where $\tilde{x}=x-\langle x \rangle$. An example of the windowing function
is the Hanning window given by \cite{Nevins2005}
\ba
H(\xi) &=& \frac{1}{2} \left [ 1 + \cos(\pi \xi) \right ], |\xi|<1 \\
       &=& 0, |\xi| \geq 1
\ea

Winters et. al. \cite{Winters2003} found from comparing 3 
realizations of shearing
box MRI simulations that the magnetic stress had a variation of
approximately $\pm 6.5\%$ after averaging over 85 orbits.  The
simulations we show here were averaged over 15 orbits, so
extrapolating from \cite{Winters2003} one might expect the
uncertainties to be larger by a factor of $\approx \sqrt{85/15}
\approx 2.4$.  This is consistent with the typical error bars we
report in Table~\ref{Ch4tab:tab2} and Figure~\ref{Ch4fig:figure7}.

\chapter{Entropy condition for an ideal gas}
\label{app:app5}
\label{app:entropy} The entropy for an ideal gas is given by $S =
nVk \ln(T^{1/(\gamma-1)}/n) + \mbox{const.}$, where $n$ is the number density, V the
volume, $T$ the temperature, and $\gamma$ the ratio of specific
heats ($=5/3$ for a 3-D mono-atomic gas). The change in entropy that results from 
adding an amount of heat $dQ$ to a uniform gas is
$$
d S = \frac{n V k}{\gamma - 1} \frac{dT}{T} = \frac{dQ}{T}.
$$
We measure temperature in energy units, so $k=1$ from now on.  
The rate of change of entropy of a system where number
density and temperature can vary in space (density is assumed to be constant in time) 
is given by \be
\dot{S} \equiv \frac{\partial S}{\partial t} =   - \int dV
\frac{{\bf \grad \cdot q}}{T} = - \int dV \frac{{\bf  q \cdot \grad}
T}{T^2} = \int dV n \chi \frac{|\nabla_\parallel T|^2}{T^2} \geq 0, \ee 
where we use an anisotropic heat flux, ${\bf q}=-n\chi {\bf \hat{b}\hat{b} \nabla } T$, 
and the integral is evaluated over the whole space with
the boundary contributions assumed to vanish. The
local entropy function is defined as $\dot{s}=-{\bf q
\cdot \grad}T/T^2$ can be integrated to calculate the rate of change of
total entropy of the system.

In Chapter \ref{chap:chap5} we use a related function (the
entropy-like function $\dot{s}^*$) defined as $\dot{s}^* \equiv -{\bf q
\cdot \grad}T$ to limit the symmetric methods using face-pairs, and
to prove some properties of different anisotropic diffusion schemes. The 
condition $-{\bf q \cdot \grad}T \geq 0$ means that heat always flows from
higher to lower temperatures.

\end{document}